\documentclass[10pt]{article}

\usepackage[dvips]{epsfig} 
\usepackage{amssymb,amsmath,amsthm,mathrsfs} 
\usepackage{graphicx}
\usepackage{lineno}

\usepackage[authoryear,sort&compress]{natbib}
\usepackage{url}
\usepackage{enumerate}
\usepackage{colordvi,color,pspicture}
\usepackage{psfrag}
\usepackage{rotating}

\newcommand{\mZ}{\mathcal{Z}}
\newcommand{\U}{\mathcal{U}}

\newcommand{\I}{\mathbf{I}}

\newcommand{\E}{\mathbf{E}}
\newcommand{\Var}{\mathbf{Var}}
\newcommand{\Cov}{\mathbf{Cov}}

\newcommand{\matclust}{\text{MatClust}}

\newcommand{\poisson}{\text{Poisson}}

\DeclareMathOperator{\BIN}{BIN}

\theoremstyle{plain}
\newtheorem{theorem}{Theorem}[section]

\theoremstyle{definition}

\theoremstyle{remark}

\newtheorem{remark}[theorem]{Remark}

\begin{document}


\title{Technical Report \# KU-EC-13-1:\\
Segregation Indices for Disease Clustering}
\author{
Elvan Ceyhan\thanks{Department of Mathematics, Ko\c{c} University, Sar{\i}yer, 34450, Istanbul, Turkey.}
}
\date{\today}
\maketitle


\begin{abstract}
\noindent
Spatial clustering has important implications in various fields.
In particular, disease clustering is of major public concern in epidemiology.
In this article,
we propose the use of two distance-based segregation indices to
test the significance of disease clustering among subjects
whose locations are from a homogeneous or an inhomogeneous population.
We derive their asymptotic distributions and
compare them with other distance-based disease clustering tests
in terms of empirical size and power by extensive Monte Carlo simulations.
The null pattern we consider is the random labeling (RL)
of cases and controls to the given locations.
Along this line,
we investigate the sensitivity of the size of these tests to the underlying background pattern
(e.g., clustered or homogenous) on which the RL is applied,
the level of clustering and number of clusters,
or differences in relative abundances of the classes.
We demonstrate that differences in relative abundance has the highest impact
on the empirical sizes of the tests.
We also propose various non-RL patterns as alternatives to the RL pattern
and assess the empirical power performance of the tests under these alternatives.
We illustrate the methods on two real-life examples from epidemiology.
\end{abstract}

\noindent
{\small {\it Keywords:}
cell-specific tests, Cuzick-Edwards' tests, empirical power, empirical size,
nearest neighbor contingency table, overall test, random labeling, spatial clustering

\vspace{.25 in}

$^*$corresponding author.\\
\indent {\it e-mail:} elceyhan@ku.edu.tr (E.~Ceyhan) }




\section{Introduction}
\label{sec:intro}
Recently,
spatial clustering has become a topic of extensive study
in many fields such as geography, ecology, astronomy, and epidemiology.
The relevant methodology is discussed in many books such as \cite{ripley:2004} and \cite{diggle:2003};
even special issues of journals are devoted to this topic,
see, e.g., \cite{banerjee:2012} and \cite{lesage:2009}.
In particular,
the significance of disease clustering in human or other populations
has received considerable attention (\cite{waller:2004}, \cite{lawson:2002} and \cite{rogerson:2006}).
Roughly speaking,
a \emph{disease cluster} is a region or neighborhood where the number of cases substantially exceeds
the expected number of cases at a specific time or for a specific time period (\cite{gomez:2003}).
There are many tests available for testing the significance of disease clustering.
Among them are tests for deviation from homogeneity like the usual Pearson's chi-square statistic for quadrat data
or Potthoff-Whittinghill's test (\cite{potthoff:1966}).
Clustering methods for detection of disease clustering can be grouped into four categories:
(i) methods based on regional count data,
(ii) individual point data (e.g., case-control data in epidemiology),
(iii) adjacencies of high count regional data,
and
(iv) distance-based methods.
Cuzick-Edwards' $k$-nearest neighbor (NN) test (\cite{cuzick:1990})
is an example of category (ii) and has been frequently employed in epidemiology
so that it is suggested in the appendix of  guidelines for disease clustering
(\cite{center:1990}).

Regional count method is the procedure in which a square grid is overlaid over the region of interest
and the number of events in each quadrat is counted.
Assuming the points are from a homogeneous Poisson process (HPP),
which is the null pattern,
the quadrat counts would be distributed as Poisson variates,
and their departure from the null case can be tested
using an index of dispersion (like ratio of variance to mean),
or $\chi^2$ test for heterogeneity of the cell counts.
This method has various shortcomings,
especially for disease clustering.
For example, the quadrats would not be square cells,
but administrative units determined by geographical limitations or human intervention.
This problem can somewhat easily be overcome by extending the
quadrat method to other shapes or administrative units by simply
using the chi-square goodness-of-fit test
using the observed and expected numbers in each region.
Other main problems are the arbitrariness in the choice of grid size
and obtaining correct expected quadrat counts on a sufficiently fine grid structure (\cite{cuzick:1990}).

Statistical methodology based on NN (or distance-based) methods
include at least six different groups (\cite{dixon:EncycEnv2002}).
Each of these methods assumes as a premise that similarity or dissimilarity between a point and its NN
provides useful information for statistical inference.
The most straightforward dissimilarity measure is the distance between a point and its NN,
while other methods could be based on classifying the types of points and their NNs.
Spatial clustering of points from one class can also be investigated by distance indices.
For example,
\cite{perry:1998} introduced
SADIE (Spatial Analysis by Distance IndicEs) to detect deviations from randomness for planar data.
These indices were designed to utilize count data
and two new indices were proposed and a comparative survey was provided for several other indices
in the same reference.
The two-class version of the methodology was also developed (\cite{perry:1997}),
and with this form of SADIE,
it would be possible to detect spatial association or interaction between two species or classes.
\cite{cuzick:1990} proposed a distance-based method which is applicable in a case-control setting and
accounts for geographical inhomogeneity in the population density
and also overcomes various confounding factors by appropriately choosing the controls (\cite{tango:2007}).
Furthermore,
in literature,
there are spatial clustering tests based on nearest neighbor contingency tables (NNCTs)
due to \cite{pielou:1961} and \cite{dixon:1994} in a two-class setting,
and due to \cite{dixon:NNCTEco2002} in a multi-class setting.
Various new segregation tests were also proposed by \cite{ceyhan:corrected} based on NNCTs.
These tests comprise of an overall test, a compound measure of deviation from the null pattern,
and cell-specific tests for pairwise comparisons after a significant overall test.
Also a `coefficient of segregation' was introduced by \cite{pielou:1961} in a two-class setting,
and `segregation indices' were proposed by \cite{dixon:NNCTEco2002} in a multi-class setting.
However,
these indices were merely introduced in passing and not studied in detail
(e.g., their asymptotic distributions were not derived)
nor they were applied for inferential purposes.
In this article,
we study their distributional properties and also propose their use for
testing spatial clustering (especially of cases compared to controls).

Disease clustering methods can also be classified as
\emph{general} or \emph{focused} (\cite{besag:1991} and \cite{tango:1995}).
In the former, presence of any cluster over the entire region is of interest,
while in the latter presence of a cluster in the vicinity of a given point is investigated.
In this article, we are concerned with the first type of clustering.
See \cite{tango:2009} for a review on existing disease clustering methods,
their advantages and disadvantages.
Several indices measure spatial autocorrelation in a given data which could suggest clustering of a disease,
e.g., Moran's I statistic (\cite{moran:1948}) and Geary's $c$ statistic (\cite{geary:1954}).
These indices were employed to assess spatial patterns of disease rates
and are compared with another index called rank adjacency statistic, $D$, by \cite{walter:1993}.
Furthermore, there are methods which provide a general clustering measure for the entire study area,
such as Whittemore's statistic (\cite{whittemore:1987}).
However, this statistic was shown to be inadequate by \cite{tango:1999},
who also proposed a corrected version of it.
As the general clustering methods fail to identify localized clusters,
the so called \emph{scan statistics} are developed.
In these methods, the region is scanned by a rectangular or
circular window to detect any anomaly in disease occurrence or intensity.
Examples of scan methods are Openshaw's GAM (\cite{openshaw:1987}),
Besag and Newell's method to detect clusters of size $k$,
which comprise of regions containing exactly $k$ observed cases (\cite{besag:1991}),
and Kulldorff \& Nagarwalla's scan statistic (\cite{kulldorff:1995}).
In literature,
despite the lack of a comprehensive comparison of many available geographical disease clustering tests,
an empirical comparison was performed by \cite{kulldorff:2003}
using spatial scan statistic,
the maximized excess events test,
and the nonparametric $M$ statistic.
All tests are shown to have good power for detection of disease clusters
and the first having good performance in locating disease hot spots.

In literature,
clustering not only in space but also in time is of interest,
especially with applications in climatology or ecology (\cite{eshel:2012}).
This type of clustering called \emph{spatio-temporal clustering}
is also of great merit in disease clustering research.
Tango suggested an index for disease clustering in time (\cite{tango:1984})
and this index is assessed in detail for performance to detect
disease clustering in time and space by \cite{kryscio:1991}.
Several other indices were also proposed in literature to capture spatial patterns
and their evolution in time.
See, \cite{woillez:2007} for an example in marine biology which measures
spatial patterns of fish populations,
and \cite{li:1993} for an example in landscape ecology,
where a new contagion index was proposed that also corrects for an existing index.
Several other indices such as
Camargo's index of evenness, Simpson's index of evenness,
Lloyd's index of mean crowding, Smith-Wilson index, and dispersion index,
a variant of the Shannon diversity index
were employed mainly to quantify biodiversity,
but were also possibly usable for measuring spatial patterns from evenness to patchiness (\cite{payne:2005}).
Gini-style indices were used in health economics to assess
the spatial patterns of health workers (\cite{brown:1994})
and to evaluate residential segregation in a social context (\cite{dawkins:2004}).

In this article,
we propose the use of Pielou's coefficient of segregation and Dixon's segregation indices
in detecting disease clustering against the RL of cases and controls to a set of given spatial locations.
Dixon's segregation indices are not bounded for all possible types of NNCTs,
hence we also suggest corrected versions of Dixon's segregation indices
which are bounded for all cell counts (zero or positive) in a NNCT.
We derive their asymptotic distributions
(more specifically asymptotic normality)
and compare these tests with various existing tests, namely, Cuzick-Edwards' $k$-NN and combined tests,
Dixon's cell-specific and overall tests, type III cell-specific and overall tests
in terms of empirical size and power.
For the RL,
we investigate the effect of the clustering of the background points
(on which RL is performed),
including the level of clustering
and number of clusters, on the empirical sizes of the tests.
We also propose various non-RL patterns as alternatives and
investigate the power performance of the tests under these alternatives
via extensive Monte Carlo simulations.
To the author's knowledge,
these null RL patterns and alternative non-RL patterns are investigated
for the first time
and in fact the non-RL patterns are newly introduced in this article.

We present the null and alternative patterns and
construction of NNCTs in Section \ref{sec:prelim}.
The two segregation indices for spatial and disease clustering
are provided in Section \ref{sec:seg-ind-dis-clust},
where the asymptotic normality of Pielou's coefficient of segregation
and Dixon's segregation indices are derived.
Other NN-based spatial clustering tests that are used for comparative purposes
are discussed in Section \ref{sec:other-spat-tests}.
We provide an extensive empirical size comparison of these tests
under RL of points from various patterns of complete spatial randomness (CSR)
or clustering in Section \ref{sec:empirical-size},
propose four types of non-RL patterns as alternatives and provide
empirical power comparison of the tests under these alternatives in Section \ref{sec:empirical-power},
illustrate the methodology on two real life data sets from epidemiology in Section \ref{sec:example-data}.
Discussion and conclusions are provided in Section \ref{sec:disc-conc}.

\section{Preliminaries}
\label{sec:prelim}

\subsection{Null and Alternative Patterns}
\label{sec:null-and-alt}
In a case-control setting,
the null pattern we consider is
$$H_o:RL,$$
which is the pattern where the class labels (i.e., case or control labels)
are randomly assigned to a given set of locations or points.
In the two-class setting,
deviations from the null hypothesis are
towards two major directions, namely,
\emph{segregation} and \emph{association}.
{\em Segregation} is the pattern in which
NN of an individual is more likely than expected to be of the same
class as the individual than to be from a different class.
That is, the probability that this individual having a same-class NN
is higher than the relative frequency of this class (see, e.g., \cite{pielou:1961}).
On the other hand,
{\em association} is the pattern in which
NN of an individual is more likely than expected
to be from another class than to be of the same class as the individual.
That is,
the probability that this individual having a NN from another class
is higher than the relative frequency of this other class.
See \cite{ceyhan:corrected} for a more detailed discussion of the null and alternative patterns.

In a case-control setting,
segregation of cases from controls would be equivalent to clustering of cases
relative to the controls.
In other words, segregation of cases would imply a larger level of clustering
compared to the level of clustering of the healthy controls in the society.
Furthermore,
if for some reason controls are segregated,
then this would also mean an (indirect) clustering of cases,
but, here the underlying dynamics behind the disease clustering would be different.
The association of the cases and controls would mean significant lack of disease clustering;
moreover,
it would mean clustering of points from both classes (i.e., attraction of controls by cases
or vice versa).
This may not be practical either,
hence is not pursued in detail in the rest of the article.
However,
association could still be relevant to disease clustering in epidemiology in other settings.
For example,
one class could be the `sources' of a contaminant or some other pollutant or disease-causing agent,
and the other class could be the `cases'.
The accumulation of cases around the sources more often than expected would mean
clustering of a disease around these sources,
which is a form of association between the classes.
But we will not pursue this type of association in this article either.

\subsection{Construction of NNCTs}
\label{sec:construction-nncts}
The segregation indices and most of the tests
we consider for comparative purposes in this article
are in some way related to NNCTs.
We provide a brief description of NNCTs below,
for a more detailed description see, e.g., \cite{ceyhan:corrected}.
In a sample of size $n$,
there are $n$ NN pairs,
and each NN pair consists of the point labeled as ``base" point
and its ``NN" point.
According to the labels of the base and NN points,
NN pairs can be classified into various categories
and
NNCTs are constructed using these categories.
For $m$ classes,
we will have a $m \times m$ NNCT
whose rows represent class labels of base points
and columns represent class labels of the corresponding NN points.
In the NNCT,
the count in cell (or entry) $(i,j)$ is $N_{ij}$,
which is the number of times the NN of a (base) class $i$ point is from class $j$.
See also Table \ref{tab:NNCT-mxm} (left)
where $C_j$ is the sum of column $j$;
i.e., number of times class $j$ points serve as NNs
for $j\in\{1,2,\ldots,m\}$
and $n_i$ is the sum of row $i$;
i.e., number of times class $i$ points serve as base class
or size of class $i$ for $i=1,2,\ldots,m$.
In what follows,
we adopt the convention that
lower case letters represent fixed quantities
while upper case letters represent random variables.
Notice that
in a NNCT-analysis, row sums are assumed to be fixed
(i.e., class sizes are given),
while column sums are random variables depending
on the NN relationship between the classes.

In a case-control setting,
we have two classes (i.e., $m=2$)
and we reserve class label 1 for cases,
and class label 2 for controls.
Hence the case-control setting yields a $2 \times 2$ NNCT
(see Table \ref{tab:NNCT-mxm} (right)).

\begin{table}
\centering
\begin{tabular}{cc|ccc|c}
\multicolumn{2}{c}{}& \multicolumn{3}{c}{NN}& \\
\multicolumn{2}{c}{}&    class 1 &  $\ldots$ & class $m$   &   total  \\
\hline
&class 1 &    $N_{11}$    &  $\ldots$ &   $N_{1m}$    &   $n_1$  \\
\raisebox{1.5ex}[0pt]{base}& $\vdots$ & $\vdots$ & $\ddots$ & $\vdots$ & $\vdots$\\
&class $m$ &    $N_{m1}$    & $\ldots$ &    $N_{mm}$    &   $n_m$  \\
\hline
&total     &    $C_1$             &  $\ldots$ &    $C_m$             & $n$  \\
\end{tabular}
\hspace{.5 cm}
\begin{tabular}{cc|cc|c}
\multicolumn{2}{c}{}& \multicolumn{2}{c}{NN}& \\
\multicolumn{2}{c}{}&    case &  control   &   total  \\
\hline
& case &    $N_{11}$    &   $N_{12}$    &   $n_1$  \\
\raisebox{1.5ex}[0pt]{base}
& control &    $N_{21}$    & $N_{22}$    &   $n_2$  \\
\hline
&total     &    $C_1$  &    $C_2$             & $n$  \\
\end{tabular}
\caption{ \label{tab:NNCT-mxm} The NNCT for $m$ classes (left) and for two classes in a case-control setting (right).}
\end{table}

\section{Segregation Indices for Spatial and Disease Clustering}
\label{sec:seg-ind-dis-clust}

\subsection{Pielou's Coefficient of Segregation}
\label{sec:seg-ind-nncts}
In a two-class setting (i.e., for $k=2$),
Pielou's coefficient of segregation is defined as
\begin{equation}
\label{eqn:piel-seg-cf}
S_P=1-\frac{N_{12}+N_{21}}{\E[N_{12}]+\E[N_{21}]}
\end{equation}
where $\E[N_{ij}]$ is the expected value of $N_{ij}$ (\cite{pielou:1961}).
Notice that the numerator in the second part of $S_P$ is
\begin{multline*}
N_{12}+N_{21}= \sum_{i=1}^n \I(\text{point $i$ is from class $1$ with a NN from class $2$}\\
\text{ or point $i$ is from class $2$ with a NN from class $1$})
\end{multline*}
where $\I()$ stands for the indicator function.
In general a $k \times k$ contingency table may result from two multinomial frameworks:
row-wise and overall multinomial frameworks.
However for spatial data,
we have a special type of contingency table,
namely, NNCT, which requires different dependency structure compared to these frameworks
for completely mapped data,
which include the locations of all points in the region of interest.
Below, we discuss these frameworks for completeness,
and when each one would be appropriate for a NNCT-analysis.

\subsubsection{The Row-wise Multinomial Framework}
\label{sec:row-wise-multinomial}
In this framework,
the rows of a contingency table result from independent multinomial distributions.
In particular,
in the two-class case,
each row has a binomial distribution independent of the other rows
(so this framework is also referred to as the \emph{binomial framework}).

Let $\pi_{ij}$ be the probability of a point from
class $j$ serving as NN to a point from class $i$ for $i,j \in\{1,2\}$.
In this framework, we assume that $N_i=n_i$ are given and
$N_{ij} \sim \BIN(n_i,\pi_{ij})$, the binomial distribution with $n_i$
independent trials and probability of success being $\pi_{ij}$.
Hence, in the two-class case,
$(N_{11},N_{12})$ and $(N_{21},N_{22})$ are assumed to be independent
and so are the individual trials (which are base---NN pairs).
Hence this framework would be appropriate for a NNCT-analysis,
provided that we have an independent set of (base-NN) pairs,
that is,
each (base-NN) pair is independent of other pairs.
However, for completely mapped data,
this assumption does not hold,
due to the inherent spatial dependence
(for example, a base point would be more likely to be a NN of its own NN
compared to being a NN of an arbitrarily selected point).
However, if we have data obtained by sparse sampling,
this dependence would be nonexistent or negligible,
then this framework would be appropriate for the corresponding NNCT.
In what follows,
when we say data is from sparse sampling,
we also assume that (base-NN) pairs constitute an (almost) independent sample.
Thus,
under sparse sampling with the binomial framework,
we would have $(N_{11},N_{12})$ is independent of $(N_{21},N_{22})$,
and $N_{ij} \sim \BIN(n_i,\pi_{ij})$.
Under the null hypothesis of random assignment of case and control labels to any given point
proportional to the class sizes,
we would have
$N_{11} \sim \BIN(n_1,\nu_1)$,
$N_{12} \sim \BIN(n_1,\nu_2)$,
$N_{21} \sim \BIN(n_2,\nu_1)$,
and
$N_{22} \sim \BIN(n_2,\nu_2)$,
where $\nu_i$ is the population proportion of class $i$ points for $i=1,2$.
Also under $H_o$, we have
$\E[N_{12}]=n_1 \nu_2$ and $\E[N_{21}]=n_2 \nu_1$
and so
$$S_P=1-\frac{N_{12}+N_{21}}{\E[N_{12}]+\E[N_{21}]}=1-\frac{N_{12}+N_{21}}{n_1 \nu_2+n_2 \nu_1}.$$
We also have $\E[S_P]=0$,
since $\E[N_{12}+N_{21}]=\E[N_{12}]+\E[N_{21}]$ by linearity of expectation.
Furthermore,
$\Var[N_{ij}]=n_i \pi_{ij} (1-\pi_{ij})$,
so under $H_o$,
$\Var[N_{12}]=n_1 \nu_1 \nu_2$
and
$\Var[N_{21}]=n_2 \nu_1 \nu_2$,
and $N_{12}$ and $N_{21}$ are independent.
Hence
$\Var[S_P]=\frac{n \nu_1 \nu_2}{(n_1\nu_2+n_2 \nu_1)^2}$.
However,
in practice $\nu_i$ would be unknown,
and hence need to be estimated.
Given sample of size $n_i$ from class $i$,
we estimate $\nu_i$ as
$\widehat \nu_i=n_i/n$
for $i=1,2$.
Then for large $n_i$,
$\E[N_{12}] \approx n_1 n_2/n$ and $\E[N_{21}]\approx n_2 n_1/n$,
so $S_P \approx 1-\frac{N_{12}+N_{21}}{(2 \, n_1 n_2/n)}$.
Furthermore,
we have
$\Var[S_P] \approx \frac{n}{4 n_1 n_2}$.
Then for large $n_i$, $i=1,2$,
under sparse sampling,
$S_P/\sqrt{\Var[S_P]}$ approximately has $N(0,1)$ distribution
where $N(\mu,\sigma)$ stands for normal distribution with mean $\mu$ and standard deviation $\sigma$.

\subsubsection{The Overall Multinomial Framework}
\label{sec:overall-multinomial}
An alternative modeling of a contingency table in general is that
cell counts (or the entries) are assumed to arise from multinomial trials.
That is, in the two-class case,
$$\mathbf {N}=(N_{11},N_{12},N_{21},N_{22}) \sim \mathscr M(n,\nu_1\kappa_1,\nu_1\,\kappa_2,\nu_2\,\kappa_1,\nu_1\,\kappa_2)$$
where $\nu_1+\nu_2=1$ and $\kappa_1+\kappa_2=1$;
hence the name {\em overall multinomial framework}.
As in the row-wise multinomial framework,
this framework would be appropriate for a NNCT-analysis,
provided we have sparsely sampled data.
In this framework,
$N_{12} \sim \BIN(n,\nu_1 \kappa_2)$
and
$N_{21} \sim \BIN(n,\nu_2 \kappa_1)$.
Using $\kappa_2=1-\kappa_1$ and $\nu_2=1-\nu_1$,
we have
$N_{12} \sim \BIN(n,\nu_1 (1-\kappa_1))$
and
$N_{21} \sim \BIN(n,(1-\nu_1) \kappa_1)$.
Hence
$\E[N_{12}]=n \nu_1 (1-\kappa_1)$
and
$\E[N_{21}]=n (1-\nu_1) \kappa_1$,
which yields
$\E[N_{12}]+\E[N_{21}]=n (\nu_1+\kappa_1)-2n \nu_1 \kappa_1$.
Thus it follows that
$$S_P=1-\frac{N_{12}+N_{21}}{n (\nu_1+\kappa_1)-2n \nu_1 \kappa_1}.$$
Furthermore,
$$
\Var[N_{12}]=n \nu_1 (1-\kappa_1)(1-\nu_1+\nu_1 \kappa_1),~~~
\Var[N_{12}]=n (1-\nu_1)\kappa_1(1-\kappa_1+\nu_1 \kappa_1),
$$
and
$$\Cov[N_{12},N_{21}]=-n \nu_1 \kappa_2 \nu_2 \kappa_1=-n \nu_1(1-\kappa_1)(1-\nu_1)\kappa_1.$$
So
$\Var[N_{12}+N_{21}]=n(1-\nu_1-\kappa_1+2\nu_1 \kappa_1)(\nu_1+\kappa_1-2 \nu_1\kappa_1)$,
which implies
$\Var[S_P]=\frac{1-\nu_1-\kappa_1+2\nu_1 \kappa_1}{n(\nu_1+\kappa_1-2 \nu_1\kappa_1)}$.
Under $H_o$,
we have $\nu_1=\kappa_1$,
and so
$$\Var[S_P]=\frac{1}{2n}\left(\frac{\nu_2}{\nu_1}+\frac{\nu_1}{\nu_2}\right).$$
Hence for large $n_i$, $i=1,2$,
$\Var[S_P] \approx \frac{1}{2n}\left(\frac{n_2}{n_1}+\frac{n_1}{n_2}\right)$
and under sparse sampling
$S_P/\sqrt{\Var[S_P]}$ approximately has $N(0,1)$ distribution.

\begin{remark}
Both of the above multinomial frameworks require that
we have an independent sample of $n$ (base,NN) pairs,
which is approximately valid when we have sparse sampling.
However,
in a case-control setting, sparse sampling may not be a feasible procedure,
especially when the disease in question is rare.
Hence, sparse sampling in general is not advisable for detection of disease clustering.
However,
these frameworks would work when there is a substantial amount of data from both classes
in the region of interest,
and sparse sampling is a feasible practice to capture
the actual interaction between the classes.
$\square$
\end{remark}

\subsubsection{Pielou's Coefficient of Segregation under RL}
Under RL of $n_1$ cases and $n_2$ controls to $n=n_1+n_2$ given locations,
we have
$\E[N_{12}]=\E[N_{21}]=\frac{n_1 n_2}{n-1}$.
So under $H_o$,
$$S_P=1-\frac{N_{12}+N_{21}}{n_1 n_2/(n-1)}$$
and $\E[S_P]=0$.
Furthermore,
$$\Var[N_{ij}]=n\, p_{ij}+Q\,p_{iij}+(n^2-3n-Q+R)p_{iijj}-(np_{ij})^2,$$
and
$$\Cov[N_{ij},N_{ji}]=R p_{ij}+(n-R)(p_{iij}+p_{ijj})+(n^2-3n-Q+R)p_{iijj}-n^2 p_{ij}p_{ji},$$
where
$p_{ij}=\frac{n_i n_j}{n(n-1)}$,
$p_{iij}=\frac{n_i(n_i-1)n_j}{n(n-1)(n-2)}$,
$p_{ijj}=\frac{n_i n_j(n_j-1)}{n(n-1)(n-2)}$,
and
$p_{iijj}=\frac{n_i(n_i-1)n_j(n_j-1)}{n(n-1)(n-2)}$,
for $(i,j)=(1,2)$ and $(i,j)=(2,1)$,
$R$ is twice the number of reflexive pairs
and $Q$ is the number of points with shared  NNs,
which occurs when two or more
points share a NN.
Then $Q=2\,(Q_2+3\,Q_3+6\,Q_4+10\,Q_5+15\,Q_6)$
where $Q_k$ is the number of points that serve
as a NN to other points $k$ times.
Then
$\Var[N_{12}+N_{21}]=\Var[N_{12}]+\Var[N_{21}]+2\,\Cov[N_{12},N_{21}]$,
and for large $n_i$,
$S_P/\sqrt{\Var[S_P]}$ has approximately $N(0,1)$ distribution.

If we have the population proportion, $\nu_i$, for class $i$, $i=1,2$,
then we would have,
for large $n_i$,
$\E[N_{12}]=\E[N_{21}] \approx n \nu_1 \nu_2$
and
$S_P \approx 1-\frac{N_{12}+N_{21}}{n \nu_1 \nu_2}$.
Furthermore,
$p_{ij}=\nu_i \nu_j$,
$p_{iij}=\nu_i^2 \nu_j$,
$p_{ijj}=\nu_i \nu_j^2$,
and
$p_{iijj}=\nu_i^2 \nu_j^2$;
hence
\begin{equation}
\label{eqn:varNij}
\Var[N_{ij}] \approx n\, \nu_i \nu_j+Q\,\nu_i^2 \nu_j+(-3n-Q+R)\nu_i^2 \nu_j^2,
\end{equation}
and
$$\Cov[N_{ij},N_{ji}] \approx R \nu_i \nu_j+(n-R)(\nu_i^2 \nu_j+\nu_i \nu_j^2)+(-3n-Q+R)\nu_i^2 \nu_j^2,$$
for $(i,j)=(1,2)$ and $(i,j)=(2,1)$.

\subsection{Dixon's Segregation Indices}
\label{sec:dixon-seg-index}
In a multi-class setting,
\cite{dixon:NNCTEco2002} proposed the following indices
which are similar to the log odds-ratios in a NNCT:
\begin{equation}
\label{eqn:dixon-seg-ind}
S^D_{ij}=
\begin{cases}
\log \left(\frac{N_{ii}/(n_i-N_{ii})}{(n_i-1)/(n-n_i)}\right) & \text{if $i=j$,} \vspace{.15 cm}\\
\log \left(\frac{N_{ij}/(n_i-N_{ij})}{n_j/(n-n_j-1)} \right)    & \text{if $i \not= j$.}
\end{cases}
\end{equation}
For Dixon's segregation index,
we will only consider the RL as the underlying framework for the null model.
Under RL of $n_1$ cases and $n_2$ controls to $n$ given locations,
as $n_1$ and $n_2$ go to infinity,
$(N_{ii}-\E[N_{ii}])/\sqrt{\Var[N_{ii}]}$ converges in law to $N(0,1)$ distribution
and
$(N_{ij}-\E[N_{ij}])/\sqrt{\Var[N_{ij}]}$ converges in law to $N(0,1)$ distribution.
Theorem 7.7.6 in \cite{bain:1992} states that
``If $\frac{\sqrt{n}(Y_n-s)}{c} \stackrel{\mathcal L}{\sim} N(0,1)$ and if $g(y)$
has a nonzero derivative at $y=s$, i.e., $g'(s)\not=0$,
then
$\frac{\sqrt{n}(g(Y_n)-g(s))}{|c g'(s)|} \stackrel{\mathcal L}{\rightarrow} N(0,1)$"
where $\stackrel{\mathcal L}{\rightarrow}$ stands for ``converges in law".
So for $i=j$,
letting $s=\frac{n_i (n_i-1)}{n-1}$ and $g(y)=\log(y/(n_i-y))$ so that
$g'(s)=\frac{(n-1)^2}{n_i(n-n_i)(n_i-1)} \not= 0$ provided $n > 1$,
by the above theorem,
we get
$$\frac{\log \left( \frac{N_{ii}}{n_i-N_{ii}}\right)-\log \left(\frac{n_i-1}{n-n_i}\right)}
{\sqrt{\Var[N_{ii}]}\left(\frac{(n-1)^2}{n_i(n-n_i)(n_i-1)}\right)}
=
\frac{S^D_{ii}}
{\sqrt{\Var[N_{ii}]}\left(\frac{(n-1)^2}{n_i(n-n_i)(n_i-1)}\right)}$$
approximately having $N(0,1)$ distribution for large $n_i$.

Similarly
for $i\not=j$,
letting $s=\frac{n_i n_j}{n-1}$ we get
$g'(s)=\frac{(n-1)^2}{n_i n_j(n-n_j-1)} \not= 0$ provided $n > 1$.
By the above theorem,
we get
$$\frac{\log \left(\frac{N_{ij}}{n_i-N_{ij}}\right)-\log \left(\frac{n_j}{n-n_j-1}\right)}
{\sqrt{\Var[N_{ij}]}\left(\frac{(n-1)^2}{n_i n_j(n-n_j-1)}\right)}
=
\frac{S^D_{ij}}
{\sqrt{\Var[N_{ij}]}\left(\frac{(n-1)^2}{n_i n_j(n-n_j-1)}\right)}$$
approximately having $N(0,1)$ distribution for large $n_i$.

For the asymptotic approximations of Dixon's segregation indices when
the population proportion, $\nu_i$, of class $i$, $i=1,2$, are known,
see the Appendix section.

\subsection{A Correction for Dixon's Segregation Index}
\label{sec:dixon-seg-index-corr}
Dixon's segregation indices may be unbounded in either direction depending on the
cell counts in the NNCT.
Let $0<n_i<n$ for all $i$.
Then
if $N_{ii}=0$,
we get $S_{ii}^D=-\infty$ provided $n_i > 1$;
and if $N_{ij}=0$,
we get $S_{ij}^D=-\infty$ provided $n_j < n-1$.
Also,
if $N_{ii}=n_i$,
we get $S_{ii}^D=\infty$;
and if $N_{ij}=n_i$,
we get $S_{ij}^D=\infty$ provided $n_j < n-1$.
To make the segregation indices bounded for all possible cell counts,
we suggest the following corrected versions:
\begin{equation}
\label{eqn:dixon-seg-ind-corr}
S^{D,c}_{ij}=
\begin{cases}
\log \left(\frac{(N_{ii}+1)/(n_i-N_{ii}+1)}{(n(n_i-1)+(n-1))/(n_i(n-n_i)+(n-1))}\right) & \text{if $i=j$,} \vspace{.15 cm}\\
\log \left(\frac{(N_{ij}+1)/(n_i-N_{ij}+1)}{(n_i n_j+n-1)/(n_i(n-n_j-1)+(n-1))} \right)    & \text{if $i \not= j$,}
\end{cases}
\end{equation}
where denominators are chosen in this way
to have simpler asymptotic approximations for the corrected versions.

For the derivation of asymptotic distribution of these corrected versions,
see the Appendix section.

\section{Other NN-Tests for Spatial Clustering}
\label{sec:other-spat-tests}
Although there are many tests for spatial clustering
of points from one class or multiple classes in the literature (\cite{diggle:2003} and \cite{kulldorff:2006}),
one-class tests are not comparable with the segregation indices
nor very useful in disease clustering.
Some of the tests like
Moran's $I$ and Whittemore's tests are shown to perform poorly
in detection of some kind of clustering (\cite{song:2003})
and most of the tests require Monte Carlo simulation or randomization
methods to attach significance to their results.
Hence we only consider cell-specific and overall NNCT-tests due to \cite{dixon:1994} and \cite{ceyhan:corrected}
and Cuzick-Edward's $k$-NN tests and their combined versions (\cite{cuzick:1990}),
and compare the segregation indices with these tests in an extensive
Monte Carlo simulation study
in terms of size and power performance.

\subsection{Cell-Specific and Overall Segregation Tests based on NNCTs}
\label{sec:nnct-tests}
Dixon's cell-specific and overall tests (\cite{dixon:1994})
and type III cell-specific and overall tests (\cite{ceyhan:corrected}) are based on NNCTs.
These tests are discussed in detail in \cite{ECarXivCellOverall:2012};
here we only provide a brief description for completeness.
For cell $(i,j)$, \cite{dixon:1994} suggests
\begin{equation}
\label{eqn:dixon-Zij}
Z^D_{ij}=\frac{N_{ij}-\E[N_{ij}]}{\sqrt{\Var[N_{ij}]}}
\end{equation}
as the cell-specific tests,
where under RL the expected cell counts are
$ \E[N_{ij}]= n_i(n_i-1)/(n-1) \I(i=j) + n_i\,n_j/(n-1)  \I(i \not= j)$
and the variance $\Var[N_{ij}]$ is given in \cite{ceyhan:corrected}.
In the multi-class case with $m$ classes,
combining the $m^2$ cell-specific tests,
\cite{dixon:NNCTEco2002} suggests the following quadratic form as an overall test:
\begin{equation}
\label{eqn:dix-chisq-mxm}
C_D=(\mathbf{N}-\E[\mathbf{N}])'\Sigma_D^-(\mathbf{N}-\E[\mathbf{N}])
\end{equation}
where $\mathbf{N}$ is the $m^2\times1$ vector of $m$ rows
of the NNCT concatenated row-wise,
$\E[\mathbf{N}]$ is the vector of $\E[N_{ij}]$,
$\Sigma_D$ is the $m^2 \times m^2$ variance-covariance matrix for
the cell count vector $\mathbf{N}$ with diagonal entries being equal to
$\Var[N_{ij}]$ and off-diagonal entries being $\Cov[N_{ij},\,N_{kl}]$
for $(i,j) \neq (k,l)$.
The explicit forms of the variance and covariance terms are provided in \cite{dixon:NNCTEco2002}.
Also, $\Sigma_D^-$ is a generalized inverse of $\Sigma_D$ (\cite{searle:2006})
and $'$ stands for the transpose of a vector or matrix.
Then under RL,
$C_D$ approximately has a $\chi^2_{m(m-1)}$ distribution for large $n_i$.

On the other hand,
type III cell-specific test suggested by \cite{ceyhan:corrected} for cell $(i,j)$ is
\begin{equation}
\label{eqn:new-ZIIIij}
Z^{III}_{ij}=\frac{T^{III}_{ij}}{\sqrt{\Var\left[T^{III}_{ij}\right]}},
\end{equation}
where
$T^{III}_{ij}= \left(N_{ii}-\frac{(n_i-1)}{(n-1)}C_i \right) \I(i=j) + \left(N_{ij}-\frac{n_i}{(n-1)}C_j \right) \I(i \not= j)$.
The explicit forms of expectation and variance of
$T^{III}_{ij}$ are presented in \cite{ECarXivCellOverall:2012}.
We obtain the type III overall test by
combining the type III cell-specific tests.
Let $\mathbf {T^{III}}$ be the vector of $m^2$ $T^{III}_{ij}$ values, i.e.,
$$\mathbf {T^{III}}=\left(T^{III}_{11},T^{III}_{12},\ldots,T^{III}_{1m},T^{III}_{21},T^{III}_{22},\ldots,T^{III}_{2m},\ldots,T^{III}_{mm}\right)',$$
and let $\E\left[\mathbf {T^{III}}\right]$ be the vector of $\E\left[T^{III}_{ij}\right]$ values.
Note that $\E\left[\mathbf {T^{III}}\right]=\mathbf 0$
where $\mathbf 0$ is the vector of $m^2$ zeros.
As the type III overall segregation test, we use the following quadratic form:
\begin{equation}
\label{eqn:typeIIIoverall]}
C_{III}=\left(\mathbf {T^{III}}\right)'\Sigma_{III}^-\left(\mathbf {T^{III}}\right)
\end{equation}
where $\Sigma_{III}$ is the $m^2 \times m^2$ variance-covariance matrix of $\mathbf {T^{III}}$.
Under RL,
the explicit forms of the variance-covariance matrix
are provided in \cite{ECarXivCellOverall:2012}.
Furthermore,
under RL,
$C_{III}$ approximately has a $\chi^2_{(m-1)^2}$ distribution for large $n_i$.

\subsection{Cuzick-Edwards' $k$-NN and Combined Tests}
\label{sec:cuzick-edwards-tests}
For disease clustering,
\cite{cuzick:1990} suggested a $k$-NN test based on number of cases among $k$ NNs
of the case points.
Let $z_i$ be the $i^{th}$ point
and $d_i^k$ be the number cases among $k$ NNs of $z_i$.
Then Cuzick-Edwards' $k$-NN test is
$T_k=\sum_{i=1}^n \delta_i d_i^k$,
where
\begin{equation}
\label{eqn:delta-i}
\delta_i=
\begin{cases}
1 & \text{if $z_i$ is a case,} \vspace{.025 cm} \\
0 & \text{if $z_i$ is a control.}
\end{cases}
\end{equation}
Since in practice,
the correct choice of $k$ is not known,
\cite{cuzick:1990} also suggest combining various $T_k$ tests.
Let $S=\{k_1,k_2,\ldots,k_m\}$ be a set of indices for $k$,
and assume $T_k$ with $k \in S$ being a mixture of shifts all in the same direction under an alternative.
Assuming further that $T_k$ has multivariate normal distribution,
the combined test statistic is given by
\begin{equation}
\label{eqn:Tcomb}
T_S=\mathbf 1' \Sigma^{-1/2} \mathbf{T}
\end{equation}
where $\mathbf{T}=(T_{k_1},T_{k_2},\ldots,T_{k_m})'$
(i.e., $T_S$ is the test obtained by combining $T_k$ tests whose indices are in $S$),
$\mathbf 1'=(1,1,\ldots,1)$, $\Sigma=\Cov[\mathbf{T}]$ is the variance-covariance matrix of $\mathbf{T}$.
Under RL of $n_1$ cases and $n_2$ controls to the given locations in the study region,
$T_k$ approximately has $N(\E[T_k],\Var[T_k]/n_1)$ distribution for large $n_1$;
similarly, $T_S$ approximately has $N(\E[T_S],\Var[T_S])$ distribution for large $n_1$.
The expected values $\E[T_k]$ and $\E[T_S]$ and variances
$\Var[T_k]$ and $\Var[T_S]$ are provided in \cite{cuzick:1990}.

Notice that $T_1$ is identical to the count for cell $(1,1)$ in the NNCT of Table \ref{tab:NNCT-mxm} (right).
Hence the corresponding tests $(T_1-\E[T_1])/\sqrt{\Var[T_1]}$ and $Z^D_{11}$ are identical.
Hence,
we only consider $T_2$ and $T_S$ with $S=\{1,2\}$ for Cuzick-Edwards' tests in our comparisons.

\begin{remark}
Note that under $H_o$,
expected values of $S_P$, $S^D_{ii}$, $Z^D_{ii}$, $T_k-\E[T_k]$ and
$T_S - \E[T_S]$ are all zero.
However they tend to be positive under segregation
and negative under association.
On the other hand,
under segregation,
the diagonal cell counts, $N_{ii}$, would be larger,
while under association,
the off-diagonal cell counts, $N_{ij}$, with $i \not= j$,
would be larger than expected.
Hence $S^D_{ij}$ and $Z^D_{ij}$ for $i \not= j$
tend to be negative under segregation and
positive under association.
Hence all these tests can be employed to test spatial clustering in the two directions
against $H_o$ in a two-class setting.
In a case-control setting,
segregation of cases from the controls would be our primary interest.
$\square$
\end{remark}

\begin{remark}
With $m=2$ classes (or in a case-control setting),
$S_P$, $S^D_{ii}$, $Z^D_{ii}$, $C_D$, $T_1$ and $C_{III}$
can detect the spatial interaction at small scales (at around the average NN distance),
while $T_k$ with $k > 1$ can detect at larger scales (at around $k$-th NN distance),
and so can $T_S$ with $S$ having indices other than 1
(at around $\ell$-th NN distance with $\ell=\min k_i$ to $\ell=\max k_i$ for $k_i \in S$).
Hence,
$S_P$, $S^D_{ii}$, $Z^D_{ii}$, $C_D$, $T_1$ and $C_{III}$ can be used to test the same type of interaction at the same scales,
but Cuzick-Edwards' tests can be used to do the same at higher scales.
$\square$
\end{remark}

\section{Empirical Size Analysis of the Tests}
\label{sec:empirical-size}
Let $\mZ_n=\{Z_1,Z_2,\ldots,Z_n\}$ be the given set of locations for $n$ points
(called \emph{background pattern}).
We consider RL of cases and controls to points, $\mZ_n$, generated from
various homogeneous or clustered patterns.
To remove the effect of one particular realization of the $Z_i$ points on the tests,
we consider 100 different realizations of $\mZ_n$ on which RL will be applied.
For each background realization,
we label $n_1$ of the points as class $X$ (for cases) and the remaining $n_2=n-n_1$
points as class $Y$ (for controls).

\begin{itemize}
\item[] \textbf{Types of the Background Patterns:}
\item[] \textbf{Case 1:}
We generate $\mZ_n$ points independently uniformly in the unit square $(0,1)\times(0,1)$,
i.e., $Z_i \stackrel{iid}{\sim} \U((0,1)\times(0,1))$ for $i=1,2,\ldots,n$.
We consider (a) $n_1=n_2=10,20, \ldots,100$ to determine the effect of increasing but equal sample sizes,
(b) $n_1=30$ and $n_2=30,40,\ldots,120$ to determine the differences in the sample sizes with number of cases fixed
and number of controls increasing,
and
(c) $n_2=30$ and $n_1=30,40,\ldots,120$ to determine the differences in the sample sizes with number of controls fixed
and number of cases increasing.
We perform the above RL $1000$ times for each $(n_1,n_2)$ combination at each background realization.

\item[] \textbf{Case 2:}
We generate $Z_i \stackrel{iid}{\sim} \U(S^I_\delta)$ for $i=1,2,\ldots,n$
where $S^I_\delta=((0,1) \times (0,1)) \cup ((\delta,1+\delta) \times (\delta,1+\delta))$.
We consider $\delta = 0.2,0.4.,\ldots,2.0$,
so that as $\delta$ increases,
the level of clustering of background points increases.
We perform the above RL $1000$ times for each $\delta$
at each background realization with $n_1=n_2=100$.

\item[] \textbf{Case 3:}
We generate $Z_i \stackrel{iid}{\sim} \U(S^{II}_\delta)$ for $i=1,2,\ldots,n$
where $S^{II}_\delta=((0,1) \times (0,1)) \cup ((1+\delta,2+\delta) \times (0,1))$.
We consider $\delta = 0.0,0.2,0.4.,\ldots,1.4$,
so that as $\delta$ increases,
the level of clustering of background patterns increases.
We perform the above RL $1000$ times for each $\delta$
at each background realization with $n_1=n_2=100$.

\item[] \textbf{Case 4:}
We generate $Z_i \stackrel{iid}{\sim} \U(S_{\delta,k})$ for $i=1,2,\ldots,n$
where $S_{\delta,k}=((0,1) \times (0,1)) \cup ((1+\delta,2+\delta) \times (0,1)) \ldots \cup ((k-1)(1+\delta),k+(k-1)\delta) \times (0,1))$
which yields $k$ squares along the $x$-axis
for the support of $Z_i$,
with successive squares being $\delta$ units apart.
We consider $\delta = 0.5$, so that each square is clearly separated,
and $k=1,2,\ldots,10$ so that the sensitivity of the empirical sizes of the tests to the number distinct clusters could be assessed.
We perform the above RL $1000$ times for each $k$
at each background realization with $n_1=n_2=100$.

\item[] \textbf{Case 5:}
In this case,
we generate $Z_i$ points from Mat\'{e}rn's cluster process in the unit square,
denoted $\matclust(\kappa,r,\mu)$ (\cite{baddeley:2005}).
First we generate ``parent" points from a Poisson process with intensity $\kappa$
and then each parent is replaced by $N$ points independently uniformly generated inside the circle centered at the
parent point with radius $r$,
where $N \sim \poisson(\mu)$.
For each background realization,
we generate one realization of $\mZ_n$ from $\matclust(\kappa,r,\mu)$,
and let $n$ be the number of points in this realization.
Then we label $n_1=\lfloor n/2  \rfloor$ of these points as cases,
and $n_2=n-n_1$ as controls,
where $\lfloor x \rfloor$ stands for the floor of $x$.
Here we take $\kappa=1,2,\ldots,10$, $\mu=\lfloor 200/\kappa  \rfloor$,
and  $r=0.1$ in our simulations.
That is,
we take $(\kappa,\mu) \in \{(1,200), (2,100), (3,66) \ldots, (10,20)\}$,
so that on the average we
would have about 200 $Z$ points
of which 100 are $X$ and 100 are $Y$ points.
\end{itemize}

At each Monte Carlo replication in each of the above cases,
we compute the following test statistics:
Pielou's coefficient of segregation, $S_P$,
Dixon's segregation indices, $S^D_{ij}$, for $i,j=1,2$,
and the corrected versions, $S^{D,c}_{ij}$, for $i,j=1,2$,
Dixon's cell-specific tests, $Z^D_{ij}$, for $i,j=1,2$
type III cell-specific tests, $Z^{III}_{ij}$, for $i,j=1,2$,
Dixon's overall test, $C_D$,
type III overall test, $C_{III}$,
Cuzick-Edwards' $k$-NN tests, $T_k$, for $k=1,2$,
and combined test, $T_S$, for $S=\{1,2\}$
(which is denoted as $T_{1,2}$ in short).
However,
the case-control setting corresponds to a two-class case.
Hence in our further analysis,
we only consider and present
$S^D_{ii}$ for $i=1,2$ among Dixon's segregation indices,
since $S^D_{11}=-S^D_{12}$ and $S^D_{22}=-S^D_{21}$;
$Z^D_{ii}$ for $i=1,2$ among Dixon's cell-specific tests,
since $Z^D_{11}=-Z^D_{12}$ and $Z^D_{22}=-Z^D_{21}$;
$Z^{III}_{ii}$ for $i=1,2$ among type III cell-specific tests,
since $Z^{III}_{11}=-Z^{III}_{21}$ and $Z^{III}_{22}=-Z^{III}_{12}$.
Furthermore,
among Cuzick-Edwards' $k$-NN tests,
we only consider and present $T_2$,
and combined test for $S=\{1,2\}$,
since $(T_1-\E[T_1])/\sqrt{\Var[T_1]}=Z^D_{11}$ in the two-class case.
Also,
$S^D_{ij}$ for $i,j=1,2$,
and the corrected versions,
$S^{D,c}_{ij}$ for $i,j=1,2$ provide very similar empirical size estimates, hence only the former are presented.
In our empirical size analysis (and also in the power analysis in Section \ref{sec:empirical-power}),
we use standardized forms of Pielou's coefficient of segregation
and Cuzick-Edwards' tests.
That is,
we use $Z_P=S_P/\Var[S_P]$,
and $(T_k-\E[T_k])/\sqrt{\Var[T_k]}$ for $k=1,2$,
and $(T_S-\E[T_S])/\sqrt{\Var[T_S]}$ for $S=\{1,2\}$.

In case 1, we have the background pattern from a HPP;
i.e., each realization of $\mZ_n$ is from the CSR pattern.
In this case,
we investigate the effect of equal but increasing sample sizes,
and differences in the relative abundances (in both directions,
with fixed number of cases and increasing number of controls and vise versa).
In case 2,
we consider an increasing level of clustering along the diagonal $y=x$ with increasing $\delta$,
and for $\delta>1$, the two clusters are disjoint.
In case 3,
we already have two disjoint clusters along the $x$-axis,
and the level of clustering increases with increasing $\delta$.
Hence in cases 2 and 3,
the effect of clustering level on the empirical sizes are assessed.
In case 4,
we already have $k$ disjoint clusters with $\delta=0.5$
and assess the effect of number of clusters on the empirical sizes.
In case 5,
we have clusters where the size and location of the clusters are random
according to a Mat\'{e}rn clustering process.
In this case, we assess the effect of such clustering on the empirical sizes.

\begin{figure} [hbp]
\centering
\rotatebox{-90}{ \resizebox{2.1 in}{!}{\includegraphics{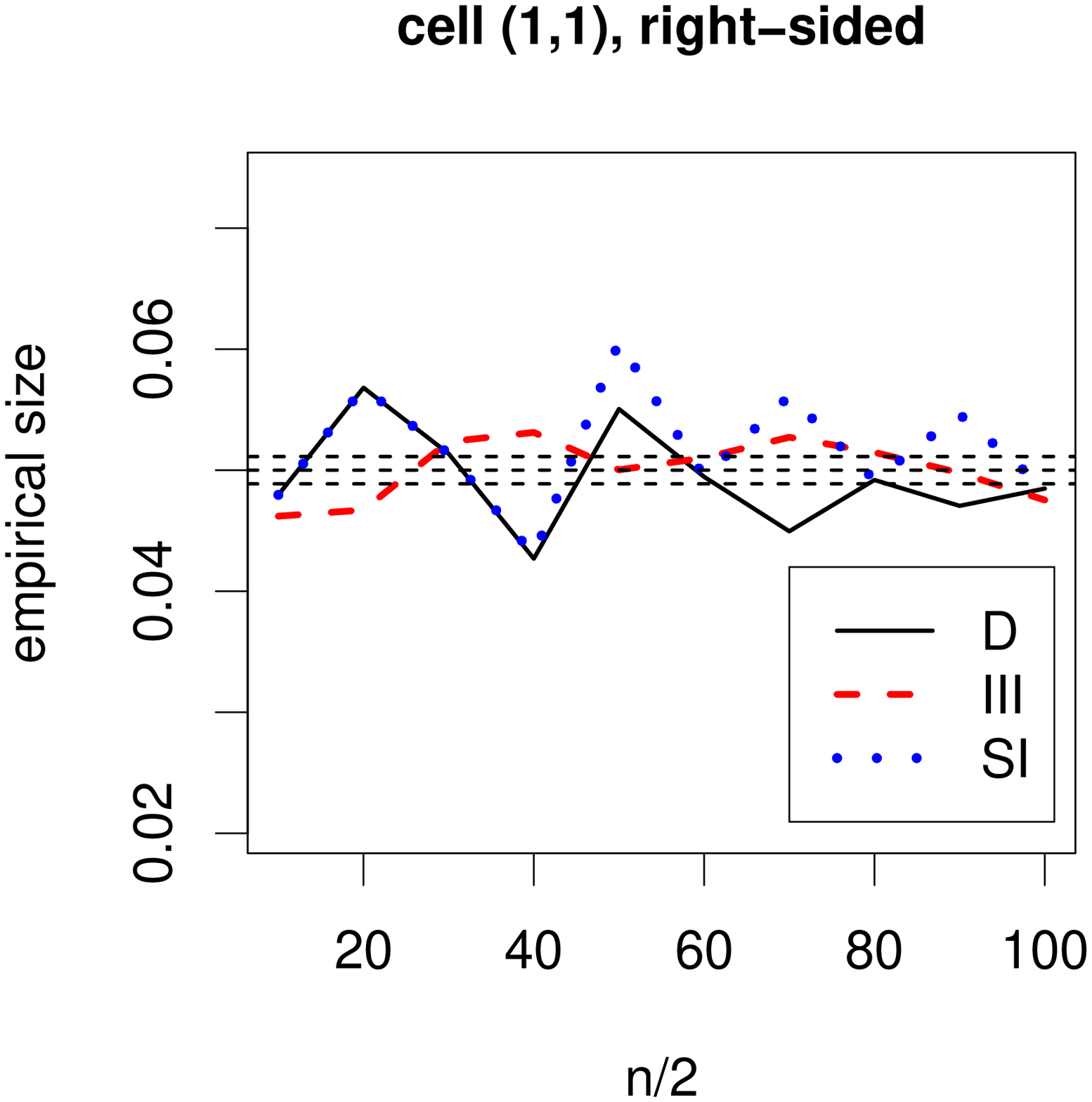} }}
\rotatebox{-90}{ \resizebox{2.1 in}{!}{\includegraphics{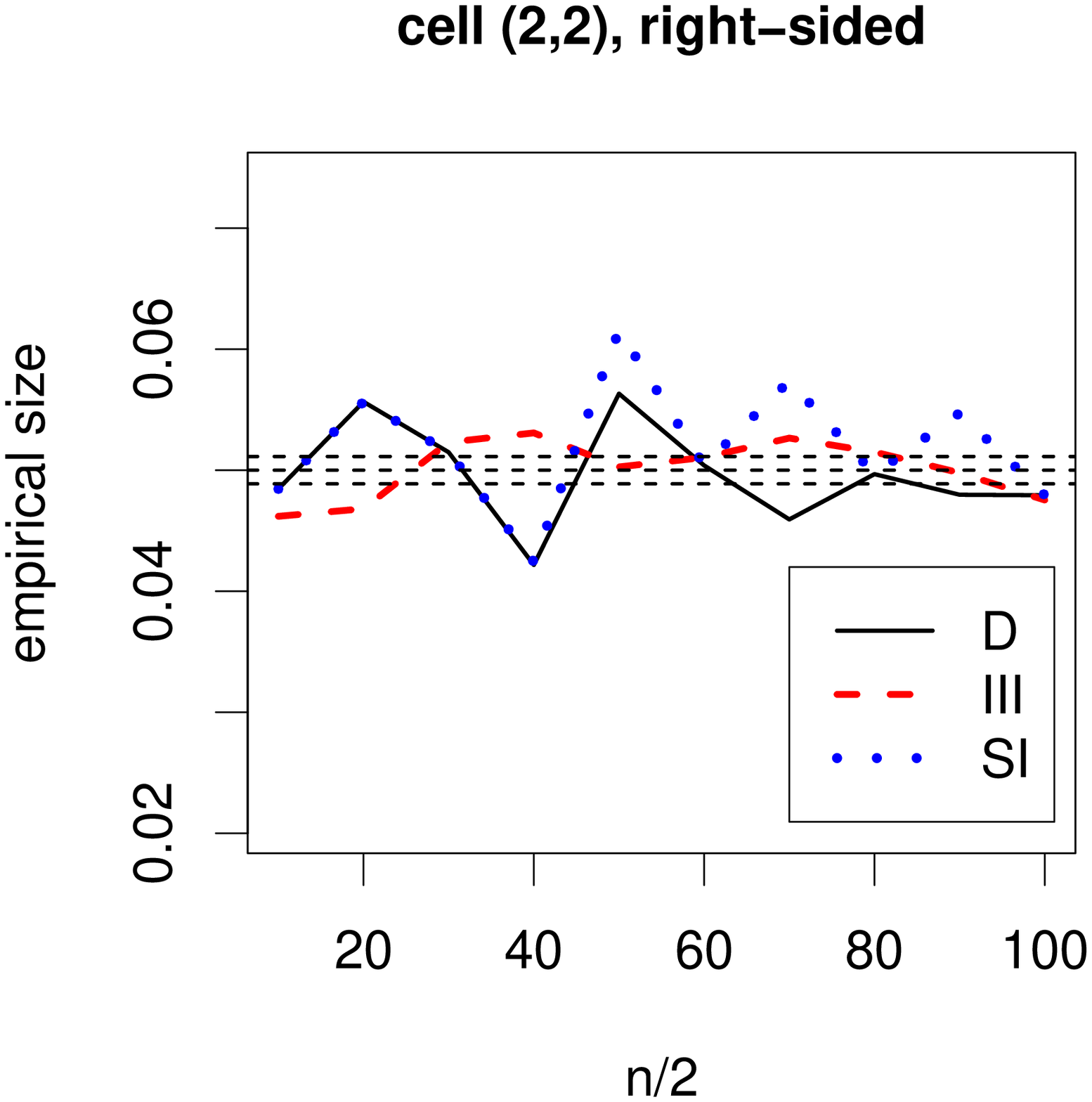} }}
\rotatebox{-90}{ \resizebox{2.1 in}{!}{\includegraphics{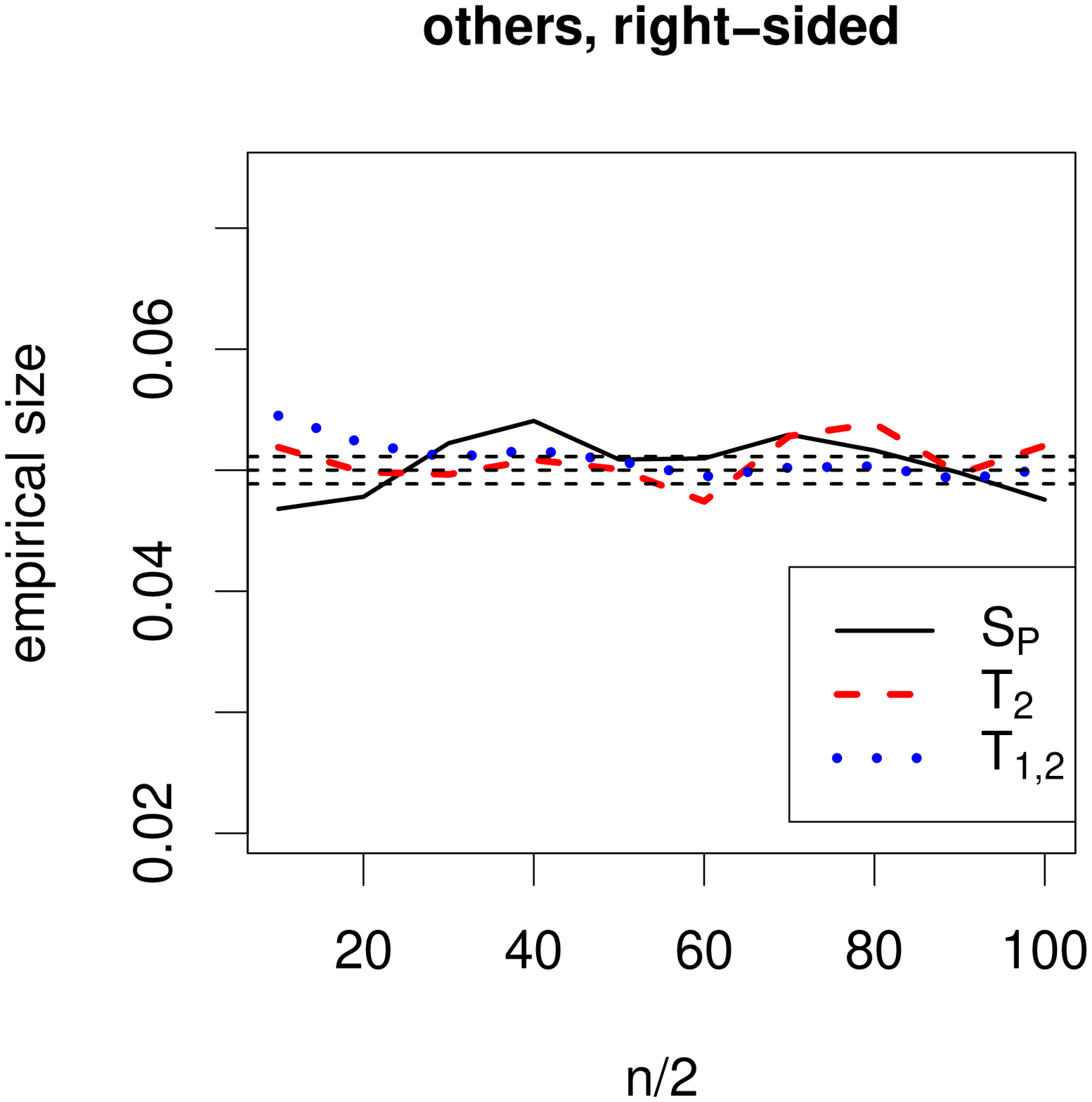} }}
\rotatebox{-90}{ \resizebox{2.1 in}{!}{\includegraphics{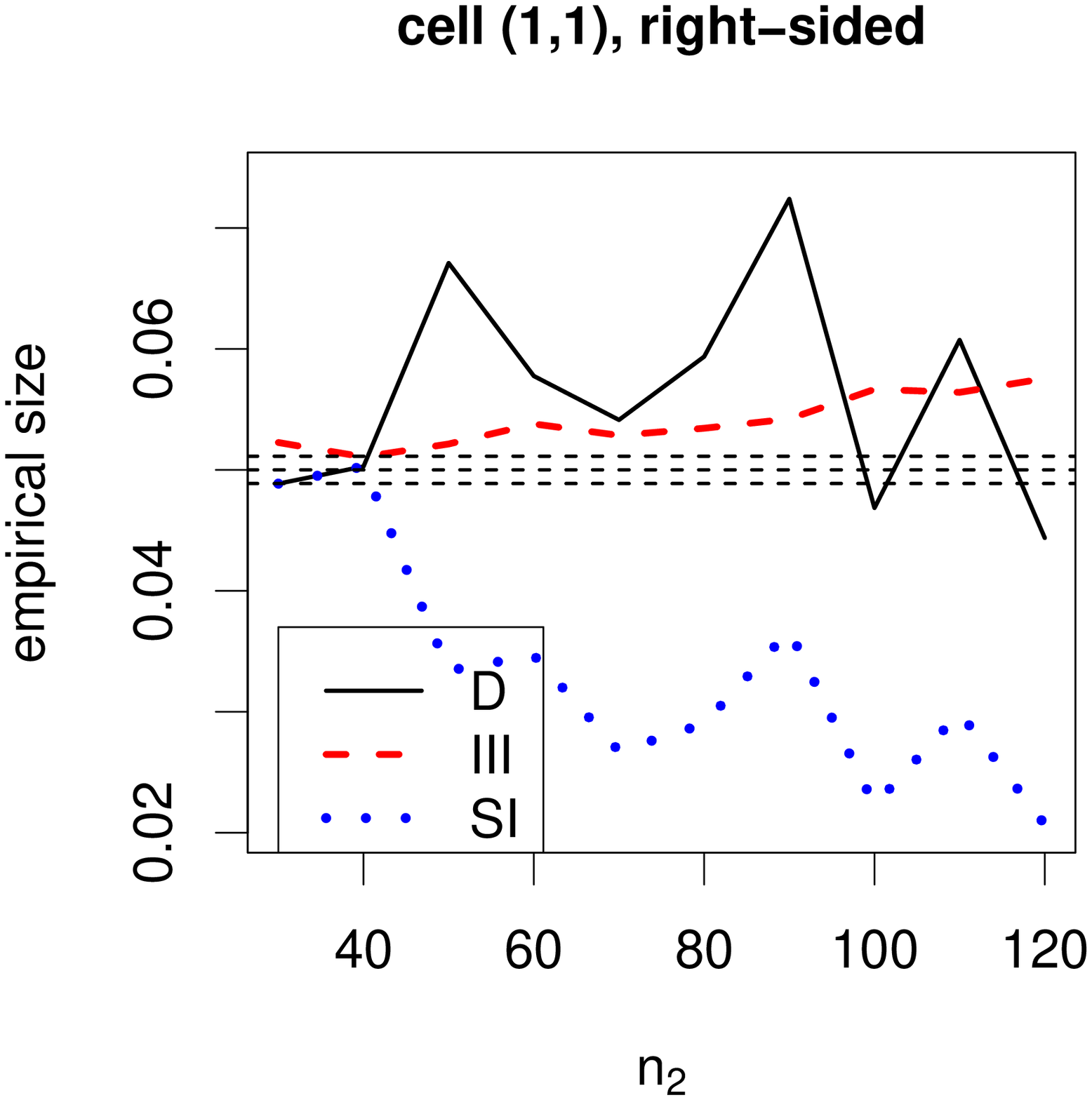} }}
\rotatebox{-90}{ \resizebox{2.1 in}{!}{\includegraphics{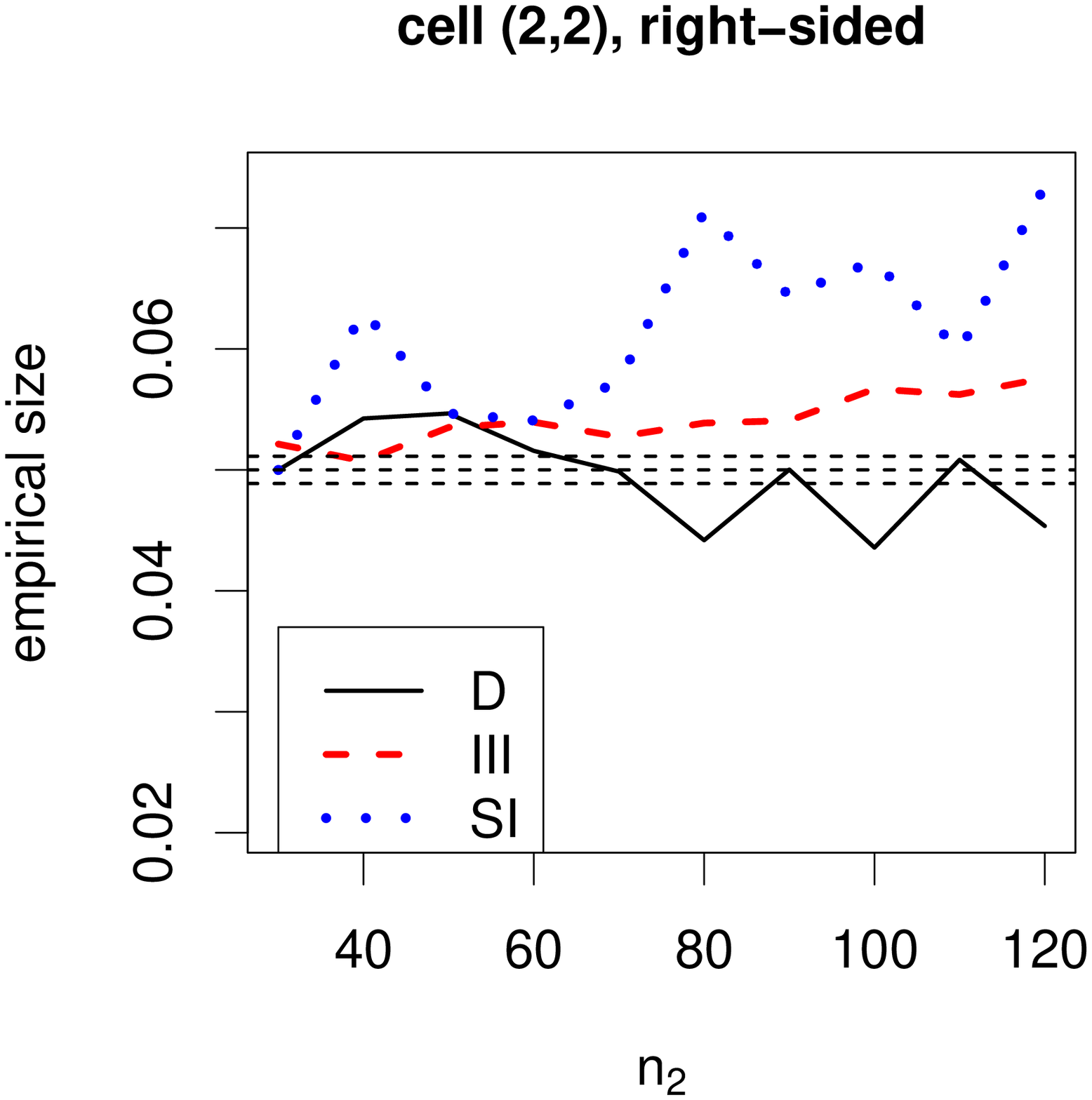} }}
\rotatebox{-90}{ \resizebox{2.1 in}{!}{\includegraphics{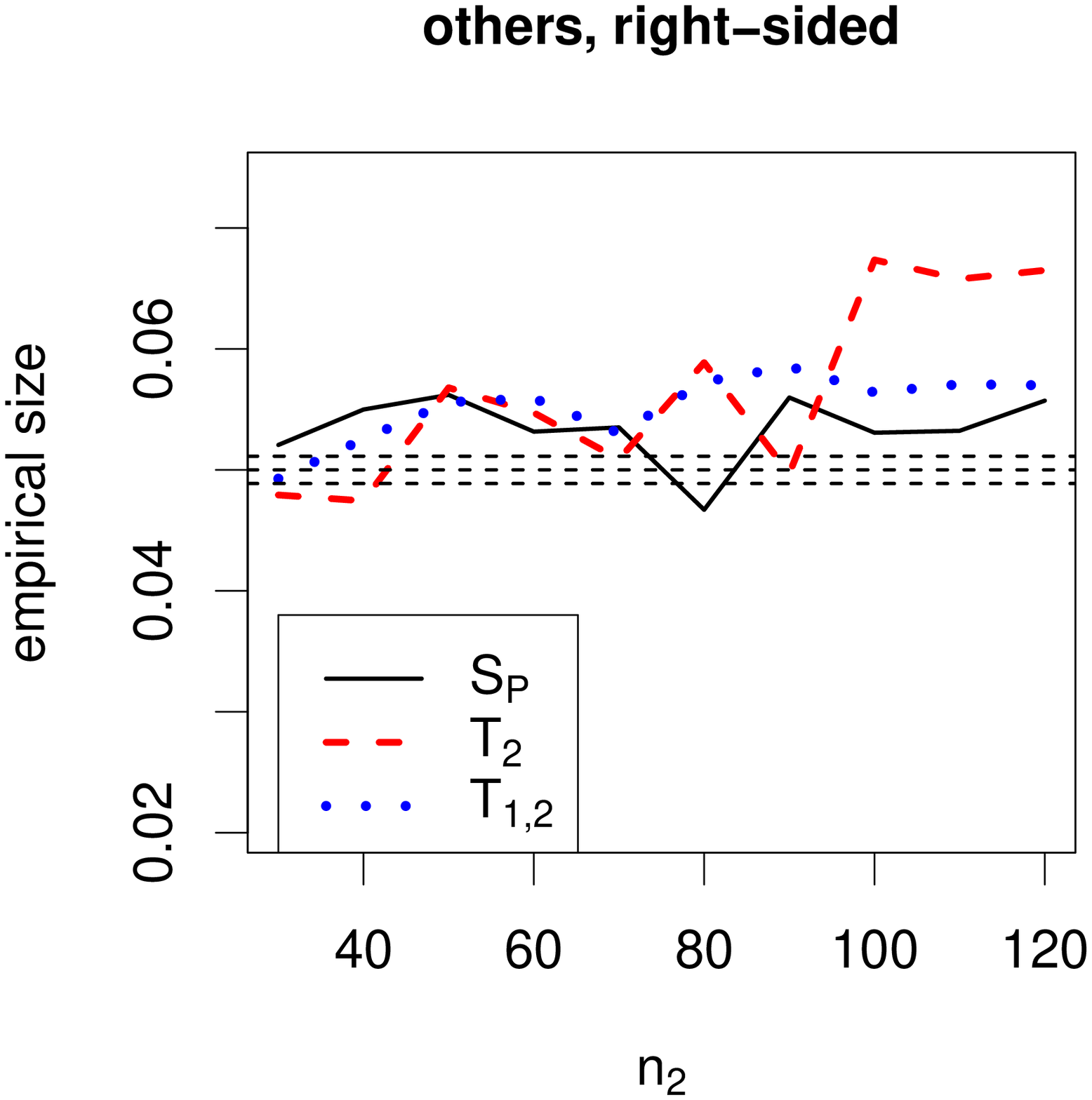} }}
\rotatebox{-90}{ \resizebox{2.1 in}{!}{\includegraphics{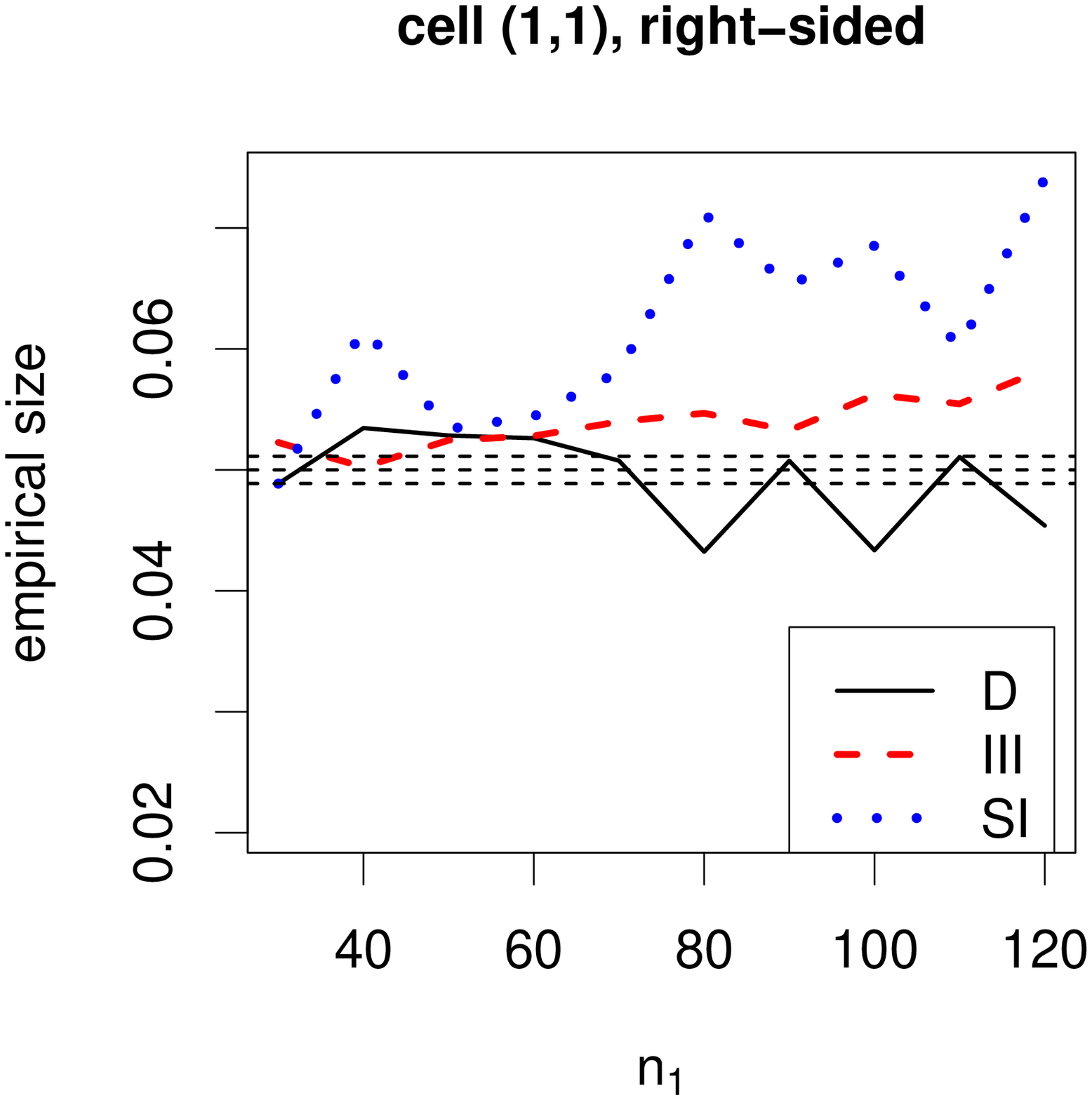} }}
\rotatebox{-90}{ \resizebox{2.1 in}{!}{\includegraphics{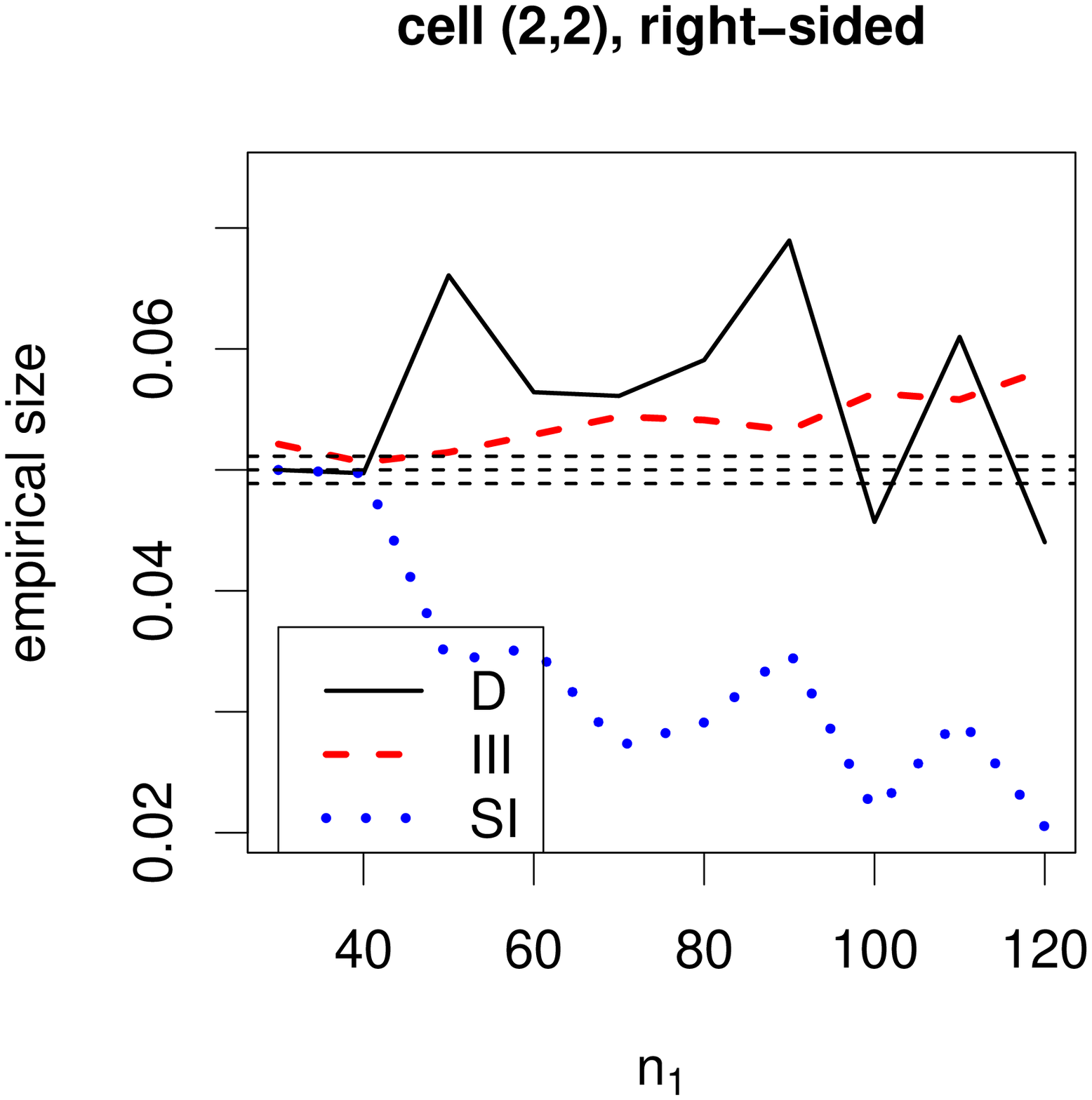} }}
\rotatebox{-90}{ \resizebox{2.1 in}{!}{\includegraphics{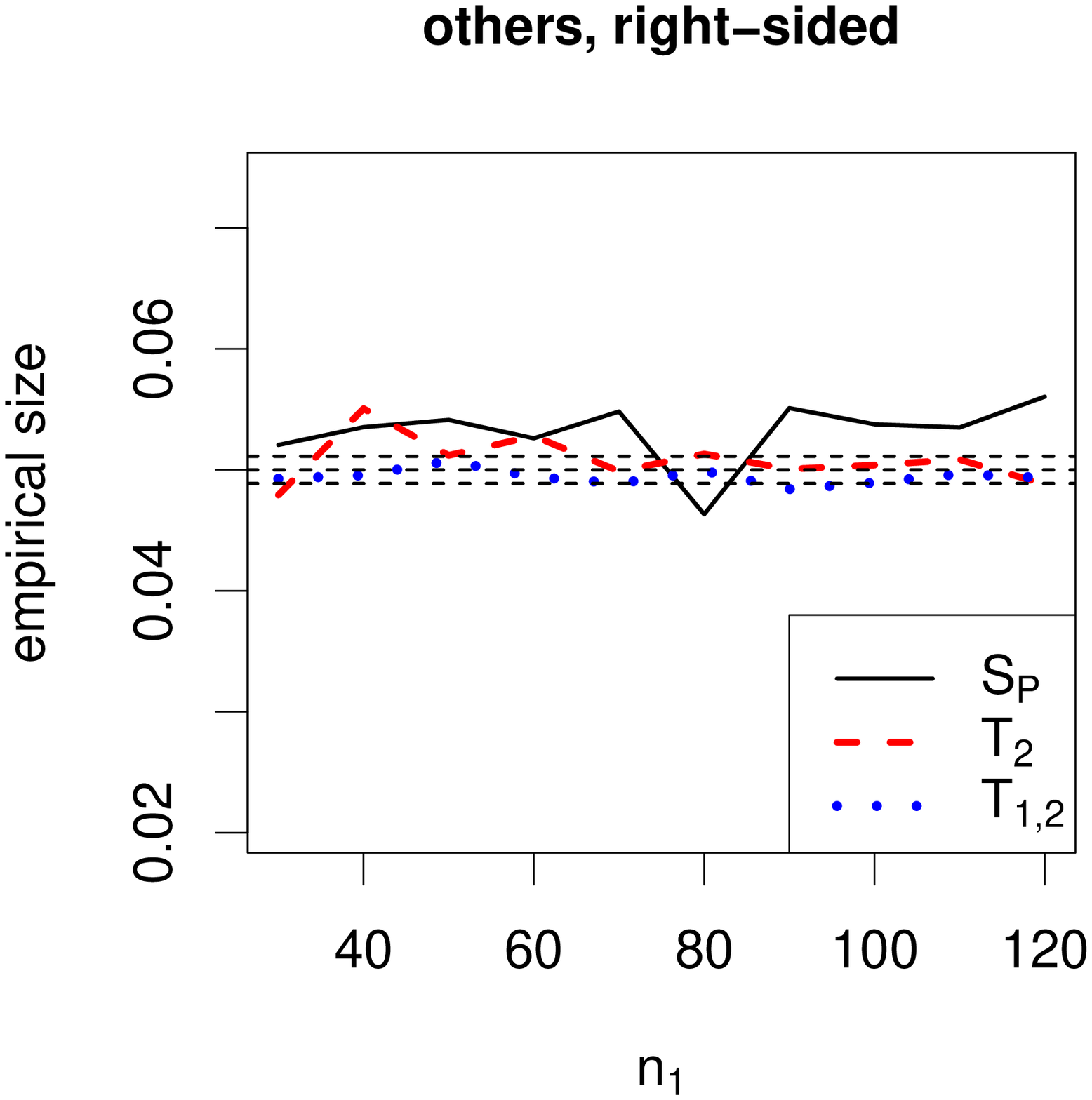} }}
\caption{
\label{fig:RS-RL-cases1a-c}
The empirical size estimates
for the tests under the RL cases 1(a)-(c)
for the right-sided alternative.
In case 1(a) (top row), we take $n_1=n_2=n/2=20,30,\ldots,100$,
in case 1(b) (middle row), we take $n_1=30$ and $n_2=30,40,\ldots,100$ and
in case 1(c) (bottom row) we take $n_1=30,40,\ldots,100$ and $n_2=30$.
In the legends,
D stands for Dixon's cell-specific tests,
III for type III cell-specific tests,
SI for Dixon's segregation indices,
$S_P$ for Pielou's coefficient of segregation,
$T_2$ for Cuzick-Edwards' $2$-NN test,
and
$T_{1,2}$ for Cuzick-Edwards' combined test, $T_S$, for $S=\{1,2\}$
The dashed horizontal lines are at .04887 and .05113,
the lower and upper bounds for significant deviation from .05.
Also, empirical size estimates for each test are joined by straight lines
for better visualization.
}
\end{figure}

\begin{figure} [hbp]
\centering
\rotatebox{-90}{ \resizebox{2.1 in}{!}{\includegraphics{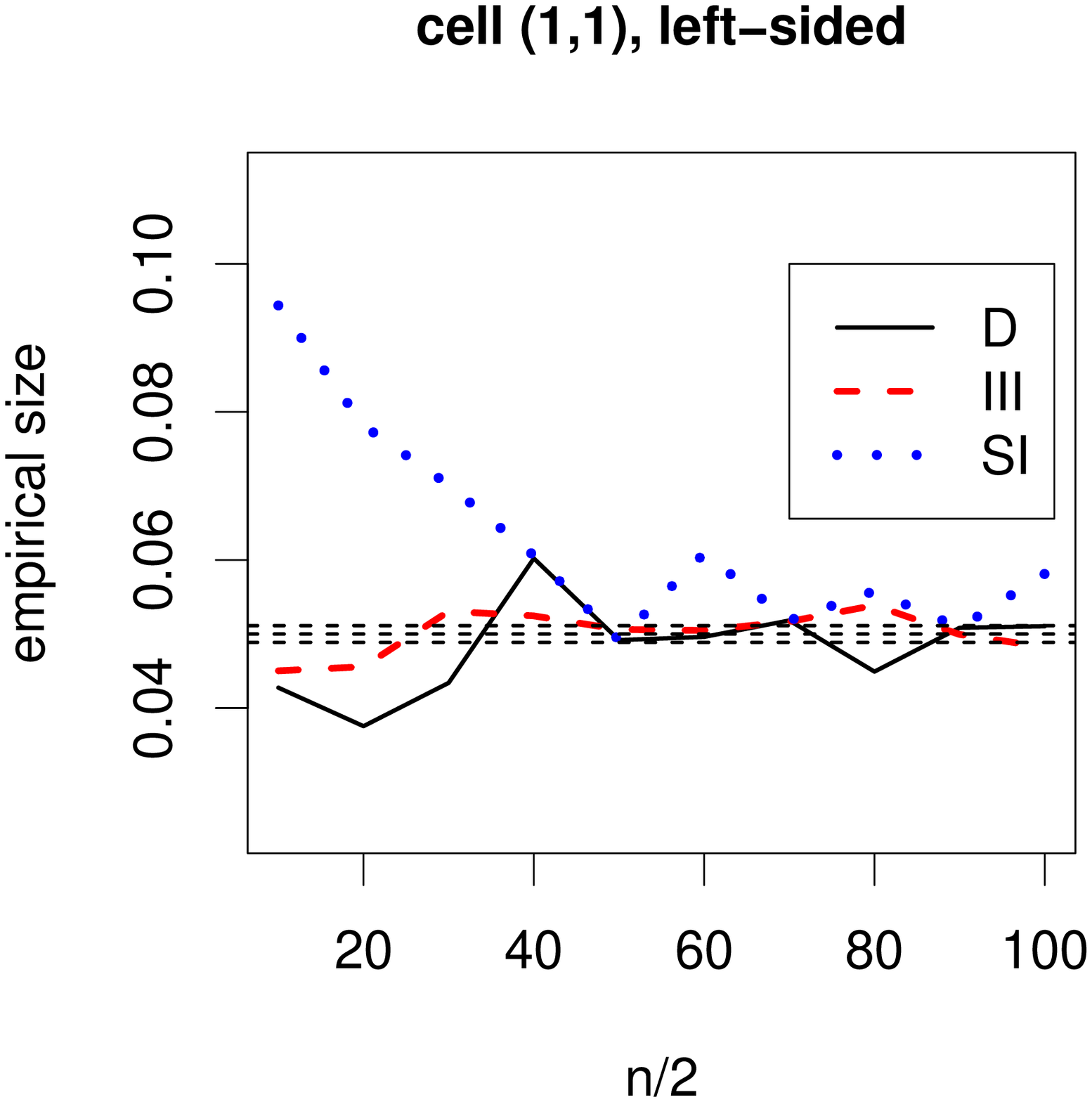} }}
\rotatebox{-90}{ \resizebox{2.1 in}{!}{\includegraphics{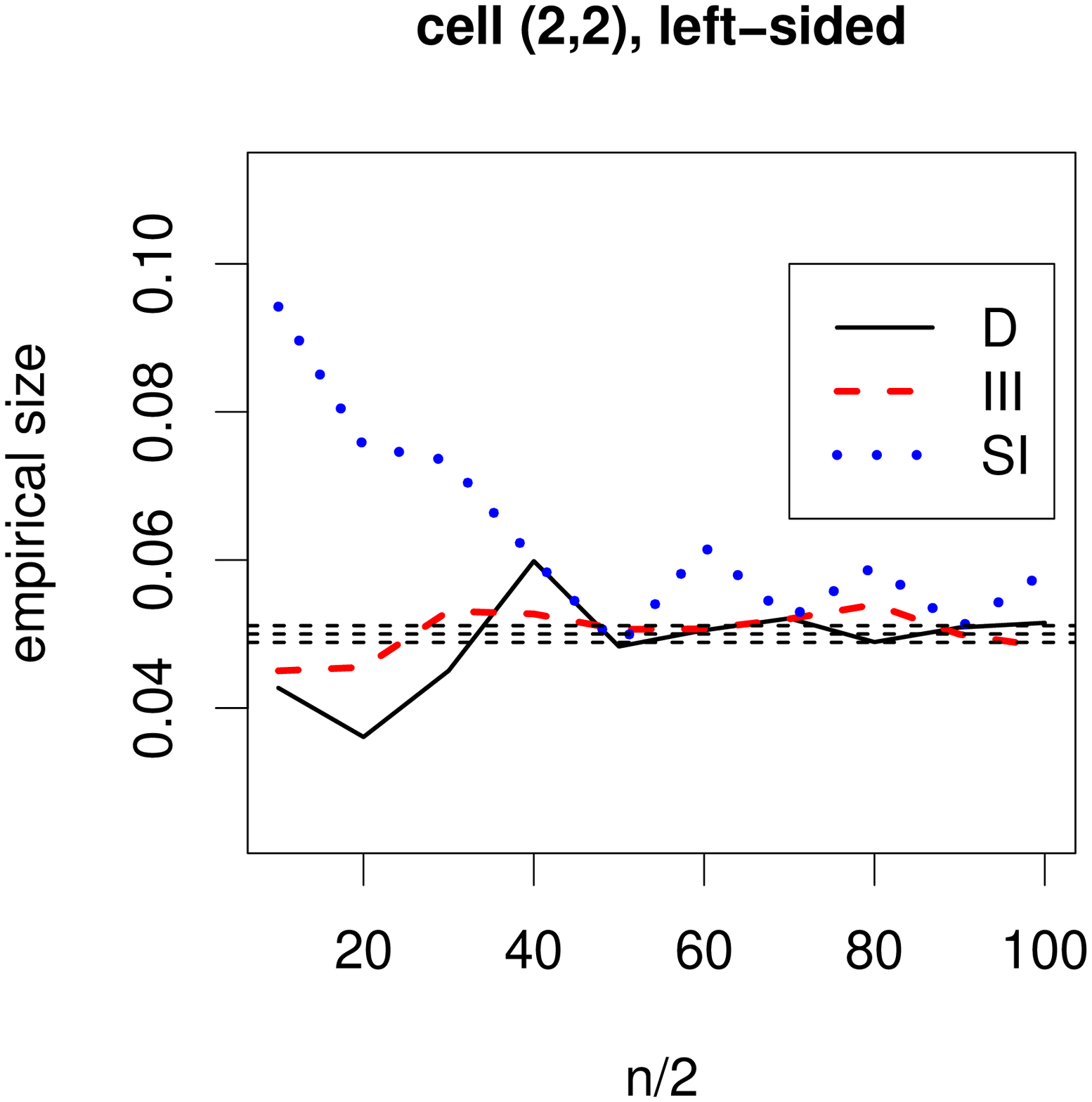} }}
\rotatebox{-90}{ \resizebox{2.1 in}{!}{\includegraphics{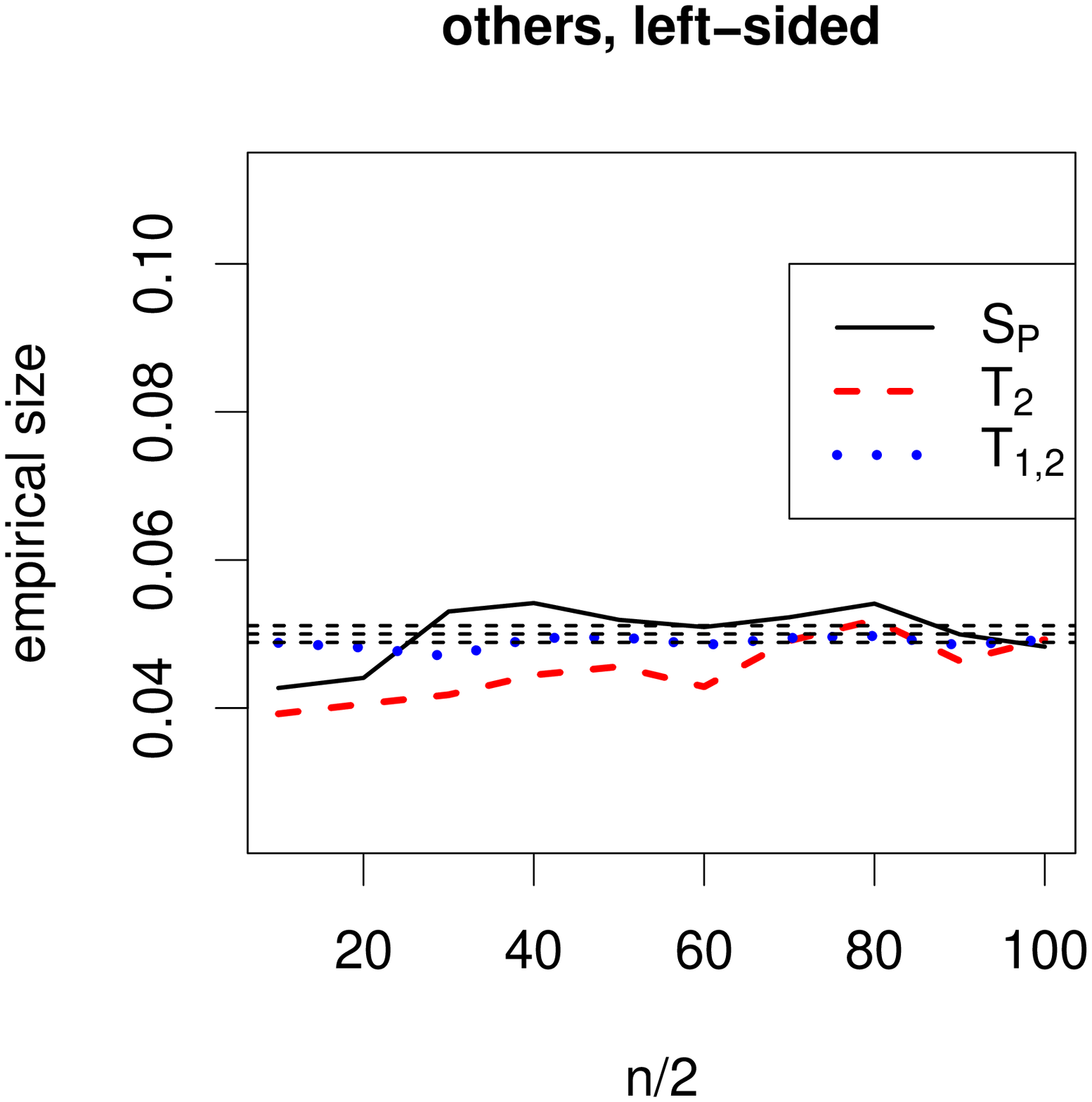} }}
\rotatebox{-90}{ \resizebox{2.1 in}{!}{\includegraphics{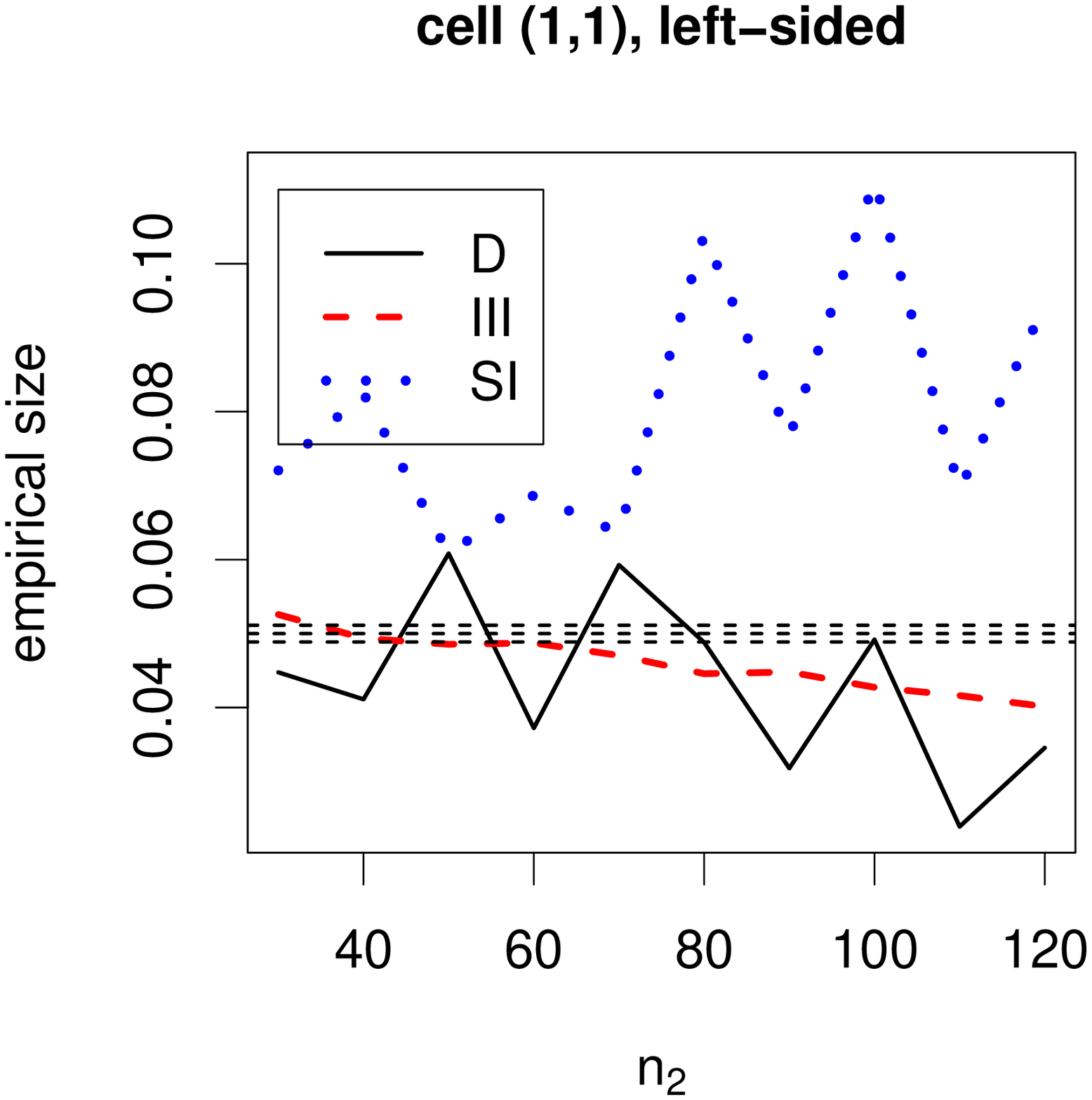} }}
\rotatebox{-90}{ \resizebox{2.1 in}{!}{\includegraphics{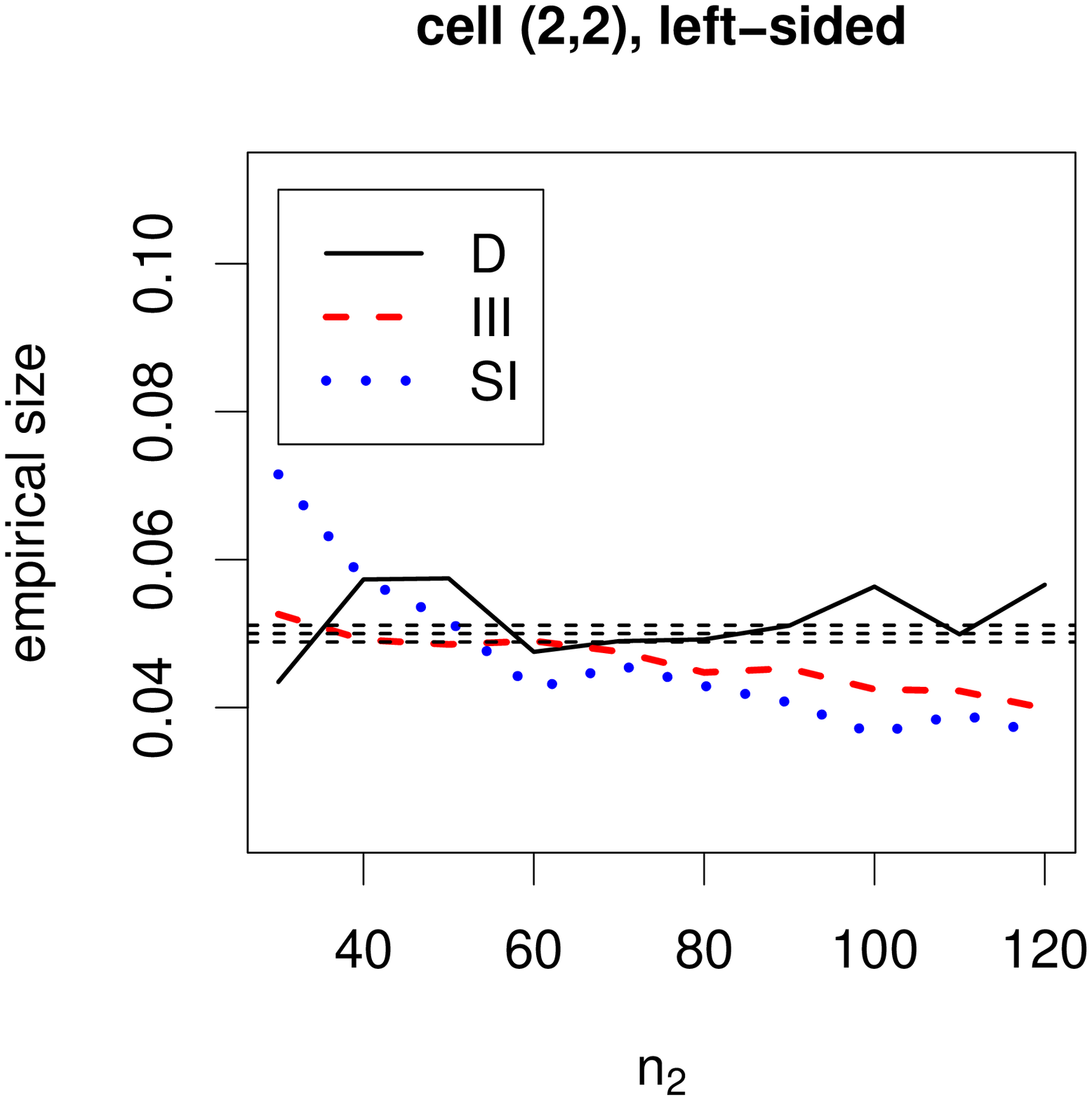} }}
\rotatebox{-90}{ \resizebox{2.1 in}{!}{\includegraphics{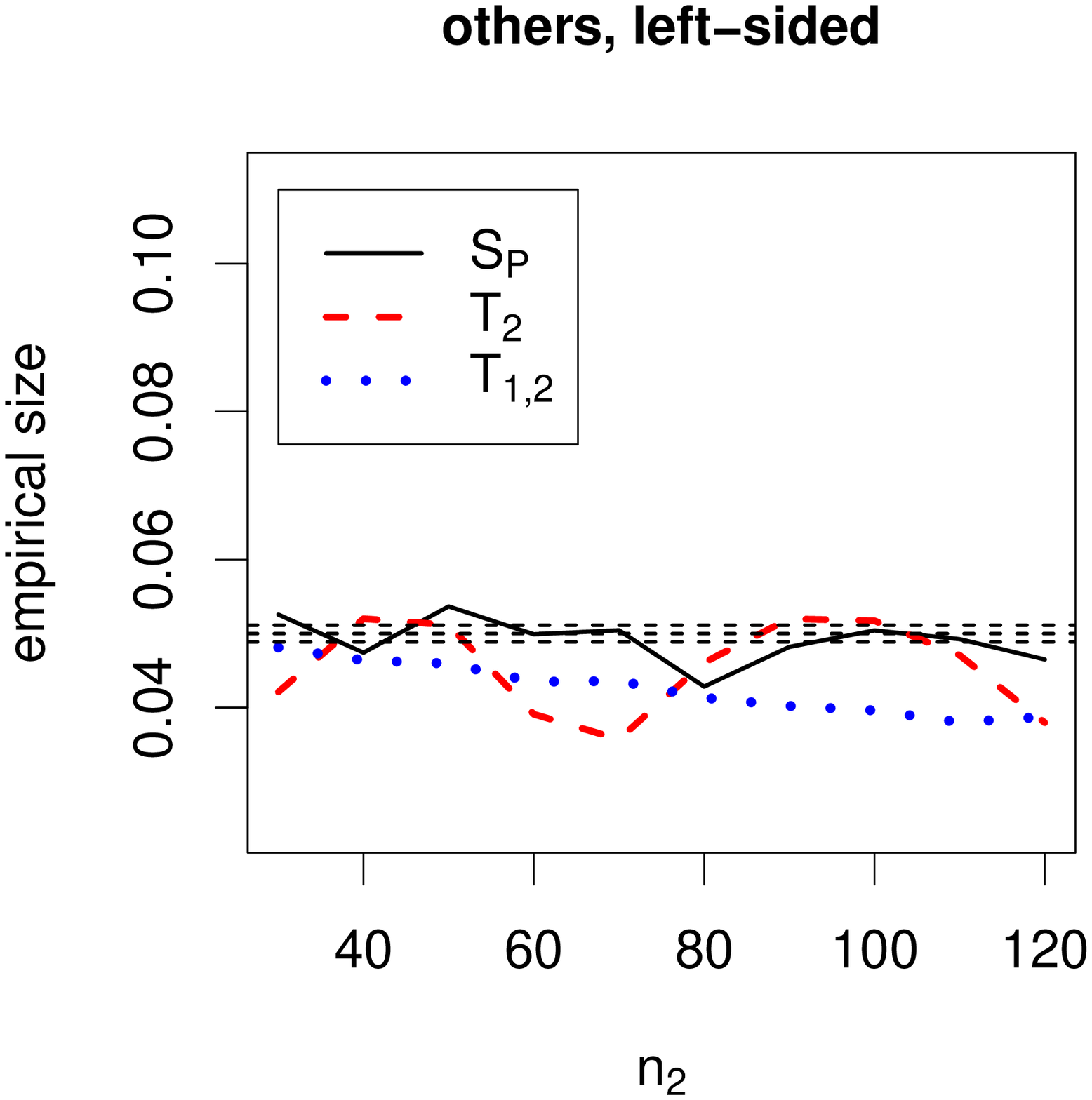} }}
\rotatebox{-90}{ \resizebox{2.1 in}{!}{\includegraphics{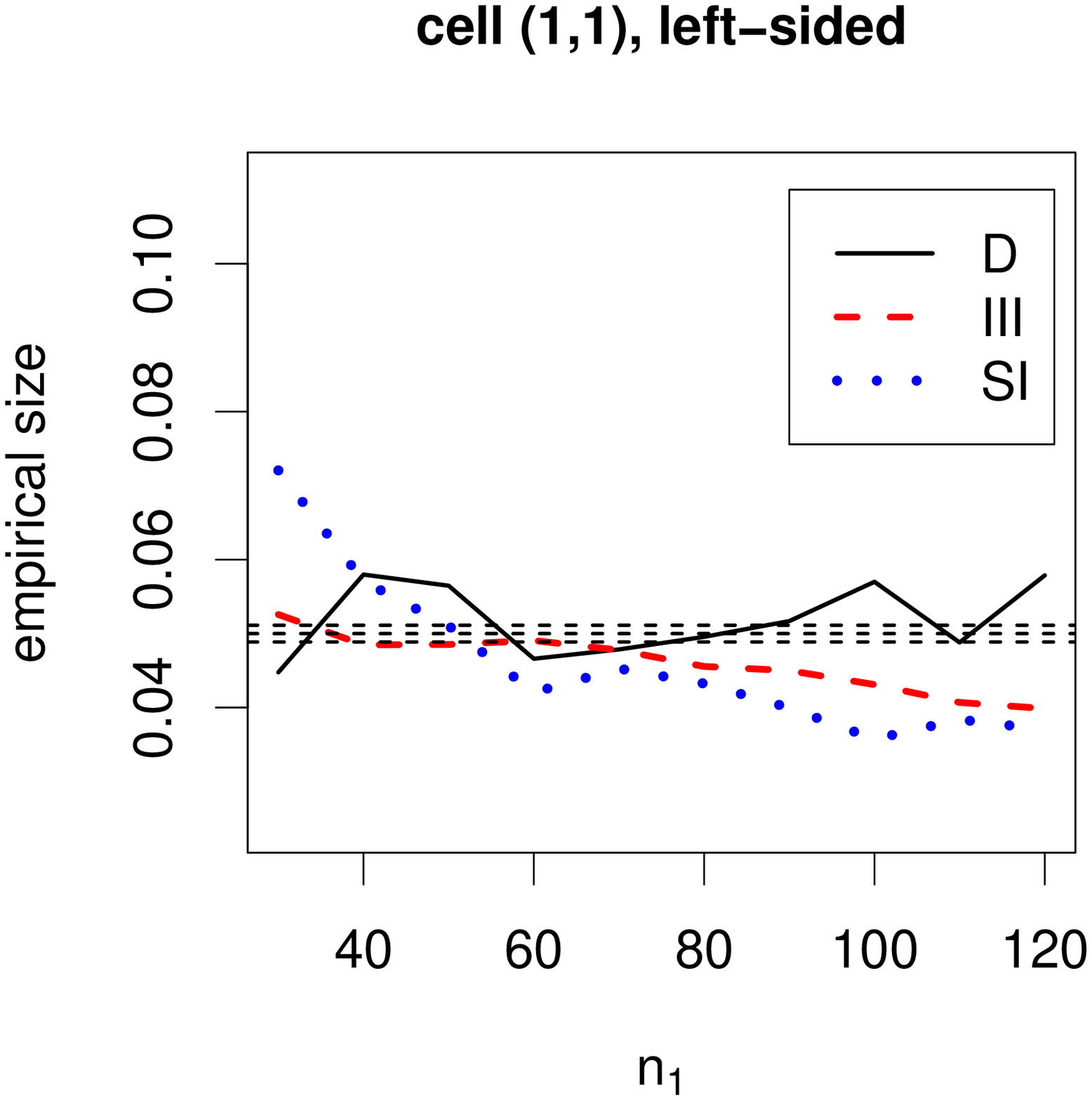} }}
\rotatebox{-90}{ \resizebox{2.1 in}{!}{\includegraphics{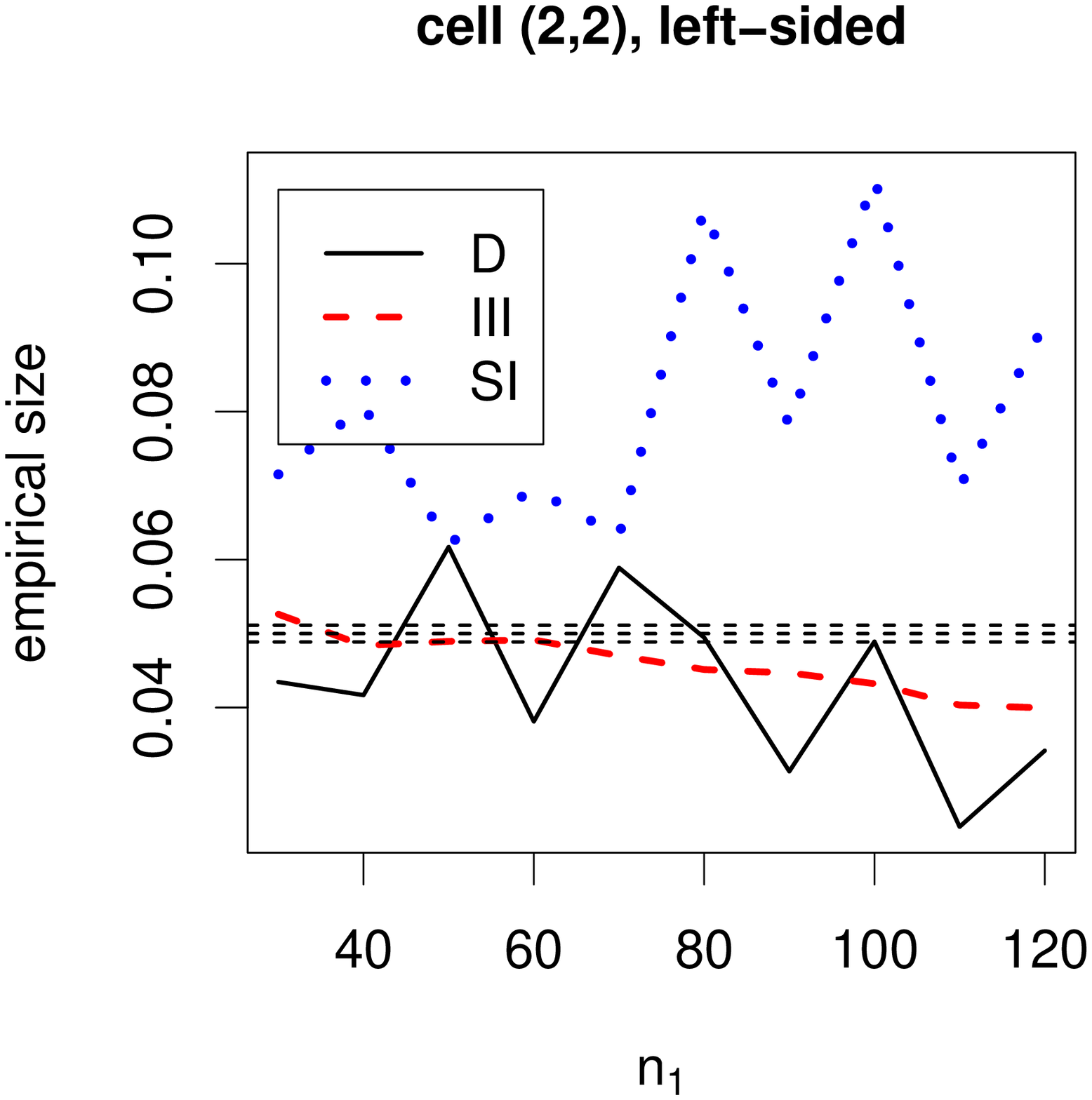} }}
\rotatebox{-90}{ \resizebox{2.1 in}{!}{\includegraphics{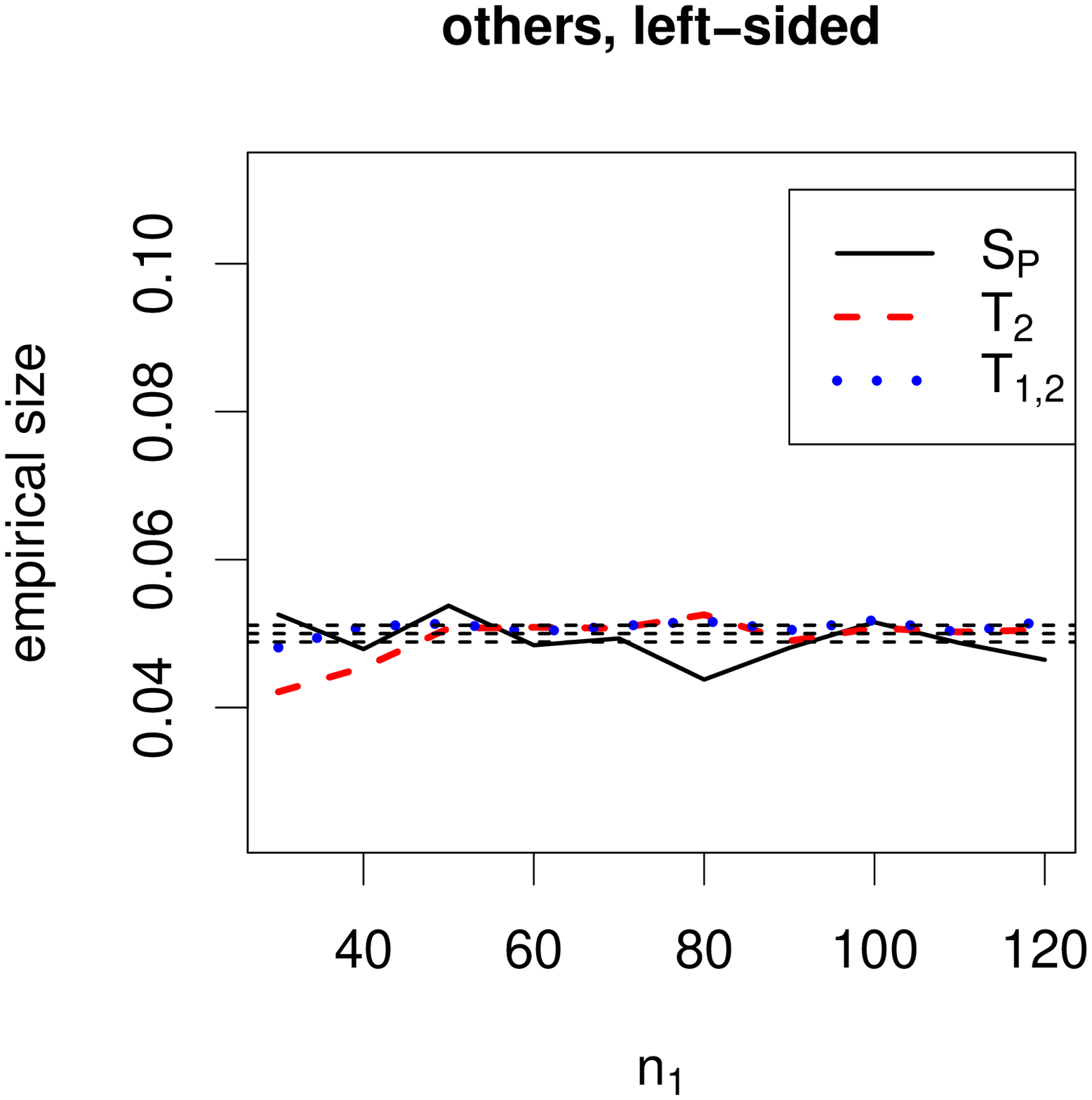} }}
\caption{
\label{fig:LS-RL-cases1a-c}
The empirical size estimates
for the tests under the RL cases 1(a) (top row), 1(b) (middle row) and 1(c) (bottom row)
for the left-sided alternative.
The horizontal lines and legend labeling are as in Figure \ref{fig:RS-RL-cases1a-c}.
}
\end{figure}

\begin{figure} [hbp]
\centering
\rotatebox{-90}{ \resizebox{2.1 in}{!}{\includegraphics{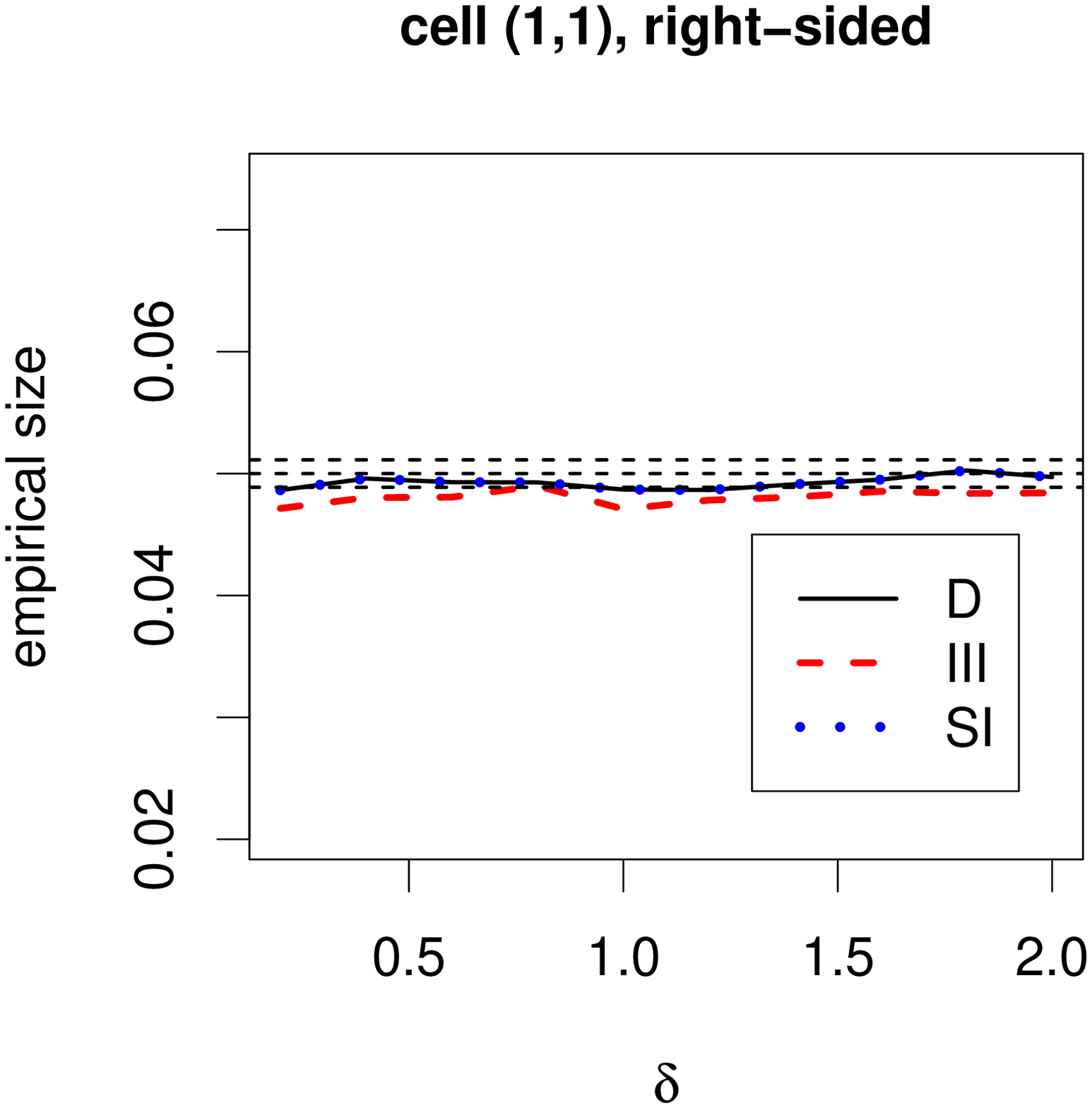} }}
\rotatebox{-90}{ \resizebox{2.1 in}{!}{\includegraphics{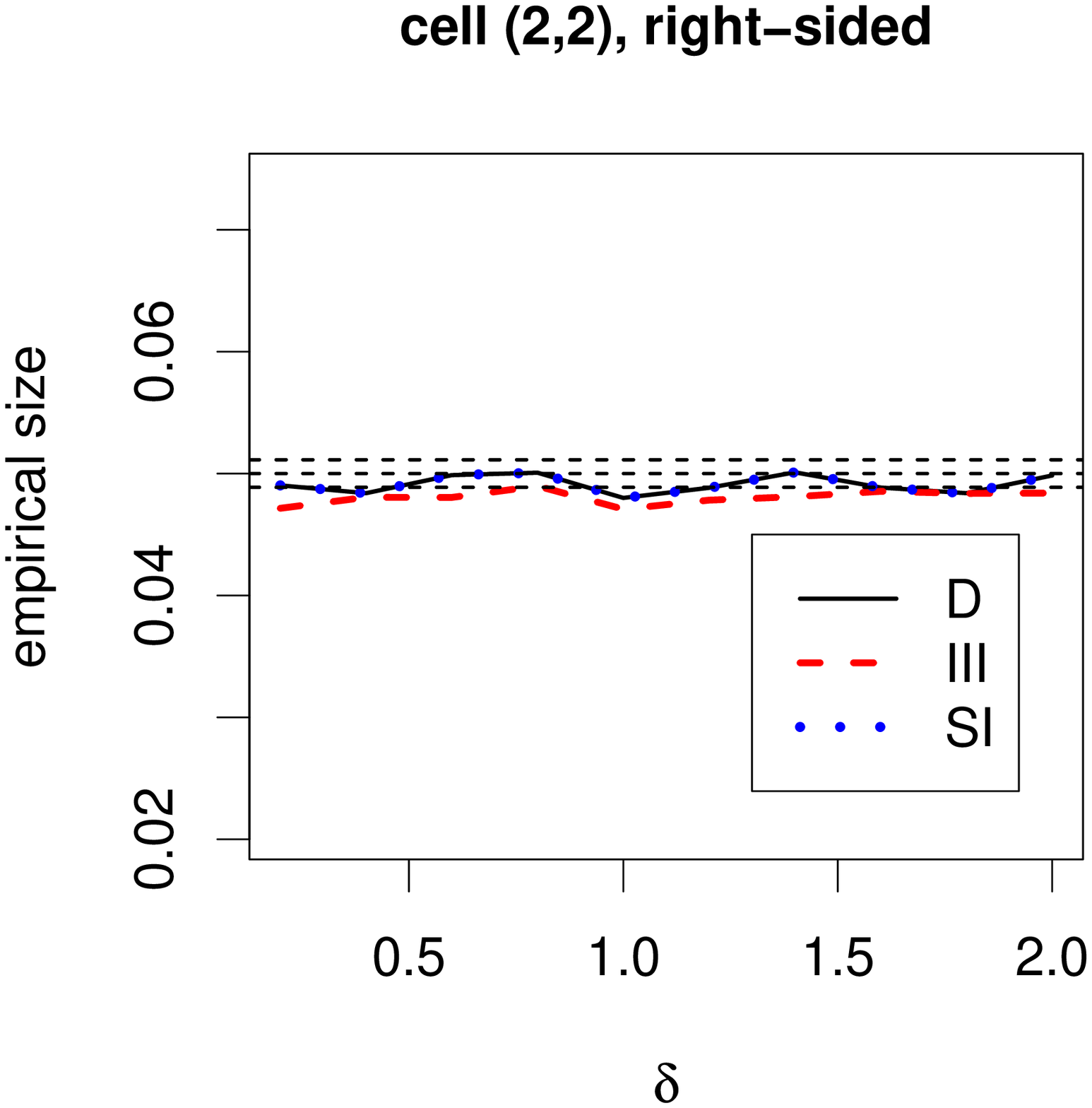} }}
\rotatebox{-90}{ \resizebox{2.1 in}{!}{\includegraphics{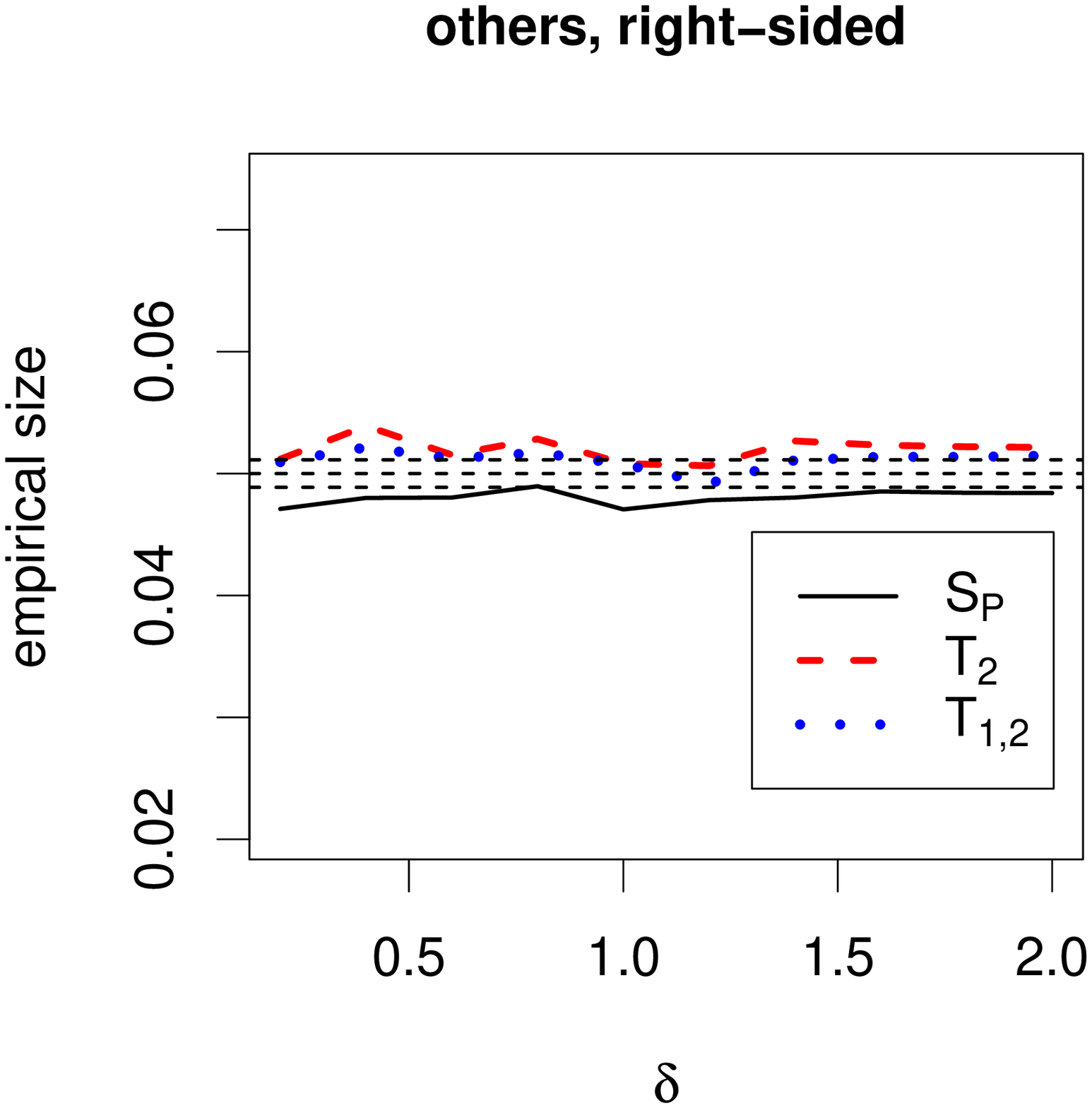} }}
\rotatebox{-90}{ \resizebox{2.1 in}{!}{\includegraphics{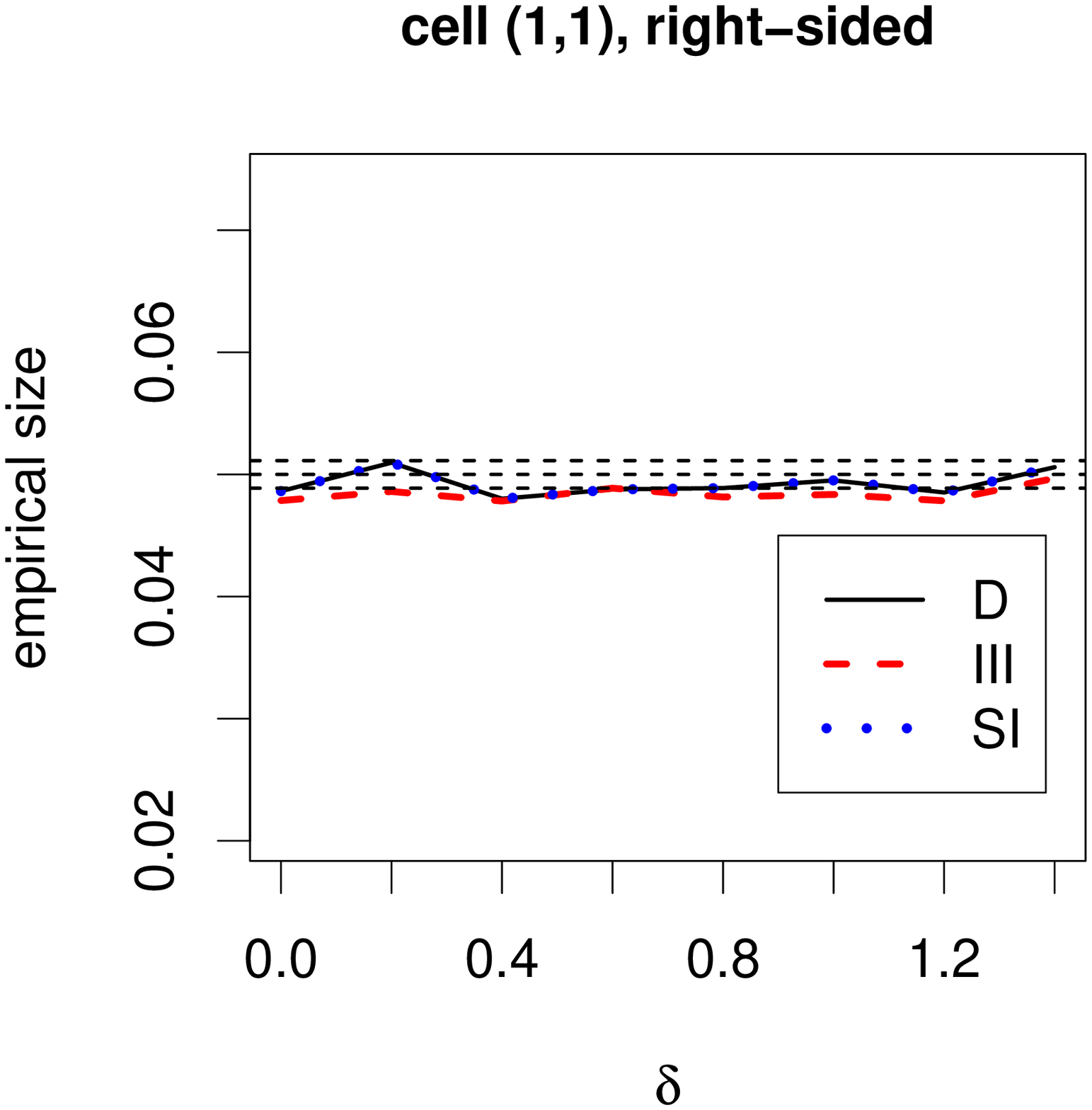} }}
\rotatebox{-90}{ \resizebox{2.1 in}{!}{\includegraphics{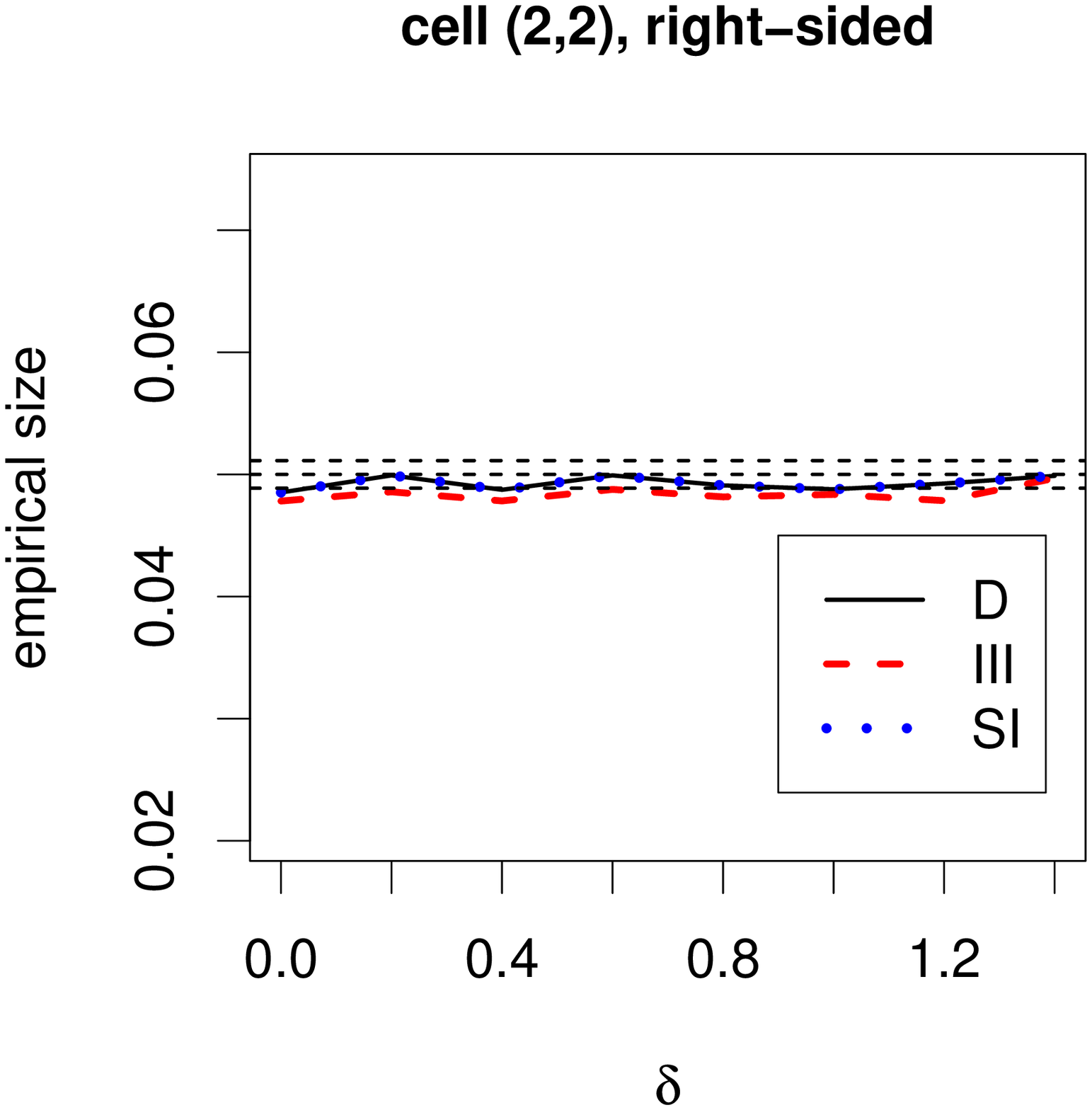} }}
\rotatebox{-90}{ \resizebox{2.1 in}{!}{\includegraphics{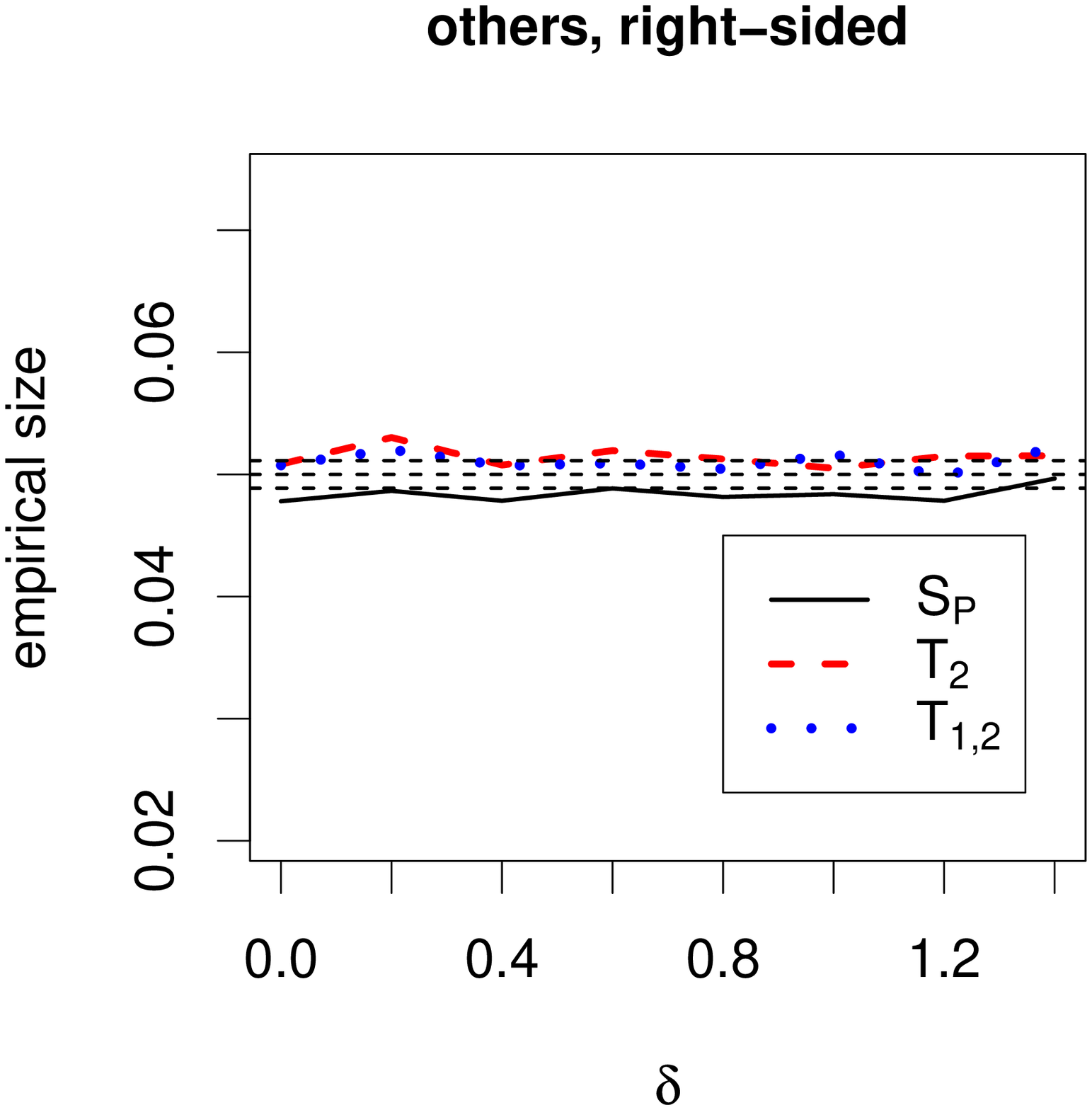} }}
\rotatebox{-90}{ \resizebox{2.1 in}{!}{\includegraphics{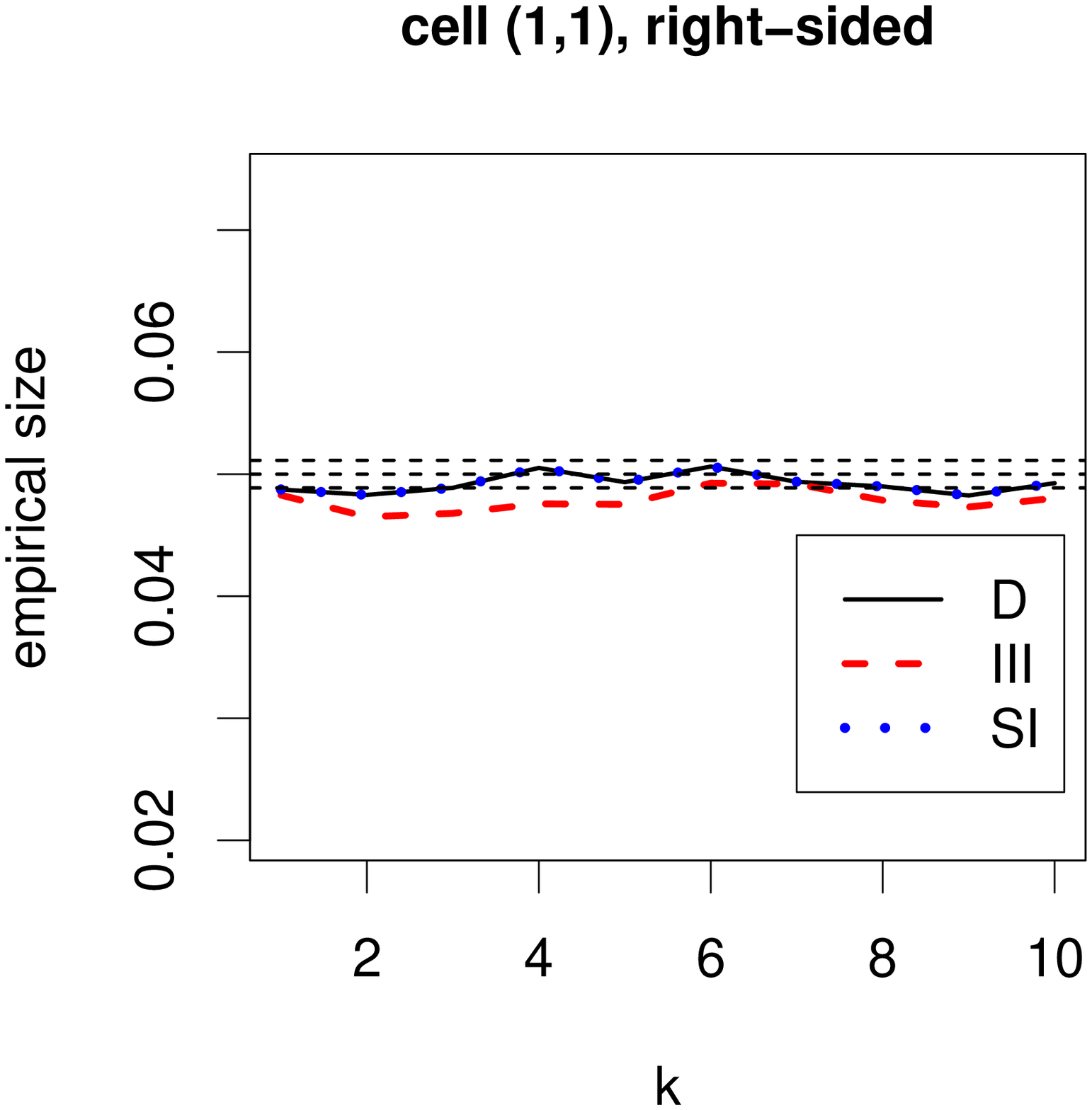} }}
\rotatebox{-90}{ \resizebox{2.1 in}{!}{\includegraphics{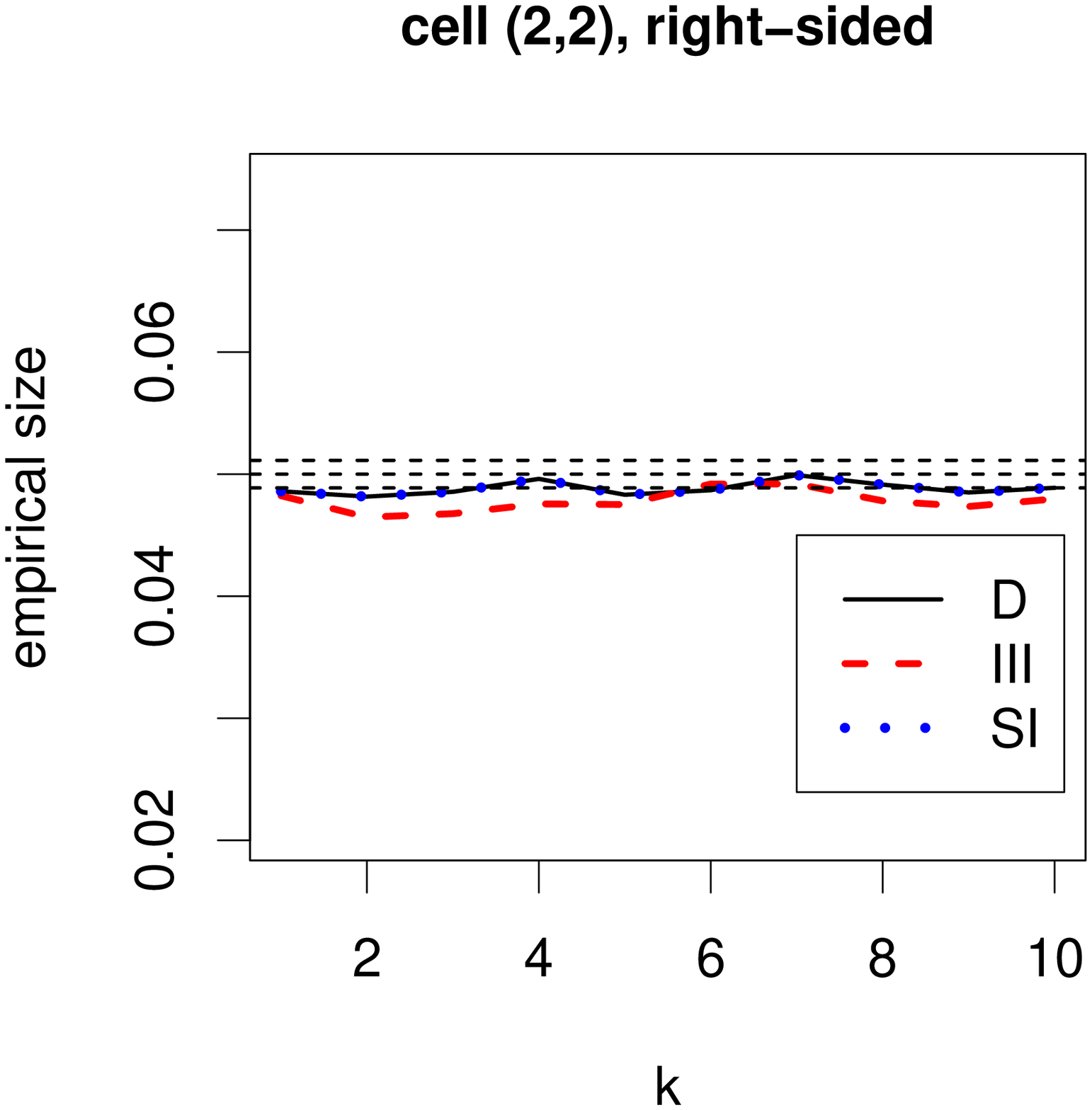} }}
\rotatebox{-90}{ \resizebox{2.1 in}{!}{\includegraphics{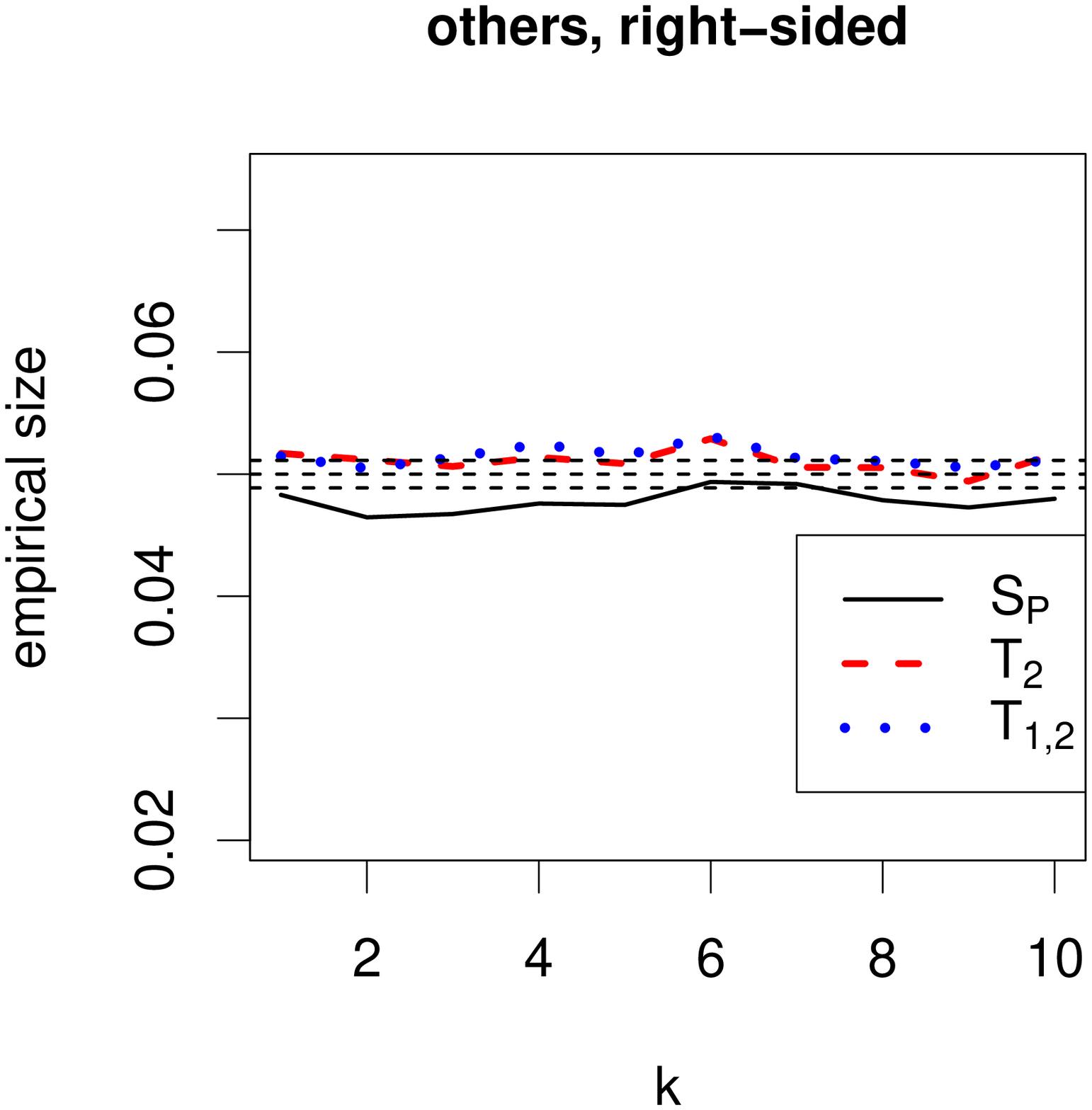} }}
\caption{
\label{fig:RS-RL-cases2-4}
The empirical size estimates of the tests for the right-sided alternative
under the RL cases 2-4 with $n_1=n_2=100$.
In case 2 (top row), we take $\delta=0.2,0.4,\ldots,1.4$,
in case 3 (middle row), we take $\delta=0.2,0.4,\ldots,1.0$ and
in case 4 (bottom row) we take $\delta=0.5$ and $k=1,2,\ldots,5$.
The dashed horizontal lines and legend labeling are as in Figure \ref{fig:RS-RL-cases1a-c}.
}
\end{figure}

\begin{figure} [hbp]
\centering
\rotatebox{-90}{ \resizebox{2.1 in}{!}{\includegraphics{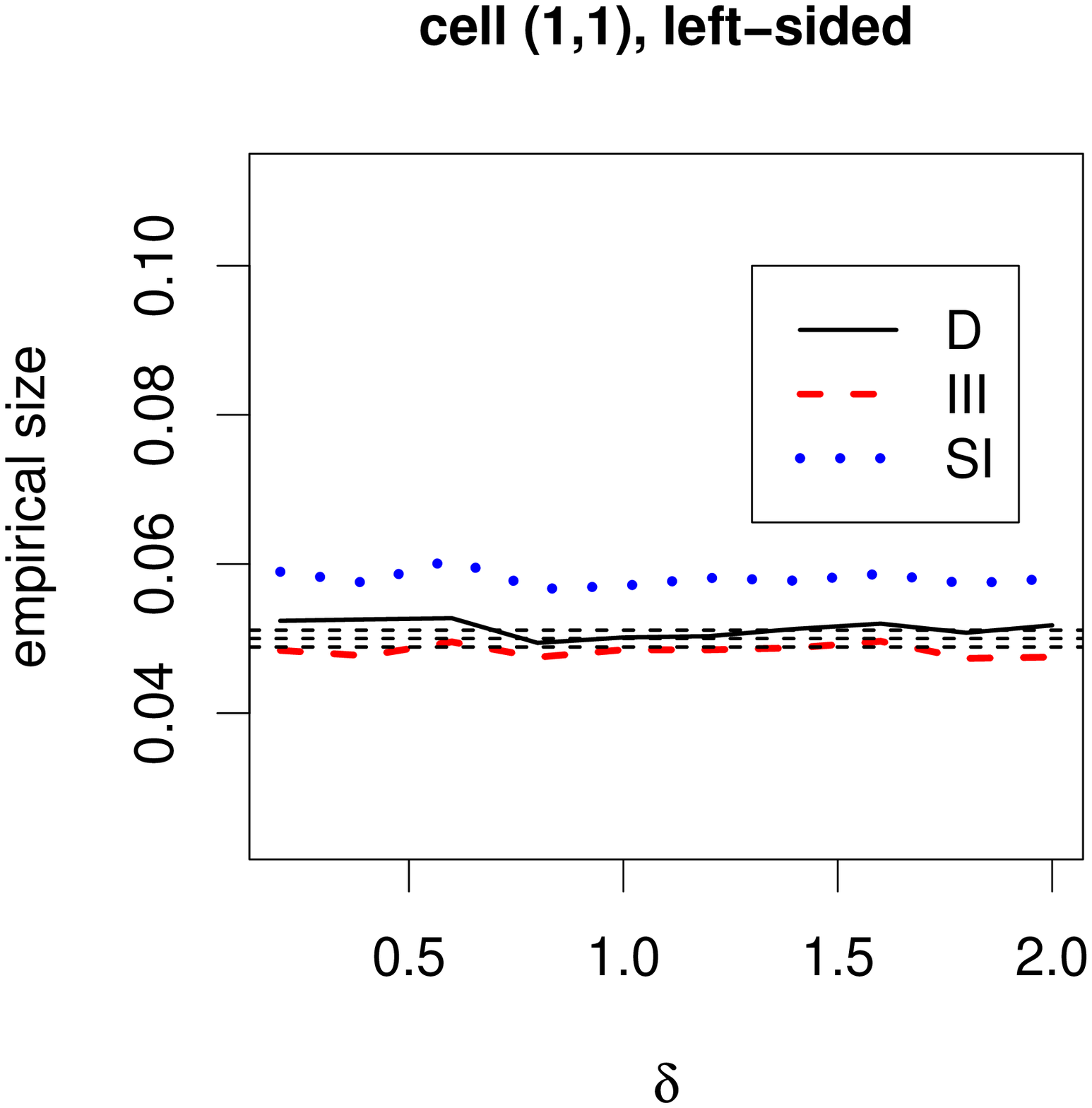} }}
\rotatebox{-90}{ \resizebox{2.1 in}{!}{\includegraphics{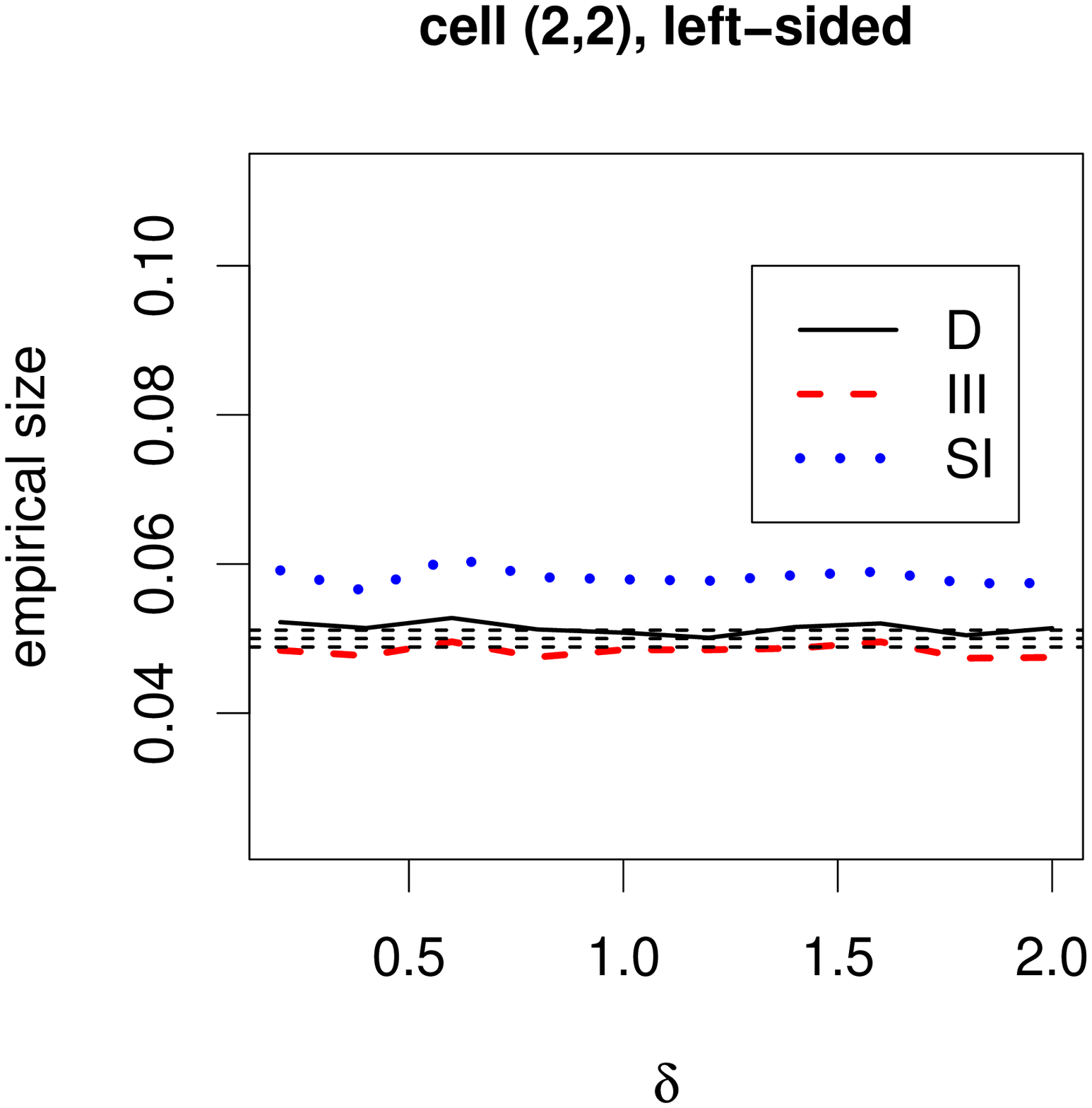} }}
\rotatebox{-90}{ \resizebox{2.1 in}{!}{\includegraphics{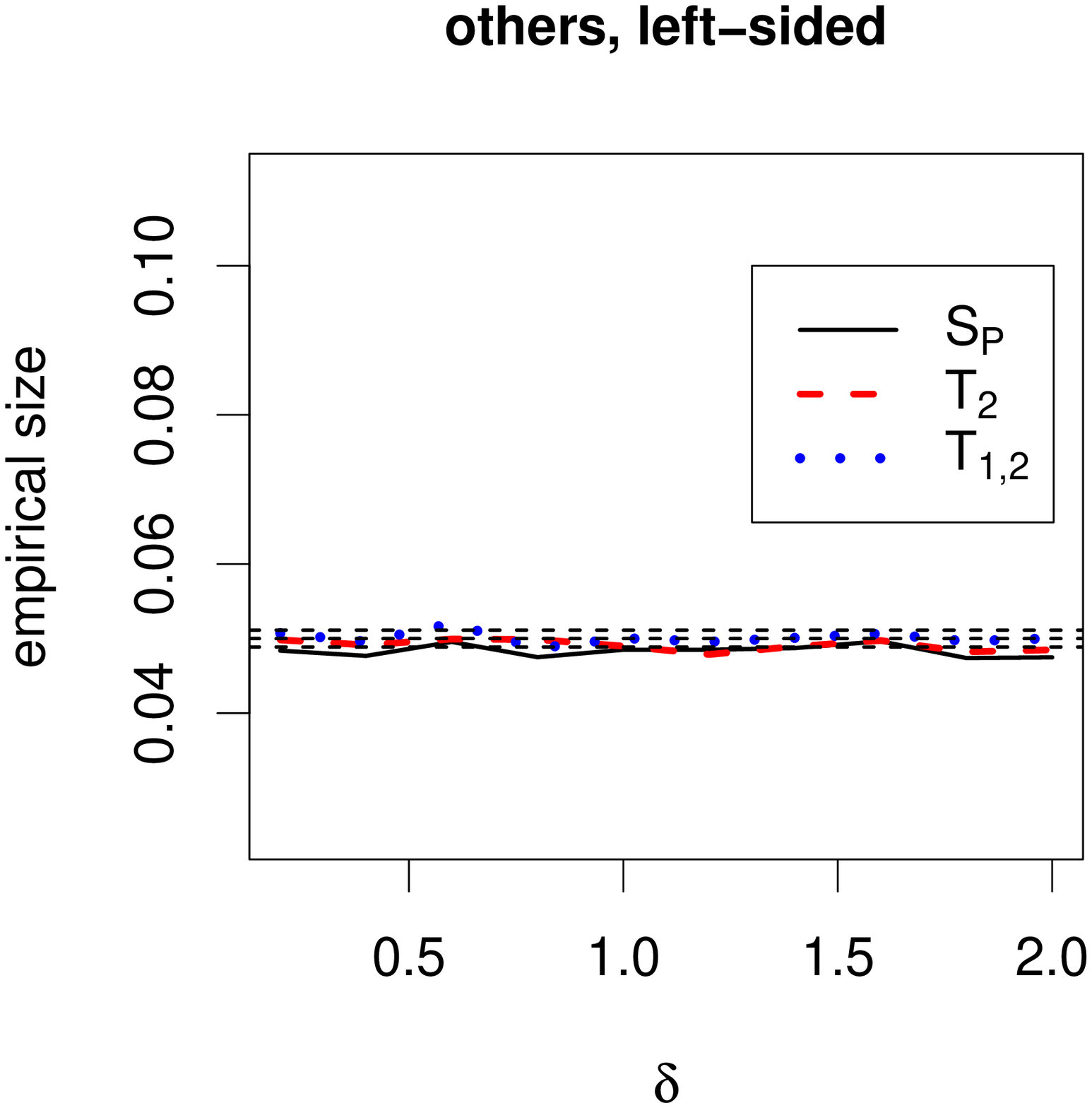} }}
\rotatebox{-90}{ \resizebox{2.1 in}{!}{\includegraphics{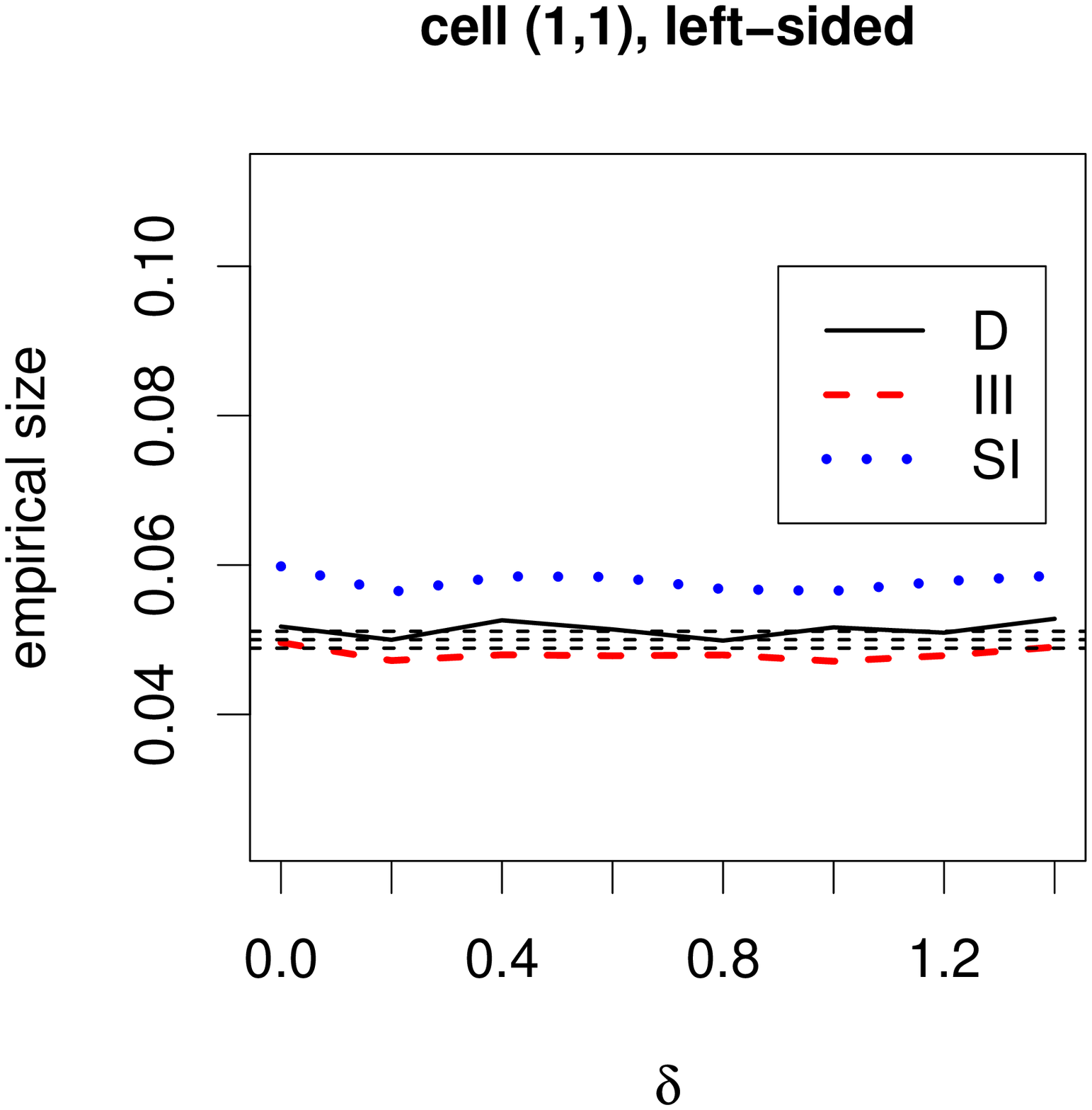} }}
\rotatebox{-90}{ \resizebox{2.1 in}{!}{\includegraphics{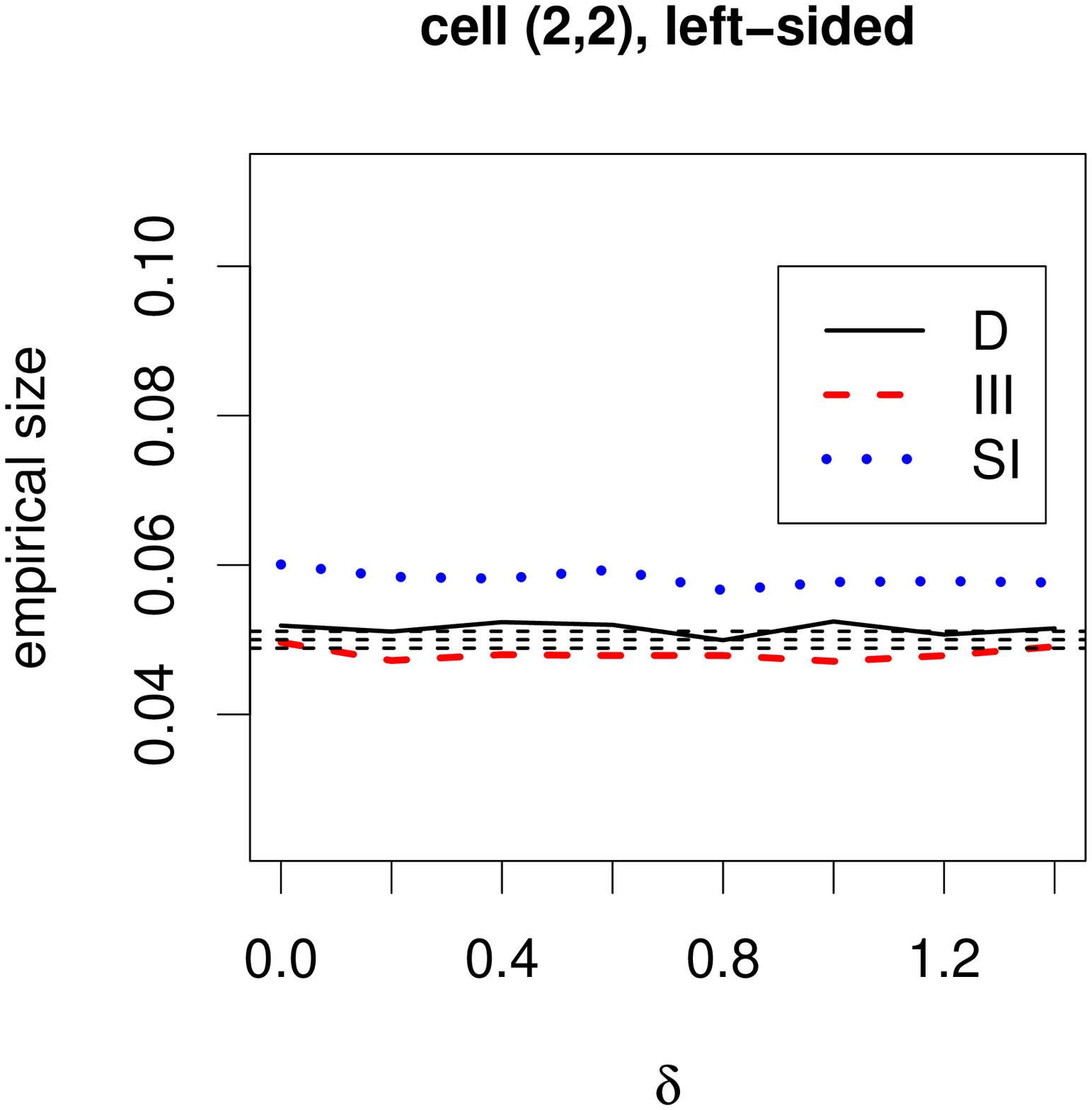} }}
\rotatebox{-90}{ \resizebox{2.1 in}{!}{\includegraphics{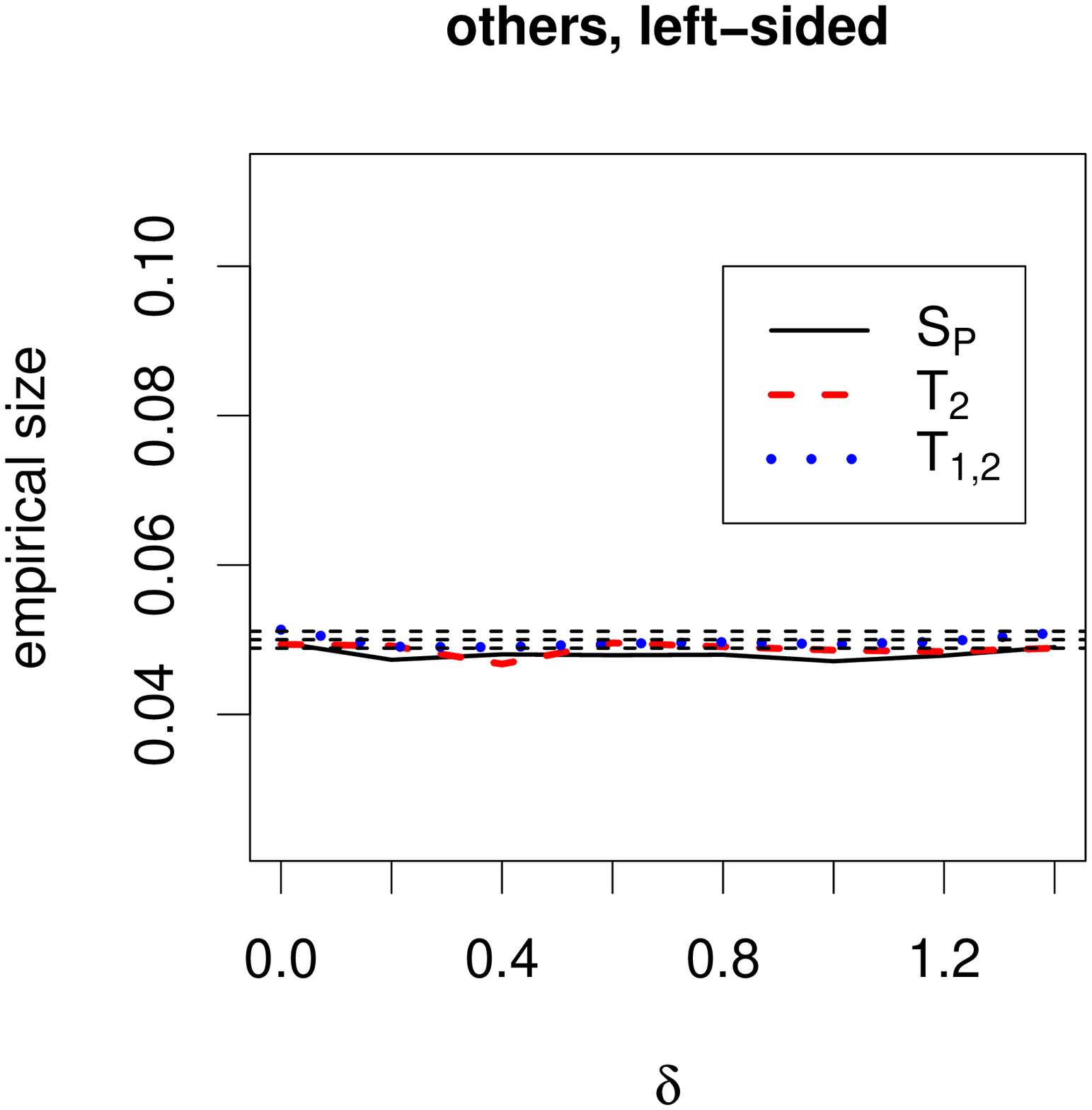} }}
\rotatebox{-90}{ \resizebox{2.1 in}{!}{\includegraphics{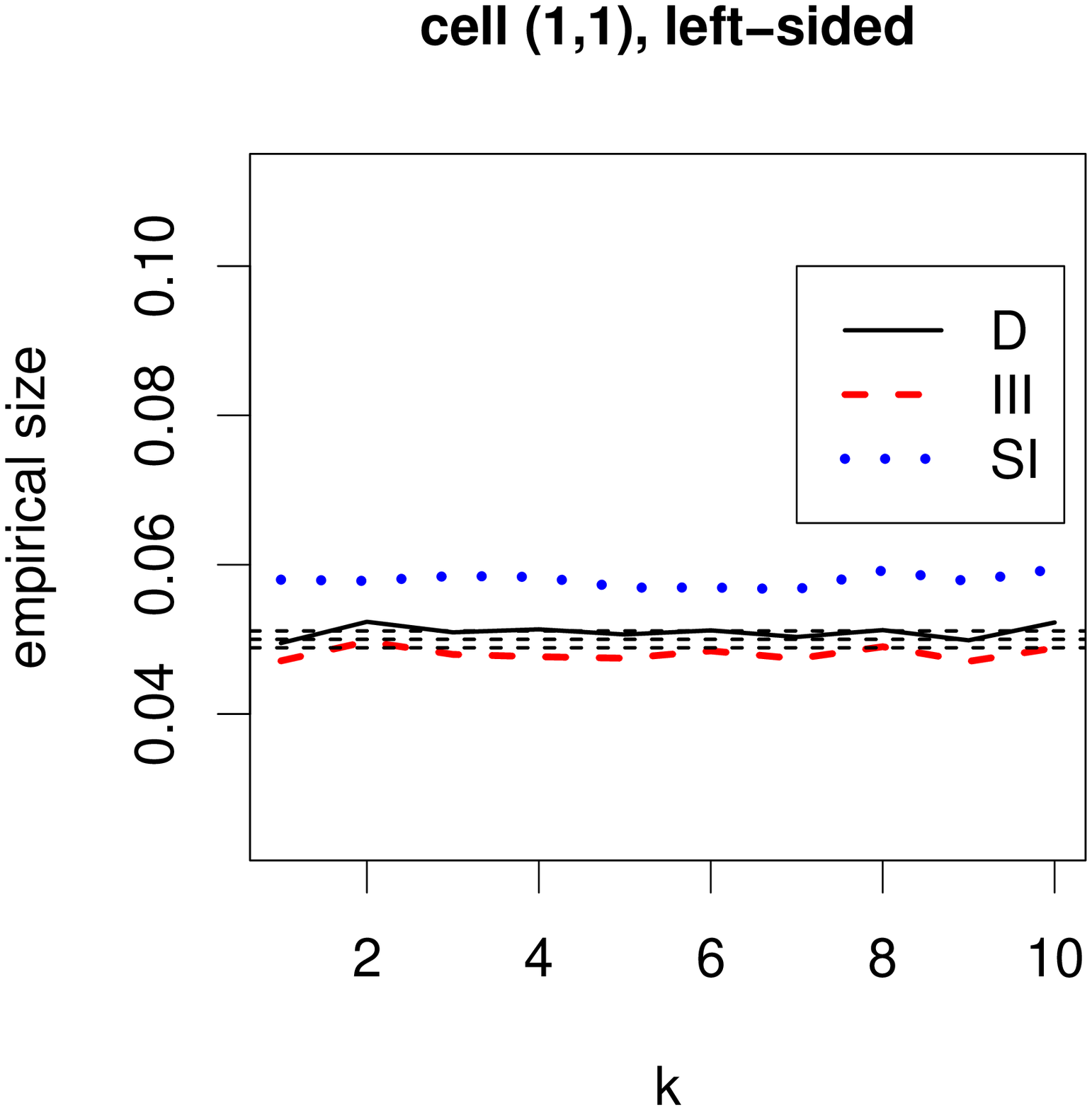} }}
\rotatebox{-90}{ \resizebox{2.1 in}{!}{\includegraphics{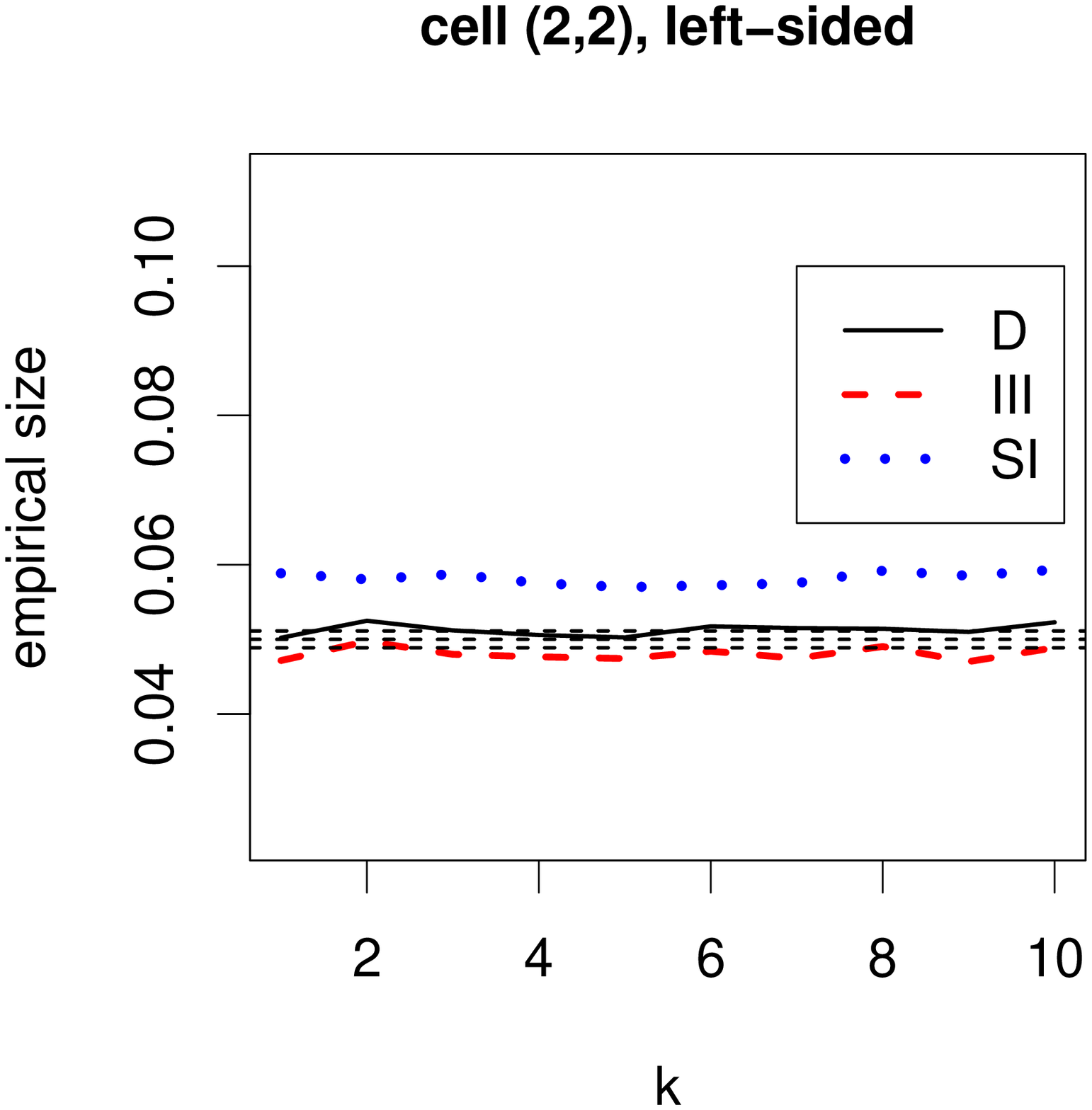} }}
\rotatebox{-90}{ \resizebox{2.1 in}{!}{\includegraphics{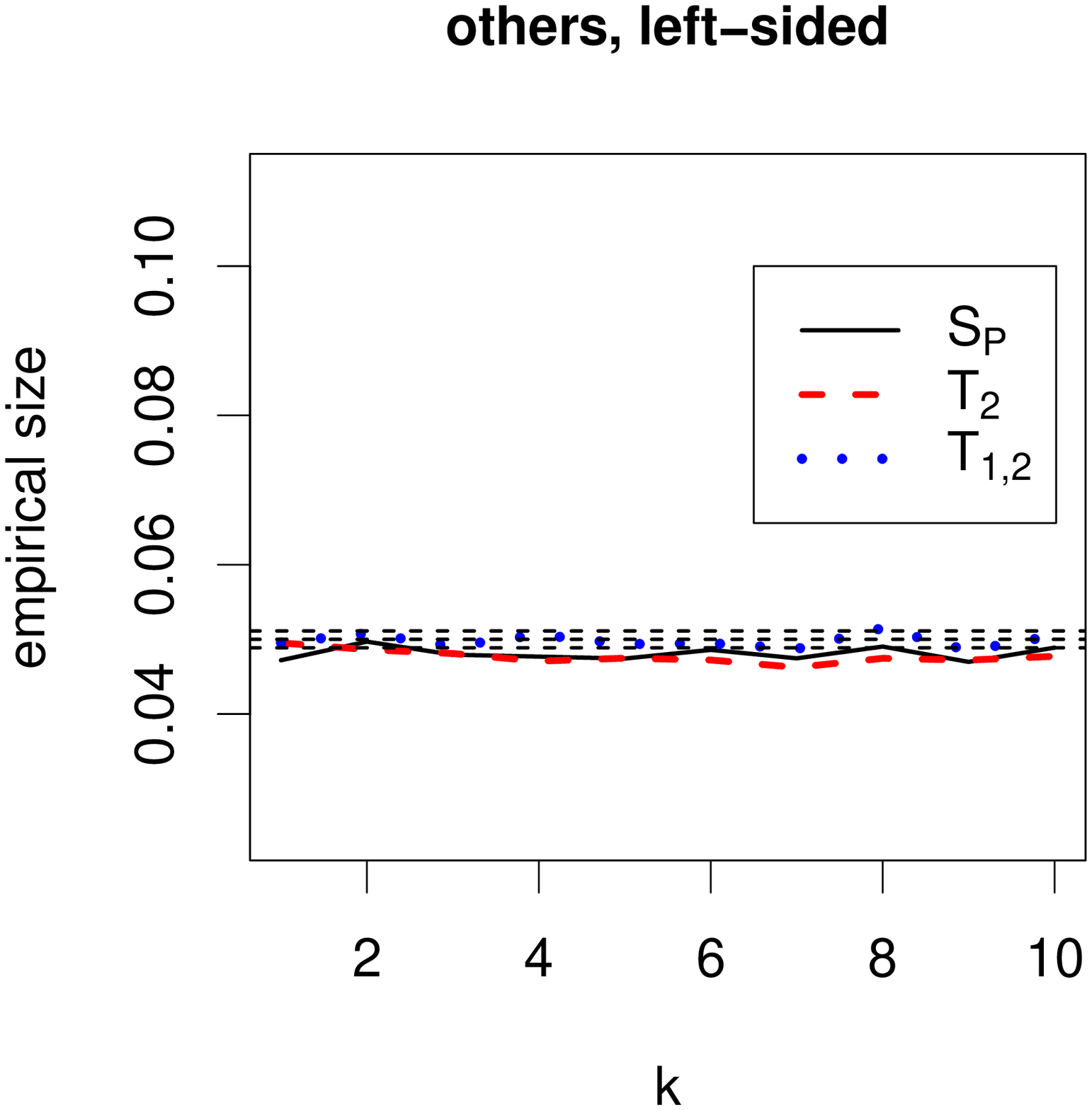} }}
\caption{
\label{fig:LS-RL-cases2-4}
The empirical size estimates of the tests for the left-sided alternative
under the RL cases 2 (top row), 3 (middle row), and 4 (bottom row).
The dashed horizontal lines and legend labeling are as in Figure \ref{fig:RS-RL-cases1a-c}.
}
\end{figure}

\begin{figure} [hbp]
\centering
\rotatebox{-90}{ \resizebox{2.1 in}{!}{\includegraphics{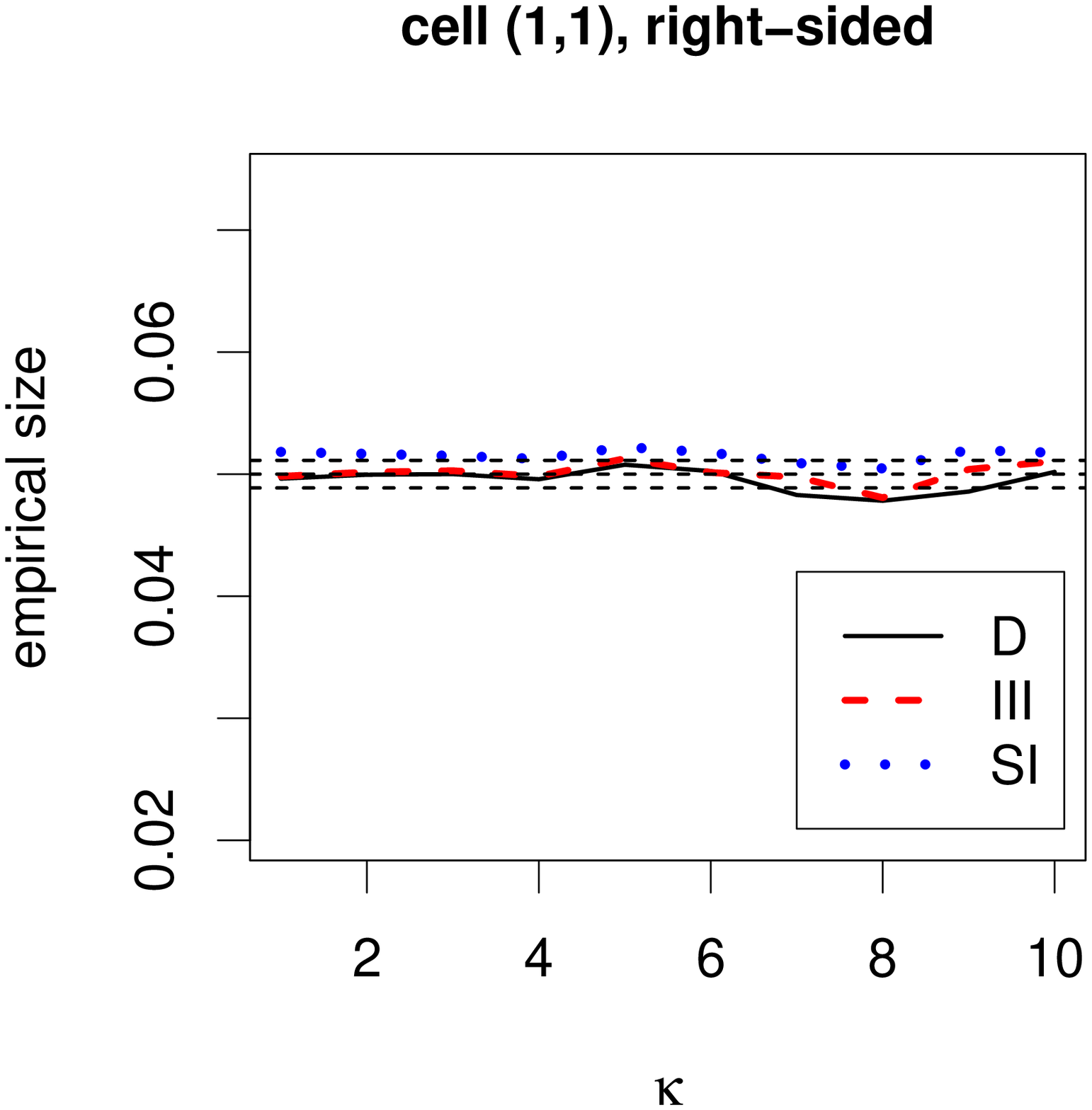} }}
\rotatebox{-90}{ \resizebox{2.1 in}{!}{\includegraphics{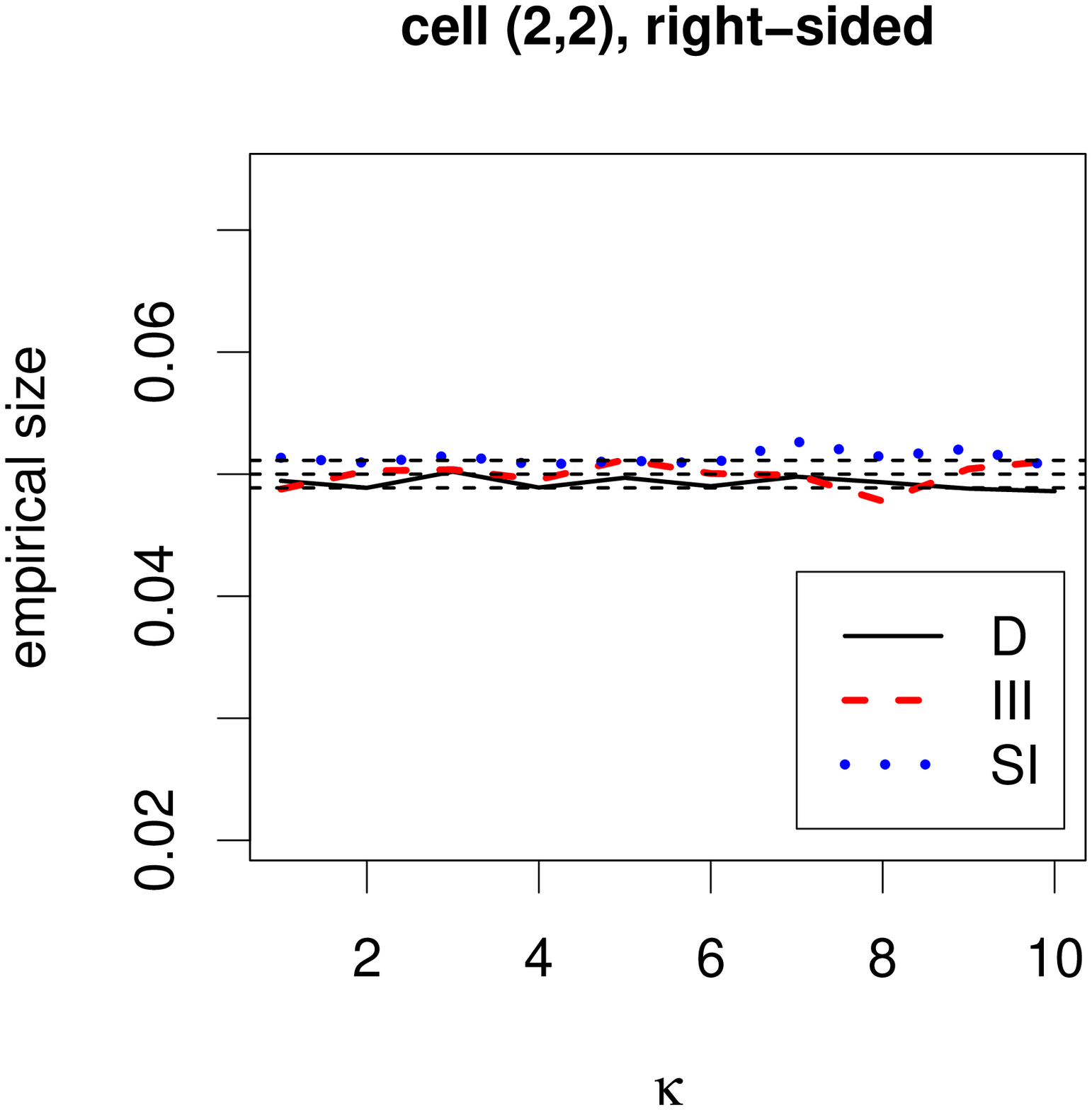} }}
\rotatebox{-90}{ \resizebox{2.1 in}{!}{\includegraphics{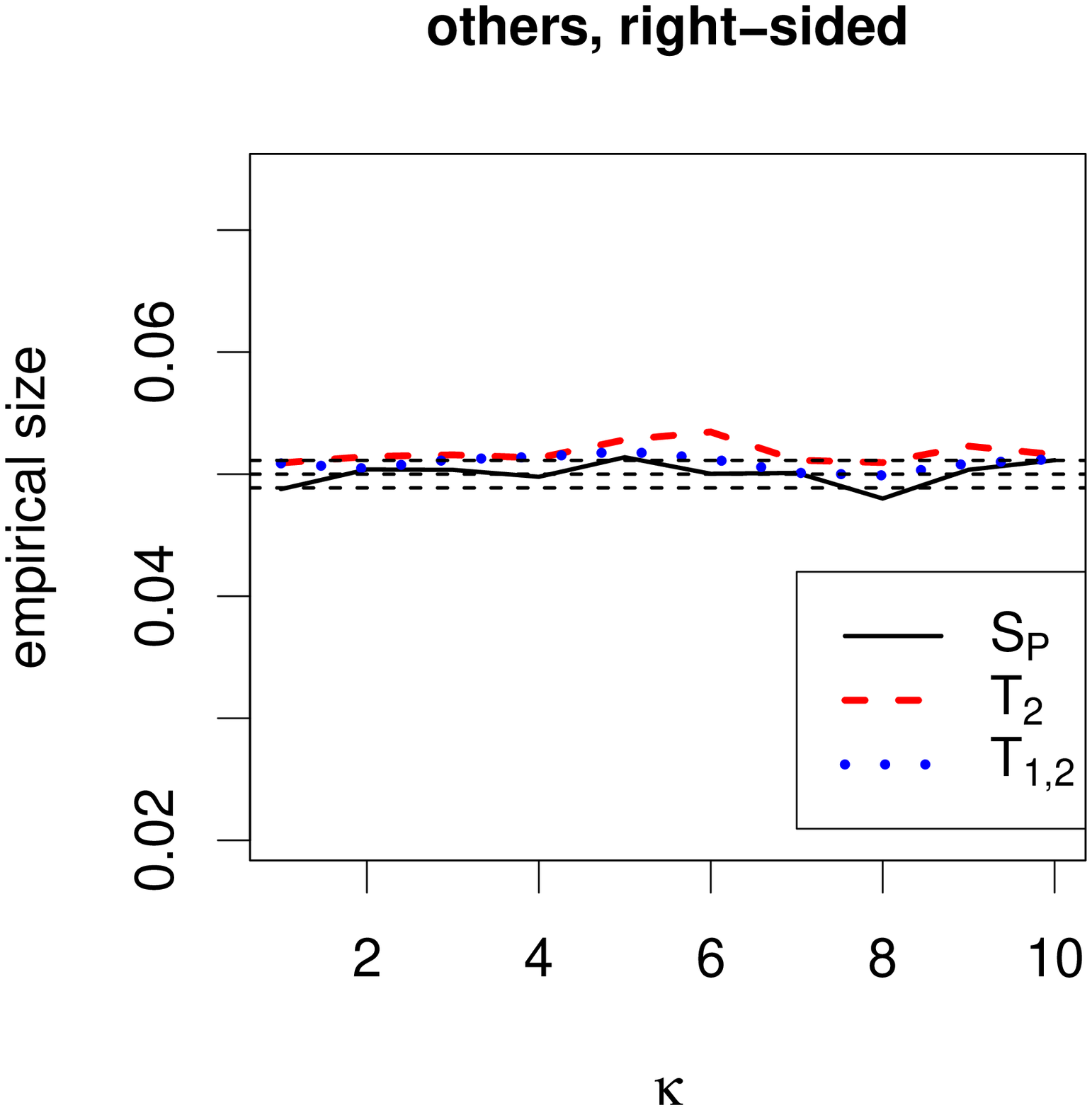} }}
\rotatebox{-90}{ \resizebox{2.1 in}{!}{\includegraphics{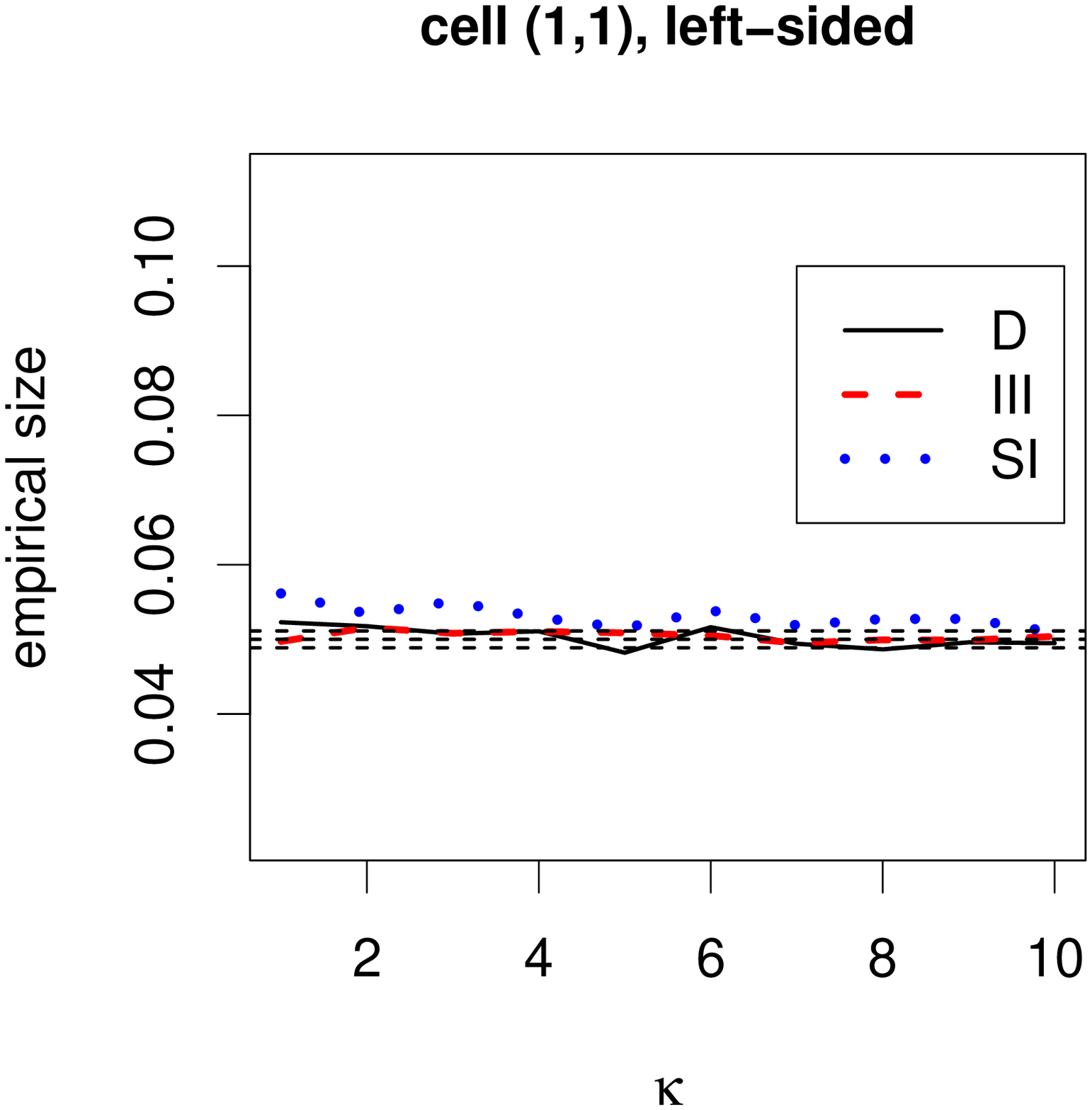} }}
\rotatebox{-90}{ \resizebox{2.1 in}{!}{\includegraphics{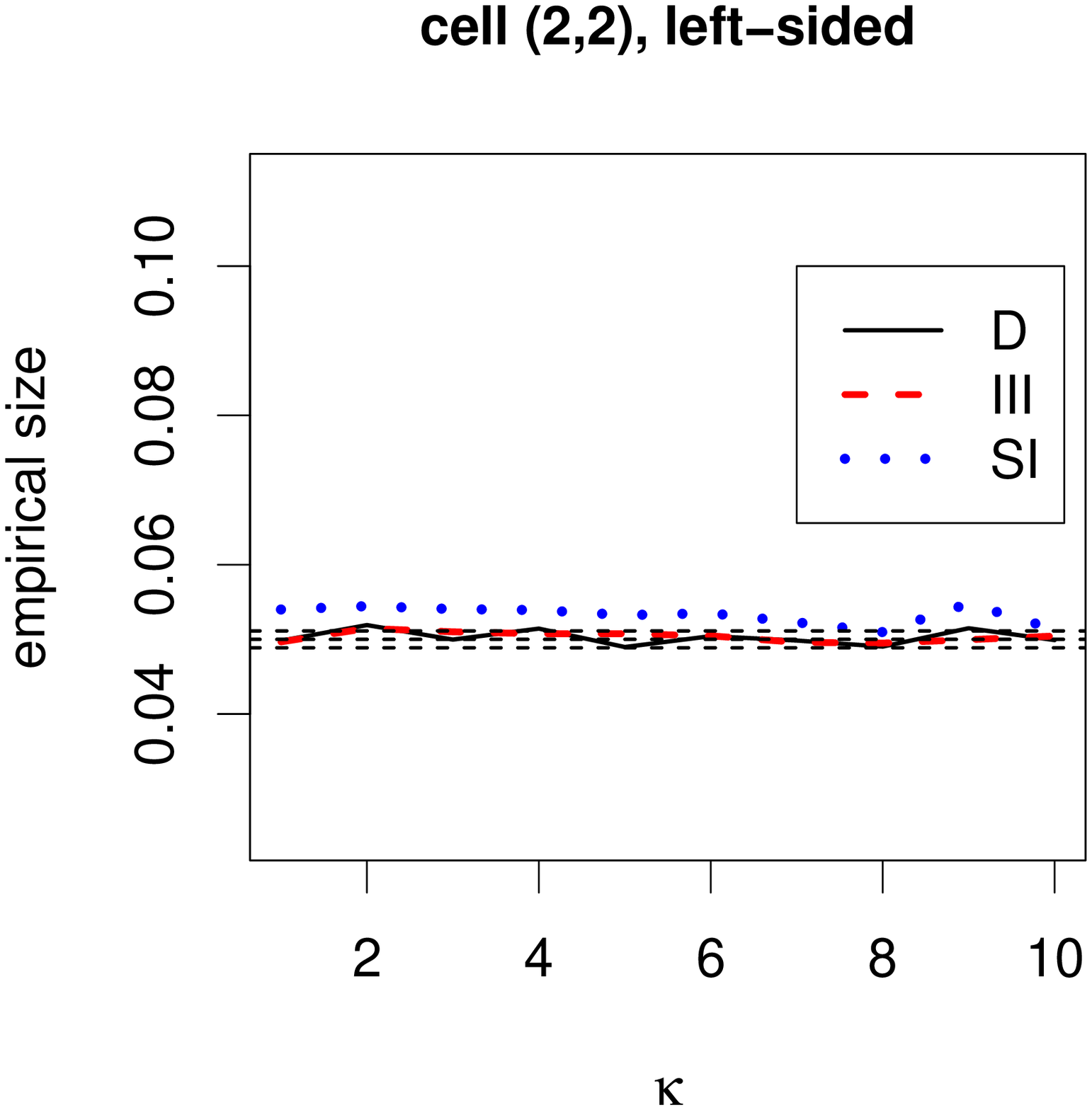} }}
\rotatebox{-90}{ \resizebox{2.1 in}{!}{\includegraphics{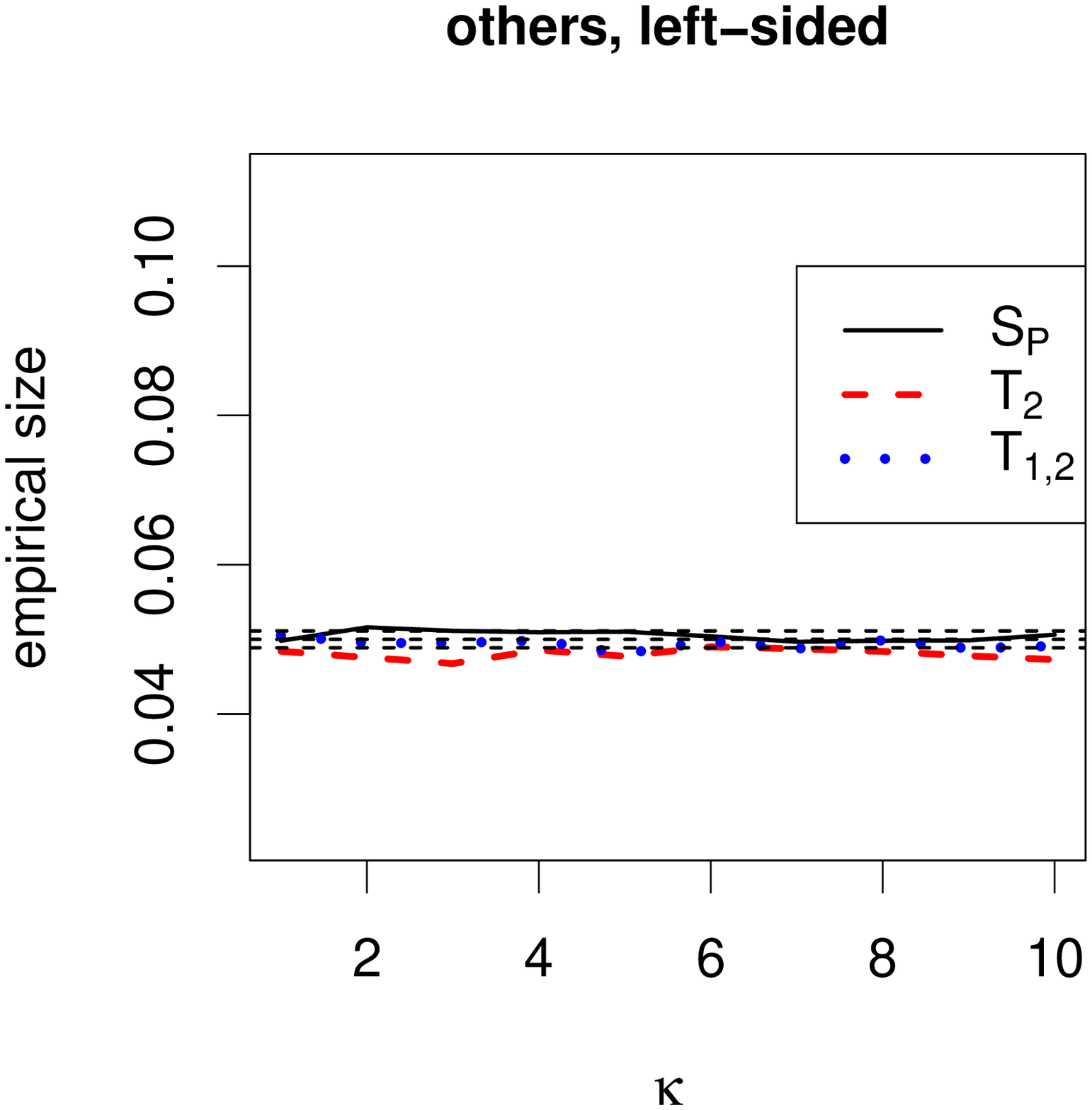} }}
\caption{
\label{fig:RS-LS-RL-case5}
The empirical size estimates of the test statistics
for the right-sided (top) and left-sided (bottom) alternatives
under the RL case 5 with $n_1$ and $n_2$ being about half the number of generated points
from the Mat\'{e}rn cluster process.
We use $\kappa=1,2,\ldots,5$.
The dashed horizontal lines and legend labeling are as in Figure \ref{fig:RS-RL-cases1a-c}.
}
\end{figure}

In Figures \ref{fig:RS-RL-cases1a-c}-\ref{fig:RS-LS-RL-case5},
we present the empirical size estimates
for the right-sided alternative (i.e., towards segregation)
and
for left-sided alternative (i.e., towards association).
The empirical size estimates are computed as follows.
For each Monte Carlo replication,
test statistics are computed and the size is estimated based on the asymptotic critical values.
For Pielou's coefficient of segregation,
Dixon's segregation indices,
Dixon's cell-specific tests,
type III cell-specific tests,
and
Cuzick-Edwards' $k$-NN and combined tests
we use the critical value $z_{.95}=1.96$ for the right-sided (clustering or segregation) alternative
and
$z_{.05}=-1.96$ for the left-sided (association) alternative.
For example, the empirical size of $S_P$
is calculated for the right-sided alternative as $\sum_{i=1}^{N_{mc}} \I(Z_{P,i} > 1.96)$
where we have $1000$ Monte Carlo replications for each of background realizations,
and since there are 100 different realizations,
we would have $N_{mc}=100000$
and $Z_{P,i}$ is the standardized version of Pielou's coefficient of segregation.
On the other hand,
for Dixon's overall test,
we use 95th percentile of $\chi^2_1$ distribution, which is $\chi^2_{1,.95}=3.84$
and
for type III overall test,
we use $\chi^2_{2,.95}=5.99$.

The empirical significance levels under cases 1(a)-(c)
for the right-sided and left-sided alternatives
are presented in Figures \ref{fig:RS-RL-cases1a-c} and \ref{fig:LS-RL-cases1a-c},
respectively.
In case 1(a), we have equal but increasing sample sizes (i.e., $n_1=n_2=n/2=10,20, \ldots,100$),
and as expected the size performance gets better
(i.e., empirical size approaches to the nominal size of 0.05)
as $n$ increases.
Furthermore, all the tests have empirical size estimates around the null region
(i.e., between .04887 and .05113).
These bounds for the null region are estimated as follows.
With $N_{mc}=100000$,
an empirical size estimate larger than .05113 is deemed liberal,
while an estimate smaller than .04887 is deemed conservative
at .05 level (based on binomial critical values with $n=100000$ trials
and probability of success 0.05).
Among the cell-related tests (i.e., cell-specific tests and segregation indices),
size estimates of type III test are closer to the nominal level of 0.05,
when all the tests considered
type III tests, Pielou's test and Cuzick-Edwards' tests have less fluctuation around 0.05,
and $T_{1,2}$ is closest to the nominal level and has the least fluctuation.
For the left-sided alternative,
(i.e., towards association)
Dixon's segregation indices are extremely liberal for $n \le 80$,
and Dixon's cell-specific tests and segregation indices fluctuate more around 0.05, compared to other tests.
Among cell-related tests, type III has the best size performance,
but all tests considered,
$T_{1,2}$ is closest to the nominal level and has the least fluctuation.

In case 1(b),
we have $n_1=30$ and $n_2=30,40,\ldots,120$,
i.e., the difference in relative abundance increases as $n_2$ increases,
and in this case, with increasing $n_2$ the disease incidence rate is decreasing.
Hence in this case, we investigate the effect of decreasing incidence rate
(starting from 50\% and decreasing to 20\%) on the empirical sizes.
For the right-sided alternatives,
among cell $(1,1)$ statistics, Dixon's test fluctuates between liberalness and the desired level,
Dixon's segregation index tends to be conservative (with level of conservativeness increasing with $n_2$),
and type III cell-specific statistic is slightly above the null region
with its size estimate increasing with $n_2$.
Among cell $(2,2)$ statistics,
Dixon's segregation indices tends to be liberal (with level of liberalness increasing with $n_2$),
and type III has the same performance as in cell $(1,1)$,
and Dixon's cell-specific is closest to the nominal level.
$S_P$, $T_1$, and $T_{1,2}$ seem to be generally above the null region,
with $S_P$ being closest to .05.
All tests considered,
Dixon's cell $(2,2)$ test and Pielou's coefficient of segregation have better performance,
with $S_P$ having slightly better performance.
For the left-sided alternatives,
among cell $(1,1)$ statistics, Dixon's test fluctuates between conservativeness and the desired level
(and tends to get more conservative with increasing $n_2$),
Dixon's segregation index tends to be extremely liberal (although fluctuating, the level of liberalness tends to increase with $n_2$),
and type III statistic is slightly below the null region
with its size estimate decreasing with $n_2$.
Among cell $(2,2)$ statistics,
Dixon's segregation indices have a decreasing trend with increasing $n_2$
(starting liberal and getting conservative eventually),
and type III has the same performance as in cell $(1,1)$,
and Dixon's cell-specific test is closest to the nominal level.
$T_1$ and $T_{1,2}$ are slightly conservative
with a clear decreasing trend in size estimate of $T_{1,2}$.
On the other hand,
$S_P$ is closest the null region.
All tests considered,
Dixon's cell $(2,2)$ test and $S_P$ have better performance,
with $S_P$ having slightly better performance.
Hence, the differences in the relative abundances increasing in favor of controls
(i.e., decreasing incidence rate of the disease) confounds most test statistics.
Among the tests considered, $S_P$ seems to be the most robust to such differences in sample sizes.

In case 1(c),
we have $n_2=30$ and $n_1=30,40,\ldots,120$,
i.e., the difference in relative abundance increases as $n_1$ increases,
and in this case, with increasing $n_1$ the disease incidence rate is increasing.
Hence in this case, we investigate the effect of increasing incidence rate
(starting from 50\% and increasing to 80\%) on the empirical sizes.
The trends in $S_P$ and type III tests are as in case 1(b), with the roles of classes switched,
the tests yield the same results for a given data.
Furthermore,
Dixon's cell $(i,i)$ statistics and
segregation indices behave similar to those for cell $(j,j)$ of case 1(b) for $i \not= j$
switching also $n_2$ with $n_1$.
For the right-sided and left-sided alternatives,
$T_2$ and $T_{1,2}$ are closest to the null region
and have better performance than the other tests
(with $T_{1,2}$ having the best performance).
Hence, the differences in the relative abundances increasing in favor of cases
(i.e., increasing incidence rate of the disease) confounds most test statistics.
Among the tests considered, $T_2$ and $T_{1,2}$ and to a lesser extent
$S_P$ seem to be the most robust statistics to such differences in sample sizes
and $T_{1,2}$ has the best performance.
The better performance of Cuzick-Edwards' tests in this case is no coincidence,
since these tests are designed to detect the clustering of cases (i.e., class 1 points),
and the number of class 1 points increases in this case.

The empirical size estimates under cases 2-4
for the right-sided and left-sided alternatives
are presented in Figures \ref{fig:RS-RL-cases2-4} and \ref{fig:LS-RL-cases2-4}, respectively.
In case 2, we have equal sample sizes with $n_1=n_2=100$,
but with increasing $\delta$,
the level of clustering of the two clusters in the background pattern increases
(in fact, with $\delta>1$, the clusters get separated).
For the right-sided alternative,
all tests are almost within the null region with Dixon's cell $(1,1)$ statistics closest to the nominal level.
Cell-related tests except Dixon's cell $(1,1)$ test and $S_P$ tend to be slightly conservative,
while Cuzick-Edwards' tests tend to be slightly liberal.
For the left-sided alternative,
all tests except Dixon's segregation indices are within the null region.
Dixon's segregation indices are liberal and has size estimates about 0.06,
Dixon's cell-specific tests are slightly liberal,
and all other tests are slightly conservative.
Hence, with sample sizes are equal and large,
most tests are unaffected seriously with increasing level of clustering in the background realizations,
and Dixon's segregation indices are most severely confounded with $\delta$.
There is no clear (increasing or decreasing) trend in the size estimates of the tests with increasing $\delta$.

In case 3, we have equal sample sizes with $n_1=n_2=100$,
but with increasing $\delta$,
the level of clustering of the two separated clusters in the background pattern increases.
For both alternatives,
the empirical size performance of the tests is similar to the performance under case 2.
In case 4, we have equal sample sizes with $n_1=n_2=100$,
and same separation length $\delta=0.5$,
but the number of clusters in the background pattern increases.
For the one-sided alternatives,
the empirical size performance of the tests is similar to the performance under case 2.
Hence we notice that with sample sizes being equal and large,
the sizes of the tests are not affected by the increasing number of clusters in the background realizations.

The empirical size estimates under case 5 for the right-sided and left-sided alternatives
are presented in Figure \ref{fig:RS-LS-RL-case5}.
In this case,
we have sample sizes $n_1=n_2=100$ on the average,
and random number of clusters $\kappa$ (with increasing $\kappa$,
the number of clusters tend to increase),
and the locations of the clusters are also random.
For the right-sided alternative,
Dixon's segregation indices, $T_2$ and $T_{1,2}$ are above 0.05,
while other tests are around 0.05.
Type III tests, Dixon's cell-specific tests and $S_P$ seem to have the best performance.
For the left-sided alternative,
Dixon's segregation indices tend to be above 0.05,
$T_2$ and $T_{1,2}$ are below 0.05,
while other tests are around 0.05.
Type III tests, $S_P$ and $T_{1,2}$ seem to have the best performance.
Hence, with randomly occurring and randomly increasing number of clusters,
most tests are not affected seriously.
Dixon's segregation indices have the worst size performance under
RL of this type of background clustering.

\begin{figure} [hbp]
\centering
\rotatebox{-90}{ \resizebox{2.1 in}{!}{\includegraphics{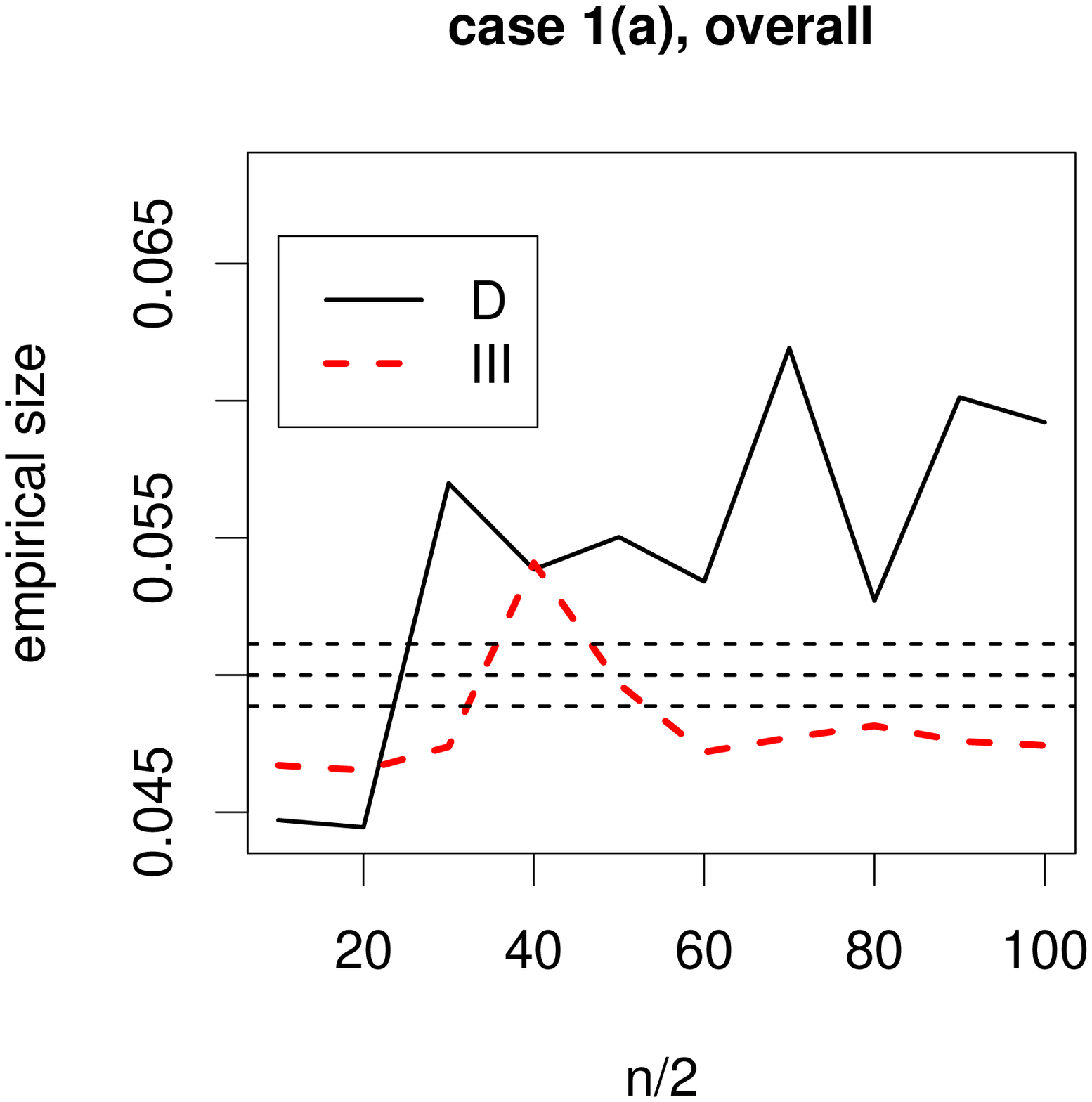} }}
\rotatebox{-90}{ \resizebox{2.1 in}{!}{\includegraphics{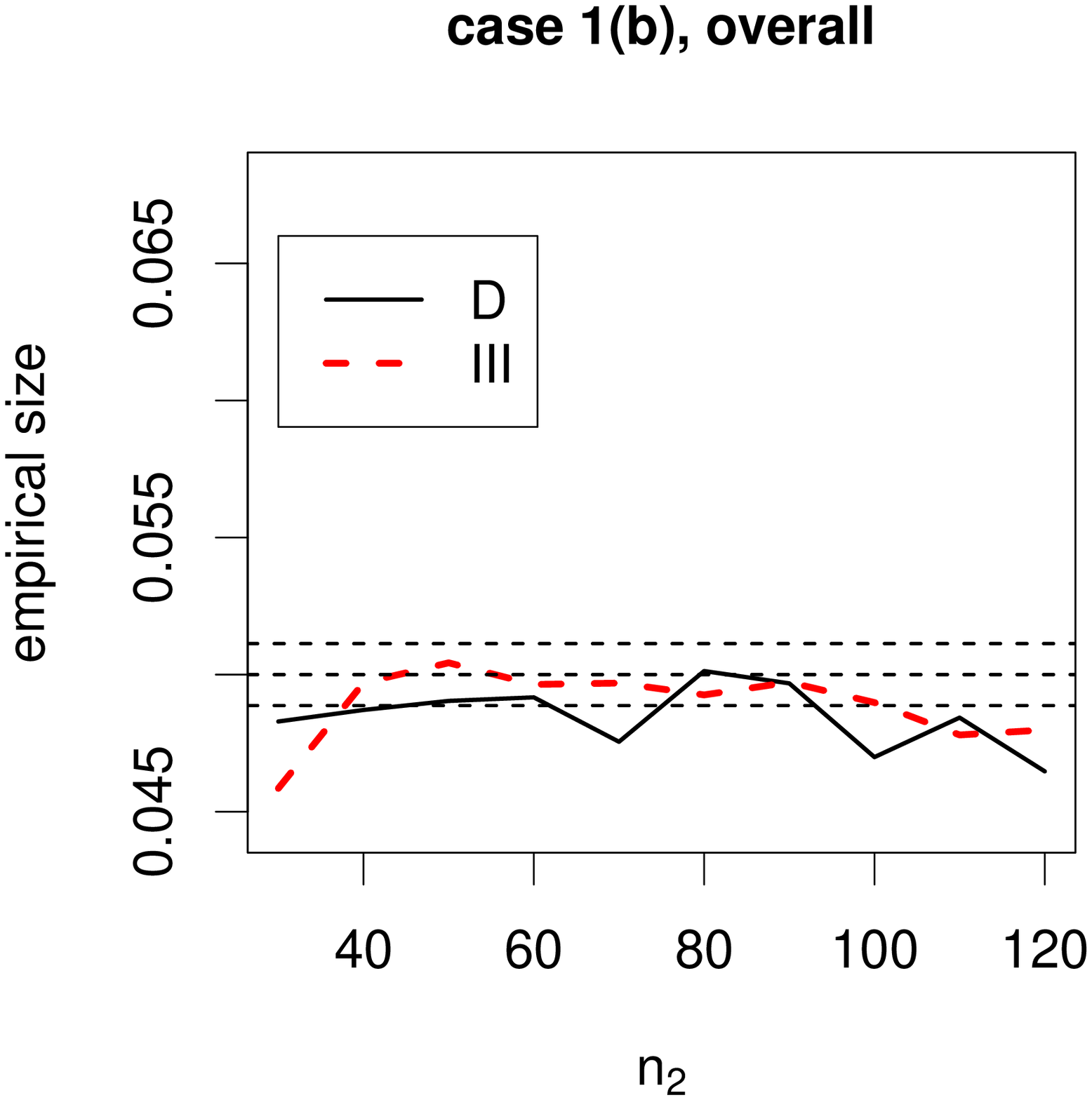} }}
\rotatebox{-90}{ \resizebox{2.1 in}{!}{\includegraphics{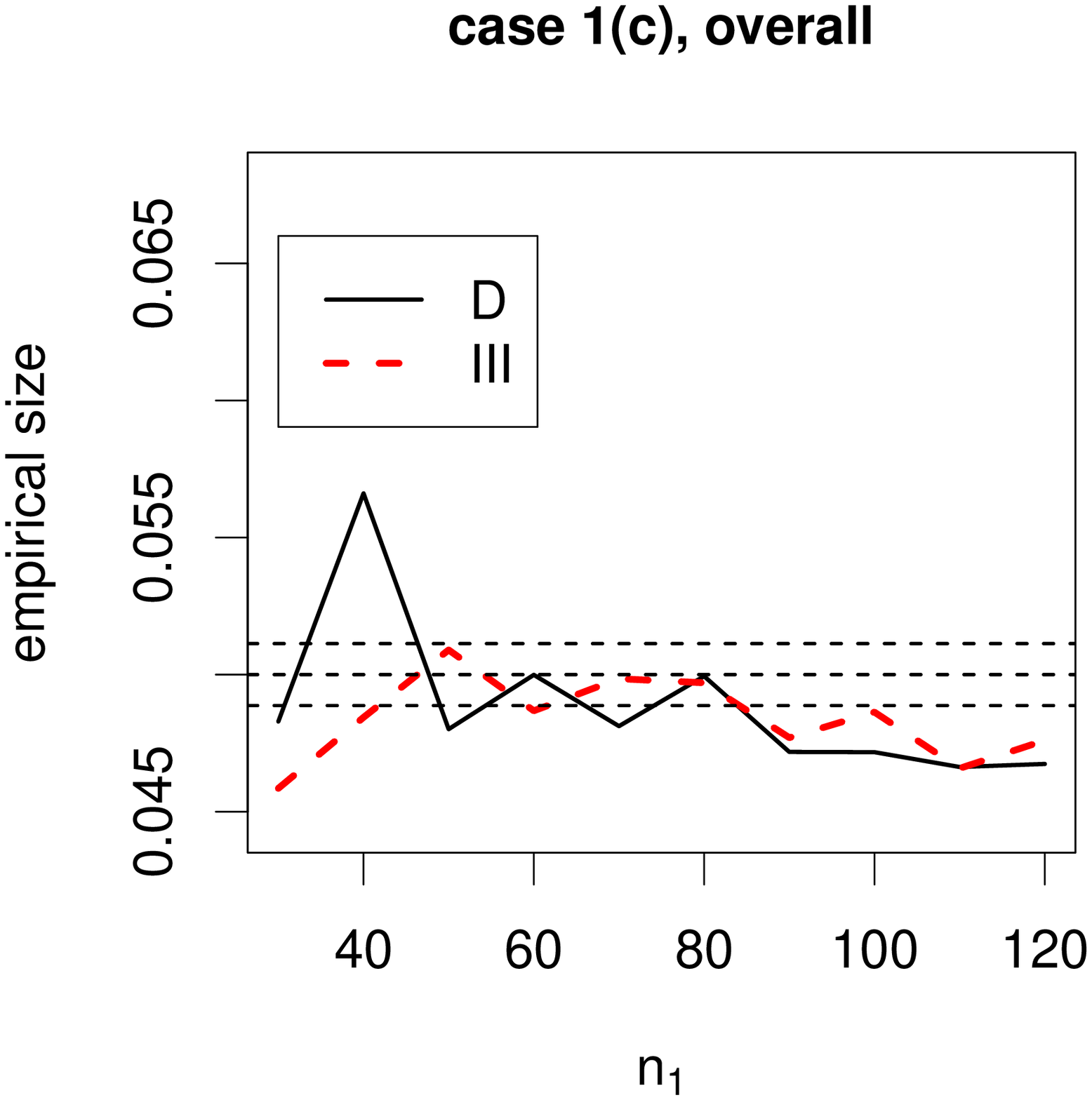} }}
\rotatebox{-90}{ \resizebox{2.1 in}{!}{\includegraphics{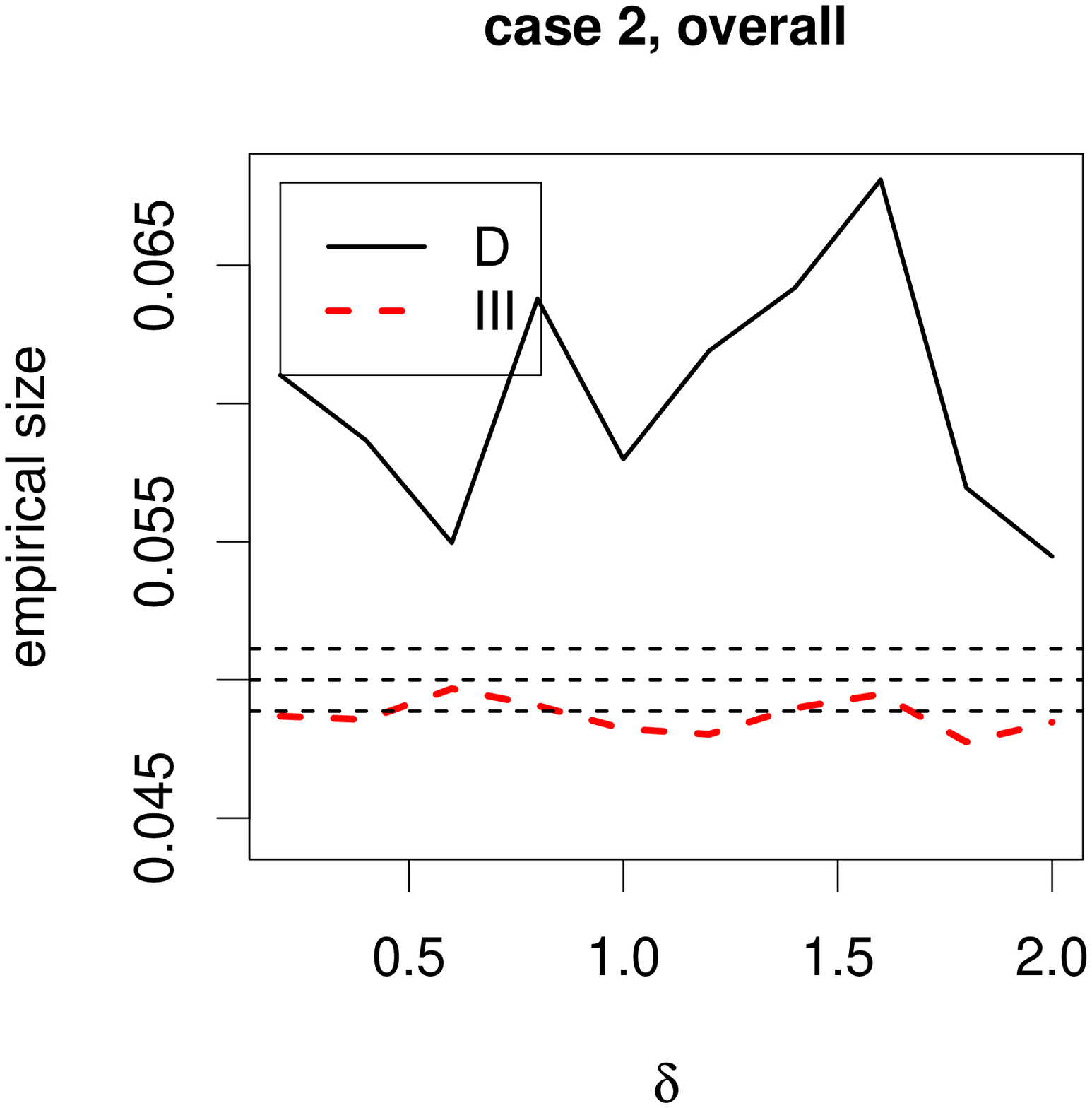} }}
\rotatebox{-90}{ \resizebox{2.1 in}{!}{\includegraphics{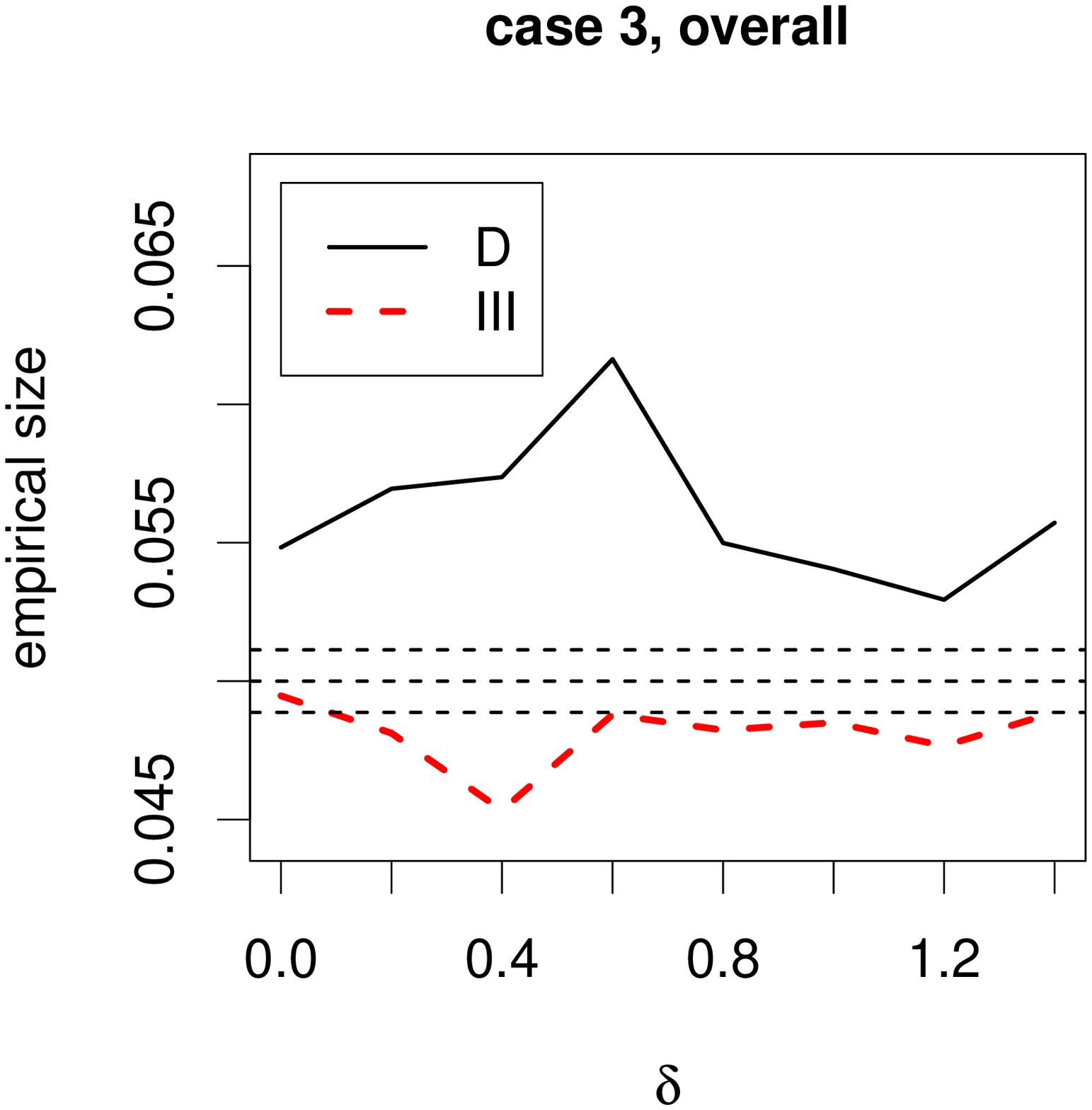} }}
\rotatebox{-90}{ \resizebox{2.1 in}{!}{\includegraphics{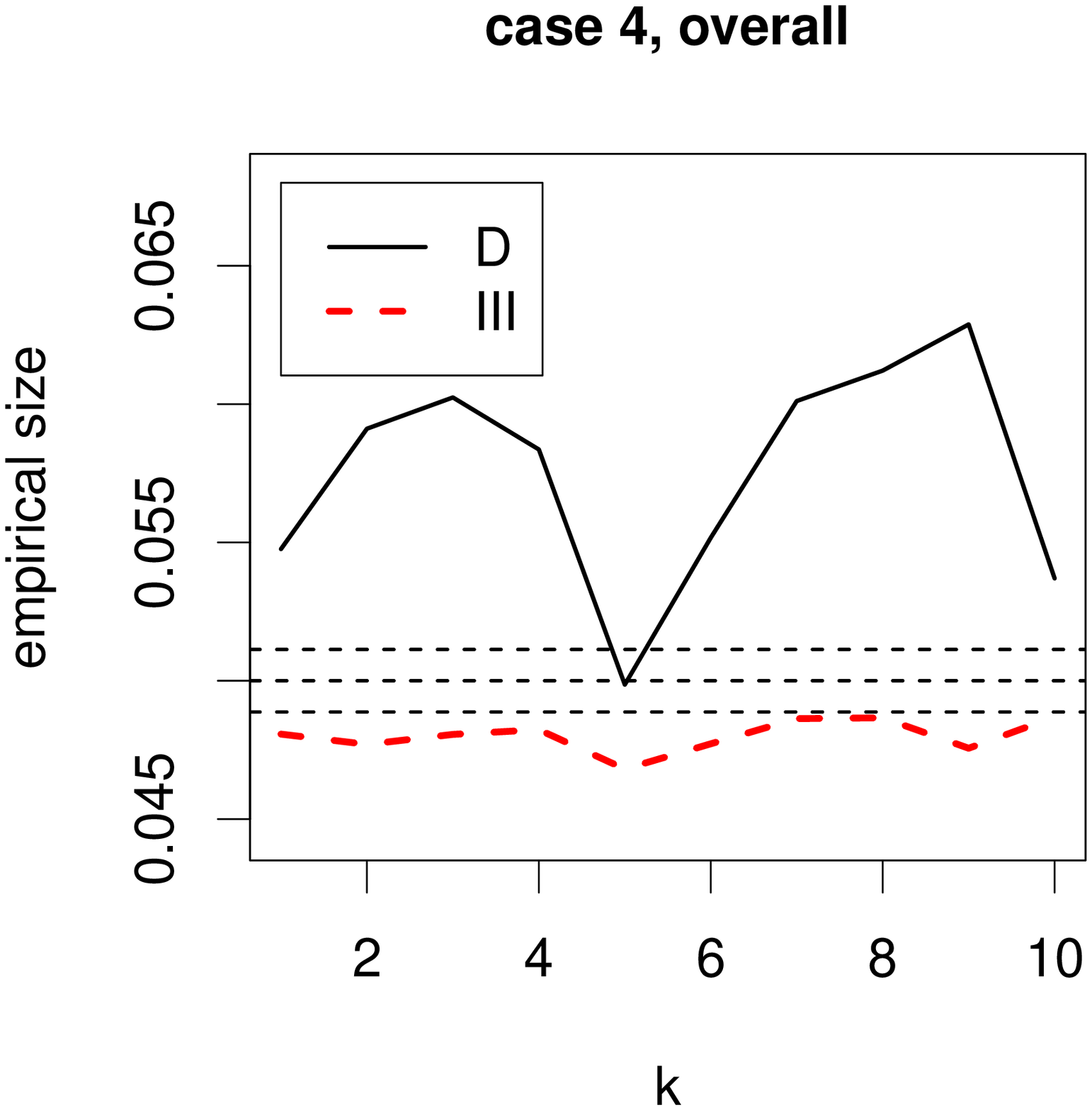} }}
\rotatebox{-90}{ \resizebox{2.1 in}{!}{\includegraphics{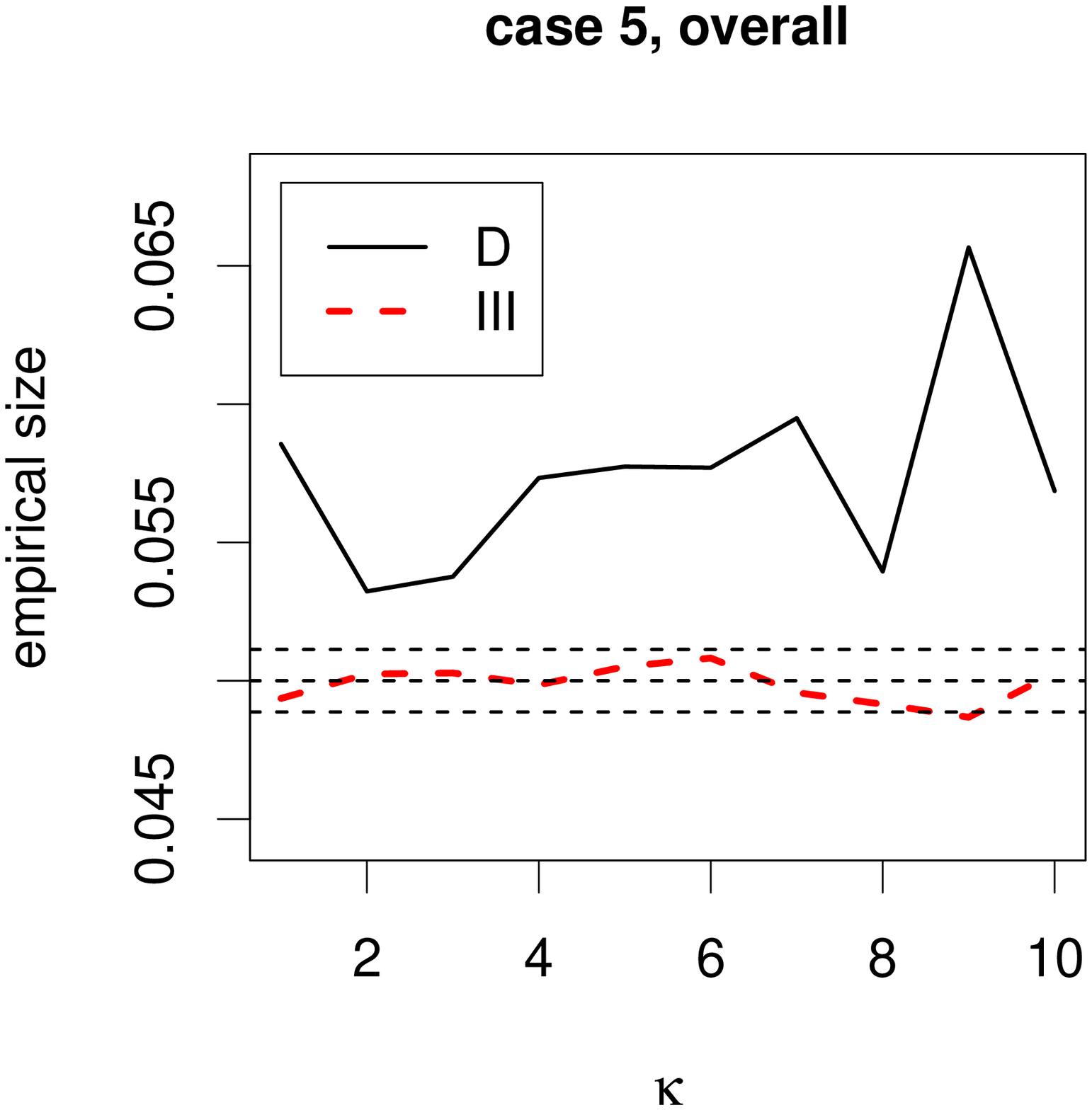} }}
\caption{
\label{fig:RL-overall}
The empirical size estimates of the overall NNCT-tests
under the RL cases 1(a)-(c) (top row in that order from left to right),
and cases 2-5 (starting at second row and ordered from left to right).
The dashed horizontal lines are as in Figure \ref{fig:RS-RL-cases1a-c},
and in the legends
D stands for Dixon's overall test,
and
III stands for type III overall test.
}
\end{figure}

The empirical size estimates for overall NNCT-tests under cases 1-5
are presented in Figure \ref{fig:RL-overall}.
In cases 1(a) and 2-5,
Dixon's overall test is mostly liberal
and around the null region in cases 1(b) and (c).
On the other hand,
type III overall test is within the null region or slightly conservative,
and has better performance compared to Dixon's overall test.
Furthermore,
there is no clear trend in the size estimates as the equal sample sizes increase,
or level and number of clusters increase.
On the other hand,
as the discrepancy between the sample sizes (i.e., differences in relative abundances)
in cases 1(b) and (c) increases,
the size estimates of the overall tests tend to decrease eventually.

\section{Empirical Power Analysis of the Tests under Non-RL Alternatives}
\label{sec:empirical-power}
We propose various non-RL alternatives
where case and control labels are assigned
(with a pattern deviating from RL pattern)
to the points generated from various homogeneous or clustering processes.
In all these alternatives
the background points in $\mZ_n$
are generated independently uniformly in the unit square $(0,1)\times(0,1)$,
i.e., $Z_i \stackrel{iid}{\sim} \U((0,1)\times(0,1))$ for $i=1,2,\ldots,n$.
To remove the effect of one particular realization of the points on the test,
we consider 100 different realizations.
We only use realizations from HPP pattern for the background,
because the level and number of clusters seem not to affect the size performance of the tests.
Hence, in the non-RL alternatives,
we only consider various non-RL schemes on the points from HPP.

\begin{itemize}
\item[] \textbf{Types of the Non-RL Patterns:}
\item[] \textbf{Case 1:}
Select a $Z_i$ randomly, assign it as a case.
Find its $k$ NNs and assign them as cases with probabilities
$\frac{n_1}{n}+\rho(1-\frac{n_1}{n}),\frac{n_1}{n}+\frac{\rho}{2}(1-\frac{n_1}{n}),\ldots,\frac{n_1}{n}+\frac{\rho}{k}(1-\frac{n_1}{n})$
until the number of cases first exceeds $n_1$.
We use
(a) $\rho=-0.2, 0.0, 0.2, 0.4, 0.6, 0.8$ and
$k=1$ which only assigns the first NN
and
(b) $\rho=0.0, 0.2, 0.4, 0.6, 0.8$ and
$k=3$ which assigns the first 3 NNs according to the above probabilities.

\item[] \textbf{Case 2:}
In this case, we have an initial proportion, $\pi_i$,
and an ultimate proportion, $\pi_u$,
with $\pi_u > \pi_i$.
First assign the initial proportion, $\pi_i$, of points as cases randomly and
pick a case among them randomly.
Then find the $k$ NNs of this case and assign them as cases with probabilities
$\rho,\rho/2,\ldots,\rho/k$.
Select a point randomly among these $k$ NNs,
find its $k$ NNs and assign them as cases with the above probabilities until
we have the proportion of cases first exceeding $\pi_u$.
We use $\pi_i=0.3$, $\pi_u=0.5$, and $\rho=0.2,0.4,0.6,0.8$ and consider
(a) $k=1$ which only assigns the first NN
and
(b) $k=3$ which assigns the first 3 NNs according to the above probabilities.

\item[] \textbf{Case 3:}
Pick a $Z_i$ randomly, mark it as a case
and label others as a case with probabilities inversely proportional to their distances to $Z_i$.
More specifically,
we use probabilities proportional to $\frac{\rho}{k_d}\left(1-\frac{d_{ji}}{d_{\max}}\right)^{k_p}$
where $d_{ji}$ is the distance from $Z_j$ to $Z_i$ for $j \not=i$,
$d_{\max}$ is the maximum of $d_{ji}$ values,
$k_p >0$ and $k_d \ge 1$.
We stop when we first exceed $n_1$ cases.
In our simulations we employ the usual Euclidean distance,
and use
(a) $\rho=0.2,0.4,.\ldots,1.0$, $k_d=1$, and $k_p=3$,
(b) $\rho=0.8$, $k_d=3,6,\ldots,15$, and $k_p=3$
and
(c) $\rho=0.8$, $k_d=1$, and $k_p=1,2,\ldots,5$.

\item[] \textbf{Case 4:}
Pick $k_0$ points $z_1^{'},z_2^{'},\ldots,z_{k_0}^{'}$ from $\mZ_n$ randomly as sources.
Let $\varphi_G$ be the pdf of $BVN(\mu,\sigma_1=\sigma_2=\sigma, \rho=0)$
where $BVN(\mu,\sigma_1,\sigma_2,\rho)$ stands for the bivariate normal
distribution with mean vector $\mu=(\mu_1,\mu_2)$,
standard deviations of univariate components are $\sigma_1$ and $\sigma_2$,
and the correlation between the components is $\rho$.
Then for each $j=1,2,\ldots,k_0$,
compute $\varphi_{G,j}(z_i)$ for all $i=1,2,\ldots,n$
where $\varphi_{G,j}$ is the pdf of $BVN(\mu=z_j^{'},\sigma_1=\sigma_2=\sigma, \rho=0)$
and add these pdf values.
That is,
find $p_G(z_i)=\sum_{j=1}^{k_0} \varphi_{G,j}(z_i)$ for each $i=1,2,\ldots,n$.
Then label the points as cases with probabilities directly proportional to
the value of the pdf sums at these points.
More specifically,
we use probabilities $\frac{1}{p_{\max}}(p_G(z_1),p_G(z_2),\ldots,p_G(z_n))$
where $p_{\max}=\max_{i=1}^n p_G(z_i)$.
We stop when we first exceed $n_1$ cases.
We use (a) $k_0=3$, $\sigma_1=\sigma_2=\sigma=0.1,0.2,\ldots,0.8$
and
(b) $k_0=1,2,\ldots,8$ and $\sigma_1=\sigma_2=\sigma=0.4$.
\end{itemize}

We simulate 1000 Monte Carlo replications for each parameterization in each case
at each background realization.
For example,
with a particular background realization,
in case 3(a) we simulate 1000 replications for
$k_p=3$, and $k_d=1$ and each of $\rho=0.2,0.4,.\ldots,1.0$ with $n_1=n_2=100$.

The empirical power estimates are computed similar to the empirical sizes.
That is,
for tests having asymptotic normality,
we use the critical value $z_{.95}=1.96$ for the right-sided alternative
and
$z_{.05}=-1.96$ for the left-sided alternative.
And for Dixon's overall test,
we use 95th percentile of the corresponding $\chi^2$ distribution,
which is $\chi^2_{1,.95}=3.84$
and
for type III overall test,
we use $\chi^2_{2,.95}=5.99$.
Furthermore,
$S^D_{ij}$ for $i,j=1,2$,
and the corrected versions,
$S^{D,c}_{ij}$ for $i,j=1,2$ provide very similar empirical power estimates, hence only the former are presented.

\begin{figure} [ht]
\centering
Power Estimates for Case 1(a) with Asymptotic Critical Values\\
\rotatebox{-90}{ \resizebox{2.1 in}{!}{\includegraphics{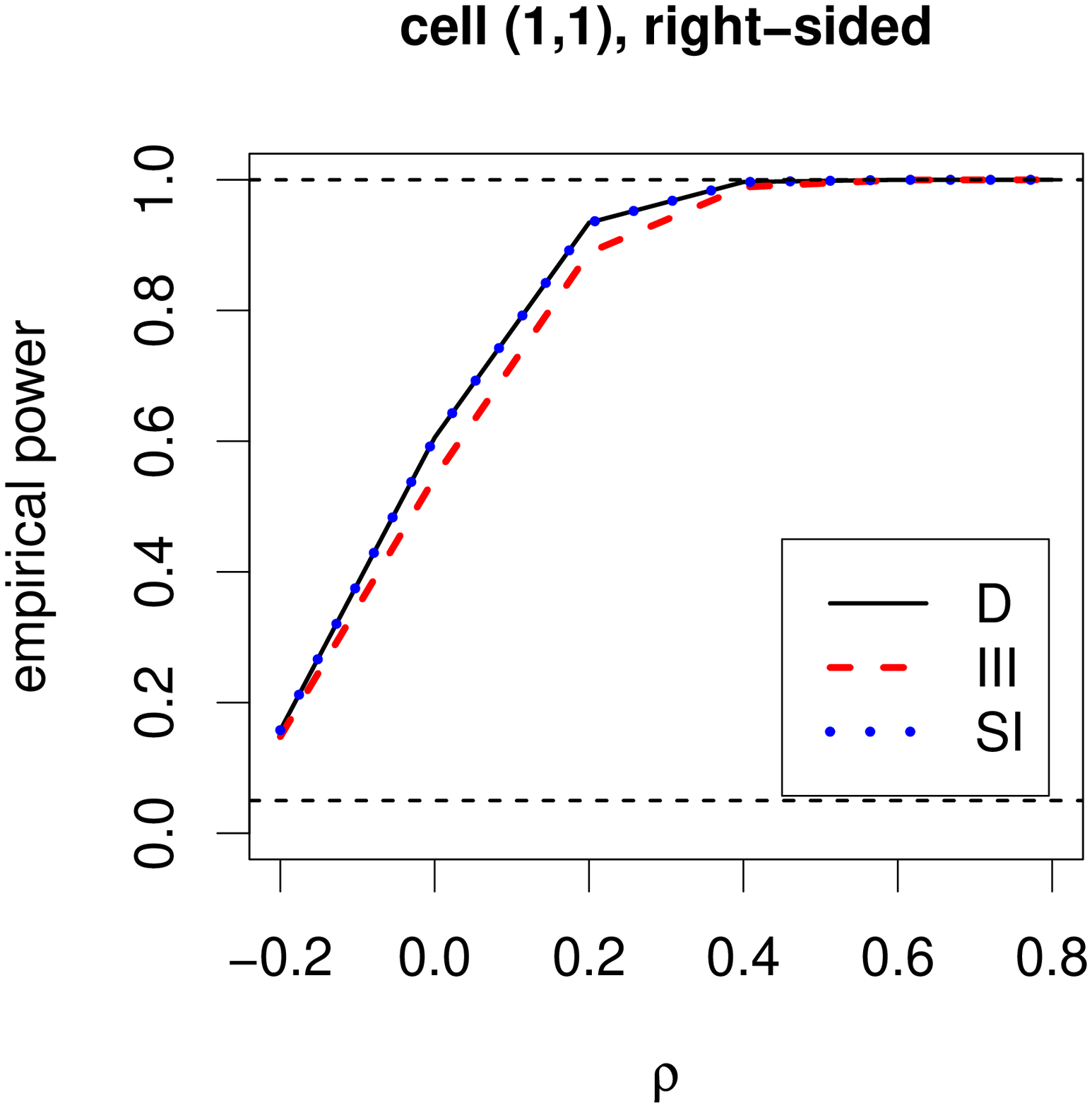} }}
\rotatebox{-90}{ \resizebox{2.1 in}{!}{\includegraphics{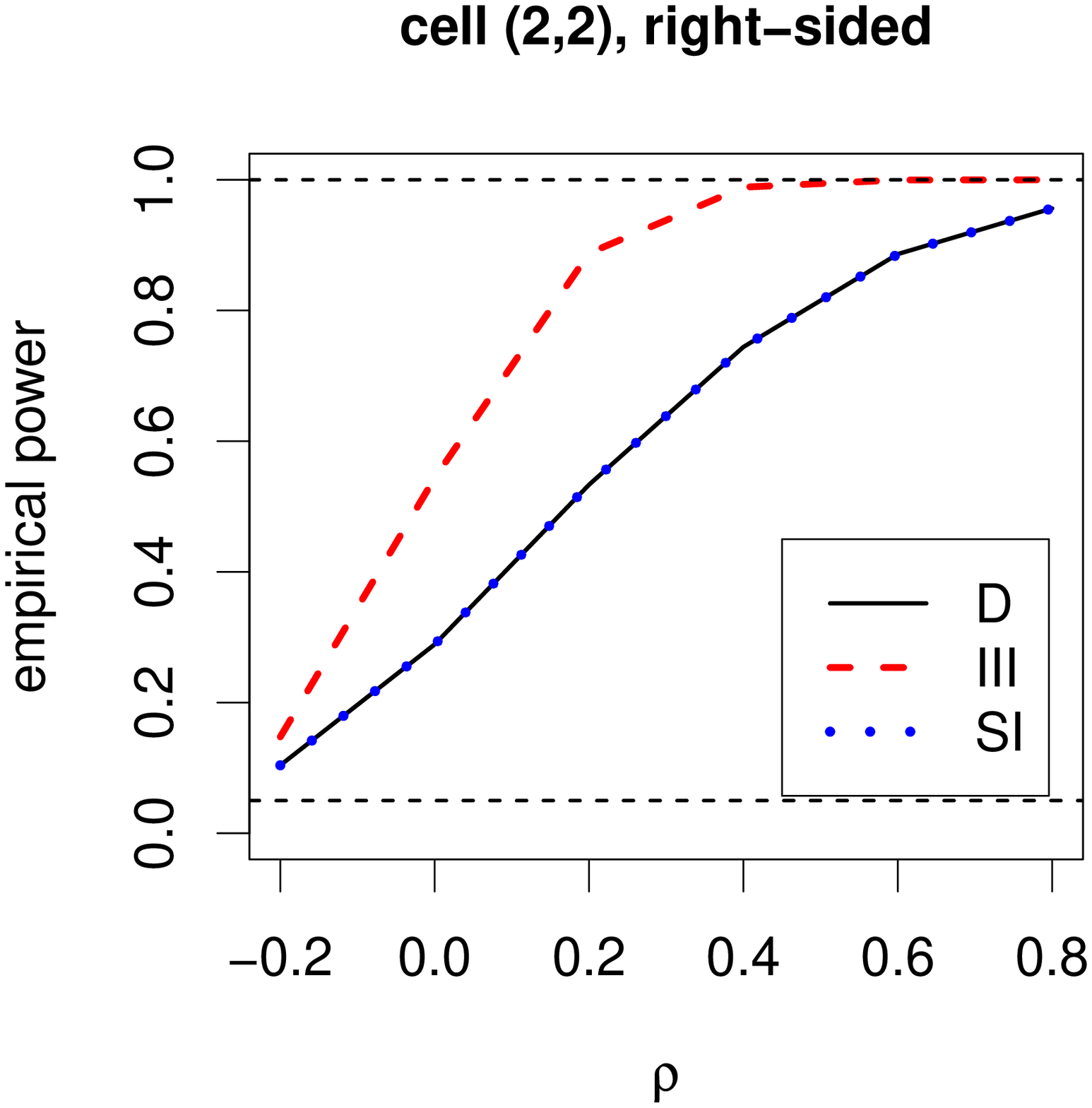} }}
\rotatebox{-90}{ \resizebox{2.1 in}{!}{\includegraphics{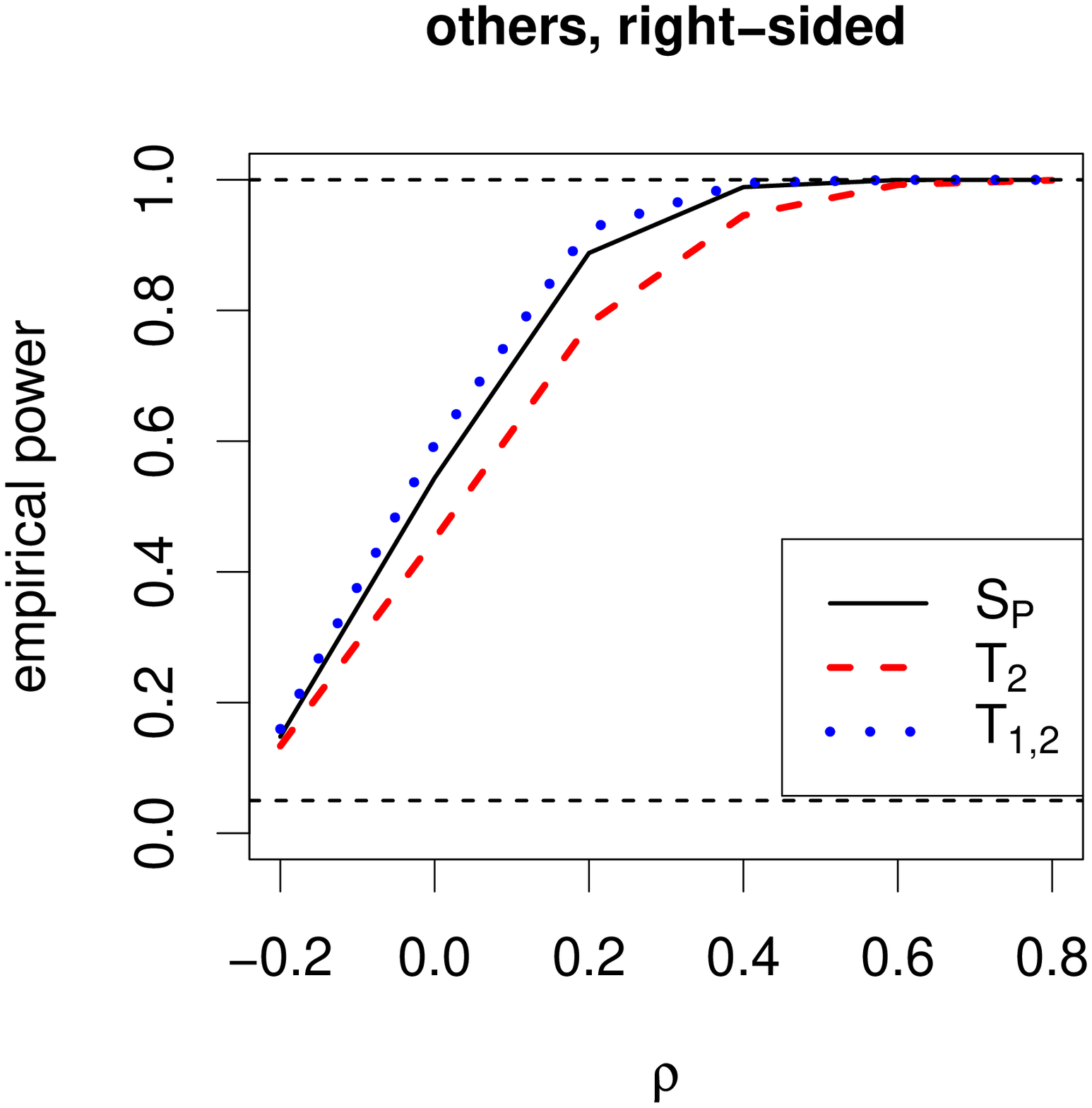} }}
Power Estimates for Case 1(a) with Monte Carlo Critical Values\\
\rotatebox{-90}{ \resizebox{2.1 in}{!}{\includegraphics{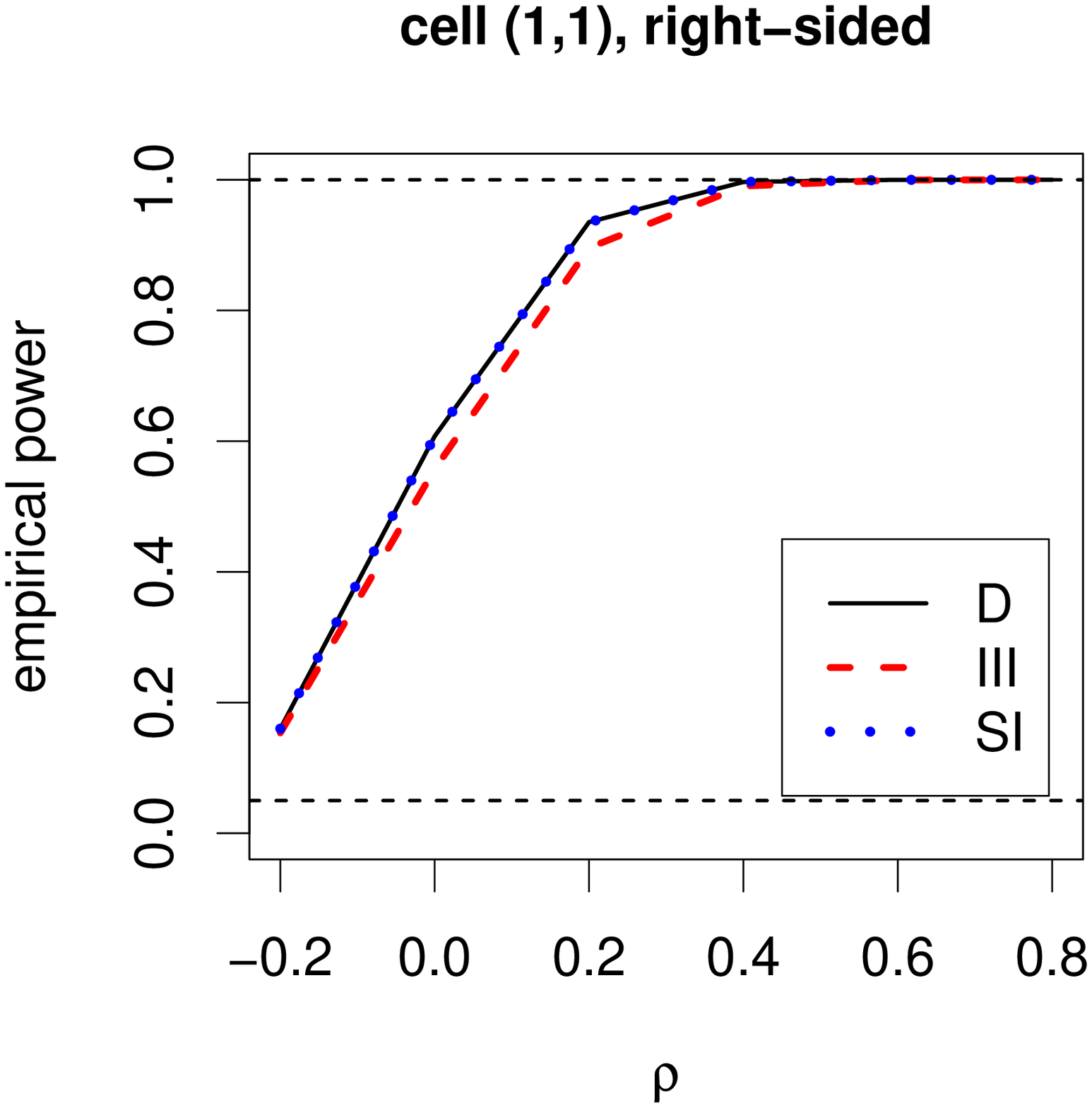} }}
\rotatebox{-90}{ \resizebox{2.1 in}{!}{\includegraphics{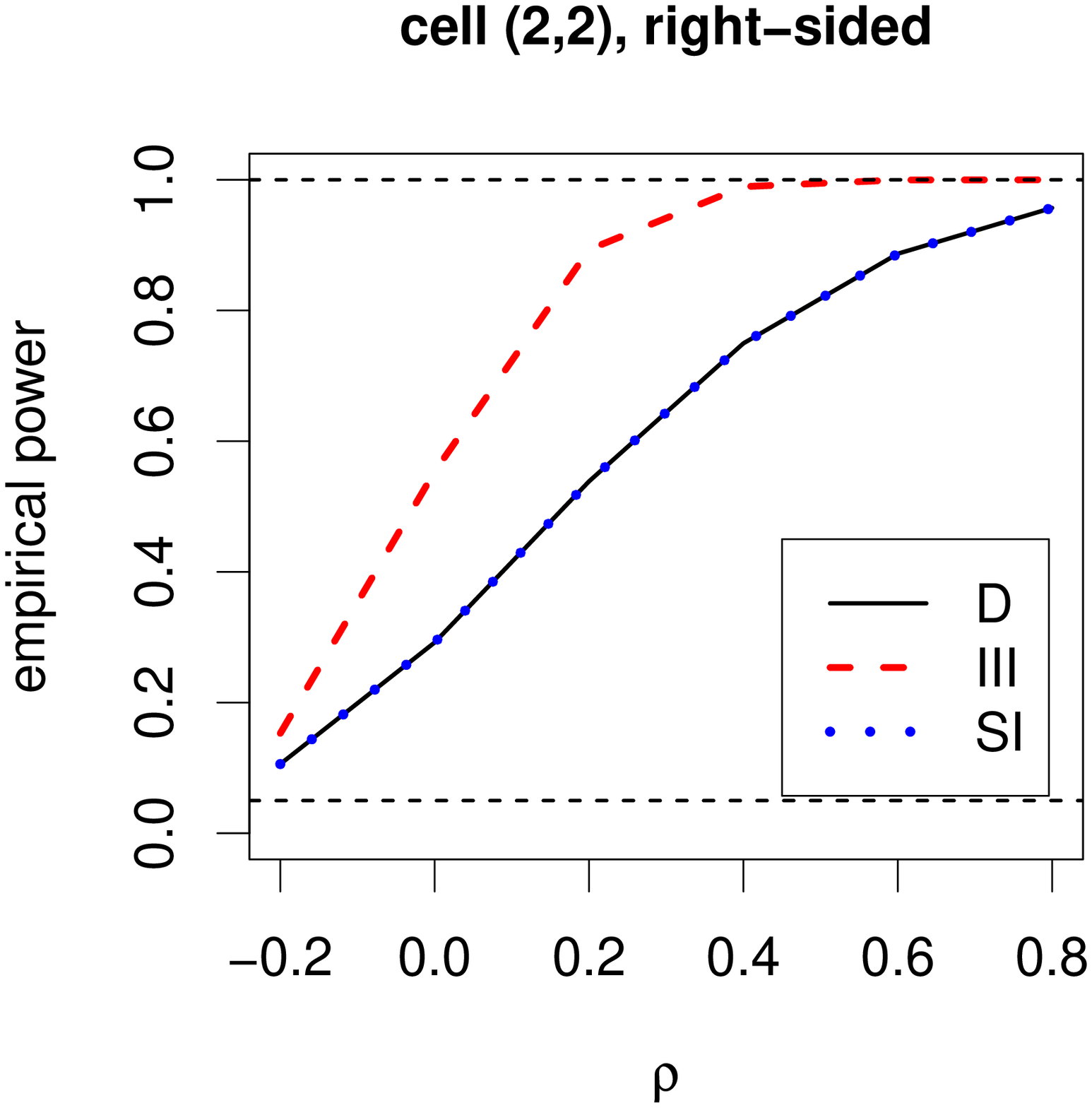} }}
\rotatebox{-90}{ \resizebox{2.1 in}{!}{\includegraphics{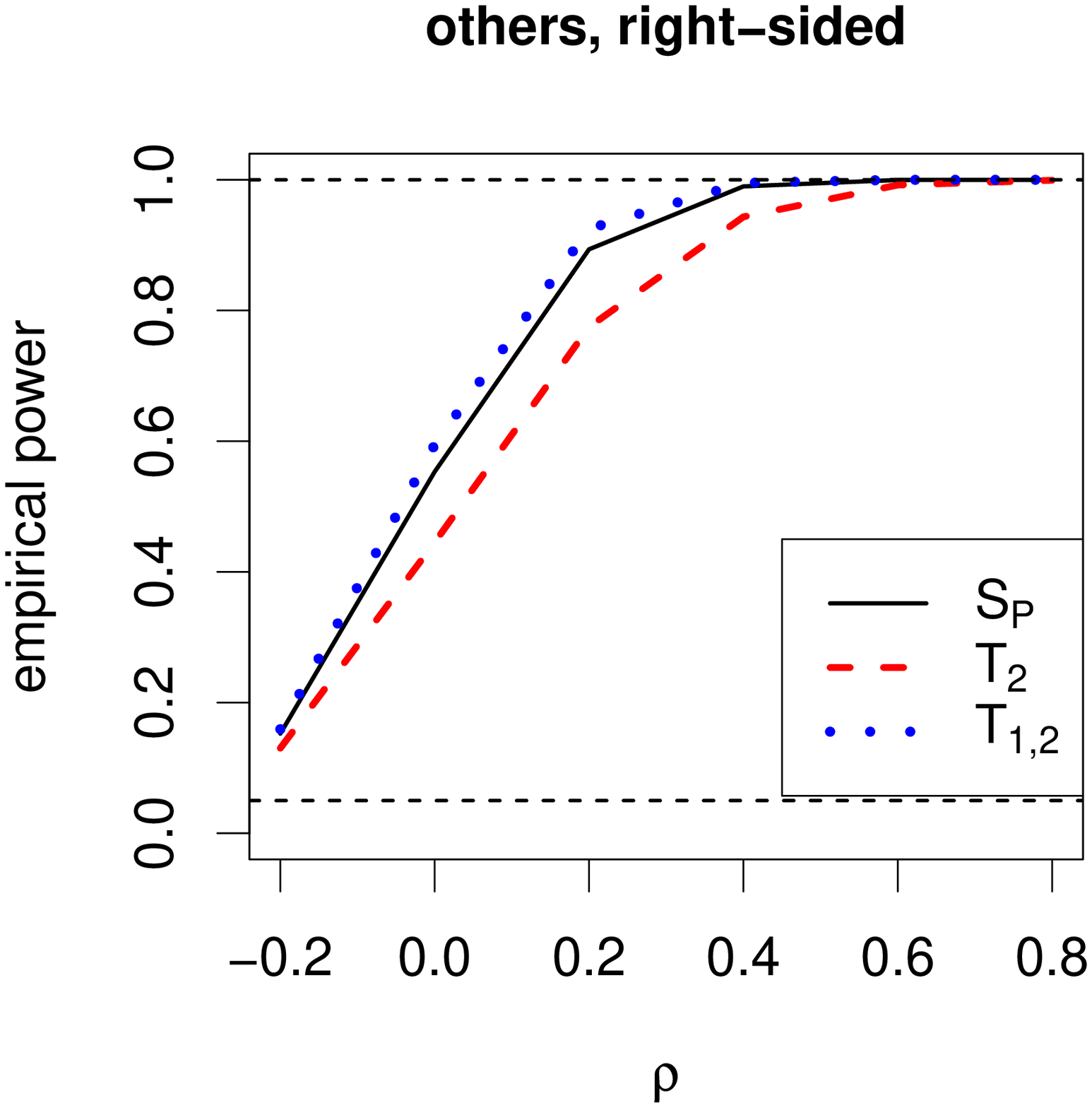} }}
Power Estimates for Case 1(b) with Asymptotic Critical Values\\
\rotatebox{-90}{ \resizebox{2.1 in}{!}{\includegraphics{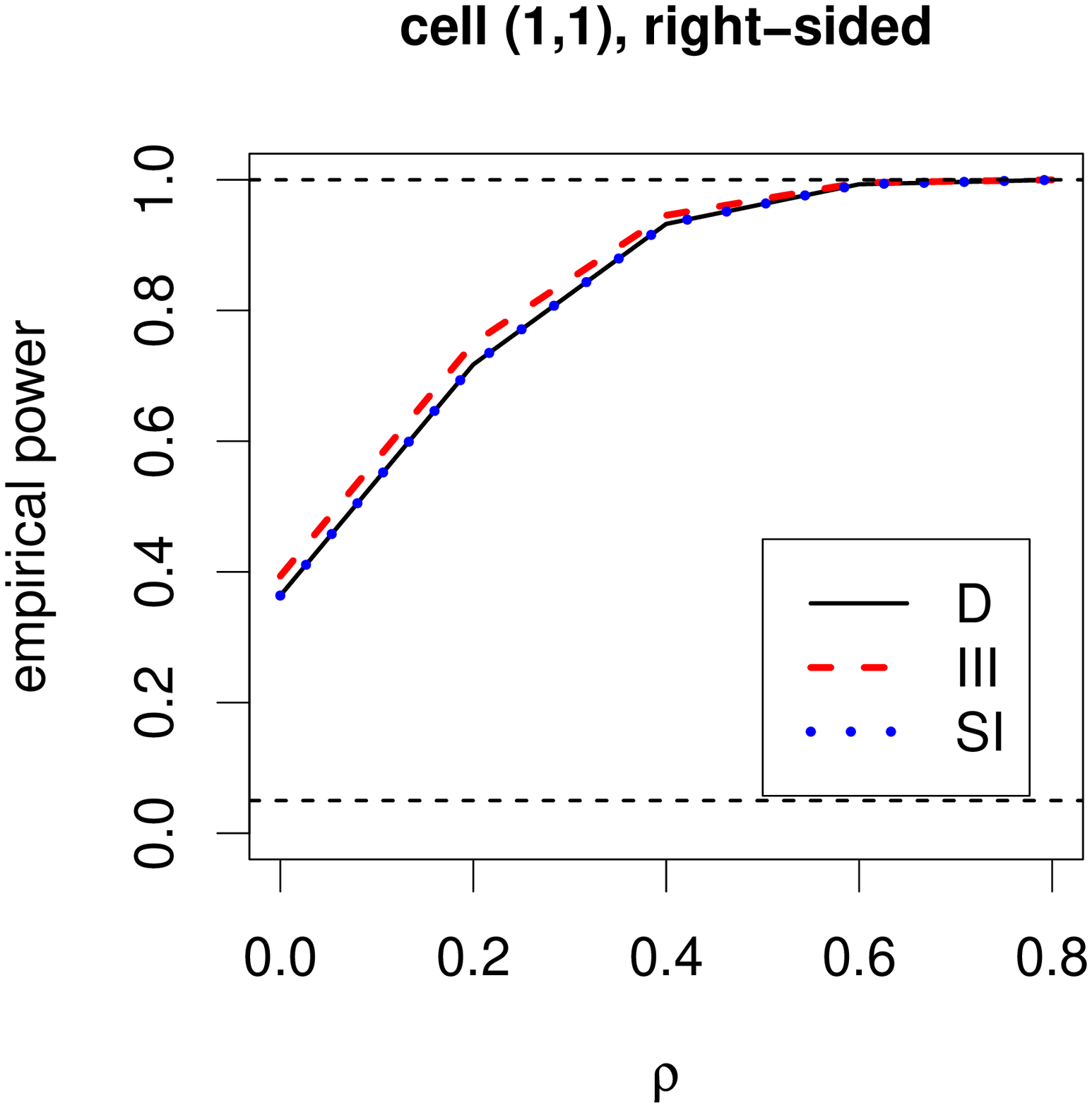} }}
\rotatebox{-90}{ \resizebox{2.1 in}{!}{\includegraphics{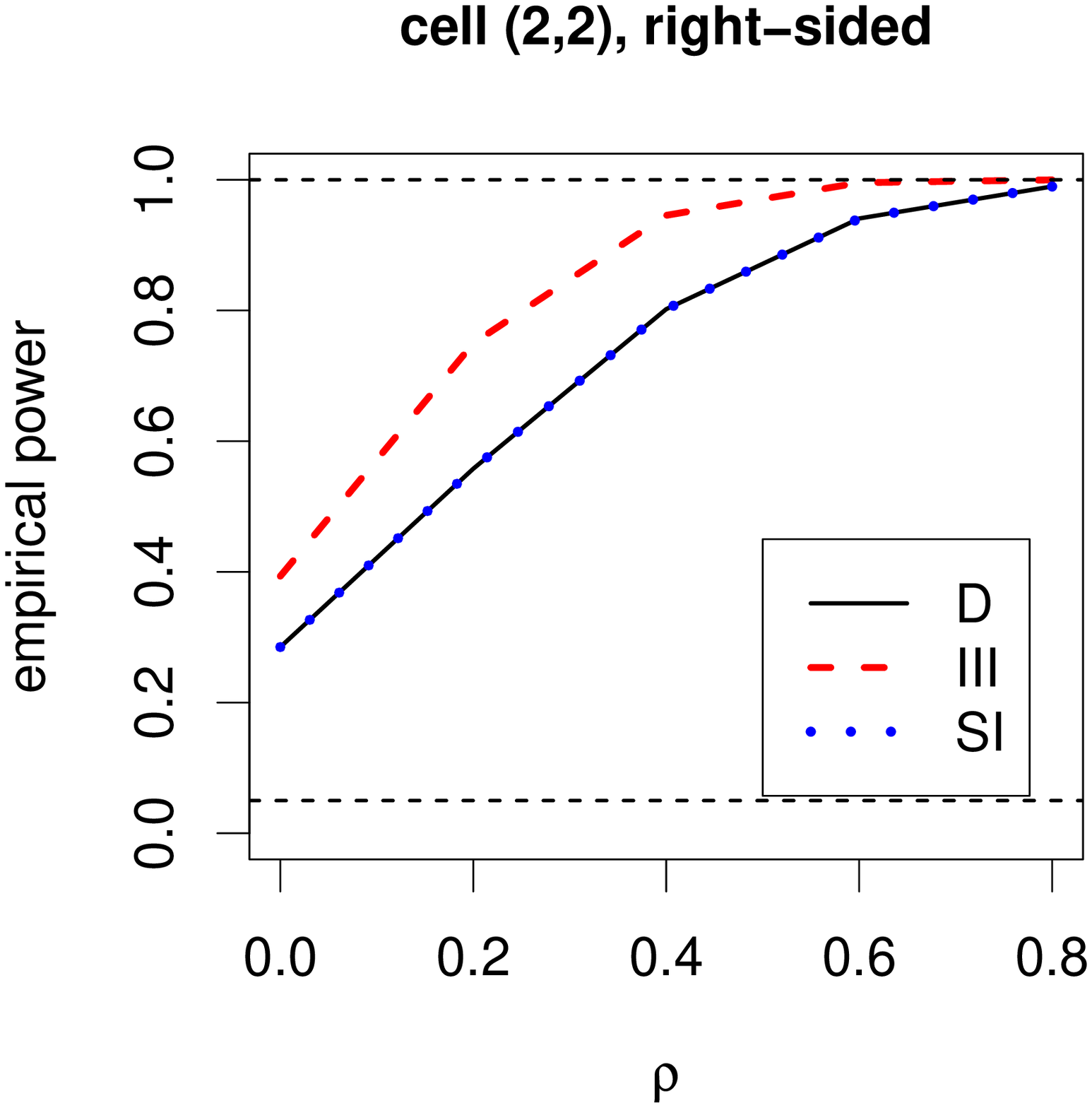} }}
\rotatebox{-90}{ \resizebox{2.1 in}{!}{\includegraphics{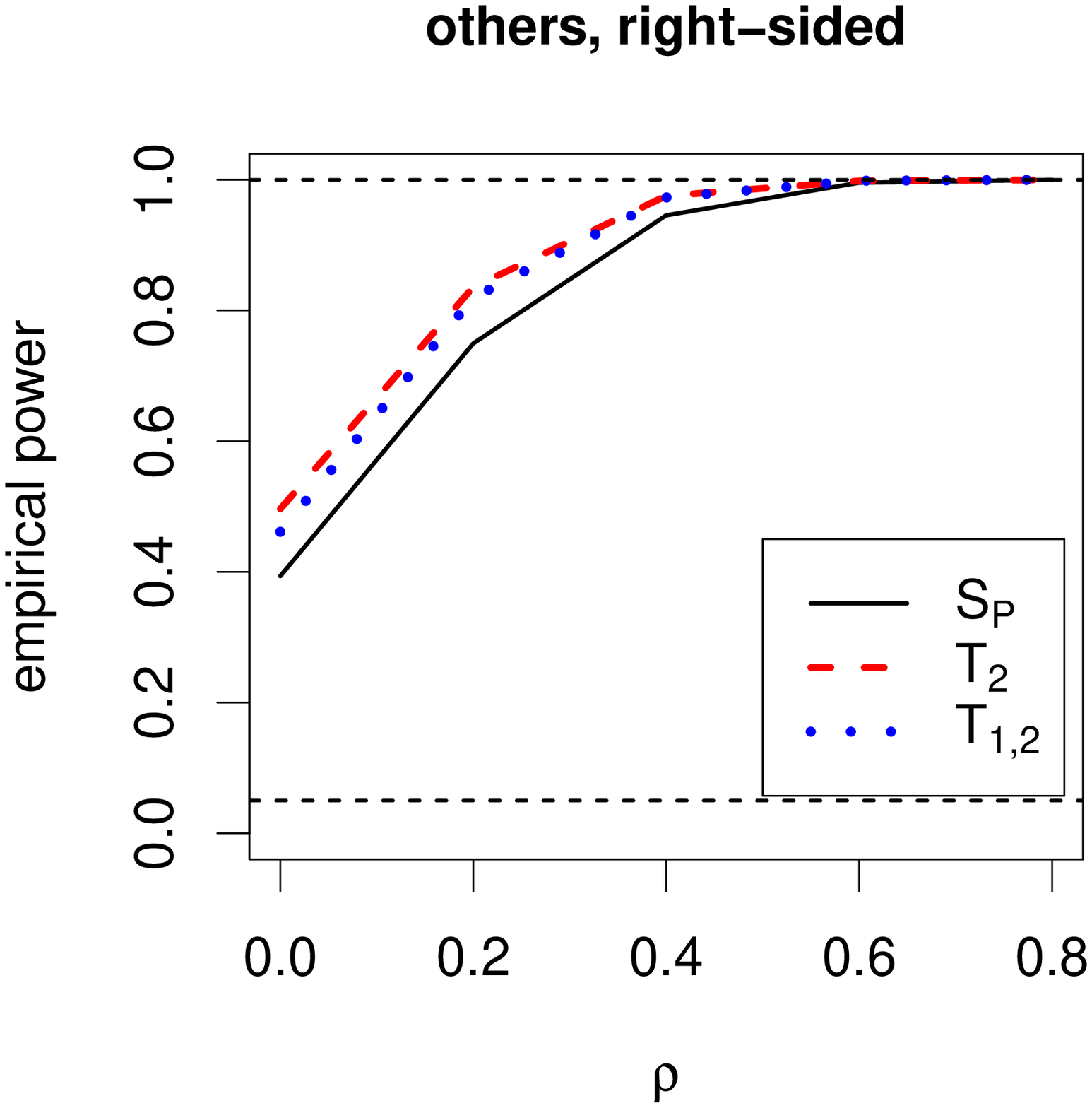} }}
\caption{
\label{fig:power-nonRL-cases1a-b}
Empirical power estimates under the non-RL cases 1(a)-(b).
In case 1(a) (top row), we take $\rho=-0.2,0.0,0.2,\ldots,0.8$ and $k=1$
and
in case 1(b) (bottom row), we take $\rho=0.0,0.2,\ldots,0.8$ and $k=3$.
The dashed horizontal lines are at 0.05 and 1.0
and legend labeling is as in Figure \ref{fig:RS-RL-cases1a-c}.
}
\end{figure}

\begin{figure} [ht]
\centering
\rotatebox{-90}{ \resizebox{2.1 in}{!}{\includegraphics{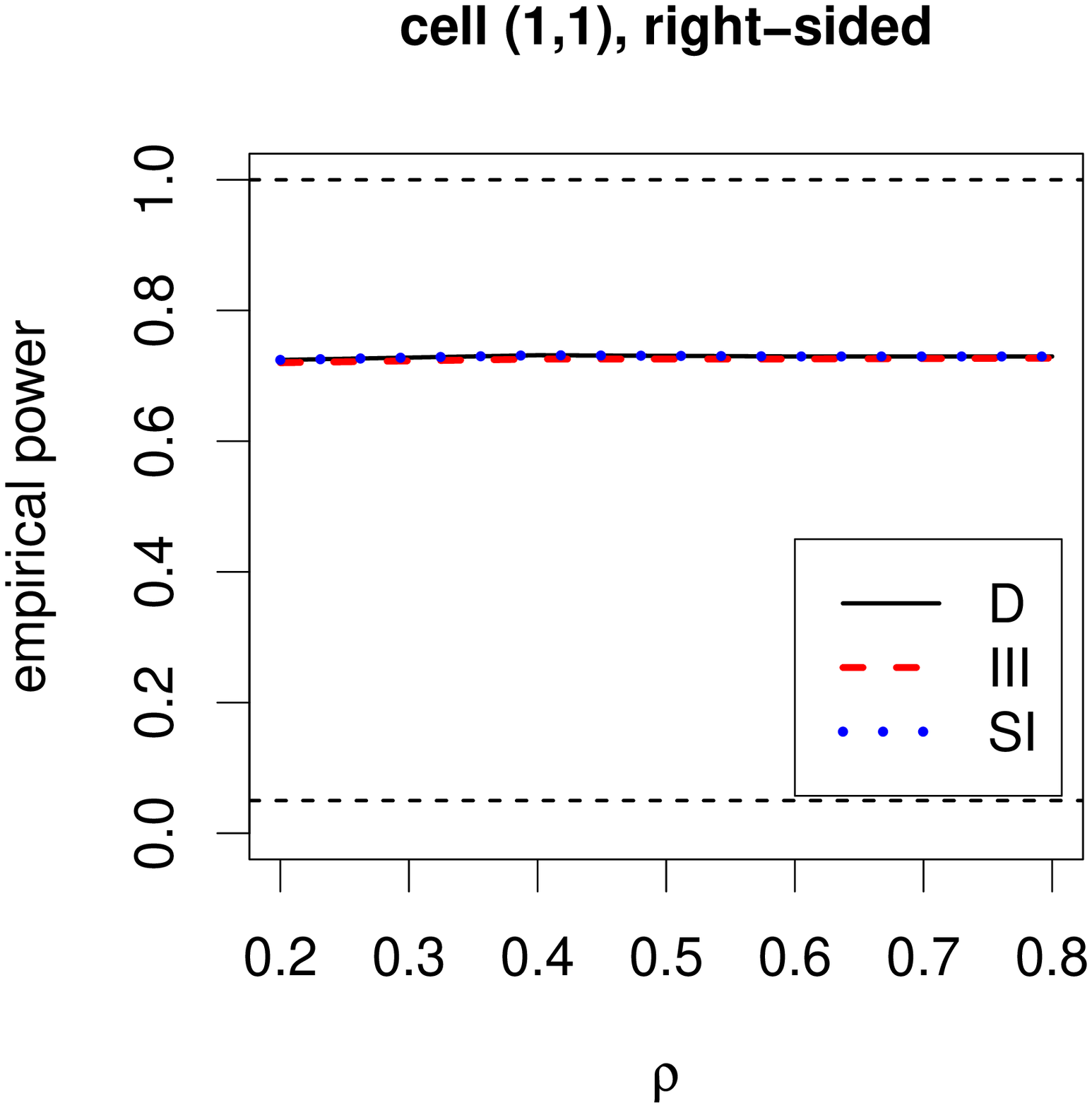} }}
\rotatebox{-90}{ \resizebox{2.1 in}{!}{\includegraphics{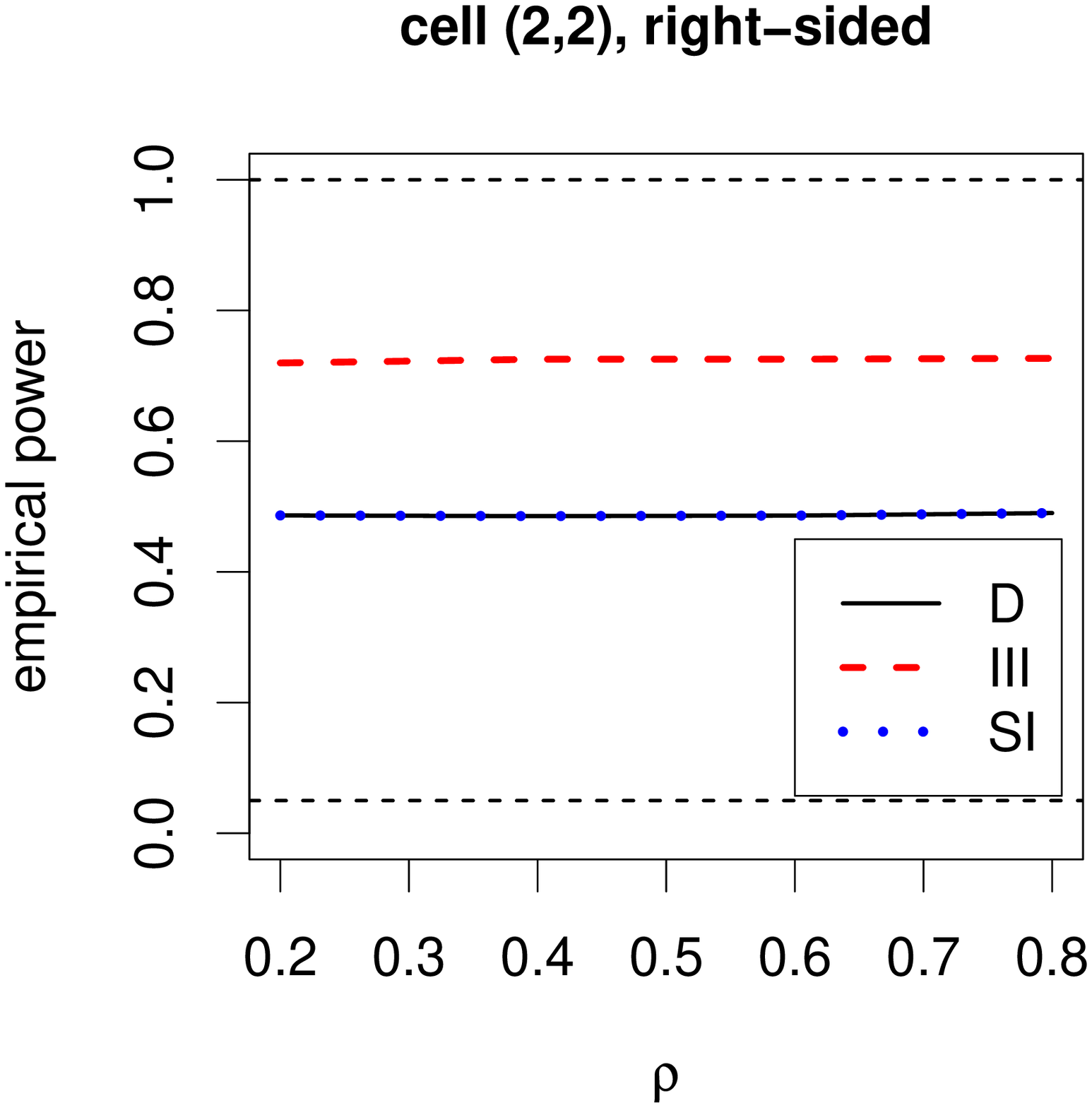} }}
\rotatebox{-90}{ \resizebox{2.1 in}{!}{\includegraphics{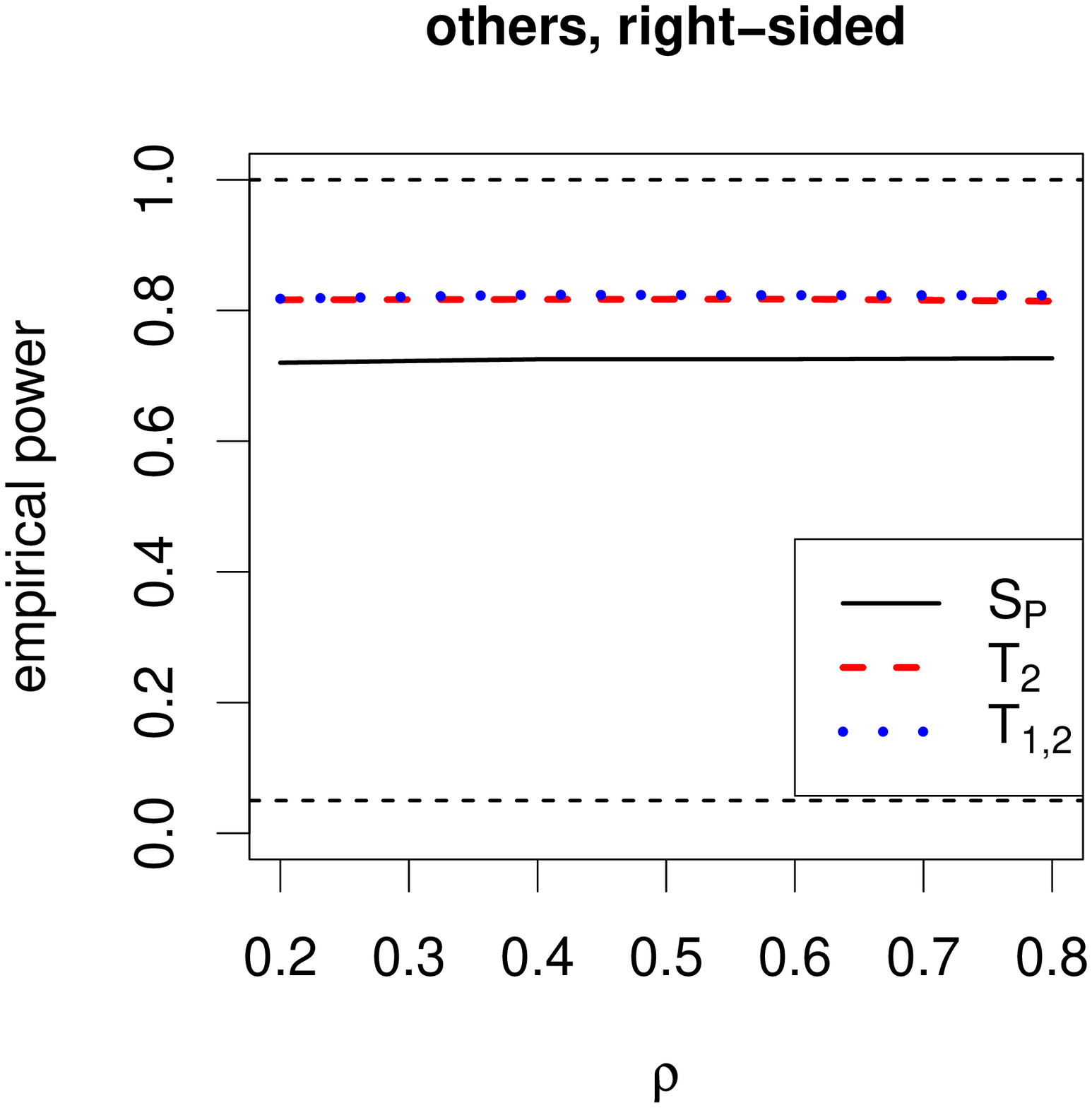} }}
\rotatebox{-90}{ \resizebox{2.1 in}{!}{\includegraphics{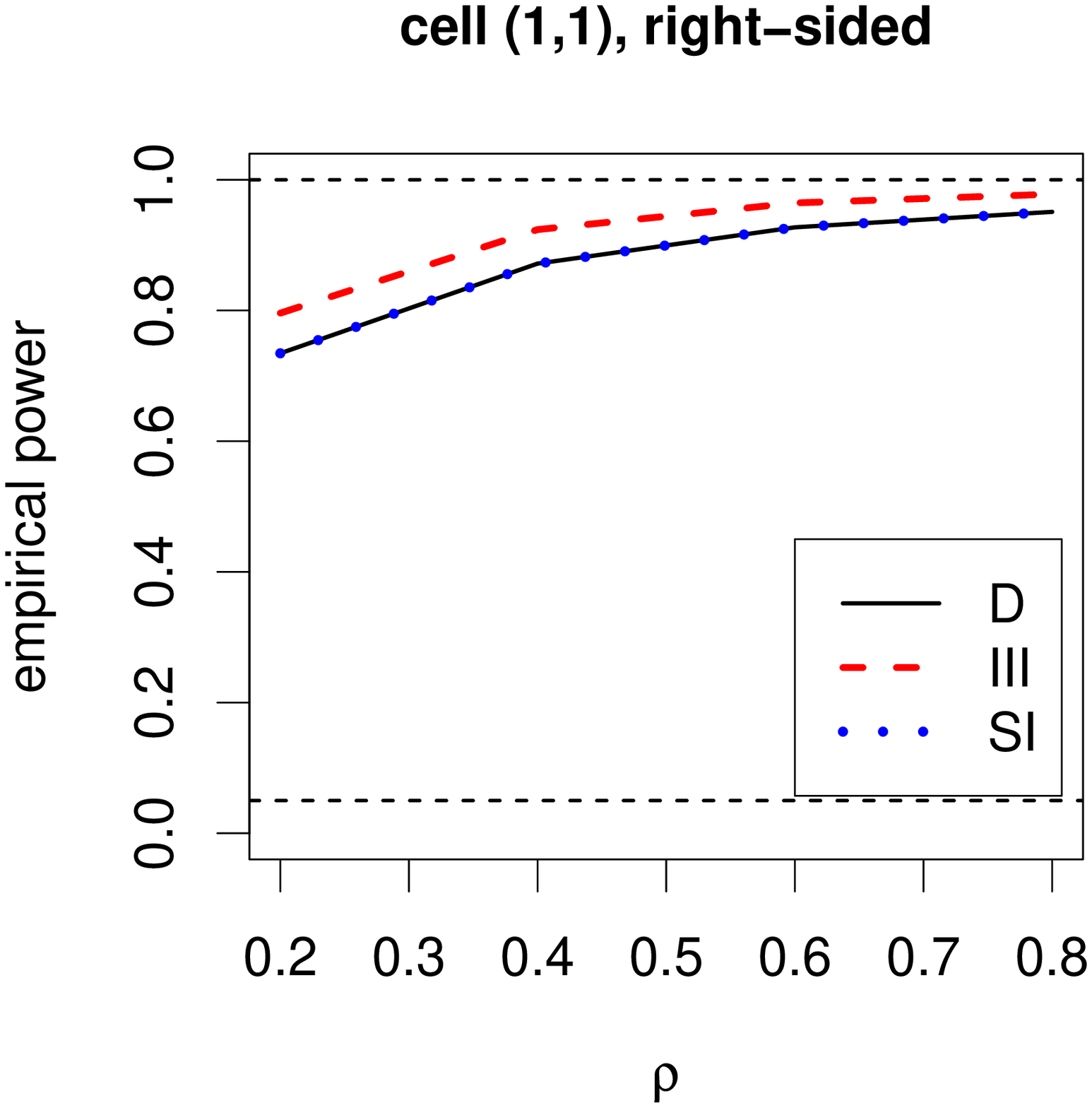} }}
\rotatebox{-90}{ \resizebox{2.1 in}{!}{\includegraphics{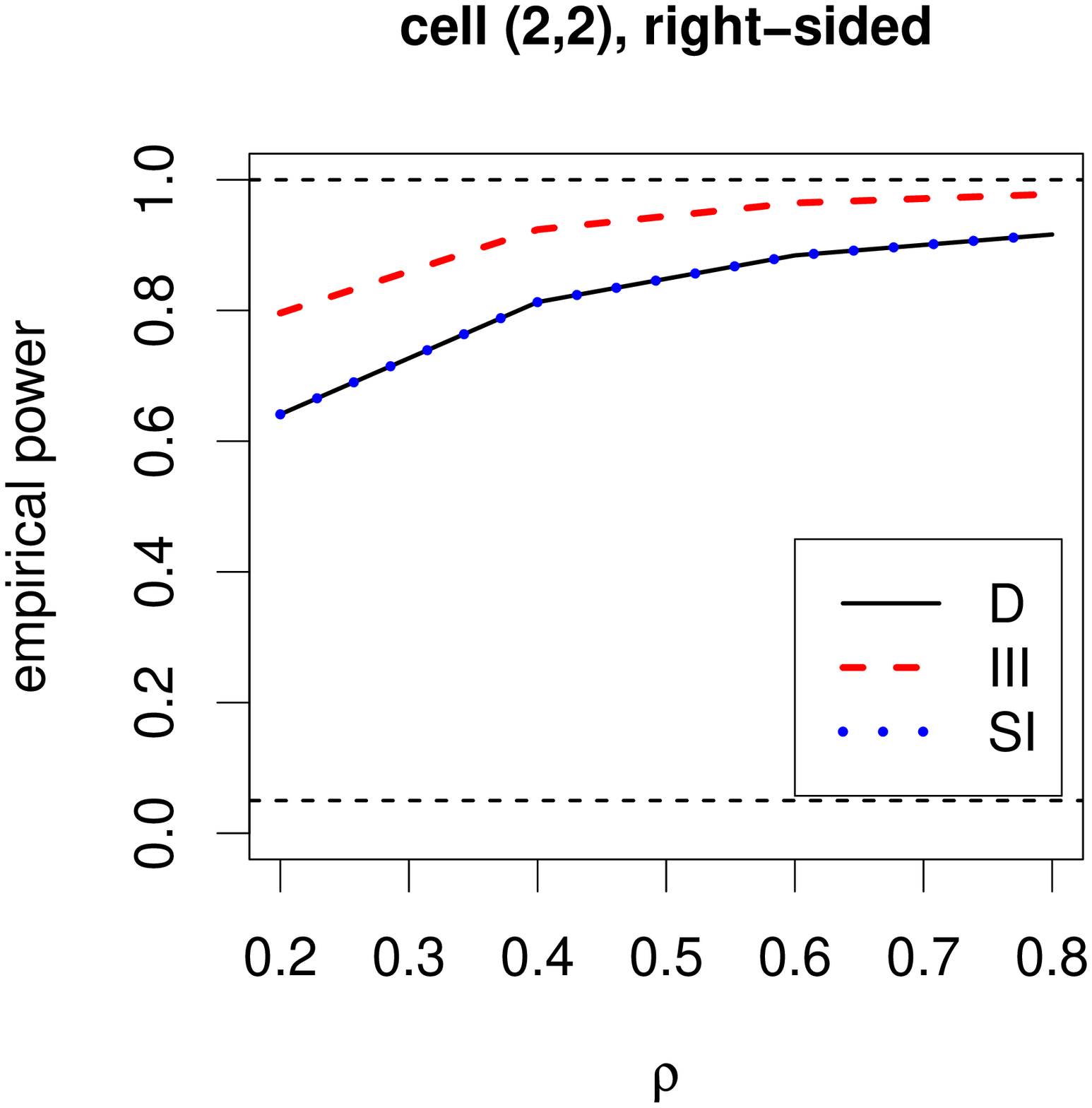} }}
\rotatebox{-90}{ \resizebox{2.1 in}{!}{\includegraphics{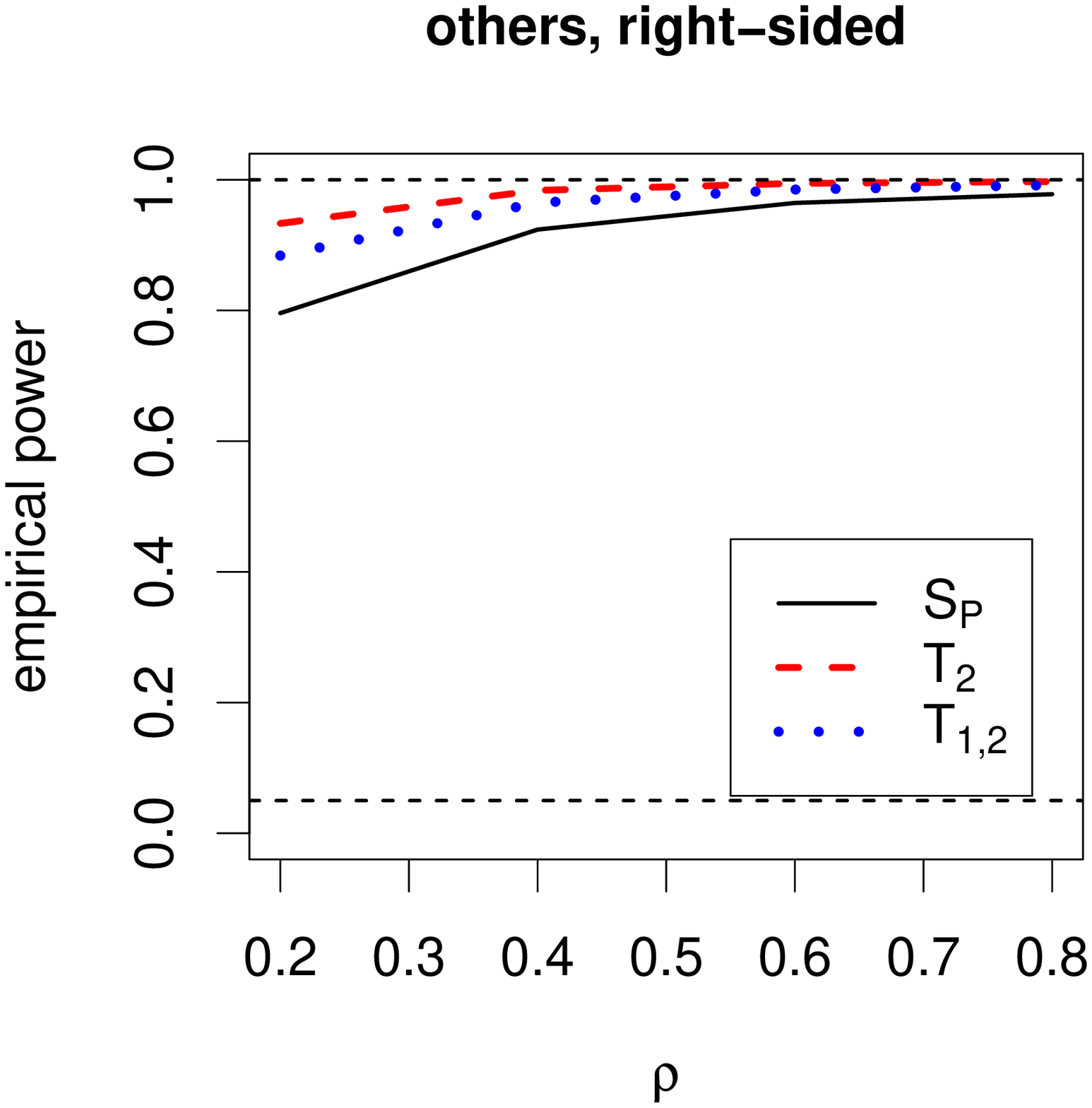} }}
\caption{
\label{fig:power-nonRL-cases2a-b}
Empirical power estimates under the non-RL case 2(a) (top row)
and case 2(b) (bottom row)
with $\pi_i=0.3$, $\pi_u=0.5$, $\rho=0.2,0.4,0.6,0.8$.
The dashed horizontal lines and  the legend labeling in both rows
are as in Figure \ref{fig:power-nonRL-cases1a-b}.
}
\end{figure}

\begin{figure} [hbp]
\centering
\rotatebox{-90}{ \resizebox{2.1 in}{!}{\includegraphics{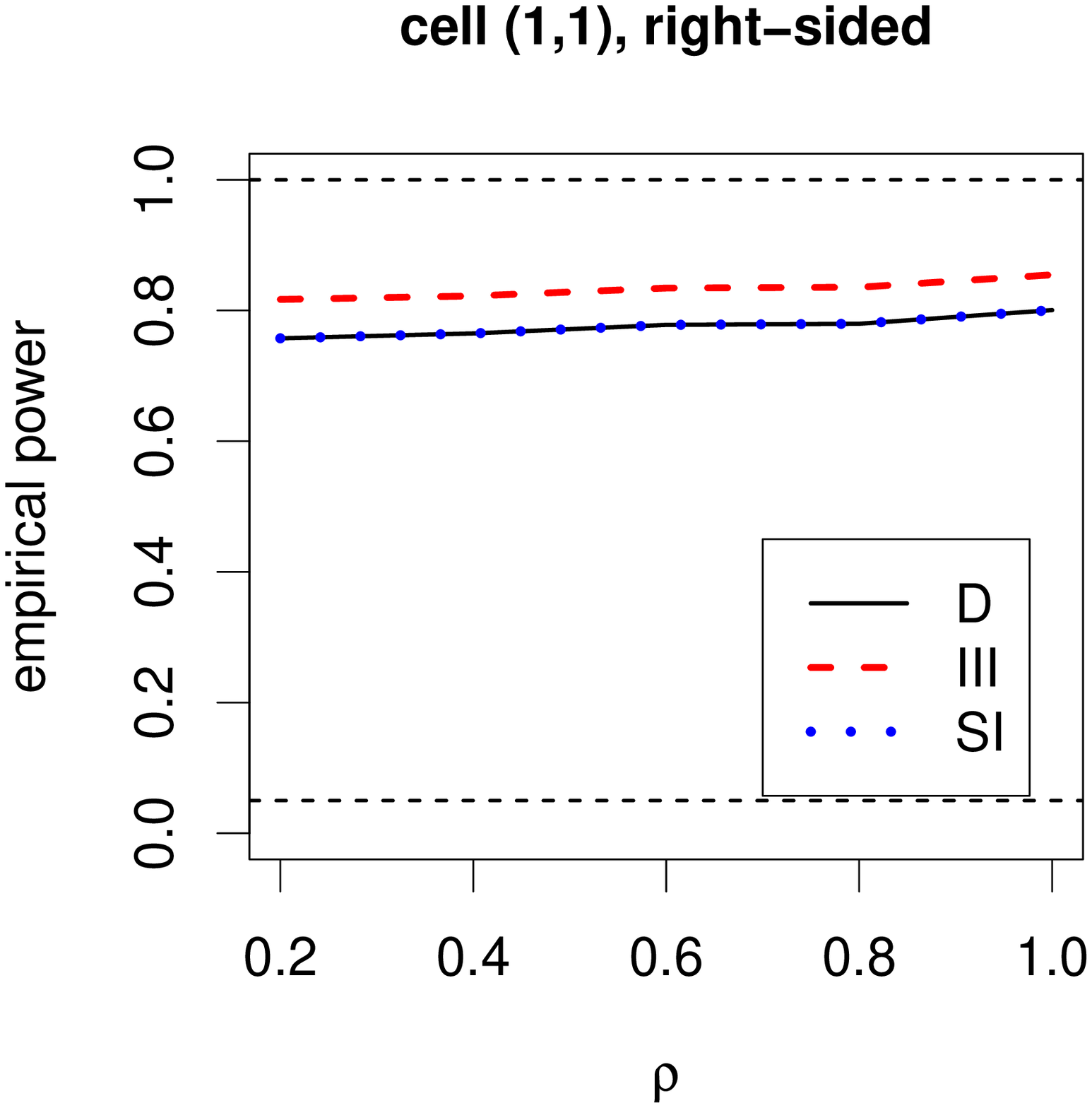} }}
\rotatebox{-90}{ \resizebox{2.1 in}{!}{\includegraphics{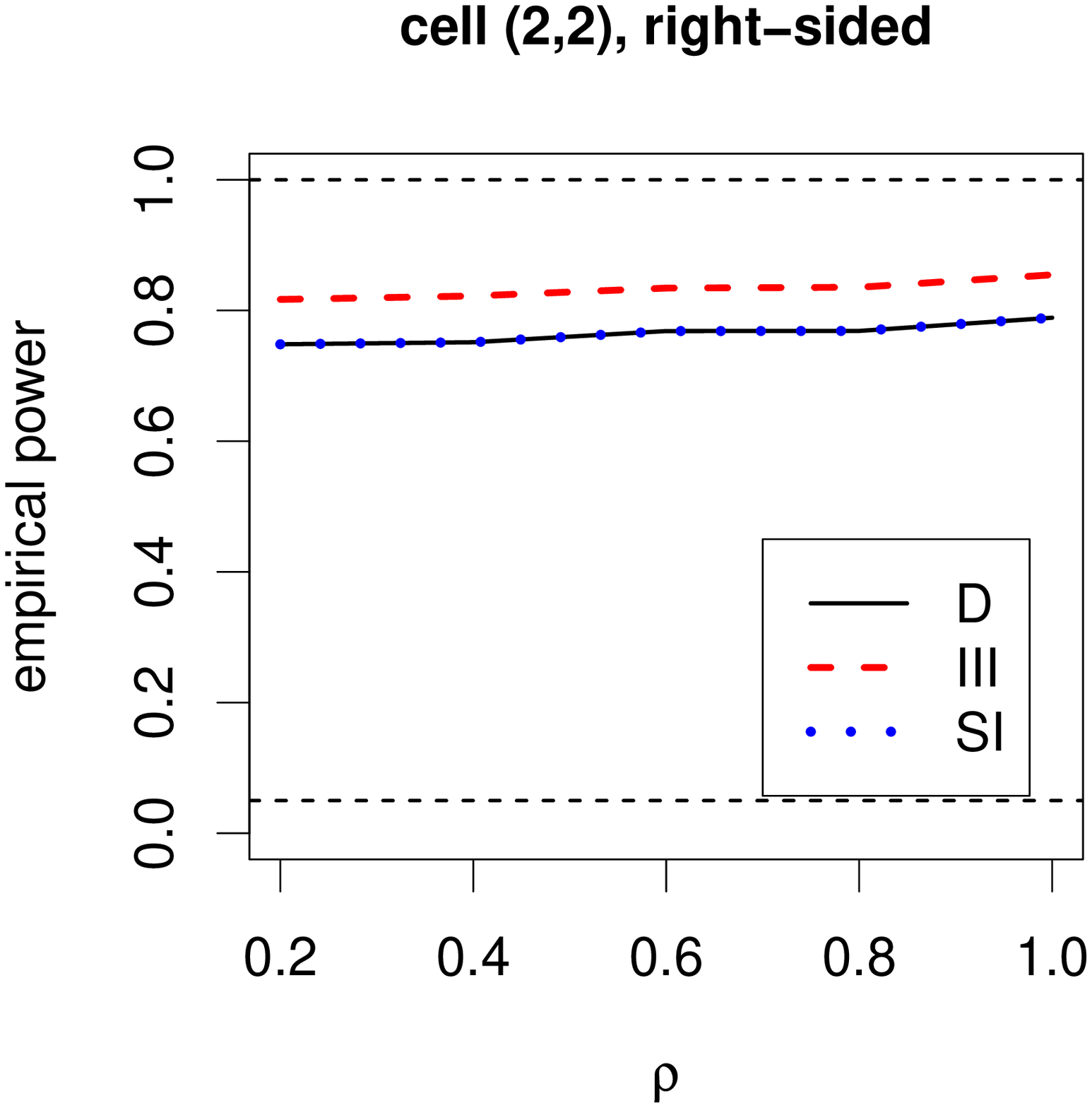} }}
\rotatebox{-90}{ \resizebox{2.1 in}{!}{\includegraphics{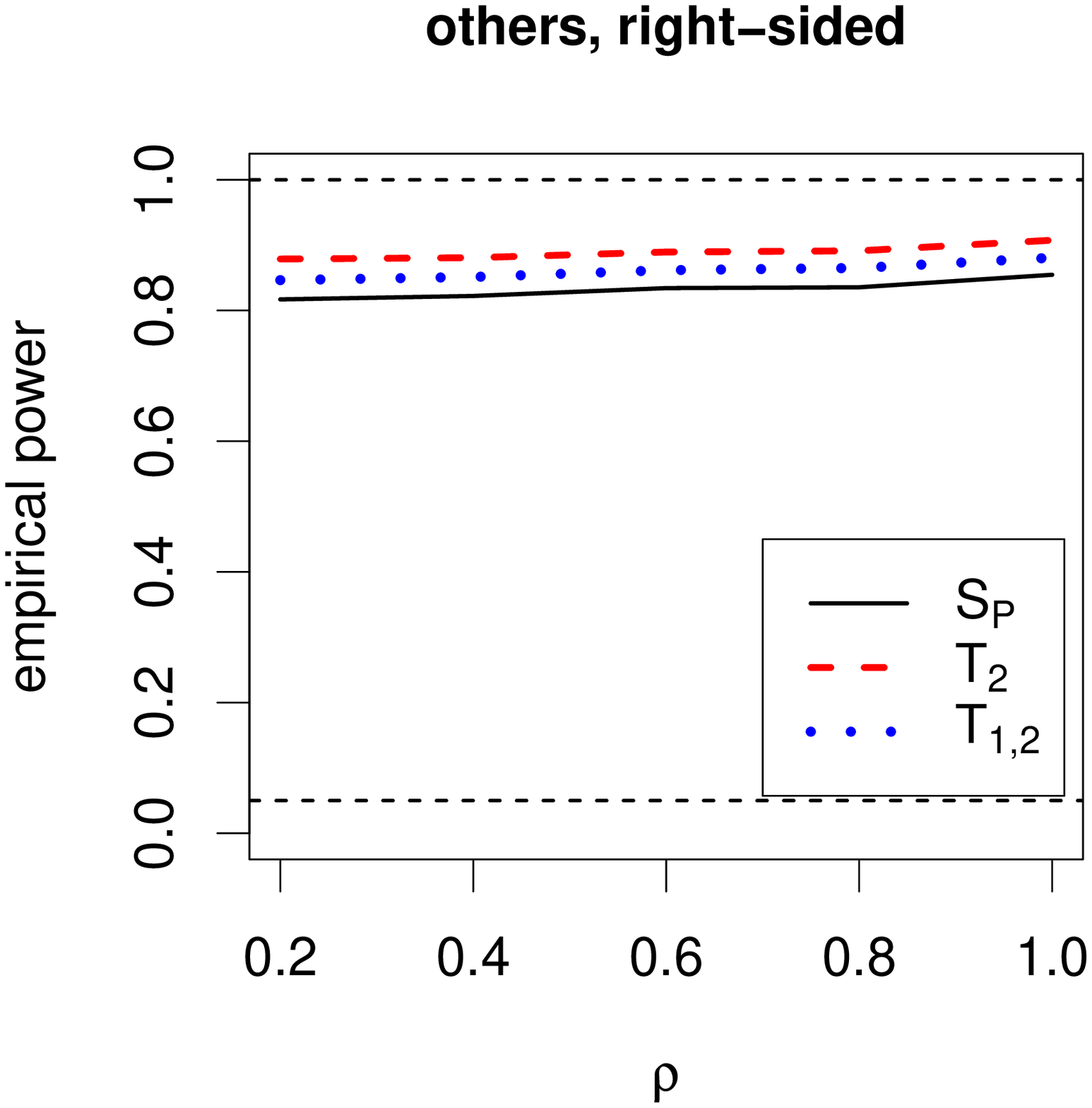} }}
\rotatebox{-90}{ \resizebox{2.1 in}{!}{\includegraphics{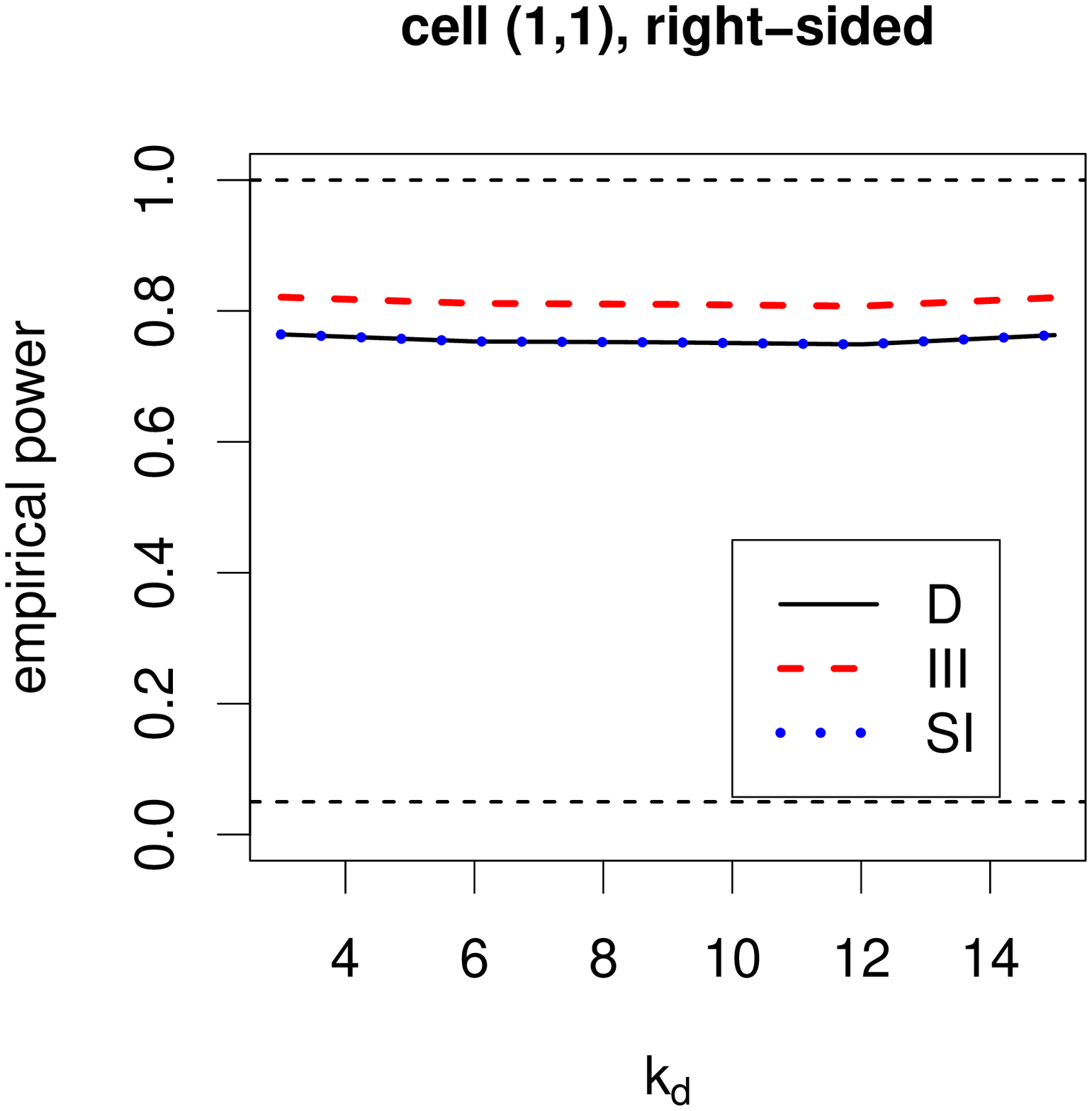} }}
\rotatebox{-90}{ \resizebox{2.1 in}{!}{\includegraphics{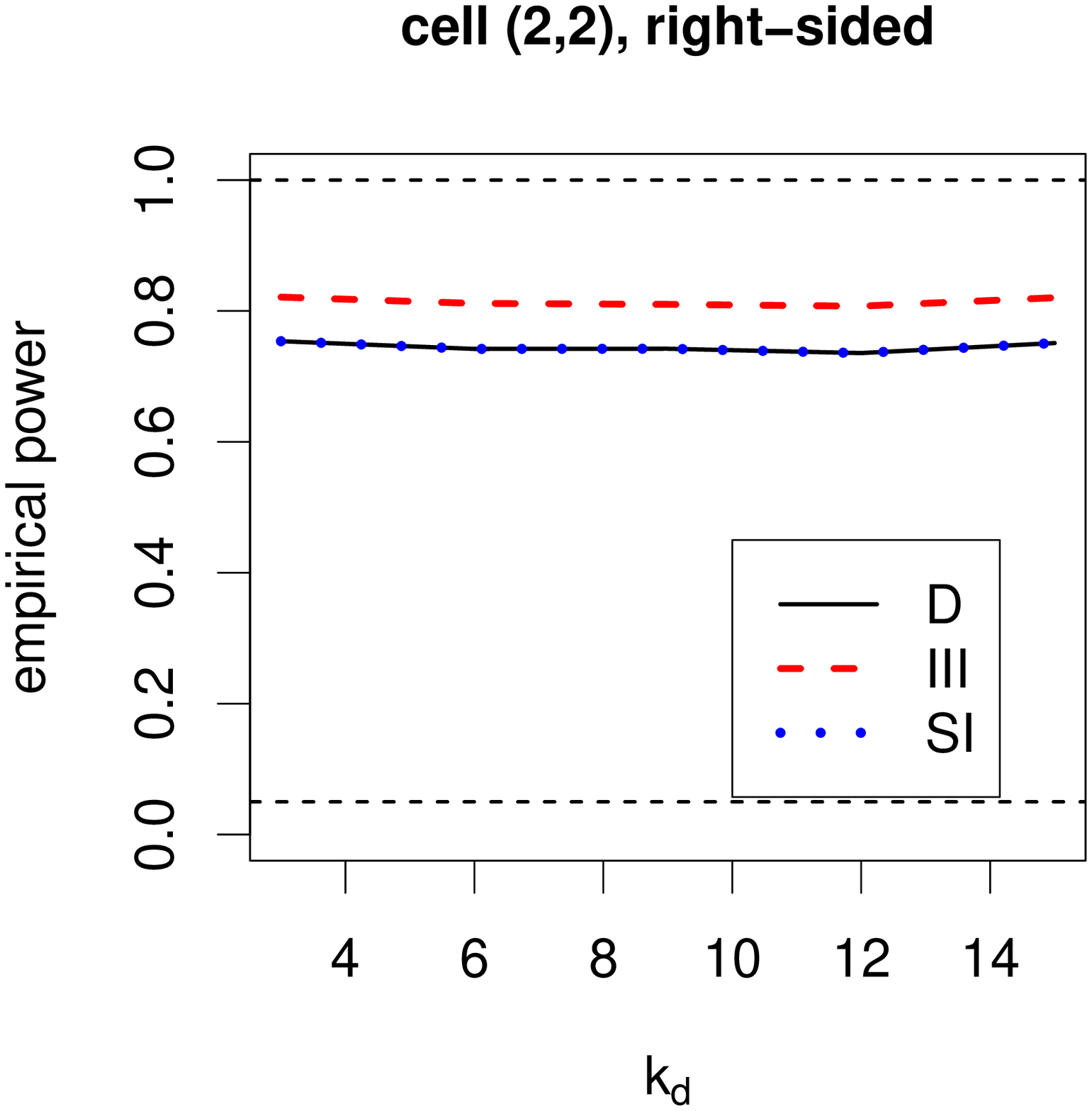} }}
\rotatebox{-90}{ \resizebox{2.1 in}{!}{\includegraphics{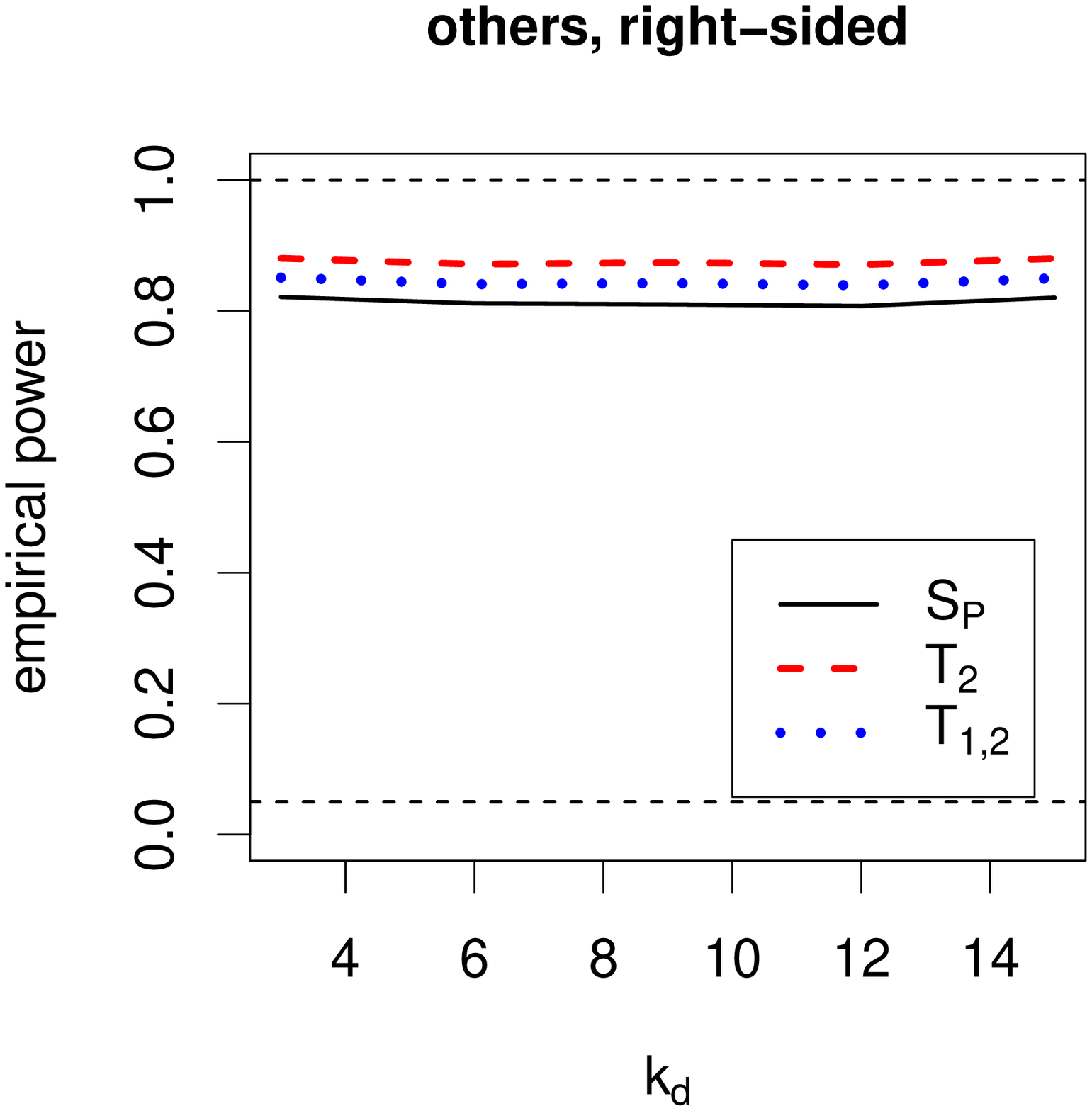} }}
\rotatebox{-90}{ \resizebox{2.1 in}{!}{\includegraphics{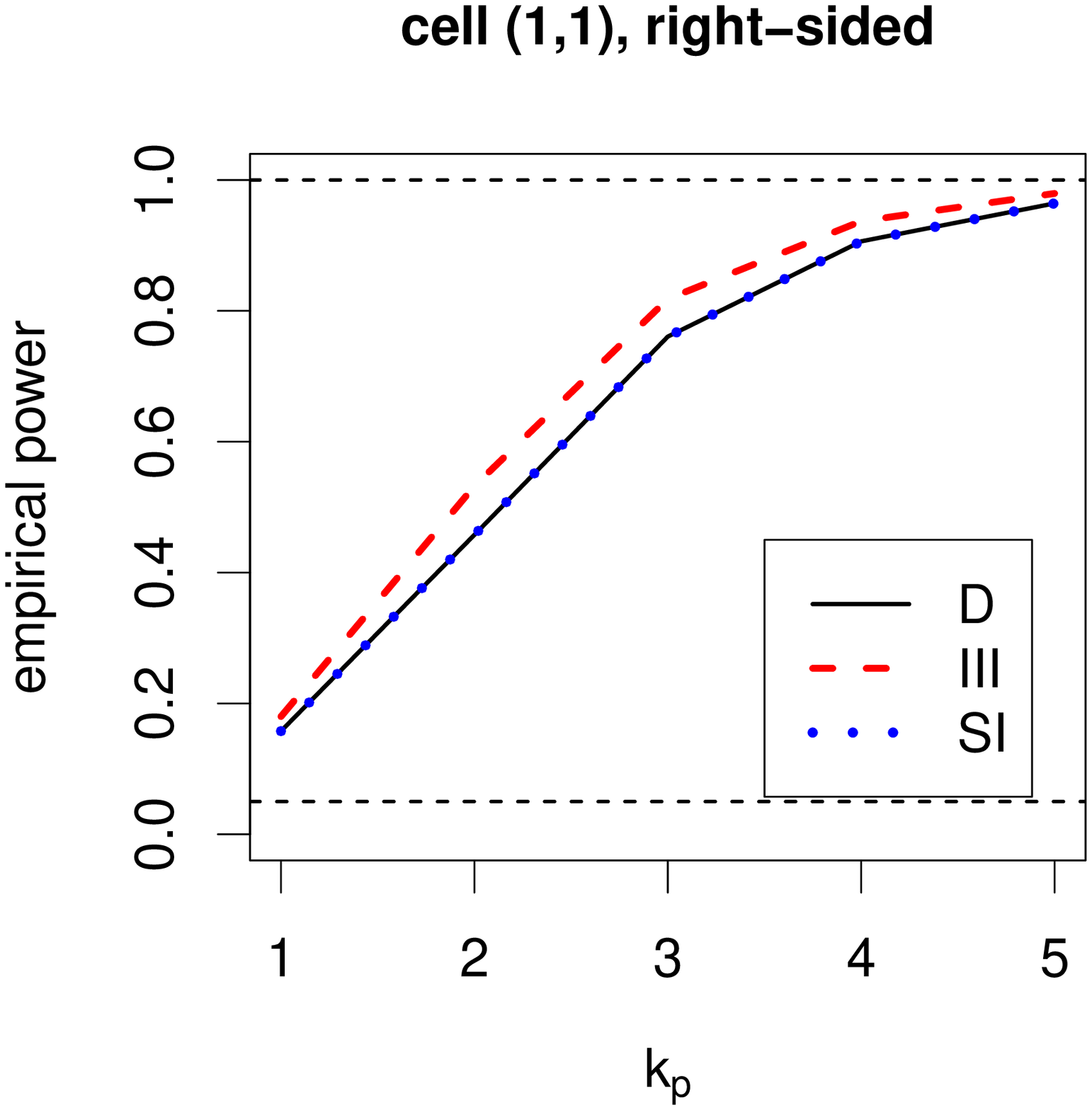} }}
\rotatebox{-90}{ \resizebox{2.1 in}{!}{\includegraphics{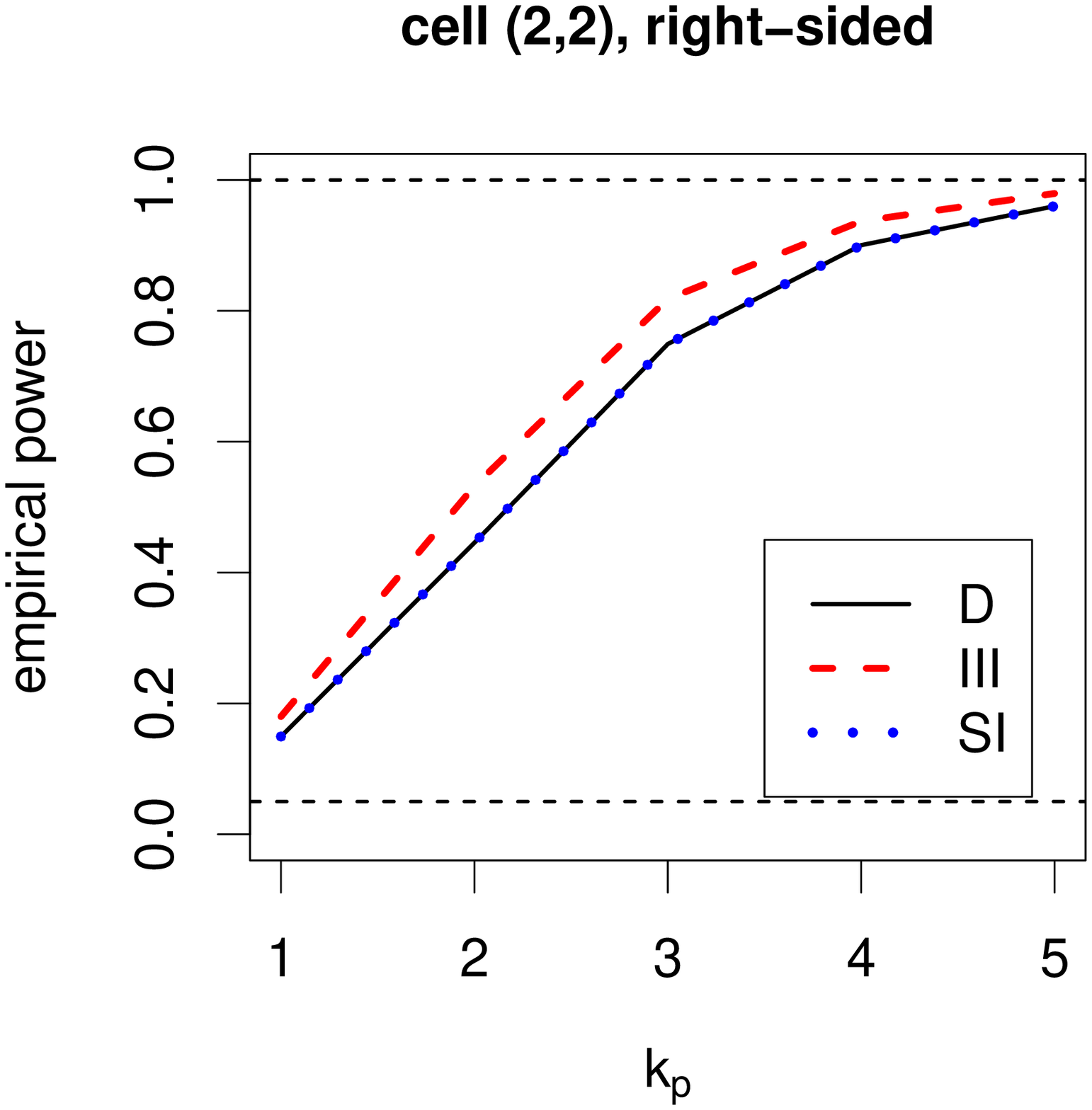} }}
\rotatebox{-90}{ \resizebox{2.1 in}{!}{\includegraphics{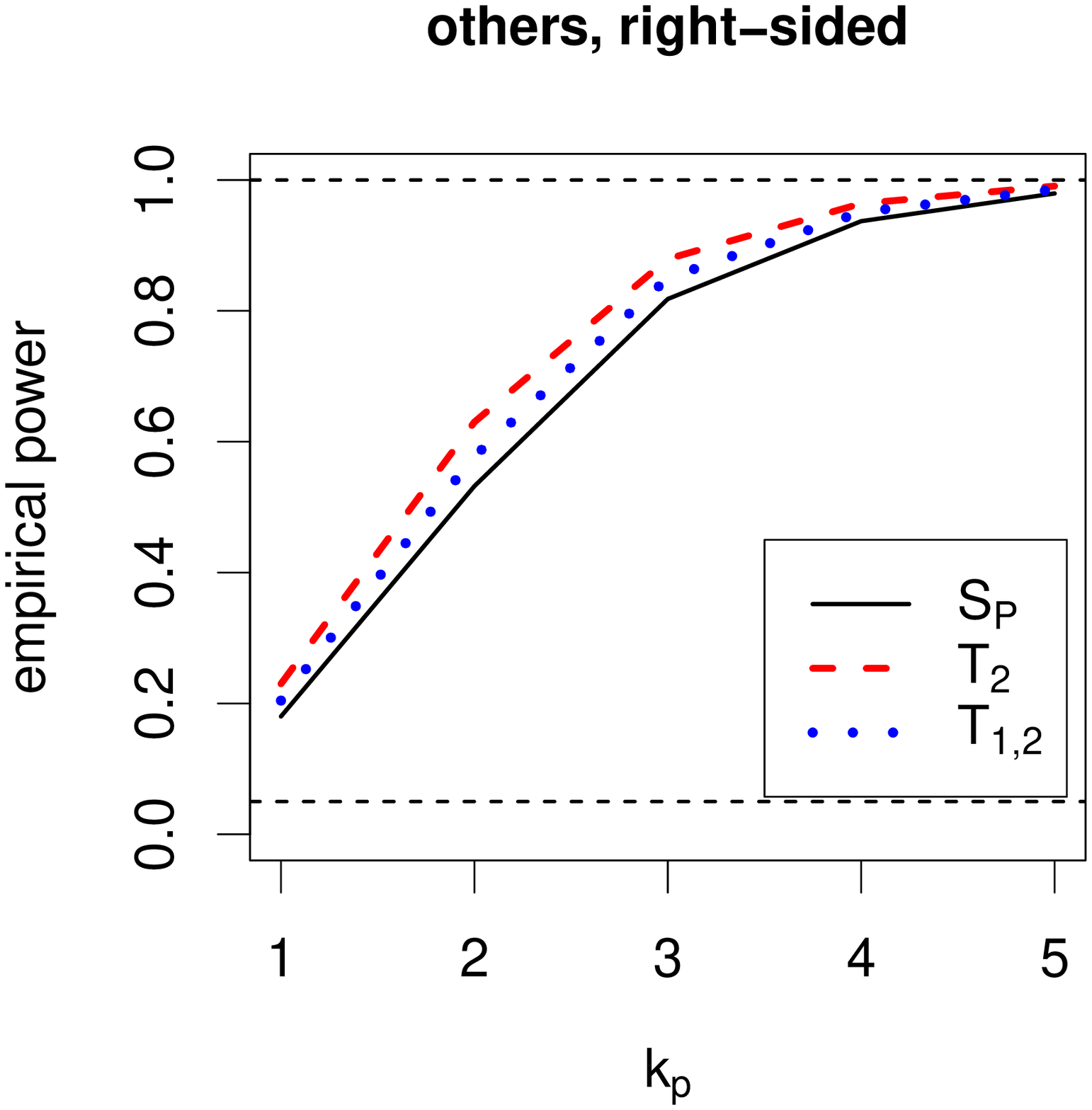} }}
\caption{
\label{fig:power-nonRL-cases3a-c}
Empirical power estimates under the non-RL cases 3(a)-(c).
In case 3(a) (top row), we take $\rho=0.2,0.4,\ldots,1.0$, $k_p=3$, and $k_d=1$,
in case 3(b) (middle row), we take $\rho=0.8$, $k_p=3$, and $k_d=3,6,\ldots,15$,
and
in case 3(c) (bottom row) we take $\rho=0.8$, $k_p=1,2,\ldots,5$, and $k_d=1$.
The dashed horizontal line and legend labeling are as in Figure \ref{fig:power-nonRL-cases1a-b}.
}
\end{figure}

\begin{figure} [ht]
\centering
\rotatebox{-90}{ \resizebox{2.1 in}{!}{\includegraphics{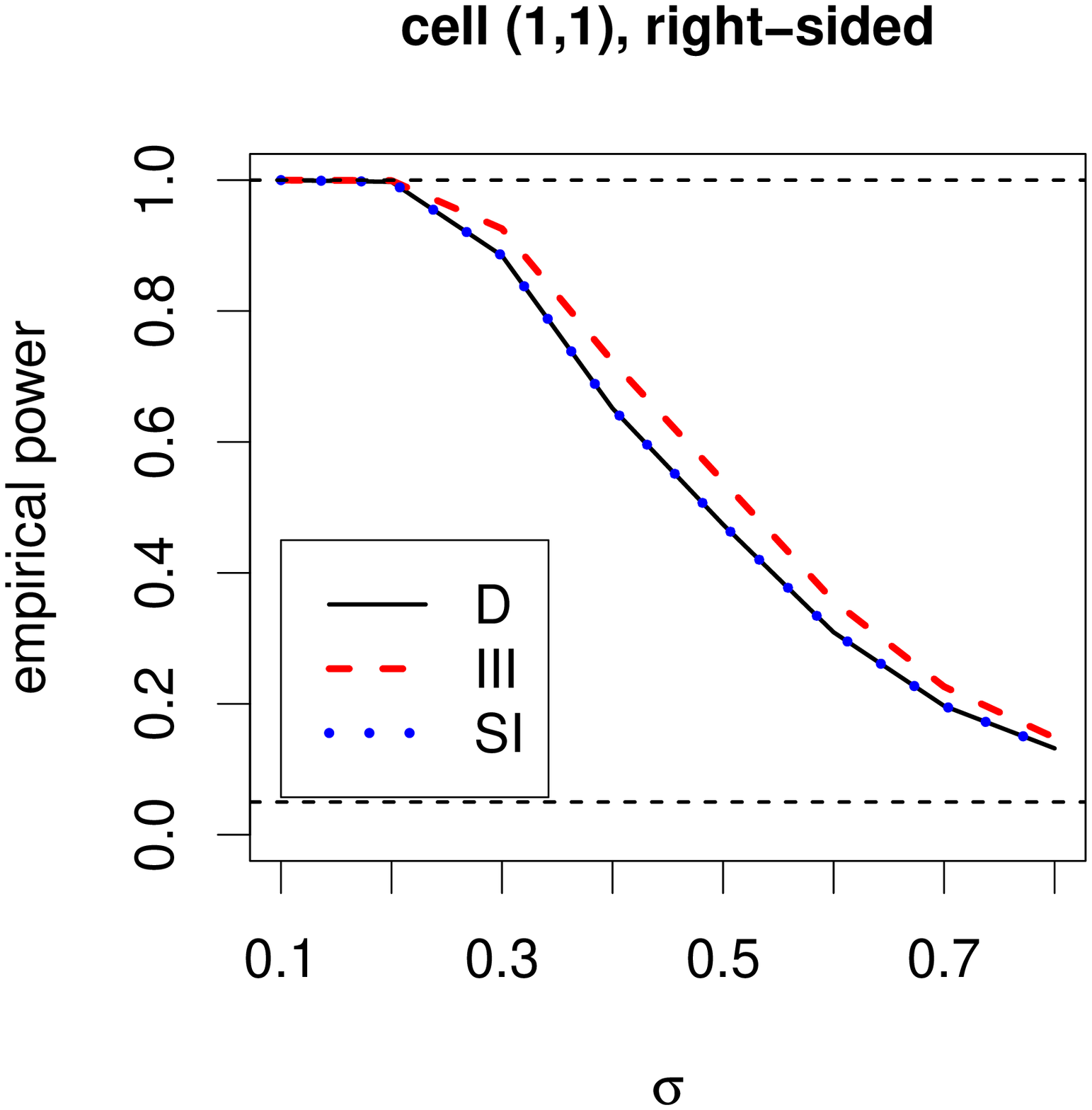} }}
\rotatebox{-90}{ \resizebox{2.1 in}{!}{\includegraphics{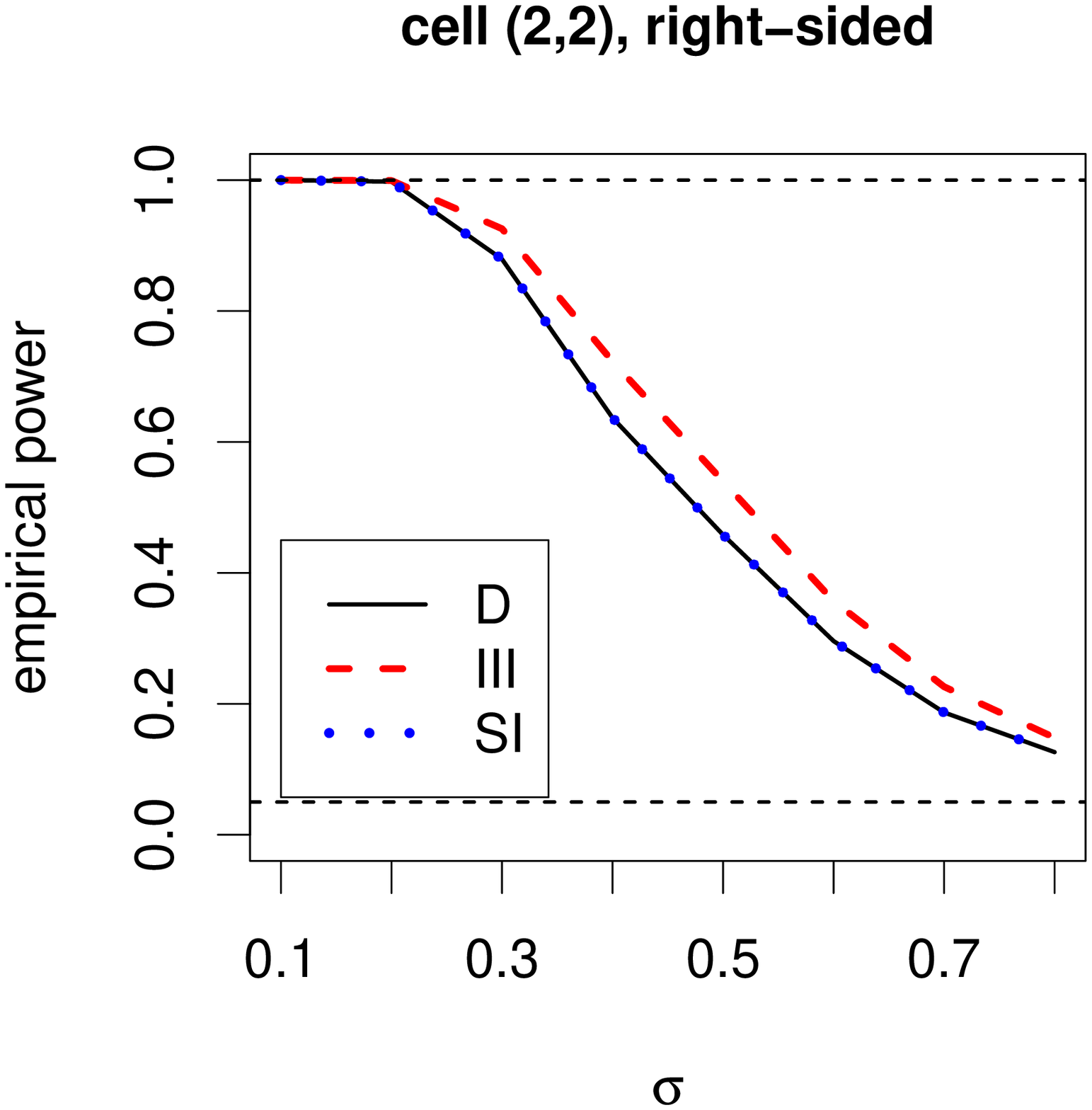} }}
\rotatebox{-90}{ \resizebox{2.1 in}{!}{\includegraphics{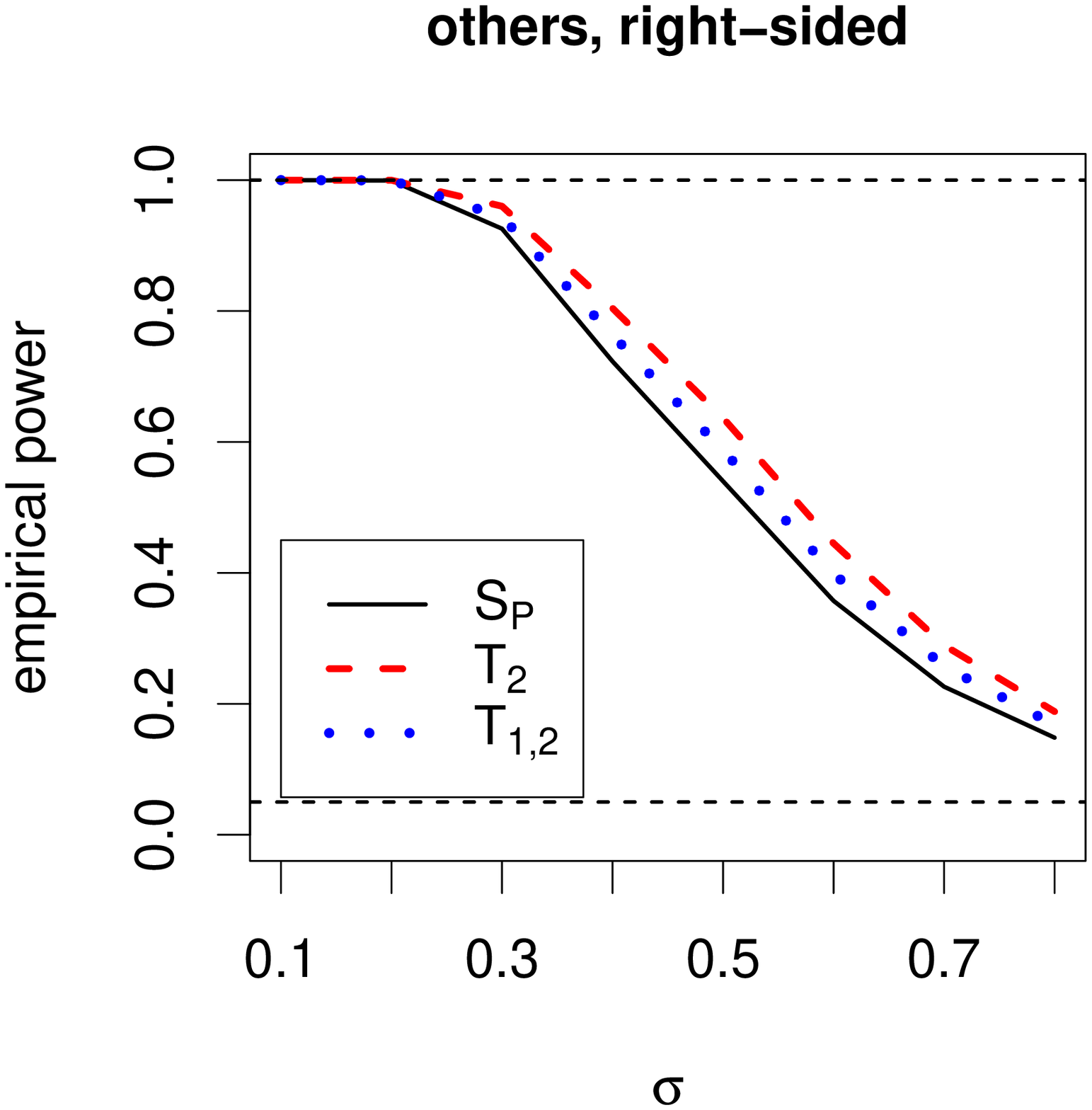} }}
\rotatebox{-90}{ \resizebox{2.1 in}{!}{\includegraphics{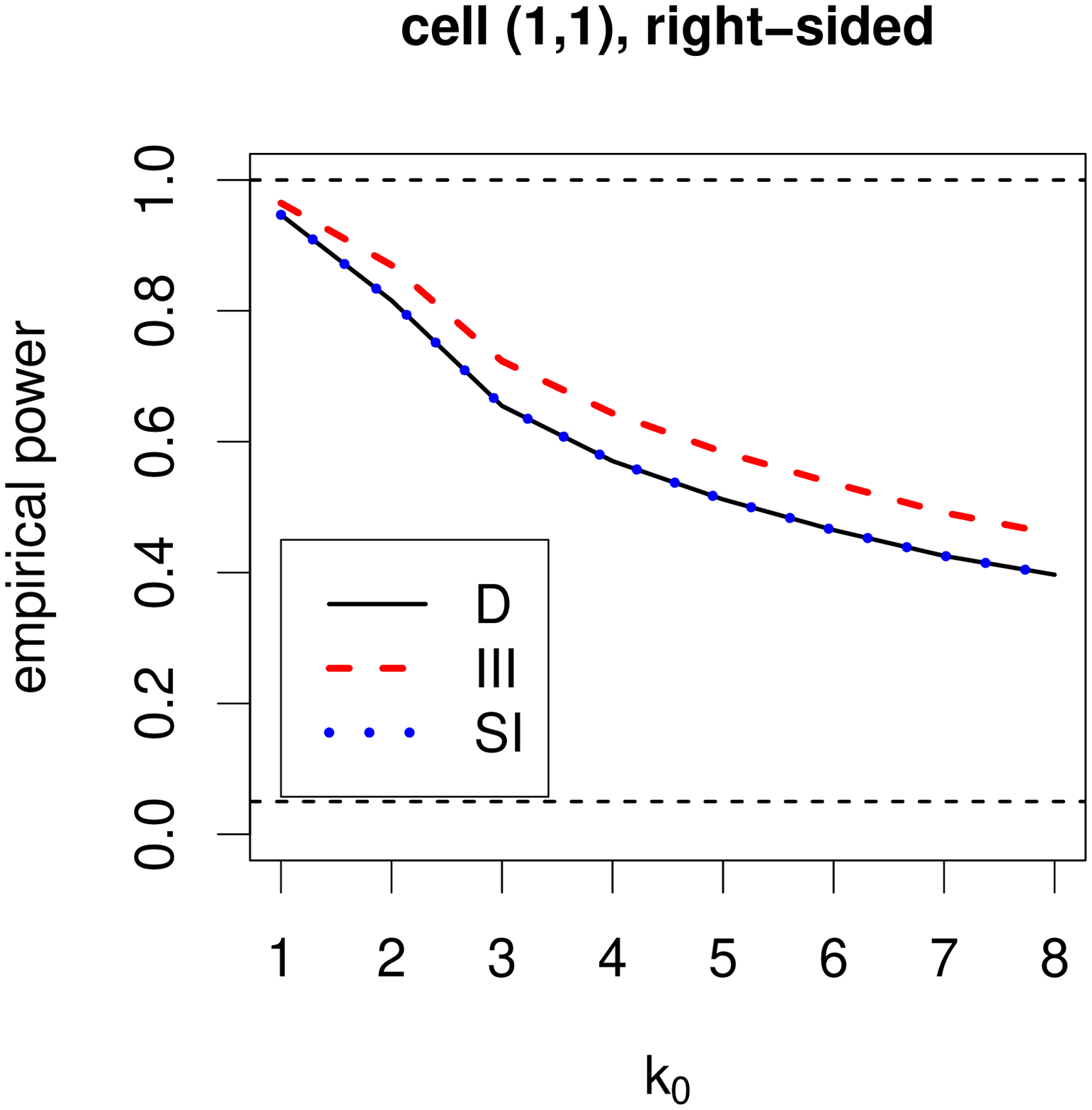} }}
\rotatebox{-90}{ \resizebox{2.1 in}{!}{\includegraphics{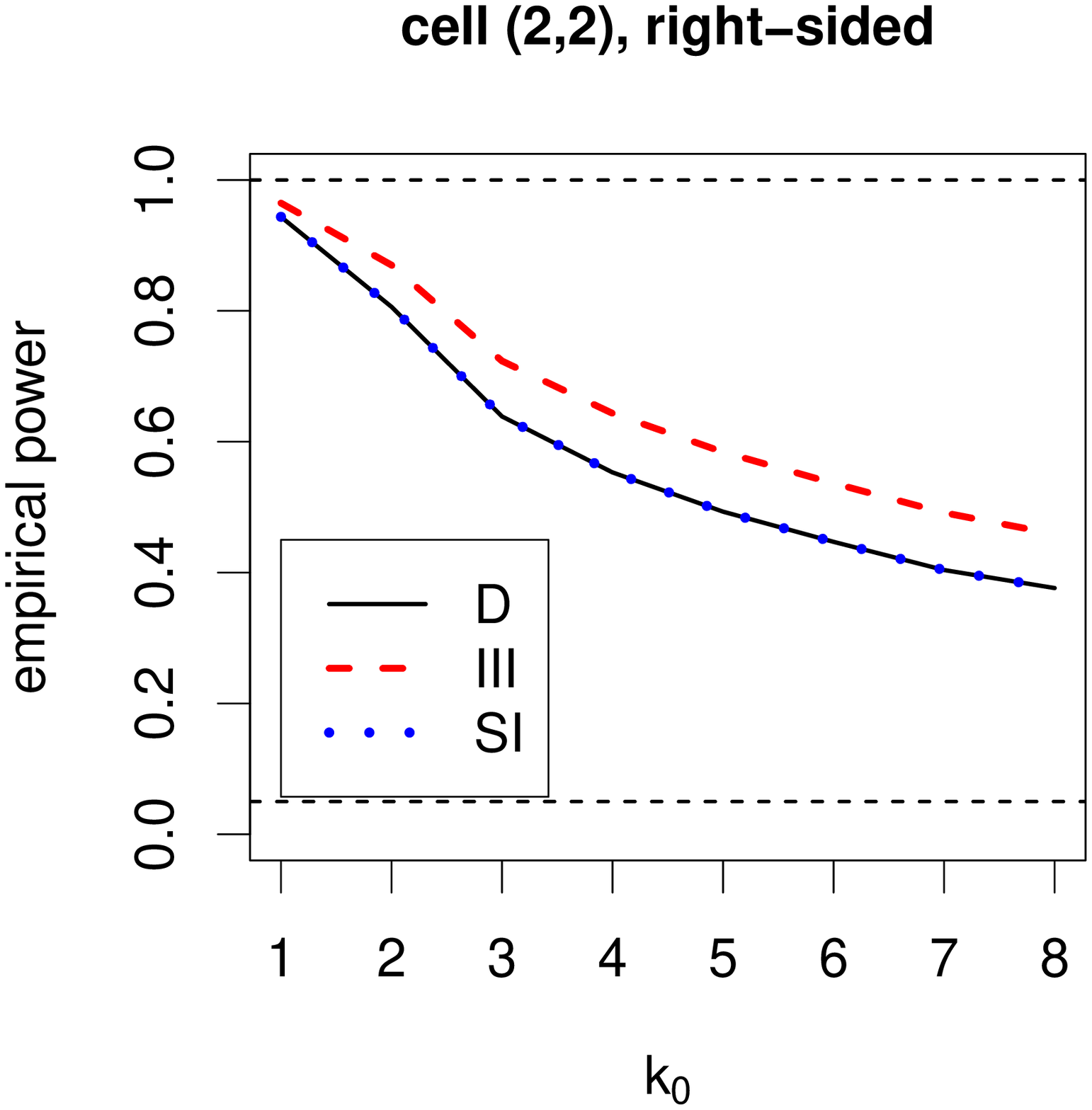} }}
\rotatebox{-90}{ \resizebox{2.1 in}{!}{\includegraphics{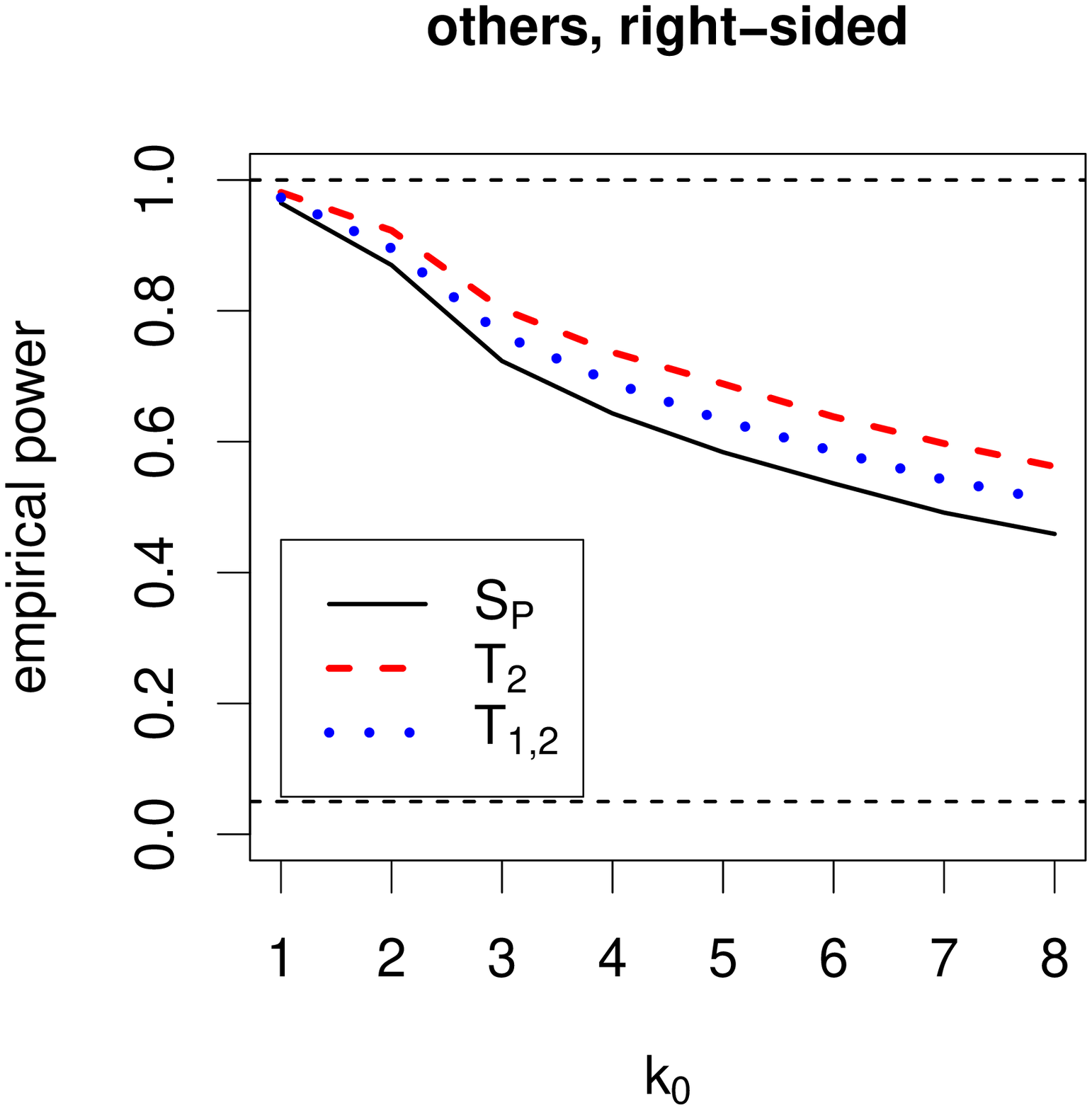} }}
\caption{
\label{fig:power-nonRL-cases4a-b}
Empirical power estimates under the non-RL cases 4(a)-(b).
In case 4(a) (top row), we take $k_0=3$ and $\sigma=0.1,0.2,\ldots,0.8$
and
in case 4(b) (bottom row), we take $k_0=1,2,\ldots,8$ and $\sigma=0.4$.
The dashed horizontal line and legend labeling are as in Figure \ref{fig:power-nonRL-cases1a-b}.
}
\end{figure}

The power estimates based on $z$-scores are plotted in Figures \ref{fig:power-nonRL-cases1a-b}-\ref{fig:power-nonRL-cases4a-b}.
In all these cases,
Dixon's cell-specific test and segregation index for cell $(i,i)$ provide very similar power estimates.
Furthermore,
we only consider right-sided alternatives,
since by design, the non-RL alternatives are for segregation (or clustering of class 1)
and the power estimates for the left-sided alternatives are virtually zero.

The empirical power estimates under cases 1(a) and (b) are presented in Figure \ref{fig:power-nonRL-cases1a-b}.
Notice that as $\rho$ increases,
the power estimates tend to increase as well.
That is,
when the probability of assigning the same label to NNs increases,
the level of segregation, hence the power of the tests increases.
Furthermore,
the power estimates are higher for $k=1$ (case 1(a)) compared to $k=3$ (case 1(b)) for each test.
Hence,
in this type of non-RL with $\rho$, $n_1$, $n_2$ being fixed,
as the number of NNs to be labeled increases,
the power estimate tends to decrease,
i.e.,
the level of segregation decreases.
In case 1(a),
among cell $(1,1)$ statistics,
Dixon's cell-specific test and segregation index have slightly higher power compared to type III cell-specific test,
among cell $(2,2)$ statistics,
type III statistics have much higher power than others,
and among other test statistics,
Pielou's segregation index and $T_{1,2}$ have higher power.
In case 1(b),
among cell-specific tests, type III test has higher power,
and among others, Cuzick-Edwards' tests, $T_2$ and $T_{1,2}$, have higher power.

We also compute empirical power estimates based on Monte Carlo critical values.
Under case 1(a) of RL pattern with $n_1=n_2=100$,
we compute the $95^{th}$ empirical percentiles of the test statistics
computed in the Monte Carlo simulations
and use these as the Monte Carlo critical values.
For example the empirical power
(based on Monte Carlo critical value) for $S_P$
is calculated for the right-sided alternative as $\frac{1}{N_{mc}}\sum_{i=1}^{N_{mc}} \I(Z_{P,i} > Z^{mc}_{P,crit})$
where we have $N_{mc}=100000$
and $Z^{mc}_{P,crit}$ is the $95^{th}$ empirical percentile of the
standardized version of Pielou's coefficient of segregation
under RL case 1(a) with $n_1=n_2=100$.
The estimates for the other tests are similar.
The empirical power estimates under non-RL case 1(a)
based on the Monte Carlo critical values are also presented in Figure \ref{fig:power-nonRL-cases1a-b}.
We observe that the power estimates using the asymptotic critical values and
those using the Monte Carlo critical values
are virtually identical (this trend persists in other cases as well).
Hence we only present the power estimates with the asymptotic critical values henceforth.

The empirical power estimates
under cases 2(a) and (b) are presented in Figure \ref{fig:power-nonRL-cases2a-b}.
In case 2(a),
the power estimates are almost constant,
with Cuzick-Edwards' tests having power around .80,
Dixon's cell $(2,2)$ test and segregation index having power around .50,
and
all others having power around .70.
In case 2(b),
the power estimates are higher compared to case 2(a),
and they increase as $\rho$ increases.
Among the cell-related tests,
type III test has higher power
(and $S_P$ has about the same power as the type III tests).
Cuzick-Edwards' tests have the higher power estimates,
with $T_2$ having the highest power.
In this type of non-RL,
the power seems not to depend on $\rho$ if only the first NN is labeled
according to the probabilities.

The empirical power estimates under cases 3(a)-(c) are presented in Figure \ref{fig:power-nonRL-cases3a-c}.
In cases 3(a) and (b),
notice that the power estimates slightly increase as $\rho$ increases,
but it seems that the power estimates
(hence the level of segregation)
does not crucially depend on $k_d$ or $\rho$.
In case 3(c),
the power estimates tend to increase as $k_p$ increases.
Hence,
as $k_p$ increases,
the probability of assigning the same label to NNs increases.
In all these cases,
among cell-specific tests,
type III test has the highest power estimates,
and among others $T_2$ has highest power.

The empirical power estimates under cases 4(a) and (b) are presented in Figure \ref{fig:power-nonRL-cases4a-b}.
In case 4(a),
notice that as $\sigma$ increases,
the power estimates tend to decrease.
That is,
when $\sigma$ decreases (with $n_1$, $n_2$, and $k_0$ being fixed),
the probability of assigning the same label to NNs of the source points increases,
hence
the level of segregation, thereby the power of the tests increases as well.
In case 4(b),
as number of source points $k_0$ increases,
the power estimates tend to decrease.
That is, when the number of source points increases
(with $n_1$, $n_2$, and $\sigma$ being fixed),
the (relative) probability of assigning the same label to NNs of the source points decreases,
hence
the level of segregation,
thereby the power of the tests decreases as well.
In both cases,
among cell tests,
type III test has higher power,
and among others $T_2$ has higher power estimates.

\begin{figure} [hbp]
\centering
\rotatebox{-90}{ \resizebox{2.1 in}{!}{\includegraphics{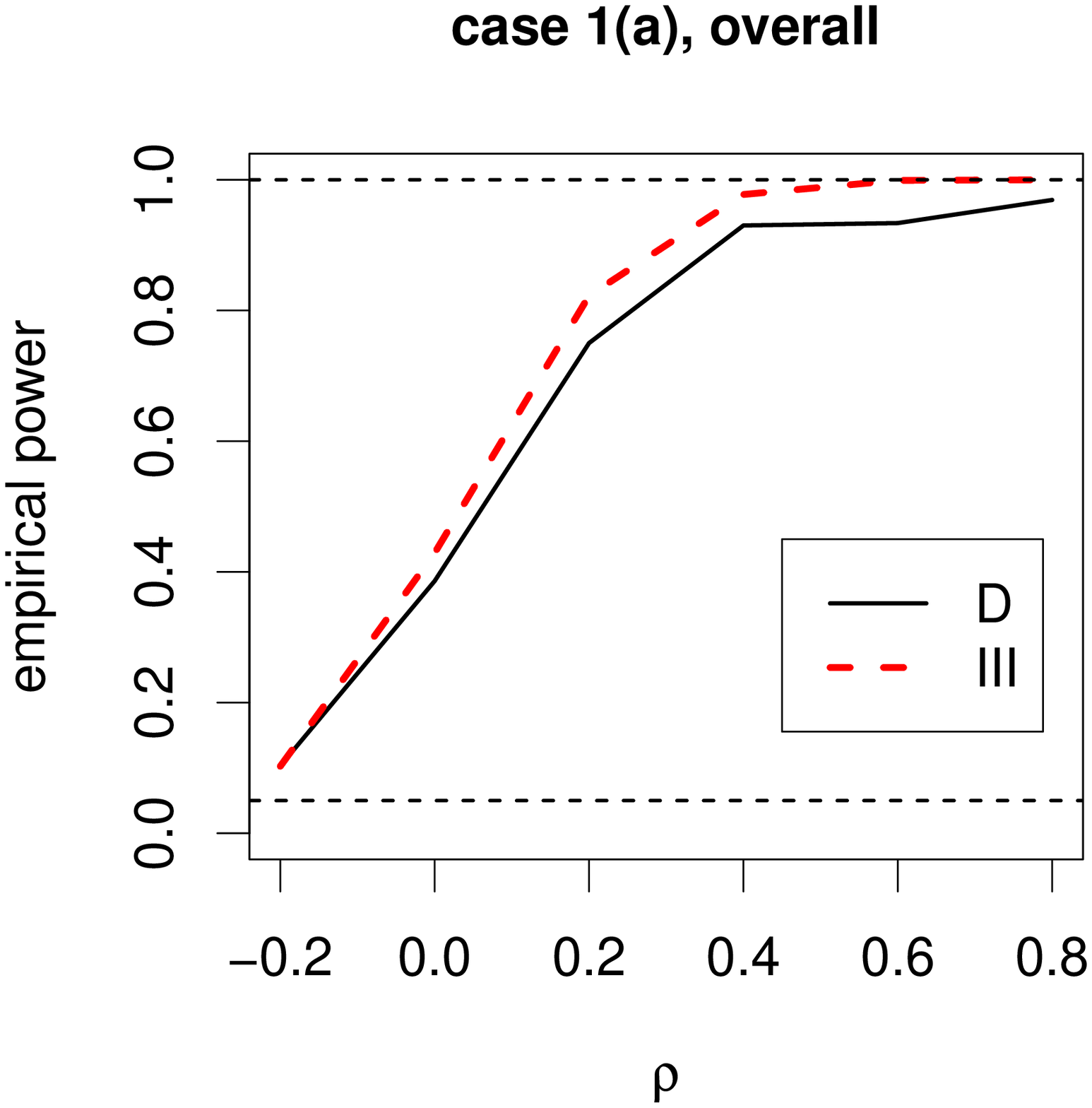} }}
\rotatebox{-90}{ \resizebox{2.1 in}{!}{\includegraphics{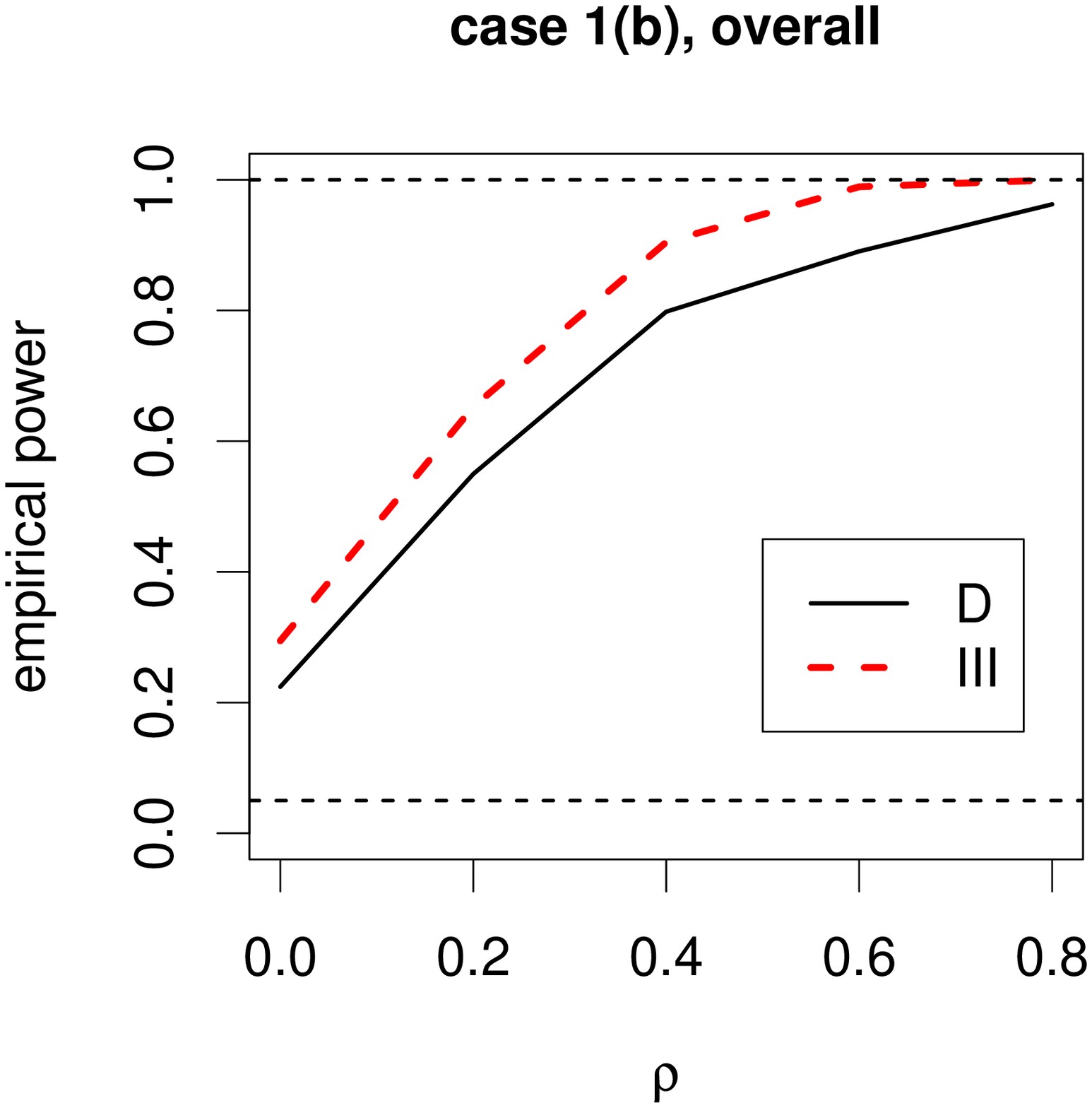} }}
\rotatebox{-90}{ \resizebox{2.1 in}{!}{\includegraphics{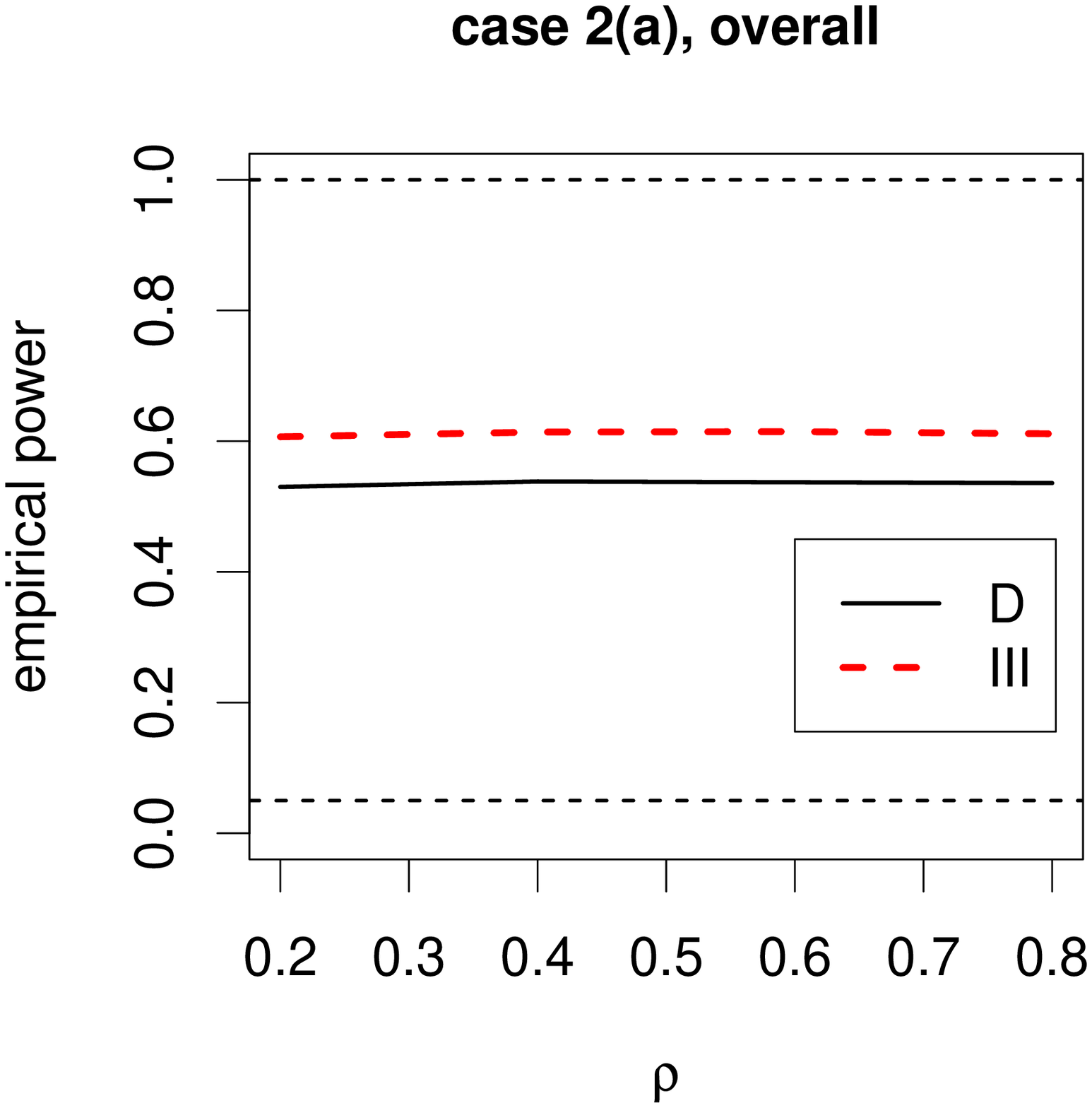} }}
\rotatebox{-90}{ \resizebox{2.1 in}{!}{\includegraphics{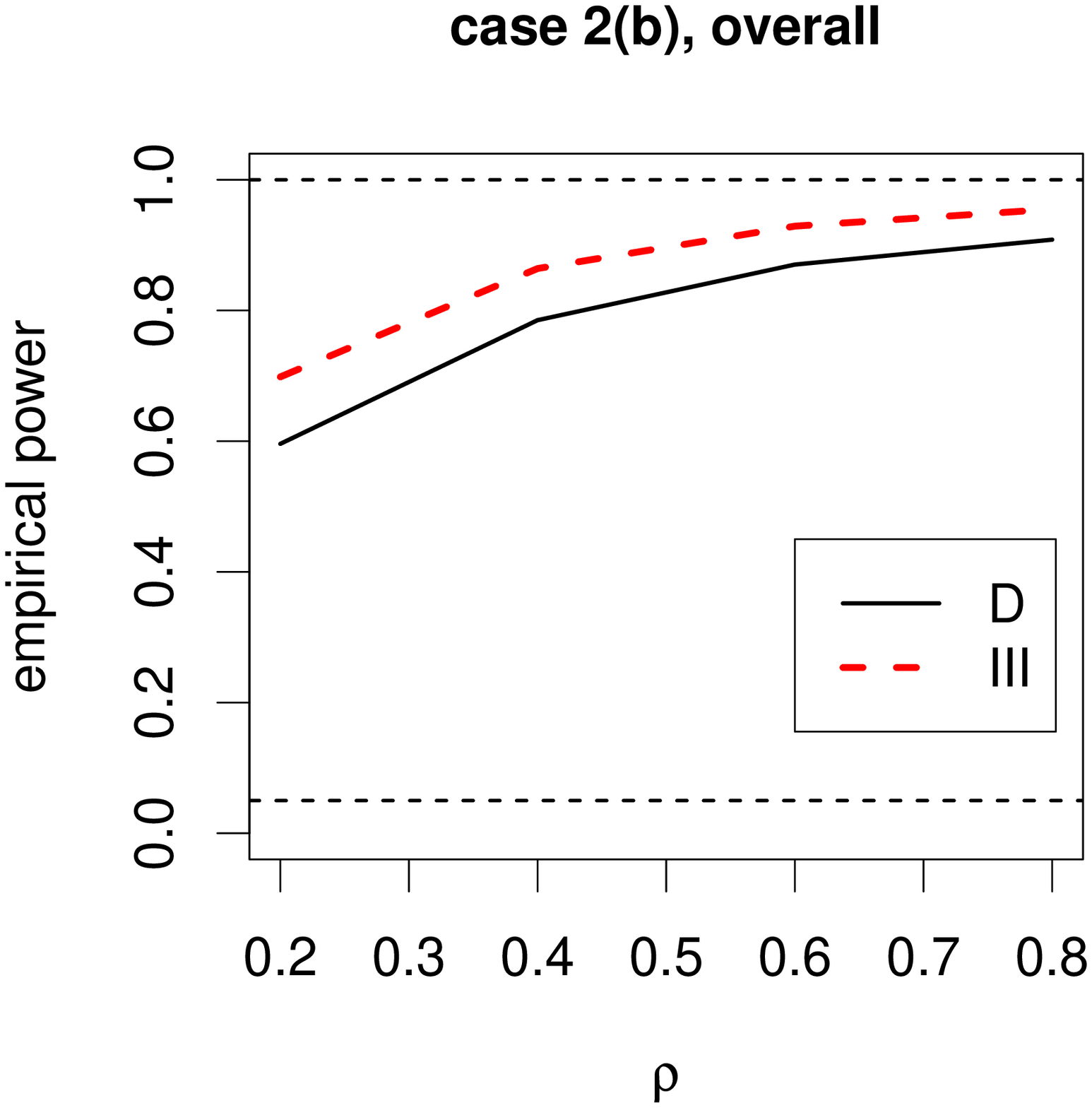} }}
\rotatebox{-90}{ \resizebox{2.1 in}{!}{\includegraphics{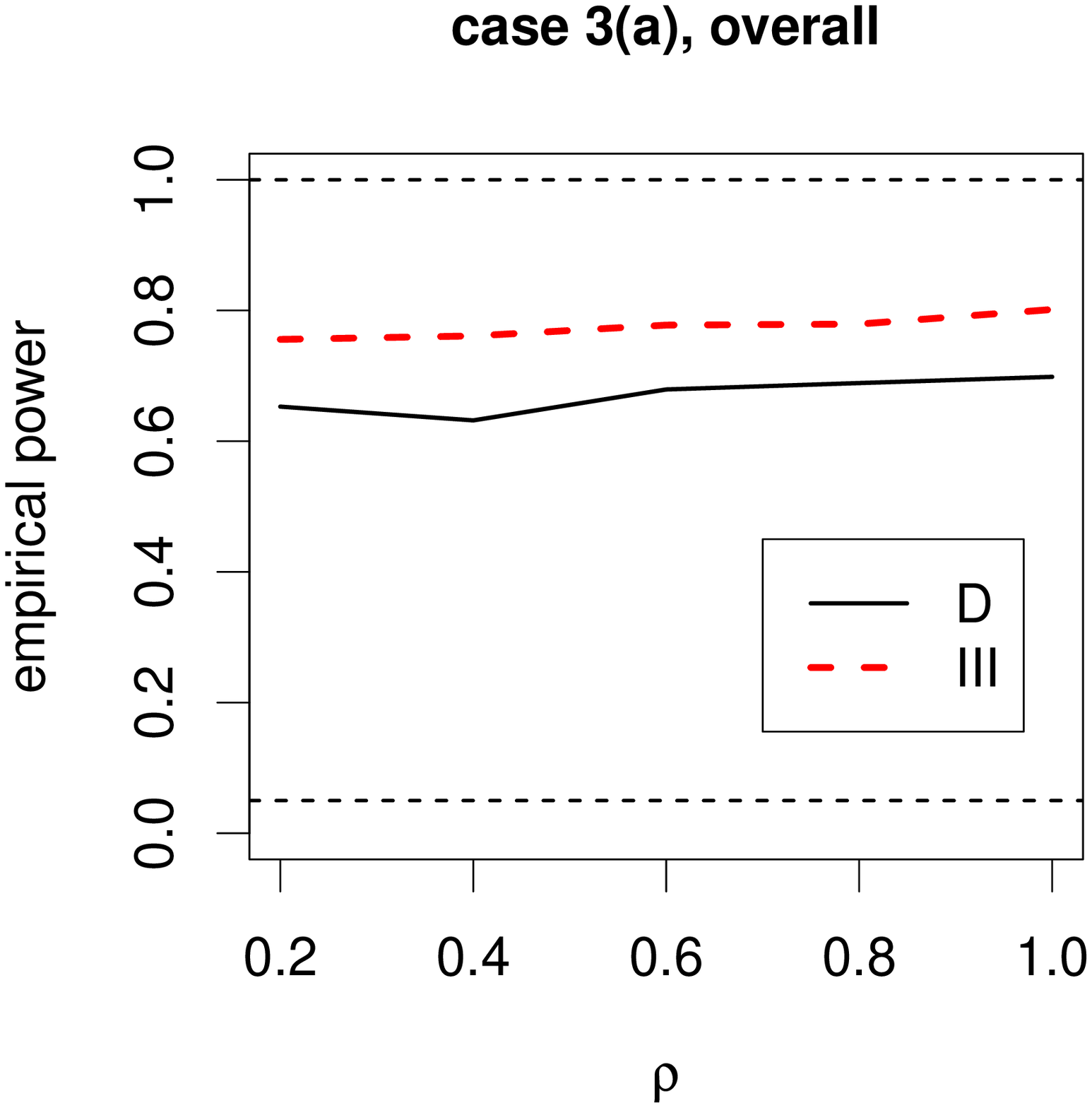} }}
\rotatebox{-90}{ \resizebox{2.1 in}{!}{\includegraphics{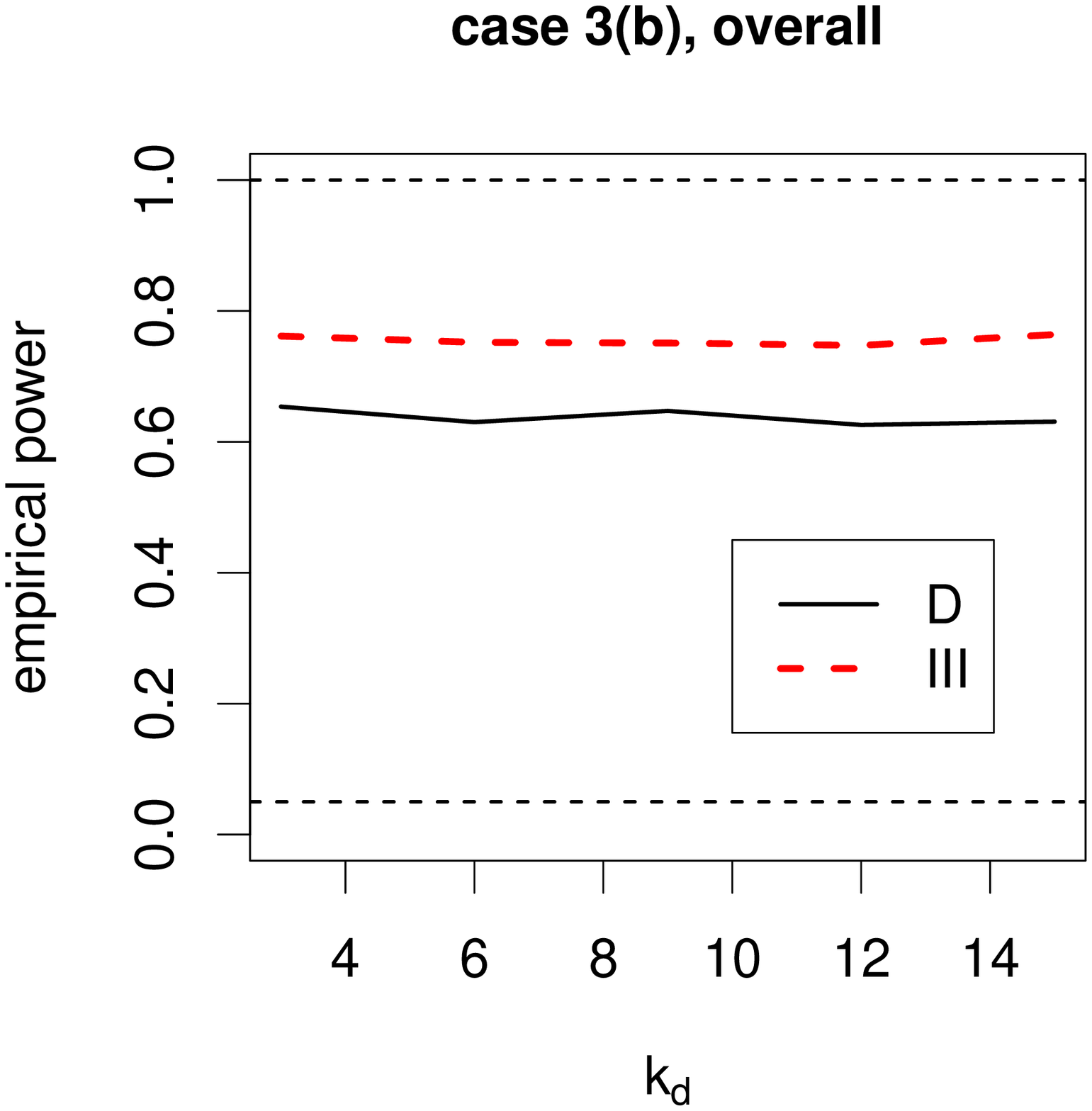} }}
\rotatebox{-90}{ \resizebox{2.1 in}{!}{\includegraphics{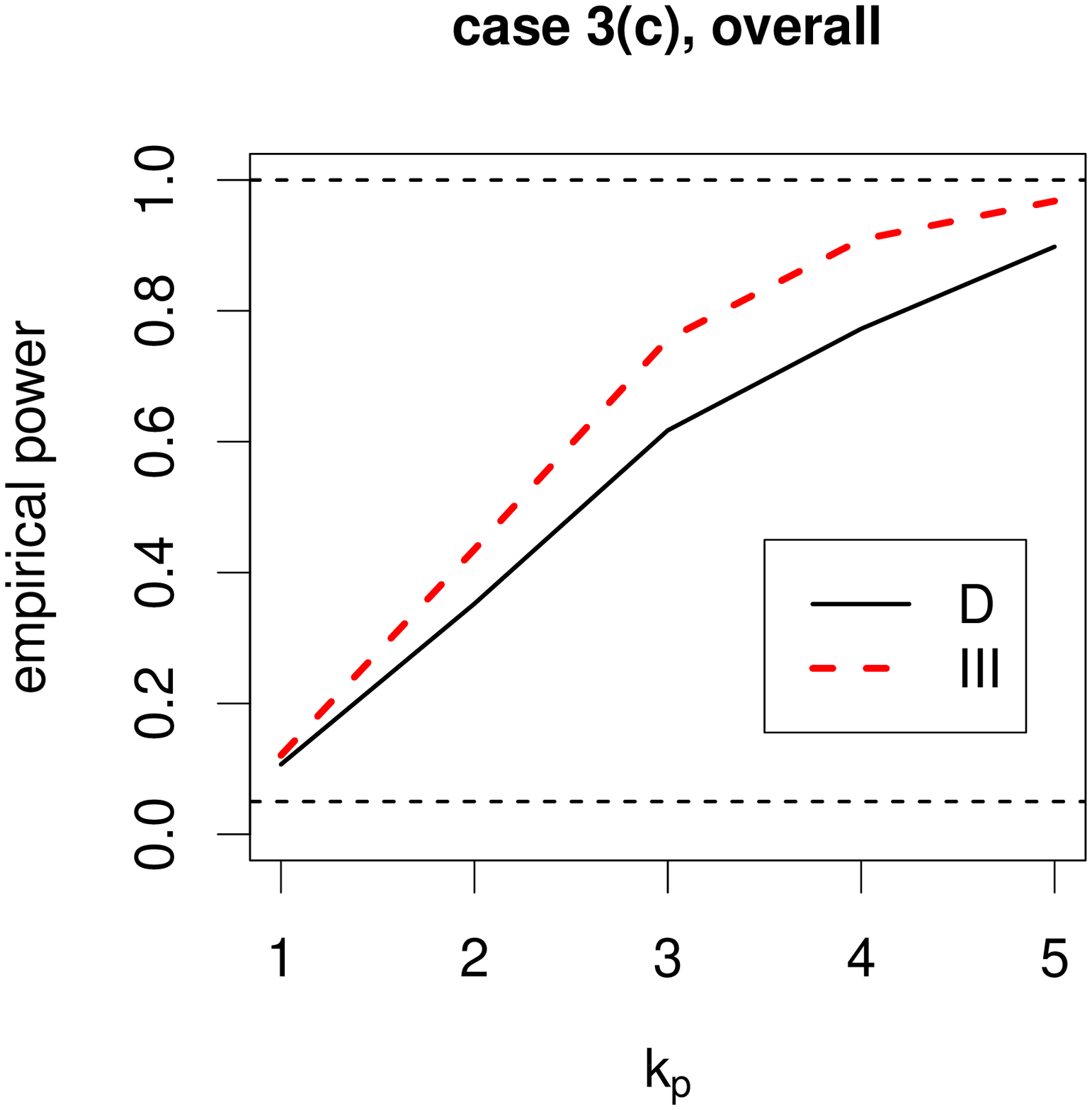} }}
\rotatebox{-90}{ \resizebox{2.1 in}{!}{\includegraphics{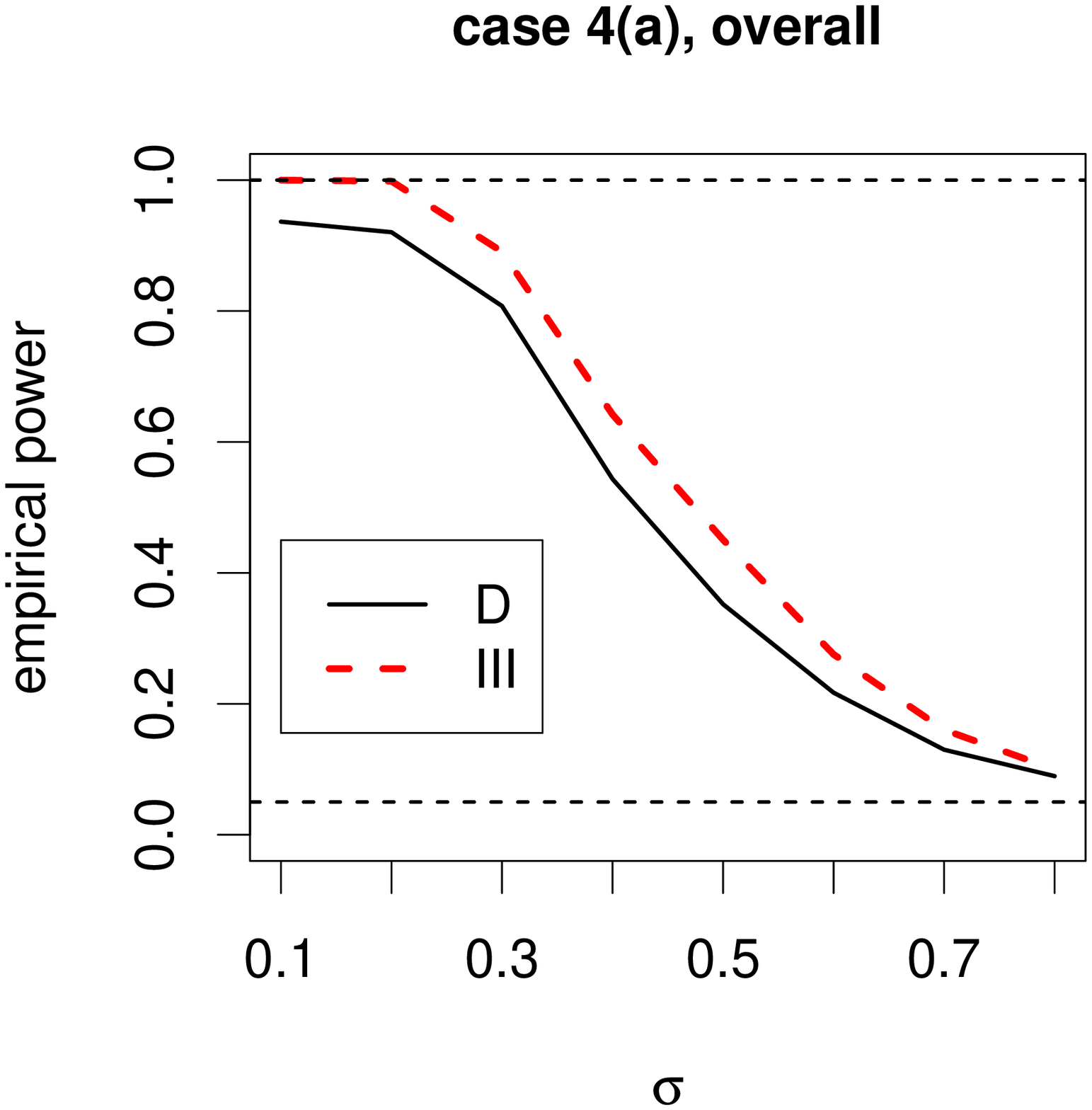} }}
\rotatebox{-90}{ \resizebox{2.1 in}{!}{\includegraphics{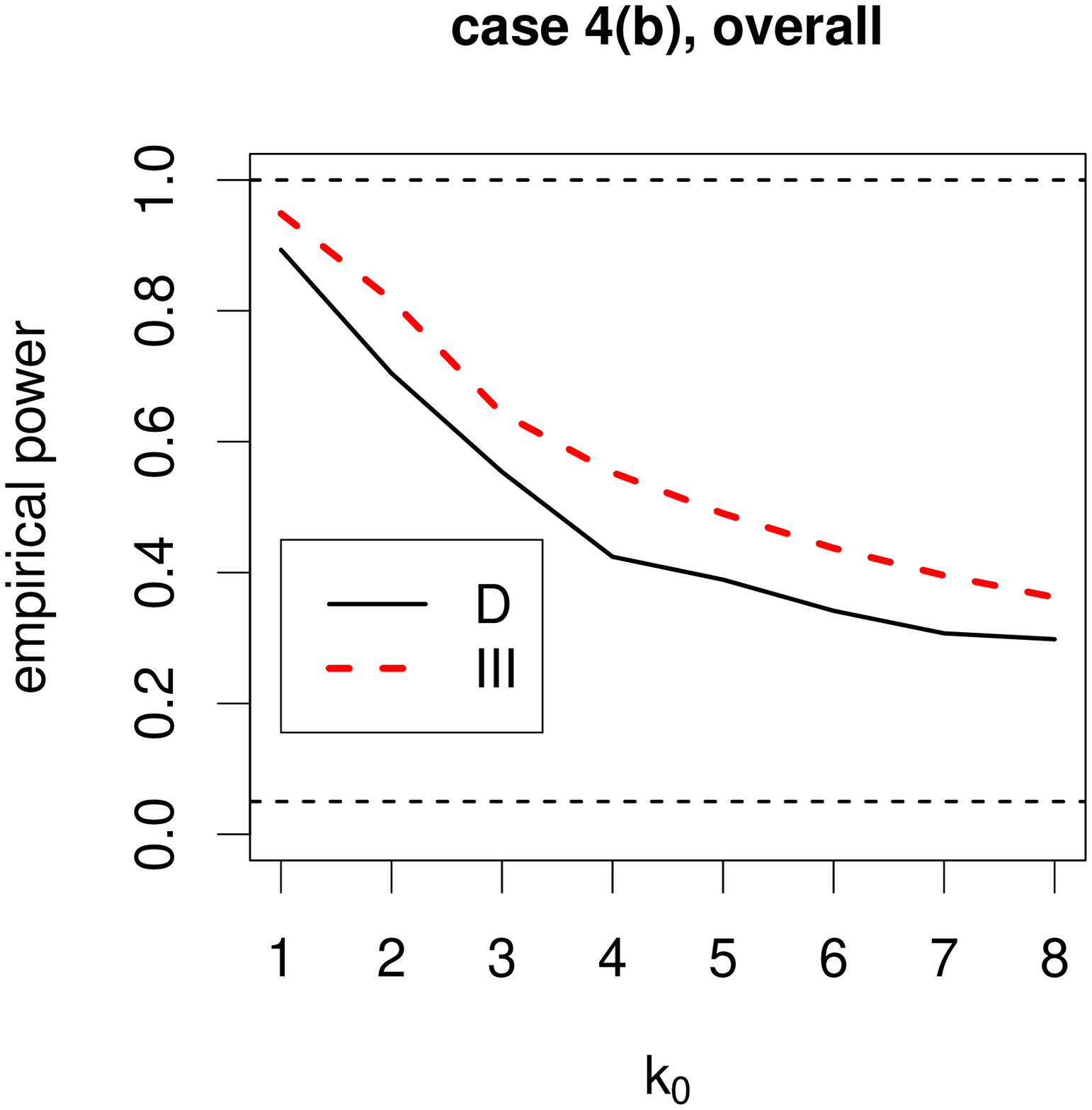} }}
\caption{
\label{fig:power-nonRL-overall}
Empirical power estimates for the overall NNCT-tests under the non-RL cases 1-4.
D stands for Dixon's overall test
and
III stands for type III overall test.
}
\end{figure}

The empirical power estimates of the NNCT overall tests under cases 1-4
are presented in Figure \ref{fig:power-nonRL-overall}.
The power estimates for cases 1(b), 2(b), and 3(b) are similar to those
for cases 1(a), 2(a), and 3(a), respectively, hence are not presented.
In all these cases,
type III overall test has higher power estimates compared to Dixon's overall test.
In cases 1(a) and (b), cases 2(a) and (b),
the power estimates increase with increasing $\rho$
(in both cases, the power estimates are higher with $k=1$ compared to $k=3$).
In case 3(a) (resp., (b))
the power estimates does not seem to depend on the parameter $\rho$ (resp., $k_d$).
In case 3 (c),
power estimates increase as $k_p$ increases,
in case 4(a) (resp., (b)) power estimates decrease as $\sigma$ (resp., $k_0$) increases.

\section{Example Data Sets}
\label{sec:example-data}

\subsection{Childhood Leukemia Data}
\label{sec:leukemia-data}
\begin{figure}[ht]
\centering
\rotatebox{-90}{ \resizebox{3. in}{!}{\includegraphics{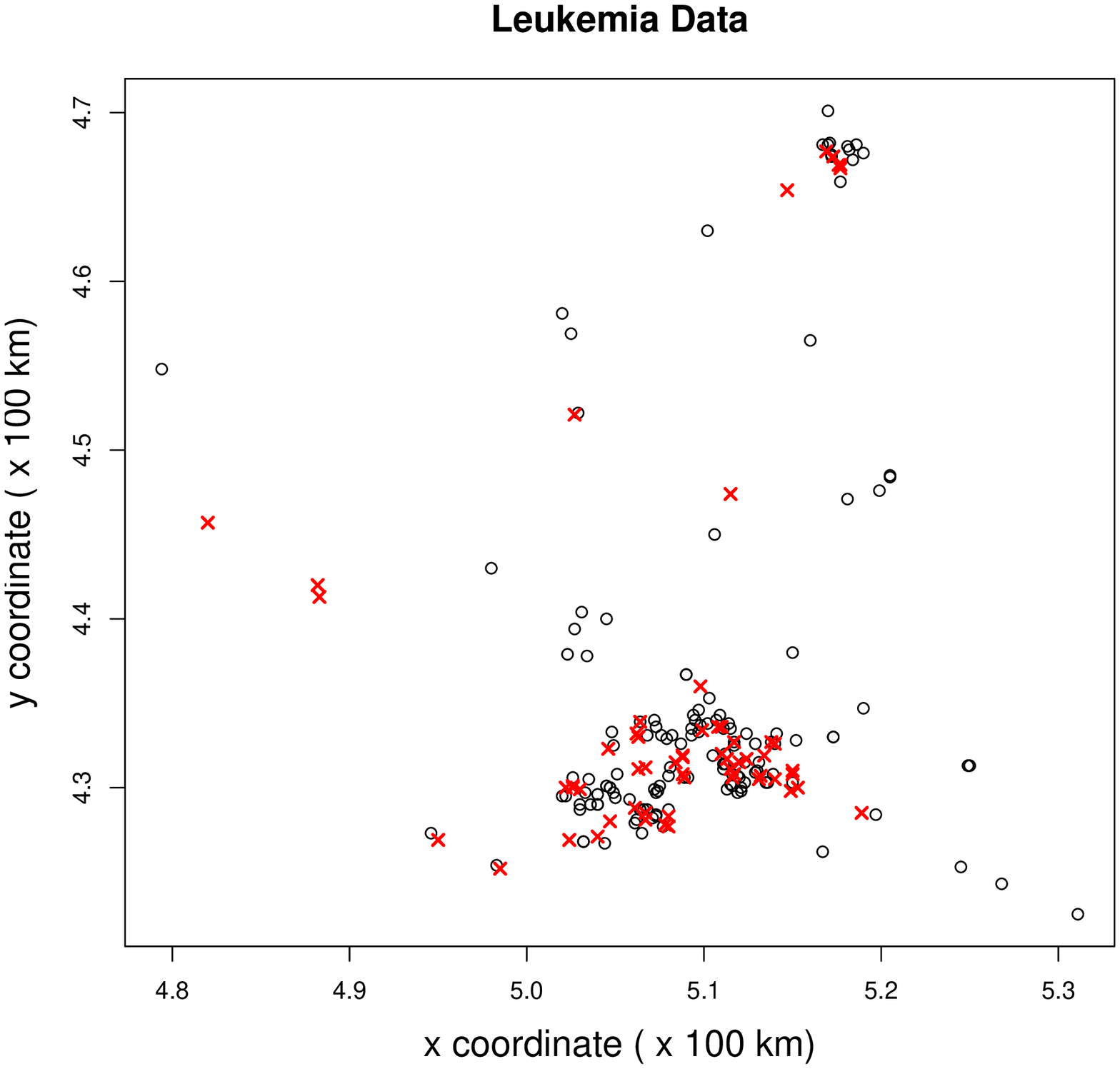} }}
 \caption{
\label{fig:leukemia}
The scatter plots of the locations of cases (crosses $\times$) and controls
(circles $\circ$) in North Humberside leukemia data set.}
\end{figure}

This data set consists of spatial locations of 62 cases of childhood
leukemia in the North Humberside region of the UK, between the years 1974 to 1982 inclusive
\cite{cuzick:1990}.
From the same region,
a random sample of 143 controls was selected using the completely randomized design.
We analyze the spatial clustering of leukemia cases with respect to controls
in this data with the tests considered above.
The locations of the points in the study region are plotted in
Figure \ref{fig:leukemia} and the segregation indices (together with standard errors)
are provided in Table \ref{tab:leukemia-test-stat}.
The figure is suggestive of mild clustering of leukemia cases,
and the indices together with their standard errors suggest only mild segregation (if any).
Here,
the indices and their standard errors are sufficient for an initial clustering assessment,
since the indices either have zero expected value (as in $S_P$)
or their expectation is approximately zero
(and tending to zero with increasing class sizes) as in Dixon's segregation indices.

\begin{table}[ht]
\centering
\begin{tabular}{|c|c|c|c|c|}
\multicolumn{5}{c}{Segregation Indices for Leukemia Data} \\
\hline
$S_P$ & $S^D_{11}$ & $S^D_{22}$ & $S^{D,c}_{11}$ & $S^{D,c}_{22}$  \\
\hline
.1348 ($\pm .090$) & .3548 ($\pm .314$) & .2362 ($\pm .175$) & .3420 ($\pm .272$) & .2317 ($\pm .251$)\\
\hline
\end{tabular}
\begin{tabular}{|c|c|c|c|c|c|c|c|c||c|c|}
\multicolumn{11}{c}{Test statistics for Leukemia Data} \\
\hline
$Z^D_{11}$ & $Z^D_{22}$ & $Z^{III}_{11}$ & $Z^{III}_{22}$ & $Z^S_{11}$ & $Z^S_{22}$ & $Z_P$& $T_2$ & $T_{1,2}$ & $C_D$ & $C_{III}$\\
\hline
1.2021 & 1.2829 & 1.4568 & 1.4590 & 1.1292 & 1.3482 & 1.4983 &  2.6263 & 2.1206 & 2.2604 & 2.1254\\
\hline
\multicolumn{11}{|c|}{associated $p$-values, with asymptotic critical values} \\
\hline
.1147 & .0998 & .0726 & .0723 & .1294 & .0888 & .0670 & .0043 & .0170 & .3230 & .1449 \\
\hline
\multicolumn{11}{|c|}{associated $p$-values, with Monte Carlo randomization} \\
\hline
.1365 & .0743 & .0784 & .0780 & .1294 & .1100 & .0726 & .0211 & .0696 & .4460 & .1462 \\
\hline
\end{tabular}
\caption{ \label{tab:leukemia-test-stat}
Pielou's coefficient of segregation and Dixon's segregation indices ($\pm$ standard errors) together with the corrected versions
and the test statistics and the associated $p$-values for the right-sided alternatives for North Humberside leukemia data.
$Z^D_{ii}$ ($Z^{III}_{ii}$) is Dixon's (type III) cell-specific test for cell $(i,i)$,
$Z^S_{ii}$ is the standardized version of Dixon's segregation indices for cell $(i,i)$, $i=1,2$,
$Z_P$ is the standardized version of Pielou's coefficient of segregation,
$T_2$ is Cuzick-Edward's $2$-NN test,
$T_{1,2}$ is Cuzick-Edward's combined test for $k=1,2$,
$C_D$ and $C_{III}$ are Dixon's and type III overall tests, respectively.
}
\end{table}

The appropriate null hypothesis is the RL pattern,
because it is reasonable to assume that some process affects a posteriori
the population of North Humberside region
so that some of the individuals get to be cases,
while others continue to be healthy (i.e., they are controls) \cite{goreaud:2003}.
In Table \ref{tab:leukemia-test-stat},
we present the test statistics and the associated $p$-values
based on asymptotic critical values
and
Monte Carlo randomization.
The latter is estimated as follows.
The test statistics for the original data are computed,
and the labels are randomly assigned to the points 10000 times.
At each random assignment,
we compute the test statistics,
and find how many times they equal or exceed the test statistics in the original data.
This number divided by 10000 yields the $p$-values based on Monte Carlo randomization.
Notice that both versions of $p$-values are similar for each test (except for $T_2$ and $T_{1,2}$).
Observe that only $T_2$ and $T_{1,2}$ are significant at .05 level,
while all others are not.
Hence, we conclude that there is no significant segregation of
cases at small scales (about the first NN-distances),
but cases tend to cluster significantly at larger scales.
The standardized versions of the corrected segregation indices are $Z^{D,c}_{11}=1.2591$ and $Z^{D,c}_{22}=1.076$
with the $p$-values for the right-sided alternative are .1040 and .1410, respectively.
The corresponding $p$-values based on Monte Carlo randomization are .1294 and .1100, respectively.

Based on the tests above, we conclude that the cases and controls
do not exhibit significant clustering (i.e., segregation) at small scales.
Based on Cuzick-Edward's tests, we find that the cases
are significantly segregated around $k$-NN distances for $k=2$.
In particular, average NN distance for leukemia data is 700 ($\pm$ 1400) m,
and the above analysis summarizes the pattern for about $t=1000$ m,
except for $T_2$ and $T_{1,2}$
where $T_2$ summarizes the pattern at about $1350$ m (since the average 2-NN distance is $1342$ m),
and $T_{1,2}$ for distances between  1000 to 1350 m.

\subsection{Liver Data}
\label{sec:liver-data}
\begin{figure}[ht]
\centering
\rotatebox{-90}{ \resizebox{3.5 in}{!}{\includegraphics{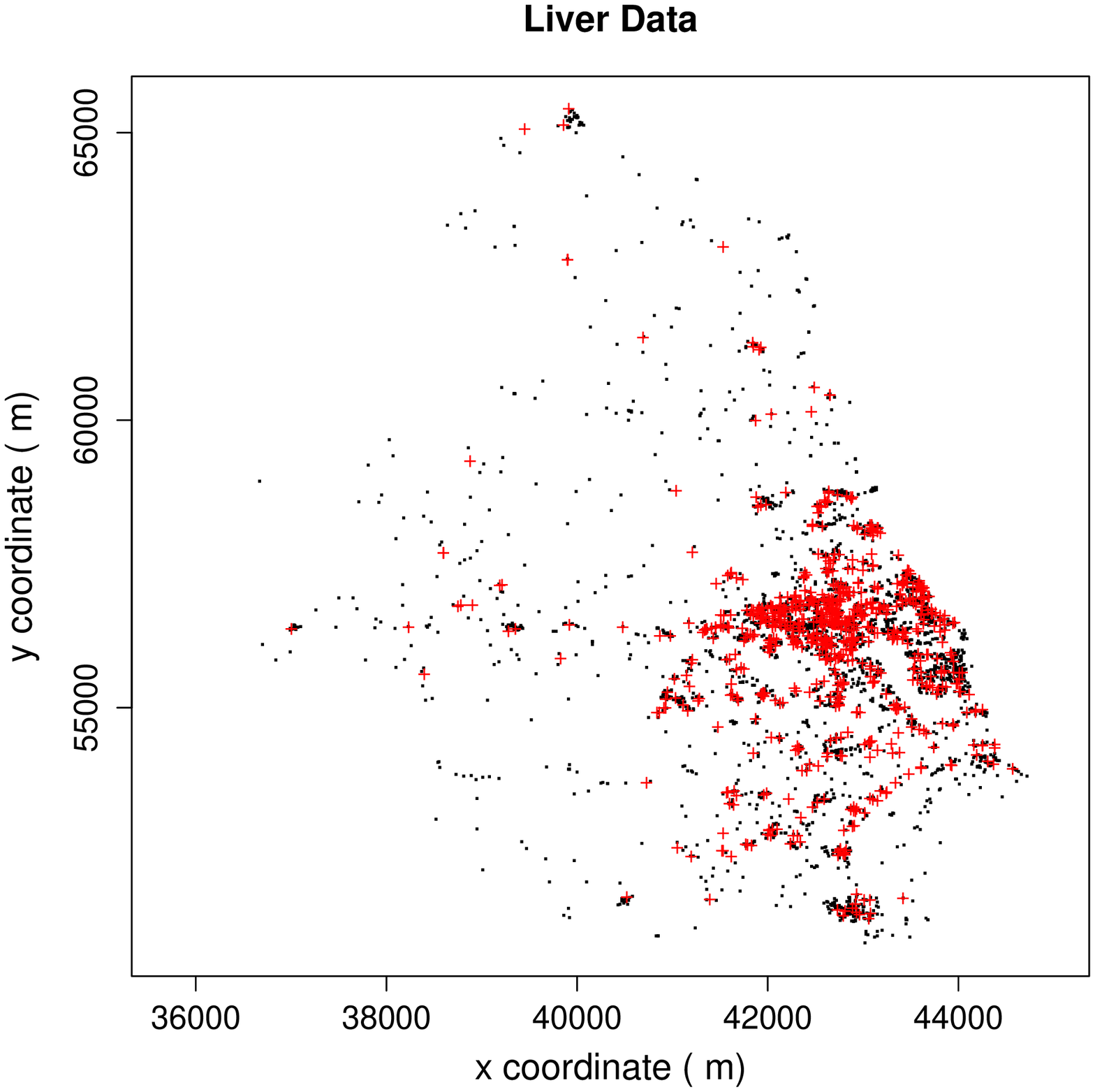} }}
 \caption{
\label{fig:liver}
The scatter plots of the locations of cases (pluses $+$) and controls
(dots $\cdot$) in Diggle's liver data set.}
\end{figure}

This data set consists of spatial locations of 761 cases of a liver disease in a region of interest
and 3044 controls in the same region \cite{diggle:2003}.
We analyze the spatial clustering of liver disease cases with respect to the healthy controls.
The locations of the points are plotted in
Figure \ref{fig:liver} and the segregation indices (together with standard errors)
are provided in Table \ref{tab:liver-test-stat}.
Observe that the plot of locations is suggestive of strong clustering of cases,
and the indices together with the standard errors support this initial assessment.

\begin{table}[ht]
\centering
\begin{tabular}{|c|c|c|c|c|}
\multicolumn{5}{c}{Segregation Indices for Liver Data} \\
\hline
$S_P$ & $S^D_{11}$ & $S^D_{22}$ & $S^{D,c}_{11}$ & $S^{D,c}_{22}$  \\
\hline
.0654 ($\pm$ .025) & .3410 ($\pm$ .117) & .0712 ($\pm$ .051) & .3393 ($\pm$ .117) & .0711 ($\pm$ .051)\\
\hline
\end{tabular}
\begin{tabular}{|c|c|c|c|c|c|c|c|c||c|c|}
\hline
\multicolumn{11}{|c|}{Test statistics for Liver Data} \\
\hline
$Z^D_{11}$ & $Z^D_{22}$ & $Z^{III}_{11}$ & $Z^{III}_{22}$ & $Z^S_{11}$ & $Z^S_{22}$ & $Z_P$& $T_2$ & $T_{1,2}$ & $C_D$ & $C_{III}$\\
\hline
3.2024 & 1.3520 & 3.2732 & 3.2729 & 2.9055 & 1.3814 & 2.5737 &  9.1854 & 7.7709 & 10.9096 & 10.7134\\
\hline
\multicolumn{11}{|c|}{associated $p$-values, with asymptotic critical values} \\
\hline
.0007 & .0882 & .0005 & .0005 & .0018 & .0836 & .0050 & $<.0001$ & $<.0001$ & .0043 & .0011 \\
\hline
\multicolumn{11}{|c|}{associated $p$-values, with Monte Carlo randomization} \\
\hline
.0004 & .0348 & .0004 & .0004 & .0004 & .0348 & .0020 & $<.0001$ & $<.0001$ & .0007 & .0007  \\
\hline
\end{tabular}
\caption{ \label{tab:liver-test-stat}
Pielou's coefficient of segregation and Dixon's segregation indices ($\pm$ standard errors) together with the corrected versions
and the test statistics and the associated $p$-values for the right-sided alternatives for Diggle's liver data.
The labels of the tests are as in Table \ref{tab:leukemia-test-stat}.
}
\end{table}

As in the leukemia data set,
the appropriate null hypothesis is again the RL pattern.
In Table \ref{tab:liver-test-stat},
we present the test statistics and the associated $p$-values
based on asymptotic critical values
and
Monte Carlo randomization
where the latter is estimated as in Section \ref{sec:leukemia-data}.
Both versions of $p$-values are similar for each test.
Observe that all tests except $Z^D_{22}$ and $Z^S_{22}$ are significant at .05 level
(but their Monte Carlo randomized versions are significant),
implying significant segregation of
cases at small scales (about the first NN-distances)
and at larger scales about the second NN-distances.
That is,
cases tend to cluster significantly at smaller scales.
The standardized versions of the corrected segregation indices are $Z^{D,c}_{11}=2.9077$ and $Z^{D,c}_{22}=1.3813$
with the $p$-values for the right-sided alternative are .0018 and .0836, respectively.
The corresponding $p$-values based on Monte Carlo randomization are .0004 and .0348, respectively.

The above tests indicate a significant segregation of cases and controls
and segregation of cases from controls seems to be much stronger compared
to that of controls from cases.
This implies a significant clustering of cases at smaller scales
around the average first NN distance.
Similarly,
Cuzick-Edward's tests also imply significant segregation
of cases and controls around $k$-NN distances for $k=2$.
In particular, average NN distance for liver data is 34.24 ($\pm$ 61.20),
and the above analysis summarizes the pattern for about $t=35$,
except for $T_2$ and $T_{1,2}$
where $T_2$ summarizes the pattern at about $50$ (since the average 2-NN distance is $52.20$),
and $T_{1,2}$ for distances between 35 to 50 units.
Notice that by construction Cuzick-Edwards' tests for $T_k$ with $k>1$
and $T_S$ with $\{1\} \subsetneq S$ provide information not available by the other tests considered.
However, this comes with a huge computational cost,
since for liver data it took about 7 hours to compute the Cuzick-Edwards tests
$T_1$, $T_2$ and $T_{1,2}$ in a HP Pavilion dv6 (Core i7 3720QM Processor 2.6GHz, 8GB RAM) laptop
but the other NNCT-based tests took only about 5 minutes.
The time difference was not that crucial for leukemia data
as Cuzick-Edwards' test took about 8 seconds,
while NNCT-tests took only about .5 seconds.
Our simulations indicate that NNCT-tests have $O(n^2)$ computing time,
but Cuzick-Edwards' tests $T_1$, $T_2$ and $T_{1,2}$ together have $O(n^{5/2})$ computing time.
Hence, when the number of cases or controls is large (more than a few hundred),
Cuzick-Edwards' tests are not computationally feasible,
but the NNCT-tests still are.

\section{Discussion and Conclusions}
\label{sec:disc-conc}
In this article,
we propose the use of two segregation indices,
namely, Pielou's coefficient of segregation \cite{pielou:1961}
and Dixon's segregation indices \cite{dixon:NNCTEco2002}
as tests to detect segregation between two classes,
in particular
to detect significance of disease clustering.
We derive their asymptotic distributions under RL of cases and controls to given locations,
and compare these tests with some other distance-based tests
(such as Dixon's and type III cell-specific and overall tests,
Cuzick-Edwards' $k$-NN and combined tests)
in terms of empirical size and power via extensive Monte Carlo simulations.
The tests related to NNCTs (i.e.,
Pielou's coefficient of segregation, Dixon's segregation indices,
Dixon's and type III cell-specific and overall tests) are for testing interaction at smaller scales
about the first NN distance
and $T_1$ is equivalent to Dixon's cell $(1,1)$ test
while $T_2$ is for the interaction at about the second NN distance,
and
$T_{1,2}$ combines the interaction information at the first and second NN distances.

We investigate the effect of the clustering (i.e., level of clustering and number of clusters) of the background points
(on which RL is applied)
and the effect of the differences in relative abundances on the size of these tests.
Our simulation results suggest that there is no increasing or decreasing trend in size
when the number of clusters or level of clustering increases.
On the other hand, the differences in relative abundances have a much stronger
influence on the size performance of the tests.
For the tests of small-scale interaction (around the first NN distance),
we observe that Pielou's coefficient of segregation,
and
type III overall tests seem to be robust to differences in relative abundances
with Pielou's coefficient of segregation being more robust.
On the other hand,
for tests of higher-scale interaction (around or up to the second NN distance),
$T_2$ and $T_{1,2}$ are both robust,
with $T_{1,2}$ being more robust.
Furthermore,
among cell-related and overall tests, type III tests have better size performance,
and when all tests are considered Pielou's coefficient of segregation and
$T_2$ and $T_{1,2}$ have better size performance.

We introduce four new non-RL algorithms yielding clustering of cases
(or segregation between the classes)
after the algorithm is executed on the background points.
With these non-RL alternatives,
we assess the power performance of the tests,
and see that type III tests and Cuzick-Edwards' tests have higher power than others
(also we notice that Pielou's coefficient of segregation
has power estimates close to Cuzick-Edwards' tests, although slightly lower).
As for the computational complexity,
Cuzick-Edwards' tests require much longer time and hence not so feasible for large sample sizes,
on the other hand the tests based on NNCTs require reasonable times
even if sample sizes are on the order of thousands.

The methodology introduced in this article
can also be used to test the deviations from CSR independence.
But in this setting,
the tests would be conditional on the values of $Q$ and $R$,
which are no longer fixed, but random quantities.
Furthermore,
the methodology is also applicable to test the spatial interaction at other contexts
(e.g., the spatial interaction between plant species in ecology).
In these contexts,
the left-sided (or association) alternative could also be of practical interest.

Our simulation study suggests that Dixon's segregation indices do not fare well
in testing spatial clustering.
Hence Dixon's segregation indices should not be employed with the asymptotic critical values in
testing spatial clustering,
but its Monte Carlo randomized version can be used.
On the other hand,
Pielou's coefficient of segregation performs similar to the best performing tests
based on NN distances (at the scale it is intended to work, i.e., at about the first NN distance).
Considering both size and power performance of the tests together,
for the interaction at small scales (around the first NN distance),
we recommend Pielou's coefficient of segregation.
In fact,
if the relative abundances of the classes are similar,
either type III tests or Pielou's coefficient of segregation can be employed;
but
if the relative abundances of the classes are different,
Pielou's coefficient of segregation is recommended.
For the interaction at higher scales,
we recommend Cuzick-Edwards' $k$-NN test with $k > 1$ and combined tests $T_S$ with $\{1\} \subsetneq S$
for testing segregation (or disease clustering) against RL
with the caveat of their computational cost in time.

\section*{Acknowledgments}
This research was supported by the research agency TUBITAK via Project \# 111T767
and the European Commission under the Marie Curie International Outgoing Fellowship Programme
via Project \# 329370 titled PRinHDD.


\section*{APPENDIX}
\subsection*{Asymptotic Approximation of Dixon's Segregation Indices when Population Proportions are Known}
If the population proportion, $\nu_i$, of class $i$, for $i=1,2$,
are known,
then for large $n_i$,
we would have the following forms of Dixon's segregation index.
For $i=j$,
$S^D_{ii}\approx \log \left(\frac{N_{ii}} {(n \nu_i-N_{ii})}\right)-\log \left(\frac{\nu_i} {1-\nu_i}\right)$.
Similarly,
for $i\not=j$,
$S^D_{ij}\approx \log \left(\frac{N_{ij}} {(n \nu_i-N_{ij})}\right)-\log \left(\frac{\nu_j} {1-\nu_j}\right)$.
Then for $i=j$,
letting $s \approx n \nu_i^2$ and $g(y) = \log(y/(n \nu_i-y))$,
we have $g'(s) \approx \frac{1}{n \nu_i^2(1-\nu_i)} \not= 0$.
By the above theorem,
for large $n_i$,
we get
$$\frac{\log \left( \frac{N_{ii}}{n \nu_i-N_{ii}}\right)-\log \left(\frac{\nu_i}{1-\nu_i}\right)}
{\sqrt{\Var[N_{ii}]}\left(\frac{1}{n \nu_i^2 (1-\nu_i)}\right)}$$
approximately having $N(0,1)$ distribution.
Similarly
for $i\not=j$,
letting $s \approx n \nu_i \nu_j$,
we get $g'(s) \approx \frac{1}{n \nu_i \nu_j(1-\nu_j)} \not= 0$.
Then for large $n_i$,
it follows that
$$\frac{\log \left(\frac{N_{ij}}{n \nu_i-N_{ij}}\right)-\log \left(\frac{\nu_j}{1-\nu_j}\right)}
{\sqrt{\Var[N_{ij}]}\left(\frac{1}{n \nu_i \nu_j(1-\nu_j)}\right)}$$
approximately has $N(0,1)$ distribution.

In the two-class case (i.e., with $k=2$),
$1-\nu_i = \nu_j$ and $1-\nu_j=\nu_i$ for $(i,j)=(1,2)$ and $(i,j)=(2,1)$.
So in the above expressions we can substitute these identities and
reduce the expressions.

\subsection*{Asymptotic Distribution of the Corrected Versions of Dixon's Segregation Indices}
Let $g(y)=\log\left(\frac{y+1}{n_i-y+1}\right)$.
Then $g'(y)=\frac{n_i+2}{(n_i-y+1)(y+1)}$.
For $i=j$,
with $s=\frac{n_i(n_i-1)}{n-1}$,
we get
$g(s)=\log \left(\frac{n_i(n_i-1)+(n-1)}{n_i(n-n_i)+(n-1)}\right)$
and
$g'(s)=\frac{(n_i+2)(n-1)^2}{(n_i(n-n_i)+(n-1))(n_i(n-n_i)+(n-1))}$.
Hence, by Theorem 7.7.6 of \cite{bain:1992},
we get
$$\frac{\log \left( \frac{N_{ii}+1}{n_i-N_{ii}+1}\right)-\log \left(\frac{n_i(n_i-1)+(n-1)}{n_i(n-n_i)+(n-1)}\right)}
{\sqrt{\Var[N_{ii}]}\left(\frac{(n_i+2)(n-1)^2}{(n_i(n-n_i)+(n-1))(n_i(n-n_i)+(n-1))}\right)}
=
\frac{S^{D,c}_{ii}}
{\sqrt{\Var[N_{ii}]}\left(\frac{(n_i+2)(n-1)^2}{(n_i(n-n_i)+(n-1))(n_i(n-n_i)+(n-1))}\right)}
$$
approximately having $N(0,1)$ distribution for large $n_i$.

Similarly
for $i\not=j$,
with $s=\frac{n_i n_j}{n-1}$,
we get
$g(s)=\log \left(\frac{n_i n_j+n-1}{n_i(n-n_j-1)+(n-1)}\right)$
and
$g'(s)=\frac{(n_i+2)(n-1)^2}{(n_i n_j+n-1)(n_i(n-n_j-1)+(n-1))}$.
Hence, by the above theorem,
we get
$$\frac{\log \left( \frac{N_{ij}+1}{n_i-N_{ij}+1}\right)-\log \left(\frac{n_i n_j+n-1}{n_i(n-n_j+1)+(n-1)}\right)}
{\sqrt{\Var[N_{ij}]}\left(\frac{(n_i+2)(n-1)^2}{(n_i n_j+n-1)(n_i(n-n_j+1)+(n-1))}\right)}
=
\frac{S^{D,c}_{ij}}
{\sqrt{\Var[N_{ij}]}\left(\frac{(n_i+2)(n-1)^2}{(n_i n_j+n-1)(n_i(n-n_j+1)+(n-1))}\right)}
$$
approximately having $N(0,1)$ distribution for large $n_i$.

If the population proportions, $\nu_i$, for $i=1,2$ are known,
the corrected and uncorrected versions are equivalent
as the class sizes tend to infinity.

\end{document}